\documentclass[8pt,pra,twocolumn,showpacs,superscriptaddress,floatfix]{revtex4-1}
\pdfoutput=1

\usepackage{amsmath,amssymb}
\usepackage{graphicx}
\usepackage{color}
\usepackage{bm}
\usepackage{hyperref}

\begin{document}

\title{Spectroscopic STM insights into Fe-based superconductors}

\author{Jennifer E. Hoffman}
\affiliation{Department of Physics, Harvard University, Cambridge, MA 02138, U.~S.~A.}


\begin{abstract}
In the first three years since the discovery of Fe-based high $T_c$ superconductors, scanning tunneling microscopy (STM) and spectroscopy have shed light on three important questions. First, STM has demonstrated the complexity of the pairing symmetry in Fe-based materials. Phase-sensitive quasiparticle interference (QPI) imaging and low temperature spectroscopy have shown that the pairing OP varies from nodal to nodeless $s\pm$ within a single family, FeTe$_{1-x}$Se$_x$.  Second, STM has imaged $C4\rightarrow C2$ symmetry breaking in the electronic states of both parent and superconducting materials. As a local probe, STM is in a strong position to understand the interactions between these broken symmetry states and superconductivity.  Finally, STM has been used to image the vortex state, giving insights into the technical problem of vortex pinning, and the fundamental problem of the competing states introduced when superconductivity is locally quenched by a magnetic field.  Here we give a pedagogical introduction to STM and QPI imaging, discuss the specific challenges associated with extracting bulk properties from the study of surfaces, and report on progress made in understanding Fe-based superconductors using STM techniques.
\end{abstract}


\maketitle
\tableofcontents

\section{\label{sec:Intro}Introduction}

The 2008 discovery of high-$T_c$ superconductivity in Fe-based materials marked an exciting turning point in the study of unconventional superconductivity\cite{KamiharaJACS2008}. Until then, cuprates were an anomalous island, hosting a zoo of confusing properties, with little outside perspective to determine which of these properties deserved deeper theoretical and experimental attention.  Furthermore, the cuprates' brittle material properties, gross electronic anisotropy, and poor normal state electrical conductivity made them challenging to incorporate into widespread technology. The Fe-based superconductors provided a foil for comparison and a fresh start for all involved in the cuprate quagmire.  Furthermore, their relative malleability, isotropy, and metallic normal state led to high hopes for useful applications.  These considerations have launched a whirlwind of research.  Three years into this new exploration, we review their properties, focusing on scanning tunneling microscopy (STM) experiments.

STM has proven to be an ideal tool to study correlated electron materials.  These materials are prone to nanoscale inhomogeneities whose effects may broaden spectral features or transitions measured by bulk techniques.  STM has been applied with great success to cuprate superconductors, addressing pairing symmetry, gap inhomogeneity, dopant placement, vortex pinning, and competing phases~\cite{FischerRMP2007}.  In recent years, STM has also been used to gain insight into momentum space, via quasiparticle interference (QPI) imaging.  QPI imaging can even provide a phase-sensitive determination of the superconducting order parameter (OP).



An early review of STM of Fe-based superconductors was written by Yin \textit{et al}\cite{YinPhysicaC2009}.  More recent, comprehensive reviews of the thousands of papers to date on Fe-based superconductors have been written by Johnston\cite{JohnstonAdvPhys2010} and Stewart\cite{StewartRMP2011}.  These longer reviews include short summaries of STM results.  Here we give a more thorough review including a pedagogical introduction to STM, a discussion of surface configurations, and several significant new results.


We start with an introduction to STM in section~\ref{sec:STM}, including a pedagogical explanation of QPI. In section~\ref{sec:Overview} we give a brief overview of Fe-based superconductivity. In section~\ref{sec:Surface}, we discuss the crystal structure and surface characterization of the several families of Fe-based superconductors. In section~\ref{sec:OP} we report on measurements of the superconducting OP, focusing on phase sensitive measurements via QPI imaging. In section~\ref{sec:Parent} we discuss the parent compound and competing electronic orders. In section~\ref{sec:Vortices}, we discuss the vortex state. In section~\ref{sec:Conclusion}, we conclude with suggestions for future STM experiments that could shed additional light on these materials.

\section{\label{sec:STM}Scanning Tunneling Microscopy}

\begin{figure}[tb]
\begin{center}
\includegraphics[width=.65\columnwidth]{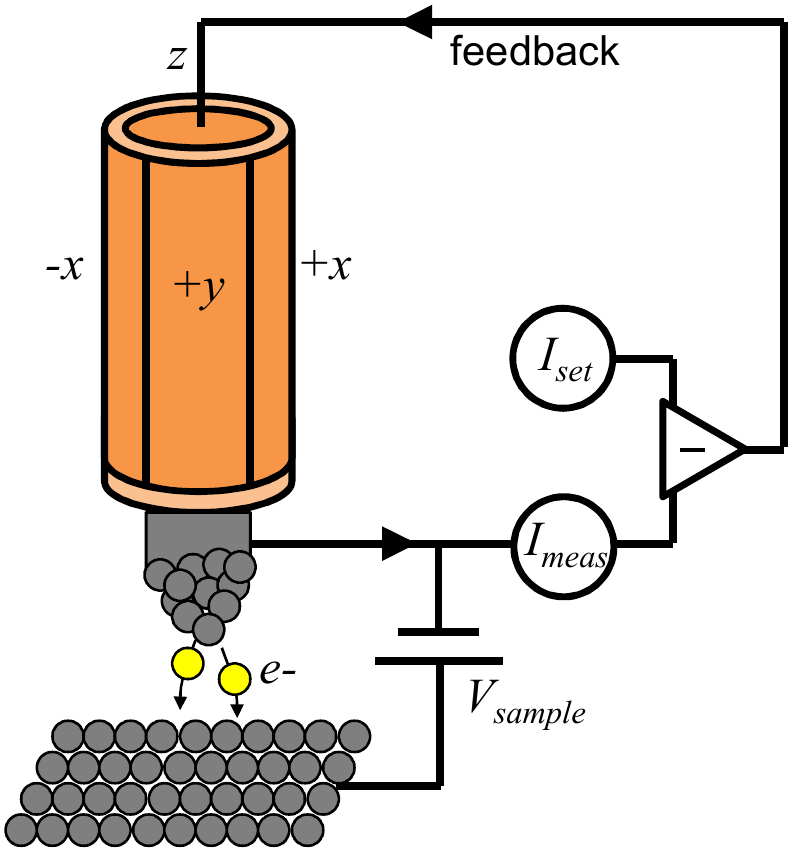}
\caption[Schematic of an STM.]{\label{fig:STMschematic} Schematic of a scanning tunneling microscope. A voltage is applied between the conducting sample surface and a sharp metallic tip, leading to a measurable tunneling current whose magnitude decays exponentially with tip-sample separation. In topographic mode, the measured tunneling current $I_{\mathrm{meas}}$ is compared to the setpoint current $I_{\mathrm{set}}$, and the difference (the error signal) is fed back to the $z$ piezo to control the tip height.}
\end{center}
\end{figure}

A scanning tunneling microscope consists of a sharp metallic tip which is rastered several {\AA}ngstroms above an electrically conducting sample surface\cite{BinnigPRL1982}. The position of the tip can be varied with sub-\AA\ precision by means of a piezoelectric scanner. When a voltage $V$ is applied between tip and sample, a current will flow. This current can be measured as a function of $(x,y)$ location and as a function of $V$. The microscope is illustrated schematically in figure \ref{fig:STMschematic}.  There are several excellent textbooks about STM\cite{ChenBook2007,StroscioBook1994,WiesendangerBook1994}.

\subsection{\label{sec:TunnelingCurrent}Tunneling Current}

\begin{figure}[tb]
\begin{center}
  {\includegraphics[width=.95\columnwidth,clip]{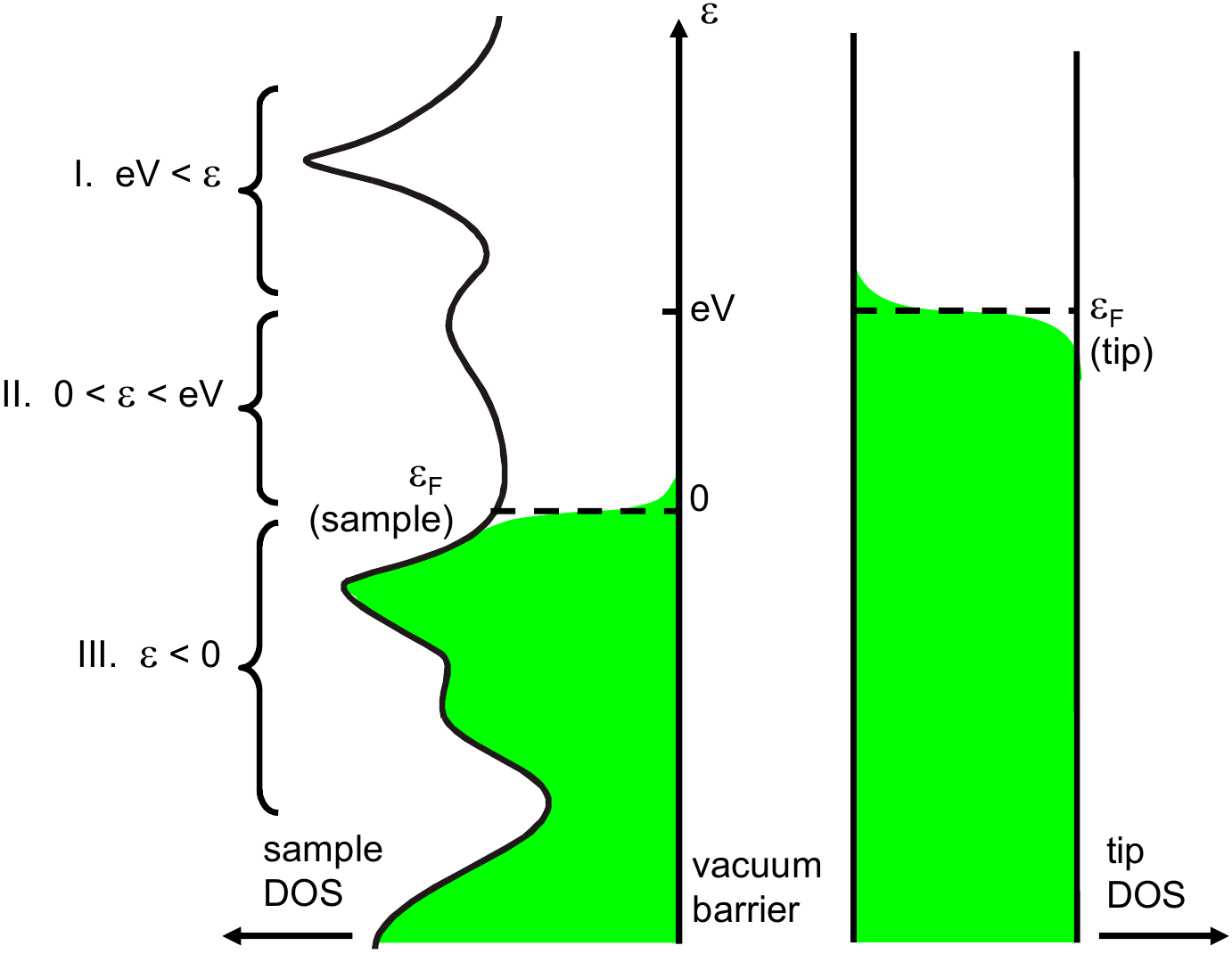}}
  \caption[Schematic of tip-sample tunneling.]
    {\label{fig:STMcurrent}
    Schematic of tip-sample tunneling.  Energy is along the vertical axis, and DOS of the sample and tip are shown along the horizontal axes.  Filled states are shown in green.  In this case, a positive bias voltage $V$ has been applied to the sample, which effectively lowers its Fermi level by energy $eV$ with respect to the Fermi level
    of the tip.  This allows for filled states from the tip (right) to tunnel into empty states in the sample (left).  The tunneling current is measured by an external circuit.}
\end{center}
\end{figure}

When a positive voltage $V$ is applied to the sample, with respect to a grounded tip, the Fermi level of the sample is lowered with respect to that of the tip.  Electrons will flow primarily from the filled states of the tip into the empty states of the sample, as illustrated in figure
\ref{fig:STMcurrent}. The elastic tunneling current from the tip to the sample, for states of energy $\varepsilon$ with respect to the Fermi level $\varepsilon_F \equiv 0$ of the sample, is given by

\begin{eqnarray}
\lefteqn{\!\!\!\!\!\!\!\! I_{t \rightarrow s}(\varepsilon) = -2e \cdot \frac{2\pi}{\hbar} |M|^2 \cdot } \nonumber \\
& & \underbrace{\rho_t(\varepsilon-eV) f(\varepsilon-eV)}_{\genfrac{}{}{0pt}{}{\mbox{{\scriptsize \# filled tip states}}}{\mbox{{\scriptsize for tunneling from}}}} \ \cdot \!\! \underbrace{ \rho_s(\varepsilon) \left[ 1 - f(\varepsilon) \right]}_{\genfrac{}{}{0pt}{}{\mbox{{\scriptsize \# empty sample states}}}{\mbox{{\scriptsize for tunneling to}}}}
\end{eqnarray}

\noindent where there is a factor of 2 for spin, $-e$ is the
electron charge, $|M|^2$ is the matrix element for the tunneling barrier,
$\rho_{s}(\varepsilon)$ is the density of states (DOS) of the sample, $\rho_{t}(\varepsilon)$ is the DOS of the tip, and $f(\varepsilon)$ is the Fermi distribution,

\begin{equation}
f(\varepsilon) = \frac{1}{1+ e^{\varepsilon/k_B T}}.
\end{equation}

Though the dominant flow of electrons for positive sample voltage $V$ will be from tip to sample, there will also be a smaller flow of electrons from sample to tip, given by

\begin{eqnarray}
\lefteqn{\!\!\!\!\!\! I_{s \rightarrow t}(\varepsilon) = -2e \cdot \frac{2\pi}{\hbar} |M|^2 \cdot } \nonumber \\
& & \underbrace{ \rho_s(\varepsilon) f(\varepsilon)}_{\genfrac{}{}{0pt}{}{\mbox{{\scriptsize \# filled sample states}}}{\mbox{{\scriptsize for tunneling from}}}} \!\!\!\!\!\! \cdot \ \ \underbrace{\rho_t(\varepsilon-eV) \left[ 1 - f(\varepsilon-eV) \right]}_{\genfrac{}{}{0pt}{}{\mbox{{\scriptsize \# empty tip states}}}{\mbox{{\scriptsize for tunneling to}}}}.
\end{eqnarray}

\noindent Summing these counter-propagating currents, and integrating over all energies $\varepsilon$, gives a net tunneling current from tip to sample

\begin{eqnarray}\label{eq:Itotal}
\lefteqn{ \!\!\!\!\!\!\!\!\!\!\!\!\!\!\!\!\!\!\!\!\!\!\!\! I_{t \rightarrow s} = -\frac{4\pi e}{\hbar} \int_{-\infty}^{\infty} |M|^2 \rho_s(\varepsilon) \rho_t(\varepsilon - eV) \cdot } \nonumber \\
 & & \!\!\!\! \left[ f(\varepsilon -eV) - f(\varepsilon) \right] d\varepsilon.
\end{eqnarray}

This expression can be simplified at low temperature, where the Fermi function cuts off very sharply at $\varepsilon_F$. (For example, at T=4.2K, the cutoff width is $\sim 2k_B T$ = 0.72 meV, which can be compared to a typical superconducting gap in Fe-based superconductors of $\sim 5$ meV.) In the approximation of a perfectly abrupt cutoff, the integrand is negligible except in the range $0 < \varepsilon < eV$.  (Likewise, a negative bias voltage $-|V|$ applied to the sample, would give an integration range of $-|eV| < \varepsilon < 0$.) The tunneling current for positive $V$ is therefore

\begin{equation}\label{eq:Isimple1}
 I_{t \rightarrow s} \approx - \frac{4\pi e}{\hbar} \int_{0}^{eV} |M|^2 \rho_s(\varepsilon)
\rho_t(\varepsilon - eV) d\varepsilon.
\end{equation}

\noindent In reality, equation \ref{eq:Isimple1} will be modified by an apparent smearing of energy features with width $\sim 4k_B T$ (composed of $\sim2k_B T$ for the sample and $\sim2k_B T$ for the tip).

A second simplification comes from the choice of a tip material with a featureless DOS near $\varepsilon_F$.  Typical choices are W, Pt, or PtIr.  If the tip DOS is flat in the energy range of interest, then $\rho_t(\varepsilon+eV)$ can be treated as a constant and taken outside the integral, giving

\begin{equation}
I_{t \rightarrow s} \approx -\frac{4\pi e}{\hbar} \rho_t(0) \int_{0}^{eV} |M|^2
\rho_s(\varepsilon) d\varepsilon.
\end{equation}

The final simplification is due to Bardeen's demonstration that under several realistic assumptions, the matrix element for tunneling will be virtually independent of the energy difference between the two sides of the barrier\cite{BardeenPRL1961}. In particular, the matrix element will remain unchanged even if one side transitions from the normal state to the superconducting state. To a reasonable approximation, the matrix element can therefore be taken outside the integral, giving

\begin{equation}
I_{t \rightarrow s} \approx -\frac{4\pi e}{\hbar} |M|^2 \rho_t(0) \int_{0}^{eV}
\rho_s(\varepsilon) d\varepsilon.
\end{equation}

The matrix element $|M|^2$ can be calculated for an $s$-wave tip\cite{TersoffPRL1983}. The tunneling probability through the vacuum barrier is approximated by $|M|^2 = e^{-2\gamma}$ with $\gamma=d\sqrt{2m\varphi / \hbar^2}$, where $m$ is the mass of the electron, $d$ is the width of the barrier (tip-sample separation), and $\varphi$ is the height of the barrier, which is some mixture of the work functions of the tip and sample, typically $\sim3-5 eV$ for metals (see table~\ref{table:work-function}). In summary, the tunneling current is fairly well approximated by

\begin{equation}
\label{eq:Ifinal} I_{t \rightarrow s} \approx -\frac{4\pi e}{\hbar} e^{-d \sqrt{
\frac {8m\varphi} {\hbar^2}}} \rho_t(0) \int_{0}^{eV}
\rho_s(\varepsilon) d\varepsilon.
\end{equation}

\begin{table}[h!]\footnotesize
\begin{center}
\begin{tabular}{ c c | c c | c c}
 \hline \hline
 Element & $\varphi$ (eV) & Element & $\varphi$ (eV) & Element & $\varphi$ (eV) \\
 \hline
 Se & 5.9 & W & 4.32-5.22 & As & 3.75 \\
 Pt & 5.12-5.93 & Te & 4.95 & Ca & 2.87 \\
 Ir & 5.42-5.76 & Fe & 4.67-4.81 & Sr & 2.59 \\
 Au & 5.31-5.47 & Bi & 4.34 & Ba & 2.52 \\
 Co & 5.0 & Pb & 4.25 & K & 2.29 \\
 \hline \hline
\end{tabular}
\caption{\label{table:work-function} Work functions of several relevant elements.  Ranges correspond to different crystalline orientations\cite{CRC2011}.}
\end{center}
\end{table}

\subsection{\label{sec:Topography}Topography}

In topographic mode (illustrated in figure~\ref{fig:STMschematic}), the tip-sample bias voltage is fixed, and a feedback loop holds the tunneling current constant by varying the tip height while it is scanned over the sample surface. Therefore, the tip follows a contour of constant integrated DOS (DOS). In the case of a homogeneous metal, the contour of constant DOS corresponds to the geometric topography of the sample surface. However, if the local DOS varies spatially, the resulting image contains a mixture of DOS and true topographic information. When the tip-sample bias voltage can be set far from the energy range of spatially inhomogeneous states, the contribution of the geometry dominates the topographic image, as desired. The recorded tip height is linear in the geometric topography, but logarithmic in the integrated DOS.

\subsection{\label{sec:Spectroscopy}Spectroscopy}

In addition to revealing the geometrical surface structure of a sample, STM can also measure the sample DOS as a function of energy, up to several eV from the Fermi level in both occupied and unoccupied states (where the upper bound is set either by catastrophic surface destruction at high fields, or by breakdown of the tunneling approximation). This is accomplished by sweeping the bias voltage $V$ and measuring the tunneling current $I$ while maintaining constant tip-sample separation $d$. By differentiating $I(V)$, the conductance $dI/dV$ is found to be proportional to the sample DOS.
\begin{eqnarray}
\frac{dI}{dV} & = & -\frac{4\pi e}{\hbar} e^{-d \sqrt{
\frac {8m\varphi} {\hbar^2}}} \rho_t(0) \frac{d}{dV}\left\{ \int_{0}^{eV}
\rho_s(\varepsilon) d\varepsilon \right\} \nonumber \\
& = & -\frac{4\pi e^2}{\hbar} e^{-d \sqrt{
\frac {8m\varphi} {\hbar^2}}} \rho_t(0) \rho_s(eV)
\end{eqnarray}
To reduce noise, a lock-in technique is typically used: a small AC modulation is summed with the DC bias voltage, and the resultant tunneling current is demodulated to yield $dI/dV$. The $dI/dV$ spectrum is a good measure of the sample DOS up to an overall constant which depends on the tip-sample bias voltage and current setpoints.
If these $dI/dV$ spectra are recorded at a dense array of locations in real space, spatial variation in the sample DOS can be extracted.


\subsection{\label{sec:QPI}QPI}

When materials are inhomogeneous and electrons are highly correlated, a full understanding of their properties requires knowledge of both the real space and momentum space behavior of quasiparticles. QPI (QPI) imaging provides some of this complementary knowledge. When quasiparticles scatter off defects or other structures within the crystal, energy-dependent standing waves form. (The special case of standing waves at the Fermi energy are known as Friedel oscillations\cite{HewsonBook1993}.)  The resulting interference patterns in the quasiparticle DOS can be imaged with scanning tunneling microscopy and spectroscopy\cite{CrommieNature1993, HasegawaPRL1993, WittnevenPRL1998, BurgiJESRP2000}.  The Fourier transform of the real-space interference patterns highlights the dominant sets of quasiparticle momenta\cite{FujitaSurfSci1999, PetersenJESRP2000, KanisawaPRL2001, HoffmanScience2002b, ZhangNatPhys2009, RoushanNature2009}. This combination of imaging and analysis is often called Fourier-transform scanning tunneling spectroscopy (FT-STS).  FT-STS is a powerful technique because it simultaneously yields energy-dependent real-space and momentum-space information on the quasiparticle wavefunctions, scattering processes, and coherence factors.  This information can be used to distinguish between candidate superconducting OPs\cite{ByersPRL1993, WangEPL2009,HanaguriScience2009,HanaguriScience2010}.\\

\begin{figure}[tbh]
\begin{center}
  {\includegraphics[width=1\columnwidth,clip]{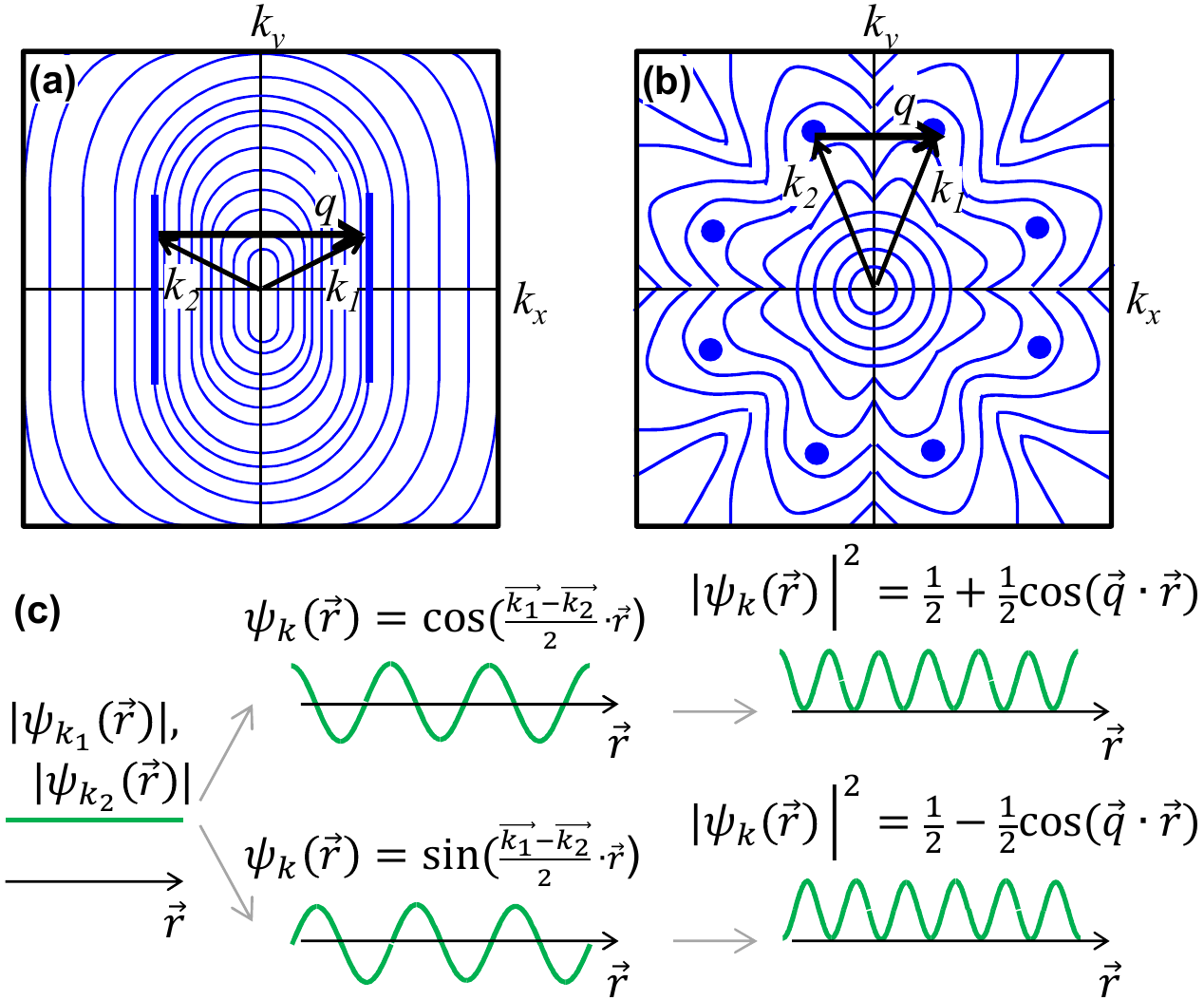}}
  \caption[Schematic of QPI.]
    {\label{fig:QPI-sketch}
    Schematic of QPI.  (a-b) BZs for two hypothetical materials show that large JDOS can arise in two different ways. The thin blue lines represent CCEs. (a) Nested regions, which will give large JDOS, are emphasized by thicker blue lines.  (b) Regions of shallow dispersion, where the CCEs are far apart, will also give large JDOS. Examples of these so-called `hotspots' are noted with blue dots. In both (a) and (b), pairs of eigenstates $\vec{k}_1$ and $\vec{k}_2$ are nested with wavevector $\vec{q}=\vec{k}_1-\vec{k}_2$. (c) Schematic showing how two different spatially homogeneous Bloch states (left) can mix to form modulations of the wavefunction (middle) and the LDOS (right). For superconducting Bogoliubov quasiparticles, the filled states (bottom right) and empty states (top right) of the LDOS are spatially out of phase with each other\cite{HanaguriNatPhys2007}.}
\end{center}
\end{figure}

\noindent \textit{\textbf{Metals}} In an ideal metal, the Landau quasiparticle eigenstates are Bloch wavefunctions characterized by wavevector $\vec{k}$ and energy $\varepsilon$. Their dispersion relation, $\varepsilon(\vec{k})$, can be measured with momentum resolved techniques such as angle resolved photoemission spectroscopy (ARPES)\cite{DamascelliPhysScript2004}.  By contrast, real space imaging techniques, such as STM, cannot directly measure $\varepsilon(\vec{k})$.  The local DOS, $\mbox{LDOS}(E,\vec{r})$, is related to the $\vec{k}$-space eigenstates $\psi_k(\vec{r})$ by

\begin{equation}
\label{eq:QI1}
 \mbox{LDOS}(E,\vec{r}) \propto \sum_k
|\psi_k(\vec{r})|^2 \delta(E-\varepsilon(\vec{k}))
\end{equation}

\noindent Substitution of a Bloch wavefunction into equation \ref{eq:QI1} shows that $\mbox{LDOS}(E,\vec{r})$ does not carry any directly observable spatial modulation at wavevector $\vec{k}$.

Spatial structures such as impurities, crystal defects, magnetic vortices, or static or fluctuating spatial orders cause elastic scattering which mixes eigenstates of different $\vec{k}$ but the same $\varepsilon(\vec{k})$.  A full treatment of this mixing was carried out using a Green's function approach for a normal metal in the presence of a weak local potential\cite{CapriottiPRB2003}.  A simpler picture can be explained as follows: elastic scattering mixes states that are located on the same quasiparticle contour of constant energy (CCE) in $\vec{k}$-space. For example, the blue lines in figures~\ref{fig:QPI-sketch}a and b show CCEs for two different hypothetical materials. When scattering mixes states $\vec{k_1}$ and $\vec{k_2}$, the result is a standing wave in the quasiparticle wavefunction $\psi_k$ of wavevector $(\vec{k_1}-\vec{k_2})/2$.  Since LDOS is proportional to $|\psi_k|^2$, the LDOS will contain an interference pattern with wavevector $\vec{q} = \vec{k_1}-\vec{k_2}$, as sketched in figure \ref{fig:QPI-sketch}c. LDOS modulations of wavelength $\lambda = 2\pi/q$ can be observed by STM as spatial modulations of the differential tunneling conductance $dI/dV$.

Although the full Green's function treatment includes additional terms, the experimentally observed amplitude of these modulations can be simply understood from Fermi's golden rule:

\begin{equation}
\label{eq:FermiGoldenRuleMetal}
 w(i \rightarrow f) \propto \frac{2\pi}{\hbar} |V(\vec{q})|^2
 n_i(E_i, \vec{k}_i) n_f(E_f, \vec{k}_f)
\end{equation}

\noindent where $E_i = E_f$ for elastic scattering, $\vec{q} =
\vec{k}_f - \vec{k}_i$, $V(\vec{q})$ is the Fourier component of the scattering potential at wavevector $\vec{q}$, and $n_i$ and $n_f$ are the density of the initial and final states.

It is apparent from equation \ref{eq:FermiGoldenRuleMetal} that a particular wavevector $\vec{q}$ can dominate the QPI at energy $E$, if the CCE contains a large joint DOS (JDOS) of $\vec{k}$-pairs connected by that $\vec{q}$. Large JDOS for a particular $\vec{q}$ can arise from nested regions where two equal-energy contours are roughly parallel and separated by $\vec{q}$, as shown in figure \ref{fig:QPI-sketch}a, or from large flat regions in $k$-space, where $\mathrm{DOS}(E) \propto 1/|\bigtriangledown_k(E)|$ is large, i.e. the CCE are farther apart, as shown in figure \ref{fig:QPI-sketch}b. These relatively flat regions of $k$-space are often called `hotspots'. (The extreme case of shallow dispersion is a van Hove singularity (vHS).  A vHS has large JDOS with all other $\vec{k}$-states, thus it will give rise to a FT-STS image which mimics the original band structure\cite{HankeArxiv1106.4217}.) The simple JDOS picture can be experimentally verified for a given material by a careful comparison between the autocorrelation of the ARPES-derived band structure (i.e. the JDOS), and the Fourier transform of the STM-derived QPI modulations\cite{MarkiewiczPRB2004, McElroyPRL2006, ChatterjeePRL2006, HashimotoPRL2011}.\\

\noindent \textit{\textbf{Superconductors}} The Bogoliubov quasiparticles in a superconductor are also Bloch states, but with dispersion

\begin{equation}
\label{eq:QI2}
 E_{\pm}(\vec{k}) = \pm \sqrt{\varepsilon(\vec{k})^2 +
|\Delta(\vec{k})|^2}
\end{equation}

\noindent where $|\Delta(\vec{k})|$ is the $\vec{k}$-dependent
magnitude of the superconducting energy gap at the Fermi surface (FS). (The FS is the CCE for $\varepsilon(\vec{k})=0$ in the normal state.) Elastic scattering of Bogoliubov quasiparticles can also result in conductance modulations, which have been treated in detail by several authors\cite{WangPRB2003,CapriottiPRB2003,ZhangPRB2003, Pereg-BarneaPRB2003, ZhuPRB2004,NunnerPRB2006,MaltsevaPRB2009,SykoraPRB2011}.  In a superconductor there is an additional complication from coherence factors\cite{WangPRB2003,Pereg-BarneaPRB2003,NunnerPRB2006,MaltsevaPRB2009,SykoraPRB2011}.  In the simple picture, Fermi's golden rule is augmented as

\begin{equation}
\label{eq:FermiGoldenRuleSC}
 w(i \rightarrow f) \propto \frac{2\pi}{\hbar}
 |u_{k_i} u_{k_f}^{*} \pm v_{k_i} v_{k_f}^{*}|^2 |V(\vec{q})|^2
 n_i(E_i, \vec{k}_i) n_f(E_f, \vec{k}_f)
\end{equation}

\noindent where the plus sign is for magnetic scatterers, the minus sign is for non-magnetic scatterers, and $|u_k|^2$ and $|v_k|^2$ are the probabilities that a pair of states with wavevectors $\pm\vec{k}$ is empty or filled, respectively\cite{TinkhamBook1996}. These Bogoliubov coefficients $u_k$ and $v_k$ are given by
\[
\label{eq:uv}
u_k=\frac{\Delta(k)}{|\Delta(k)|}\sqrt{\frac{1}{2}\left(1+\frac{\varepsilon(k)}{E(k)}\right)}; \quad v_k=\sqrt{1-|u_k|^2}
\]
\noindent Therefore, in a superconductor, a large JDOS will result in large QPI signal only when the additional constraint is satisfied that the coherence factor is non-zero.

Hanaguri\cite{HanaguriNatPhys2007} has also pointed out that for superconductors, the QPI dispersions $q_j(E)$ are particle-hole symmetric, i.e. $q_j(E) = q_j(-E)$ where $q_j$ represents a dominant scattering wavevector as exemplified in figure~\ref{fig:QPI-sketch}a and b, and $E$ is measured with respect to $\varepsilon_F$. However, for a given $q_j$, the real-space conductance modulations at $+E$ and $-E$ are spatially exactly out of phase with each other, as shown in the right panel of figure~\ref{fig:QPI-sketch}c. The QPI signal can therefore be enhanced relative to the background by computing the ratio map
\begin{equation}
\label{eq:Zmap}
Z(\vec{r},E) \equiv \frac{\frac{dI}{dV}(\vec{r},E)}{\frac{dI}{dV}(\vec{r},-E)}.
\end{equation}
\noindent The ratio also eliminates possible spurious effects from the tunneling matrix element\cite{HanaguriNatPhys2007}.\\

\begin{figure}[tb]
\begin{center}
  {\includegraphics[width=1\columnwidth,clip]{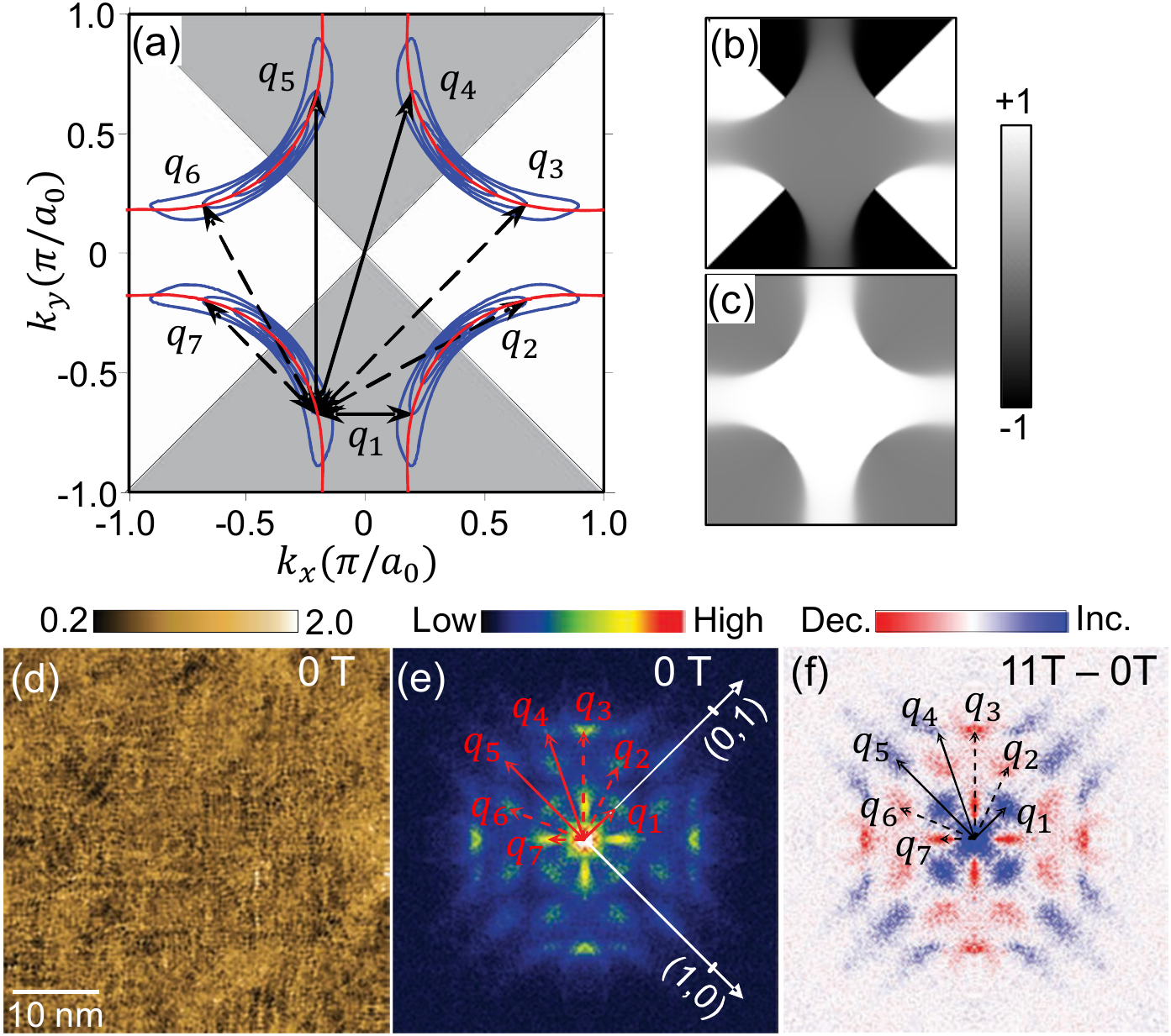}}
  \caption[QPI cuprate OP.]
    {\label{fig:Hanaguri-cuprate}
    Determination of the cuprate $d$-wave OP via QPI imaging. (a) Schematic of the cuprate BZ.  The normal state FS is shown in red.  As the material becomes superconducting, a nodal gap opens, resulting in the blue contours of constant energy (CCEs). The DOS is largest in the flat regions at the end of the banana-shaped CCEs. The QPI signal is dominated by the 7 $\vec{q}_j$'s connecting this octet of high-DOS regions. (b-c) Occupation factors $u_k$ and $v_k$ for a $d$-wave gap, mapped in the same BZ as in (a). (d) $Z(\vec{r},E=4.4\;\mathrm{meV})$ image of Ca$_{2-x}$Na$_x$CuO$_2$Cl$_{2}$ at zero field and $T$=1.6K (setup: $V_{\mathrm{sample}}=-100\,\mathrm{mV}; I_{\mathrm{set}}=100\,\mathrm{pA}$). (e) FT-STS image: 8-fold-symmetrized Fourier transform of (d), showing the 7 $\vec{q}_j$'s expected from the connections between the 8 `hotspots' in (a). (f) The difference between high-field and zero-field QPI: $Z(\vec{q},E,B=11\;\mathrm{T})-Z(\vec{q},E,B=0\;\mathrm{T})$ for $E$=4.4 meV. This figure is taken from Ref.~\onlinecite{HanaguriScience2009}.}
\end{center}
\end{figure}

\noindent \textit{\textbf{Example: Cuprate Superconductors}} Although the pairing symmetry of cuprate superconductors has long been known\cite{WollmanPRL1993,VanHarlingenRMP1995}, a recent STM study on superconducting Ca$_{2-x}$Na$_x$CuO$_2$Cl$_2$  ($x\sim$0.14 and $T_c\sim$28K) demonstrated the use of QPI imaging as a phase-sensitive probe of a superconducting OP\cite{HanaguriScience2009}.  Here we will walk through this example as preparation to better understand the determination of the FeSe OP using QPI imaging discussed later in section~\ref{sec:OP}.

In optimally doped cuprates, there is a single Cu $d$ band crossing the Fermi level, resulting in a single large hole pocket centered at the $M$ point of the single-CuO$_2$-plaquette Brillouin zone (BZ). The normal state FS is shown schematically as a red line in figure~\ref{fig:Hanaguri-cuprate}a.  Below $T_c$, a gap opens with four nodal points, resulting in the banana-shaped CCEs shown schematically in blue in figure~\ref{fig:Hanaguri-cuprate}a.  The JDOS is therefore dominated by the 7 inequivalent $q$-vectors connecting the octet of flat regions at the ends of the `bananas'\cite{HoffmanScience2002b,WangPRB2003,McElroyNature2003}.  The measured FT-STS image in figure~\ref{fig:Hanaguri-cuprate}e indeed shows all 7 of these expected dominant $\vec{q}_j$'s.

The dominant superconducting OP in cuprates is $d$-wave, with the gap $\Delta_k$ changing sign along the BZ diagonals, as denoted by the white and gray shading in figure~\ref{fig:Hanaguri-cuprate}a. The sign change in $\Delta_k$ results in sign-changing Bogoliubov coefficients $u_k$ and $v_k$, shown in figure~\ref{fig:Hanaguri-cuprate}b-c. The 7 $\vec{q}_j$'s therefore have different coherence factors $|u_{k_i}u_{k_f}^{*}\pm v_{k_i}v_{k_f}^{*}|^2$.  The question is, within a given sample, which type of scattering dominates, and therefore which sign applies within the coherence factor?  For weak scalar potential scattering, the coherence factor is suppressed for $\vec{q}_j$'s that preserve the sign of $\Delta_k$, namely $q_1$, $q_4$, and $q_5$. For scattering from magnetic impurities, the coherence factor is suppressed for $\vec{q}_j$'s that change the sign of $\Delta_k$, namely $q_2$, $q_3$, $q_6$, $q_7$.  When a magnetic field is applied, the pinned vortices serve as an additional form of gap-sign-preserving scattering\cite{MaltsevaPRB2009}, thus suppressing $q_2$, $q_3$, $q_6$, $q_7$ while preserving or enhancing $q_1$, $q_4$, and $q_5$.  This effect is observed experimentally in figure~\ref{fig:Hanaguri-cuprate}f.  The phase of the superconducting gap in cuprates is thus revealed by the evolution of the QPI upon application of the magnetic field.

\section{\label{sec:Overview}Fe-based Superconductor Overview}

\begin{figure}[tbh]
\begin{center}
  {\includegraphics[width=1.0\columnwidth,clip]{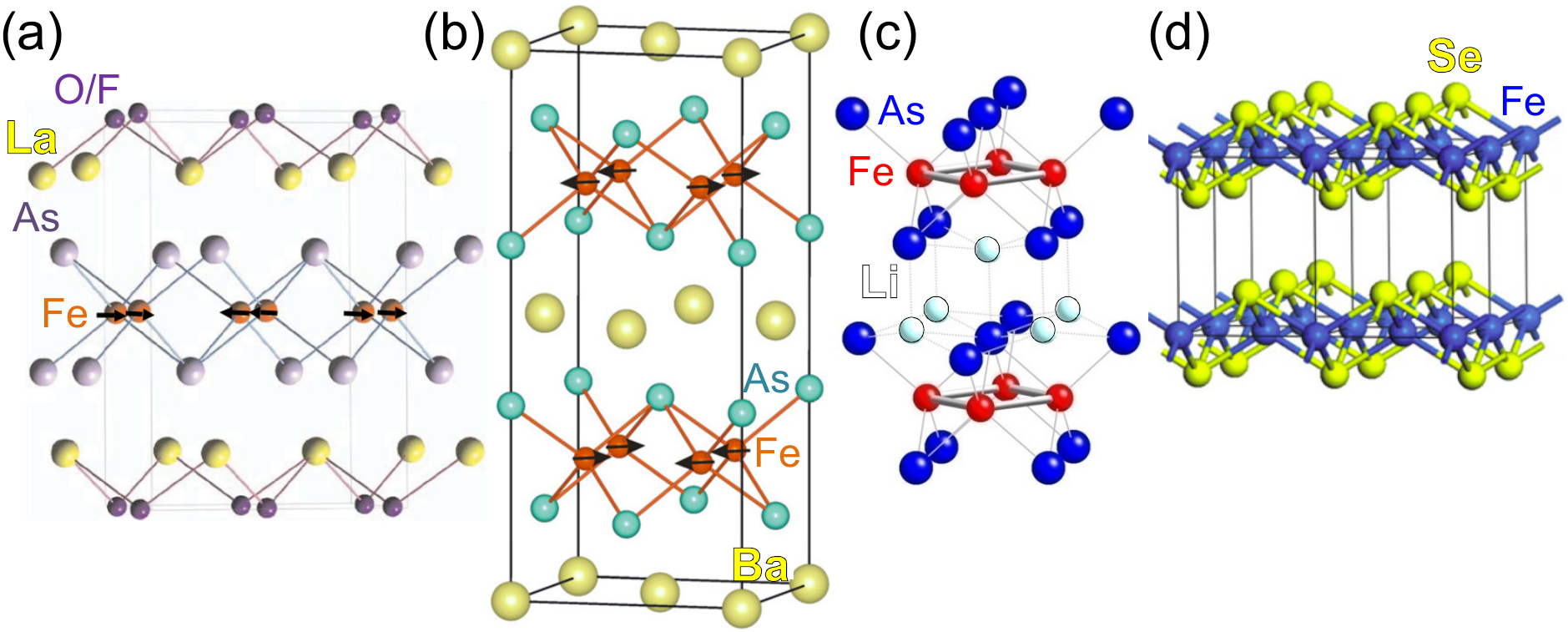}}
  \caption[Structures of Fe-based superconductors.]
    {\label{fig:structures}
    Structures of 4 families of Fe-based superconductors. (a) 1111 (specifically, LaFeAsO$_{1-x}$F$_x$), from Ref.~\onlinecite{HuangPRB2008}. (b) 122 (specifically, BaFe$_2$As$_2$), from Ref.~\onlinecite{HuangPRL2008}. (c) 111 (specifically, LiFeAs), from Ref.~\onlinecite{HanaguriAspen2011}. (d) 11 (specifically, FeTe), from Ref.~\onlinecite{SongPRB2011}}
\end{center}
\end{figure}

The common feature of all known Fe-based superconductors is an FeX square lattice, where X may be a pnictogen (As or P), or a chalcogen (Te, Se, or S).  Between these FeX layers there may be an additional oxide layer (`1111' materials), ionic layer (`122' or `111' materials), more complex intermediaries (e.g., `42622' materials), or nothing (`11' materials).  Representative crystal structures are shown in figure~\ref{fig:structures}.


\begin{figure}[tbh]
\begin{center}
  {\includegraphics[width=1.0\columnwidth,clip]{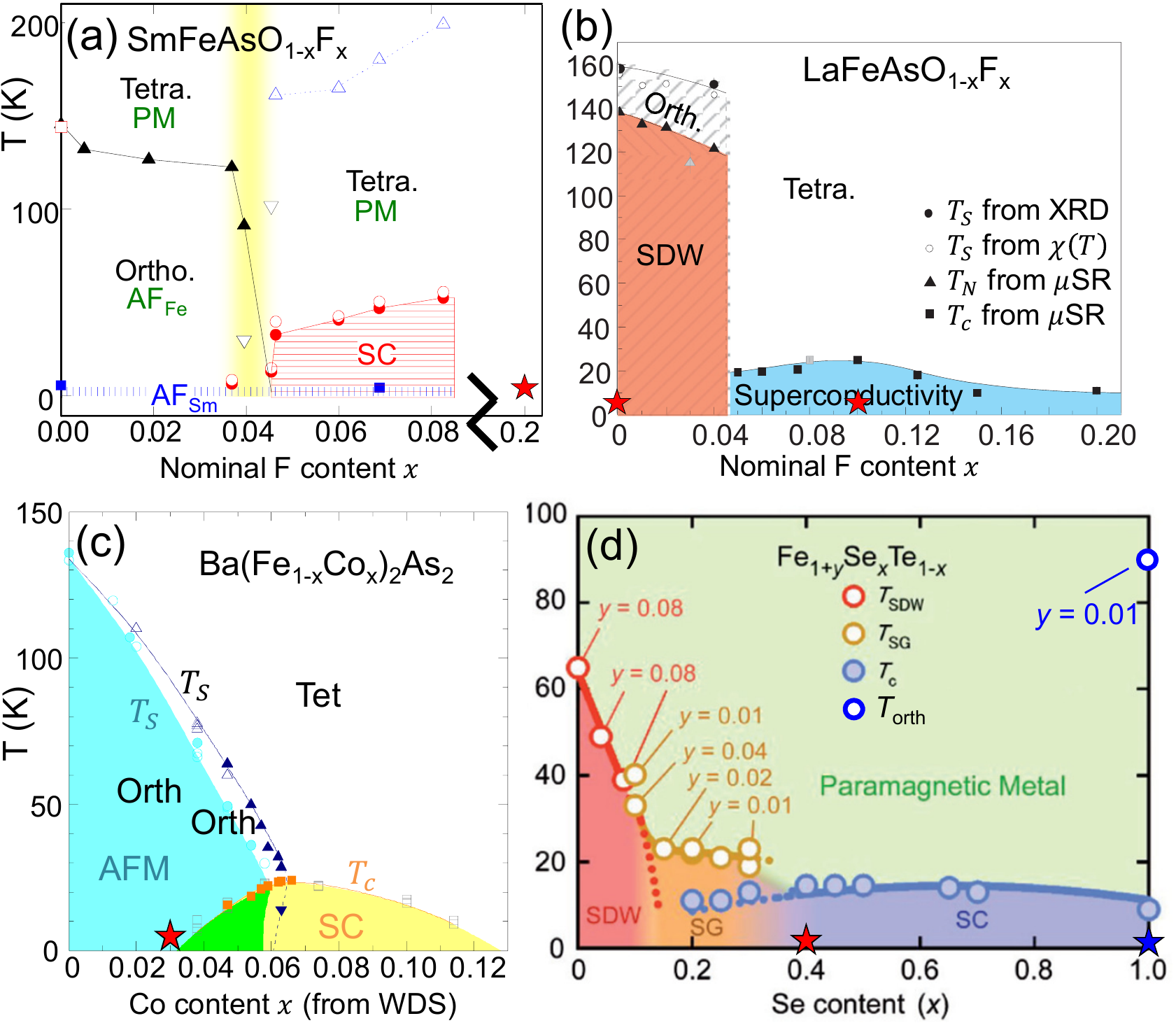}}
  \caption[Phase diagrams of Fe-based superconductors.]
    {\label{fig:phase-diagrams}
    Phase diagrams of several Fe-based superconductors, representative of the 1111, 122, and 11 families. (a) SmFeAsO$_{1-x}$F$_x$ \cite{KamiharaNJP2010}; (b) LaFeAsO$_{1-x}$F$_x$ \cite{LuetkensNatMat2009}; (c) Ba(Fe$_{1-x}$Co$_x$)$_2$As$_2$ \cite{NandiPRL2010}; (d) FeTe$_{1-x}$Se$_x$ \cite{KatayamaJPSJ2010}. The orthorhombic transition marked in (d) comes from x-ray diffraction measurements on a powder sample with stoichiometry Fe$_{1.01}$Se \cite{McQueenPRL2009}.  In each diagram, stars mark the phase space locations of STM experiments to be discussed later.}
\end{center}
\end{figure}

Fe-based superconductors, like cuprates, exhibit a dome-shaped phase diagram with an antiferromagnetic (AF) parent compound.  Additionally, they undergo a structural phase transition which is often but not always simultaneous with the magnetic phase transition.  Phase diagrams representative of three different families of Fe-based superconductors are shown in figure~\ref{fig:phase-diagrams}.

\begin{figure}[tbh]
\begin{center}
  {\includegraphics[width=1\columnwidth,clip]{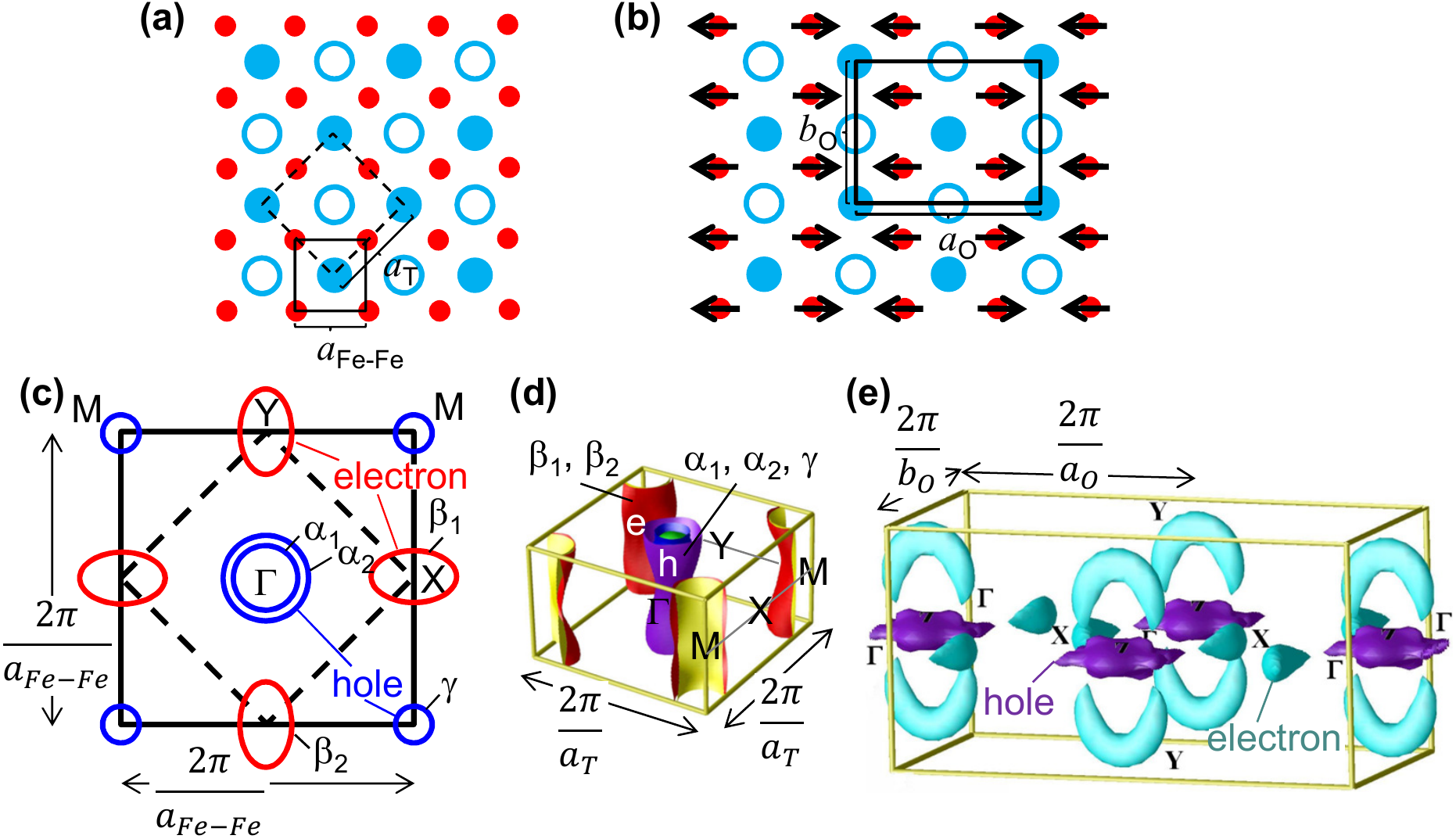}}
  \caption[Unit cells of Fe-based superconductors.]
    {\label{fig:unit-cell}
    Real space and momentum space unit cell of Fe-based superconductors. (a) Fe$_2$X$_2$ lattice, where the red dots represent Fe atoms, and the blue filled (open) circles represent X=As, P, Se, Te, or S atoms above (below) the Fe plane.  The solid square shows a unit cell with only one Fe; because of the two inequivalet X sites above and below the Fe plane, this is not the true unit cell of the crystal.  The dashed square shows the unit cell of the tetragonal structure typically seen at high temperatures. (b) The $a$ and $b$ axes of the low $T$ orthorhombic unit cell (black rectangle) typically differ by $<$1\%; the difference is exaggerated in this figure for visual clarity. (c) Unfolded BZ of the unit cell containing only a single Fe atom.  Band structure calculations typically predict as many as three inequivalent hole FSs (blue), centered at the $\Gamma$ point, and 2 inequivalent electron FSs (red), centered at the $M$ points. Variations in structure and charge doping can reduce the number of hole FSs, in some cases leaving only a single $\alpha$ surface at the $\Gamma$ point. (d) Calculated LDA FSs in the tetragonal BZ of Ba(Fe$_{0.9}$Co$_{0.1}$)$_2$As$_2$\cite{MazinPhysicaC2009}. Note that in this doubled real space unit cell and halved BZ, all hole FSs from (c) have now been folded into the central $\Gamma$ point, while the two electron FSs have been folded onto the same $M$ point. (e) Calculated LDA FS in the orthorhombic BZ for CaFe$_2$As$_2$ in the AF state\cite{MazinPRB2011}.  The real space unit cell has doubled again, and the BZ has halved again with respect to (d).}
\end{center}
\end{figure}

There are a number of points of confusion when considering the real space unit cell or the BZ (BZ) of these materials.  The common FeX layer in all of these materials (X=As,P,Se,Te,S) should be more accurately written as Fe$_2$X$_2$, because half of the X lie above the Fe plane, while half lie below, as can be seen in figure~\ref{fig:structures}. In the high temperature state, the material is tetragonal, with a unit cell twice as large (dashed square in figure~\ref{fig:unit-cell}a) and BZ half as large (figure~\ref{fig:unit-cell}d) as might be naively expected from the Fe square sublattice (solid square in figure~\ref{fig:unit-cell}a, and BZ in figure~\ref{fig:unit-cell}c).  In the underdoped region of the phase diagram, the material undergoes a tetragonal to orthorhombic phase transition as the temperature is lowered.  Then the unit cell doubles again (figure~\ref{fig:unit-cell}b), and the BZ halves again (figure~\ref{fig:unit-cell}e).  However, the difference between the new $a_O$ and $b_O$ axes in the orthorhombic state is typically $< 1\%$ (e.g. see tables A8 and A9 in Ref.~\onlinecite{JohnstonAdvPhys2010}) which is below the calibration resolution of the piezo scantubes typically used by STM.  Therefore the $a_O$ and $b_O$ axes cannot be distinguished by STM unless a twin boundary between orthorhombic domains can be found\cite{ChuangScience2010,SongScience2011}.


There is an additional source of confusion regarding the 122 materials.  The 1111, 111, and 11 materials have only a single FeX layer per unit cell.  However, the 122 materials have a double layer along the $c$-axis, so their unit cell is technically body-centered tetragonal, with a BZ that is rotated by 45$^{\circ}$ from the tetragonal BZ of the other Fe-based families.  In this review, since STM images only a single layer at a time, we will ignore this complication and we will talk about the 122 unit cell and BZ in the same tetragonal/orthorhombic notation as is used for the other Fe-based superconductors.

Fe-based superconductors are more complicated than their cuprate cousins because all five Fe $d$-bands cross the FS (FS), in contrast to the single Cu $d$ band crossing the FS in cuprates.  The schematic bands are shown in the unfolded BZ in Fig~\ref{fig:unit-cell}c.  Local density approximation (LDA) band structure calculation showed that Fe-based materials are semi-metals with hole bands at the $\Gamma$ point and electron bands at the $M$ point, in the tetragonal BZ (figure~\ref{fig:unit-cell}d)\cite{SinghPRL2008}.  The materials may be electron-doped or hole-doped, but over a large range of doping there is still usually significant nesting between the hole and electron FSs.

\section{\label{sec:Surface}Surface Considerations}

Like the cuprates, the Fe-based superconductors have a layered structure. Consequently, high-quality single crystals can be mechanically cleaved to obtain atomically flat and clean surfaces suitable for characterization by surface sensitive probes such as STM (STM) and spectroscopy (STS). With effort, samples as small as $\sim~100\,\mu$m can be cleaved and imaged (although typically 1mm is a more convenient size).  Just because a surface is clean and atomically flat does not necessarily mean it is a window into the bulk properties of the material.  Surfaces may have different carrier concentration than the bulk, and may exhibit electronic or structural reconstructions due to the reduced coordination number. In this section, we will discuss the surface characteristics for the different families of Fe-based superconductors.

\subsection{1111 Materials}
\label{sec:Surface1111}
Although LaFePO was found to superconduct with $T_c\sim$5K in 2006\cite{KamiharaJACS2006}, the Fe-based superconductors first rocketed to worldwide attention in February 2008 with the discovery of superconductivity up to $T_c$=26K in LaFeAsO$_{1-x}$F$_x$\cite{KamiharaJACS2008}.  These so-called `1111' materials have in common iron-pnictogen planes, separated by and ionically bound to oxide planes.  Because of the relatively high $T_c$, and the immediate superficial similarities between LaFeAsO$_{1-x}$F$_x$ and the cuprates (e.g. the layered structure, dome-shaped phase diagram, and surprising presence of Fe suggesting a possible spin-based superconducting mechanism), the reaction was immediate. Research labs around the world swiftly turned their attention to the new materials.  Within a few months, RFeAsO had been formulated with almost a dozen rare earth elements (R) \cite{ChenNature2008, ChenPRL2008, RenMatResInn2008, RenEPL2008, BosChemComm2008}, raising $T_c$ up to 56K\cite{WangEPL2008}.

Large single crystals proved challenging to grow, so studies of intrinsic anisotropy or inhomogeneity remained out of reach.  A few early, painstaking experiments managed to isolate single crystals as large as $\sim100\;\mu$m for transport\cite{JiaAPL2008} or ARPES\cite{KondoPRL2008}.  However, early STM experiments on polycrystalline samples\cite{MilloPRB2008,PanArxiv0808.0895} likely suffered from degraded surfaces.  The `as-grown' surface of a $\sim$100 $\mu$m single crystal of SmFeAsO$_{1-x}$F$_x$ (with nominal $x$=0.2, and measured $T_c$=45K, denoted by a red $\star$ in figure~\ref{fig:phase-diagrams}a) could not be imaged with atomic resolution, but did show a reproducible $V$-shaped gap of the approximate expected bulk value, that disappeared around the bulk superconducting $T_c$\cite{FasanoPRL2010}.  SmFeAsO$_{1-x}$F$_x$ spectra are shown in figures \ref{fig:Dynes-gap-fit}a and \ref{fig:pseudogap-DOS}a, and will be discussed further in section \ref{sec:gap-magnitude}.

In late 2009, a breakthrough allowed the growth of large single crystals of the `1111' materials using NaAs flux at ambient pressure\cite{YanAPL2009}.  Their growth method has since been replicated by other groups\cite{ChenPRB2010}. The first STM study on a cleaved single crystal `1111' material LaFeAsO$_{1-x}$F$_x$ (both parent compound, and nominal composition $x$=0.1, denoted by red $\star$'s in figure~\ref{fig:phase-diagrams}b) appeared a year later\cite{ZhouPRL2011}.  Zhou \textit{et al} were able to cleave and image atomically flat FeAs and LaO planes (figures~\ref{fig:LaFeAsO-QPI}, a and b), neither of which showed any structural reconstruction of the surface.

Although `1111' single crystals may cleave beautifully, surface-sensitive studies are expected to be problematic due to the ionic nature of the interlayer bonding, which is expected to result in a polar catastrophe at the cleaved surface\cite{NakagawaNatMat2006}. The surface electronic structure is likely to be different from that of the bulk. In the best case, there is just a rigid band shift which affects measurements of carrier density and Fermi level~\cite{HossainNatPhys2008}. In addition, atomic reconstruction, adsorption of charged contaminants, and electronic reconstruction might occur at the surface in order to compensate for the charge imbalance~\cite{HesperPRB2000}.
In fact, detailed calculations predict that several new surface bands appear in cleaved LaFeAsO$_{1-x}$F$_x$, on both La-terminated and As-terminated surfaces\cite{EschrigPRB2010}.
Zhou \textit{et al} found strong evidence for an electronic surface state on the LaO plane: the 2-dimensional electronic standing waves were so pronounced that the atomic lattice was almost completely concealed (see figure~\ref{fig:LaFeAsO-QPI}b).

The existence of dramatically different surface states in 1111 materials is further supported by ARPES measurements of the enclosed FS area on the cleaved surfaces of LaFePO~\cite{LuNature2008} and LaFeAsO~\cite{LuPhysicaC2009}.  In both cases, the ARPES-measured FS area falls short of the bulk electron count that would be expected from Luttinger's theorem.  An even more dramatic discrepancy is found in another ARPES experiment which sees the surface of NdFeAsO$_{0.9}$F$_{0.1}$ to be hole-doped, although the bulk is known to be electron-doped\cite{KondoPRL2008}. A follow-up ARPES\cite{LiuPRB2010} study found an additional large hole-like, $\Gamma$-centered FS on LaFeAsO, which was determined to be a surface state by its lack of $k_z$ dispersion, and by comparison with theory. When the LaFeAsO was electron-doped through the introduction of F, the 2-dimensional surface FS was seen to develop a large superconducting gap, likely via the proximity effect from the bulk superconducting state. The surface state superconducting gap may have different magnitude and additional symmetry components, which may confuse surface investigations of the bulk superconductivity.

Although large single crystals of LaFePO have been available since 2008, there have been no STM studies to date, likely because of its low $T_c$, and the surface challenges associated with all 1111 materials.

\subsection{\label{sec:Surface122}122 Materials}

In June 2008, a second family of Fe-based superconductors was discovered, including the same FeAs plane, but only a single layer of intervening A (alkaline earth metal) ions separating the FeAs layers, giving formula AFe$_2$As$_2$~\cite{RotterPRL2008}.  Particular members of this `122' family are referred to as Ba122, Sr122, Ca122, etc.

The 122 parent compounds become superconducting upon the introduction of hole or electron dopants, or the application of chemical or physical pressure.  For example, BaFe$_2$As$_2$ becomes a hole-doped superconductor upon replacement of Ba$^{2+}$ by K$^+$ \cite{RotterPRL2008}, or an electron-doped superconductor upon replacement of Fe$^{2+}$ by Co$^{3+}$ or Ni$^{4+}$ \cite{SefatPRL2008, LiNJP2009}, or a pressure-induced superconductor upon replacement of As by the isovalent but smaller atom P \cite{KasaharaPRB2010}, or application of physical pressure\cite{AlirezaJPCM2009}. The emergence of superconductivity through chemical substitution directly into the superconducting layer\cite{SefatPRL2008, LiNJP2009,KasaharaPRB2010} is in stark contrast with the cuprates, where the substitution of impurity elements for even a small percentage of the Cu atoms can destroy superconductivity~\cite{TarasconPRB1987}.

Because the 122 materials were successfully grown as large single crystals soon after their discovery \cite{NiPRB2008}, they were the first target of serious STM study among Fe-based superconductors.  Many papers have reported on their surface properties, yet the structure of the cleaved surface remains controversial.

The 122 materials are expected to cleave with FeAs layers intact, but A ions may end up on either of the two cleaved FeAs surfaces.  If the A ions do not divide evenly, each resulting surface will be polar, just as in the 1111 family. If the A ions do divide evenly, the resulting half-A surfaces will be nonpolar, and possibly ordered into one of several superstructures. Atomically resolved images of cleaved 122 surfaces typically show 1/2 of the atoms which would be expected for a complete As or Ba/Sr/Ca layer.  The STM groups who have achieved these images are sharply divided into two camps: those who believe that the observed structures result from a 1/2-layer of Ba/Sr/Ca, vs. those who believe that the observed structures result from a reconstruction of a complete As layer, i.e. all As atoms are present but 1/2 of these atoms are `invisible' to STM. In the following sections, we will present the arguments of both camps. \\

\begin{figure}[tbh]
\begin{center}
  {\includegraphics[width=1.0\columnwidth,clip]{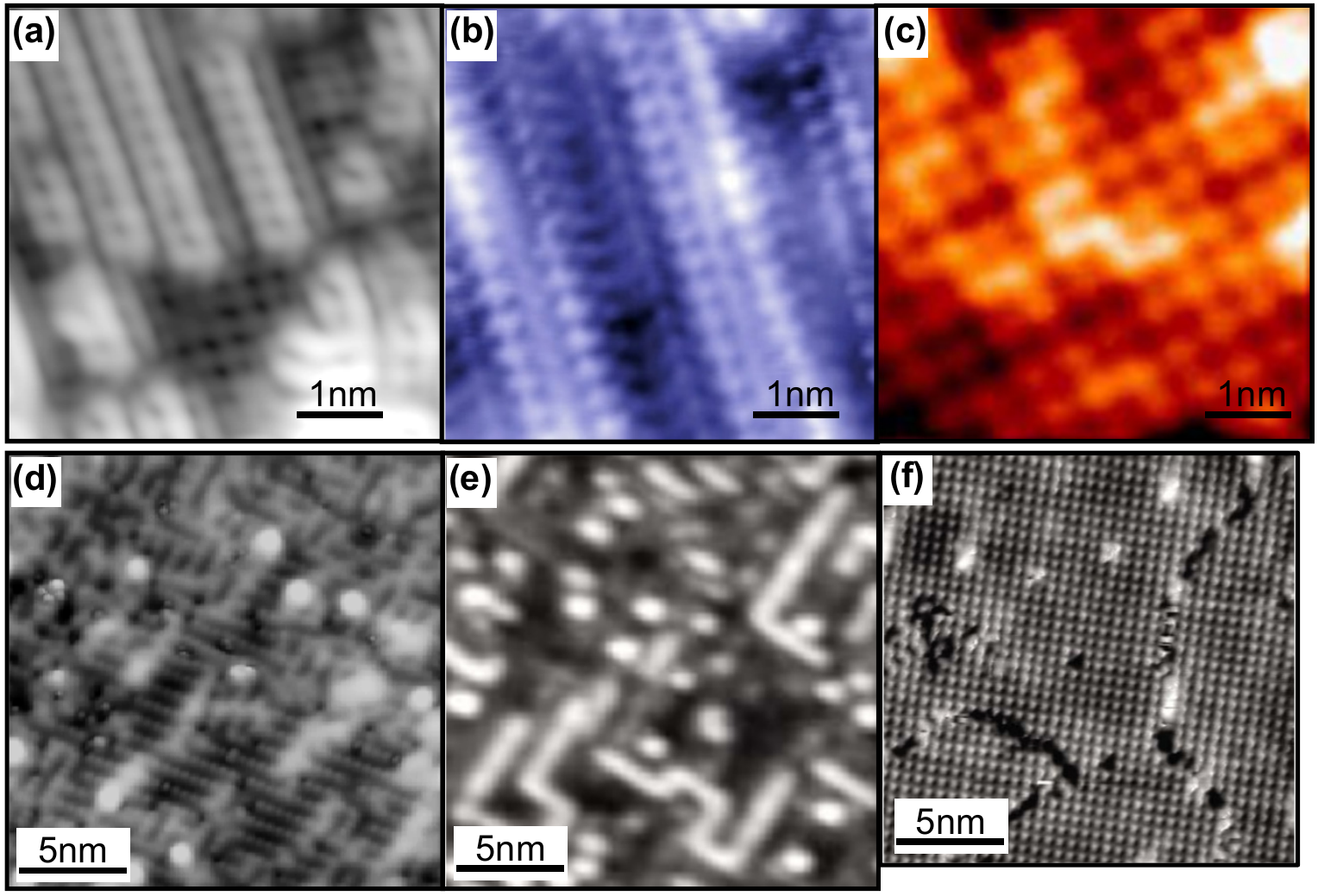}}
  \caption[Ba/Sr/Ca-terminated 122 surfaces]
    {\label{fig:122-Ba-topos}
    Various topographic images of cleaved 122 surfaces, presumed to be Ba/Sr-terminated. (a-c) $5\times5$ nm$^2$ images of cold-cleaved surfaces. (a) Sr$_{1-x}$K$_x$Fe$_2$As$_2$ ($T_c$=32K) cleaved at $\sim$10K and imaged at 5.3K. This is believed to be a nearly-complete Sr layer, with a glimpse through to the As layer below in one region\cite{BoyerArxiv0806.4400} (setup: $V_{\mathrm{sample}}=-100\,\mathrm{mV}; I_{\mathrm{set}}=200\,\mathrm{pA}$). (b) Ba(Fe$_{0.9}$Co$_{0.1}$)$_2$As$_2$ ($T_c$=25K) cleaved at $\sim$25K and imaged at $\sim$6K\cite{YinPRL2009} This surface is believed to show every other row of Ba, with the intervening As just barely visible beneath\cite{YinPhysicaC2009} (setup: $V_{\mathrm{sample}}=-20\,\mathrm{mV}; I_{\mathrm{set}}=40\,\mathrm{pA}$). (c)  Ba(Fe$_{0.915}$Co$_{0.085}$)$_2$As$_2$ cleaved at 120K, imaged at 5K with very small junction resistance to allow atomic resolution\cite{ZhangPRB2010} (setup: $V_{\mathrm{sample}}=20\,\mathrm{mV}; I_{\mathrm{set}}=2\,\mathrm{nA}$). (d-f) $20\times20$ nm$^2$ images of warm-cleaved surfaces, showing greater disorder. (d) SrFe$_2$As$_2$, cleaved at room temperature, imaged at 40K\cite{HsiehArxiv0812.2289}. (e) Ba(Fe$_{0.93}$Co$_{0.07}$)$_2$As$_2$ cleaved at room temperature, imaged at 4.2K\cite{MasseePRB2009b} (junction resistance $R_J \sim 0.75\,\mathrm{G}\Omega$). (f) Ba(Fe$_{0.93}$Co$_{0.07}$)$_2$As$_2$ cleaved at room temperature, imaged at 4.2K\cite{MasseePRB2009b} ($R_J \sim 0.75\,\mathrm{G}\Omega$).}
\end{center}
\end{figure}

\noindent \textit{\textbf{Ba/Sr/Ca Surface}} The first atomically resolved STM images of a Fe-based superconductor were reported by Boyer \textit{et al} on hole-doped Sr$_{1-x}$K$_x$Fe$_2$As$_2$ ($T_c$=32K), cleaved at $T=10$K and imaged at $T=5.3$K\cite{BoyerArxiv0806.4400}.  The low temperature cleave exposed a flat surface with average atomic spacing $\sim$4\AA\, as would be expected for either a complete Sr/K or As layer. These images also showed a $2 \times 1$ stripe reconstruction at $45^{\circ}$ to the orthorhombic lattice. Patches of missing atoms showed windows through to an underlying square lattice, also with atomic spacing $\sim$4\AA. Figure~\ref{fig:122-Ba-topos}a shows that the underlying square lattice is laterally shifted by half a unit cell from the topmost $2\times 1$ surface, and lies beneath it by 2.8(4)\AA, close to the expected 2\AA\ vertical distance between Sr and As layers in the bulk. The observed atomic separation of $\sim$4\AA\ in both striped region and square lattice region are not consistent with the Fe-Fe distance. Boyer concluded that the upper striped region was a reconstruction of the complete Sr/K layer, while the underlying square lattice patch was the bare As layer.

Yin \textit{et al} reported atomically resolved images of optimally electron-doped Ba(Fe$_{1-x}$Co$_x$)$_2$As$_2$ (nominal $x$=0.1, and measured $T_c=25.3$K) which showed a $2 \times 1$ reconstruction\cite{YinPRL2009} (see figure \ref{fig:122-Ba-topos}b). They found no step edges, but occasional 1/2-period lateral shifts.  Upon inverse transforming the four brightest spots in the Fourier transform of this topography, a pattern emerged which suggested that the surface consisted of half of the Ba atoms, arranged into stripes of single atom width, allowing a glimpse between stripes to the As layer beneath \cite{YinPhysicaC2009}. A Ba half-layer would imply a nonpolar surface of bulk-like carrier concentration, which is consistent with the observation of a ubiquitous superconducting gap, and other expected features of bulk superconductivity such as magnetic vortices\cite{YinPRL2009}.

Hsieh \textit{et al} reported atomically resolved images of the parent SrFe$_2$As$_2$, cleaved at room temperature and cooled to $T=40$K\cite{HsiehArxiv0812.2289}.  The surface (figure~\ref{fig:122-Ba-topos}d) was only partially ordered into a similar $2\times1$ reconstruction as shown by Boyer and Yin.  Upon heating to 200K and reimaging, the long range order of the $2\times1$ reconstruction was lost.  Upon recooling, the surface sometimes reordered, but sometimes remained disordered\cite{GomesPrivate2011}.  This observation suggests that the ordered $2\times1$ surface reconstruction may be a metastable ordering of a mobile half-Sr layer, resulting from the low temperature cleaving process, easily destroyed upon heating, and not always reformed upon recooling. It is harder to imagine that a clean, complete As layer would disorder so extremely on heating to 200K, and fail to reorder upon recooling.




\begin{figure}[tbh]
\begin{center}
  {\includegraphics[width=0.95\columnwidth,clip]{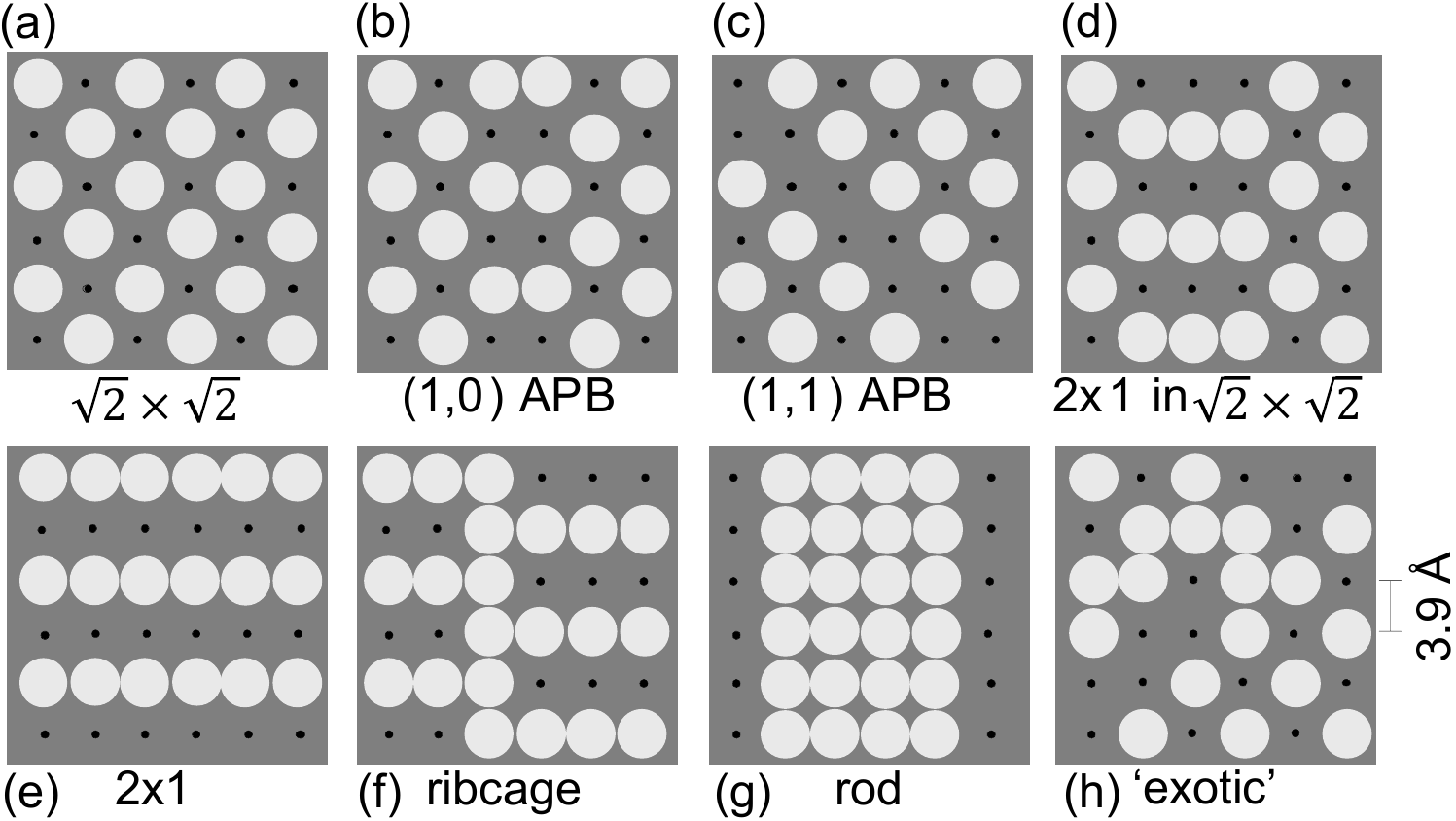}}
  \caption[Partial Ba surface configurations]
    {\label{fig:Massee-surface-schematic}
    Sketch of various possible partial-Ba surface configurations\cite{MasseePRB2009b}.  Large (small) circles indicate the presence (absence) of a Ba atom.}
\end{center}
\end{figure}

Massee \textit{et al} showed a large number of topographies of Ba(Fe$_{0.93}$Co$_{0.07}$)$_2$As$_2$, cleaved either at room temperature or at $T<80$K \cite{MasseePRB2009b} (see figure~\ref{fig:122-Ba-topos}, e and f).  At low $T$, they often observed the same $\sim$8 \AA\ stripes seen previously\cite{BoyerArxiv0806.4400,YinPRL2009}.  Like Yin, they observed occasional half-period stripe shifts, resulting in a `ribcage' structure (figure~\ref{fig:Massee-surface-schematic}f).  They also saw less regular, larger rodlike features of up to 20\AA\ width, which evolved smoothly without step edges from the more commonly observed 8\AA\ stripes.  In room temperature cleaves, they saw disordered 8\AA\ stripes (figure~\ref{fig:122-Ba-topos}e), similar to those seen by Hsieh \textit{et al.}  They also sometimes saw a 5.5\AA\ $\sqrt{2}\times\sqrt{2}$ structure, with meandering antiphase boundaries (figure~\ref{fig:122-Ba-topos}f). In total, Massee observed at least 8 different surface structures\cite{MasseePRB2009b}, although $2\times 1$ and $\sqrt{2} \times \sqrt{2}$ were found to be dominant\cite{MasseePrivate2011}.  These 8 surface structures are sketched in figure~\ref{fig:Massee-surface-schematic}. It is hard to imagine that a single clean As layer would display so many different surfaces structures.

Zhang \textit{et al} imaged Ba(Fe$_{1-x}$Co$_x$)$_2$As$_2$ with six different nominal dopings: $x$=0, 0.04, 0.085, 0.10, 0.125, and 0.16. The latter four were found to be superconducting with $T_c$=9K, 25K, 22K, and 9K, respectively\cite{ZhangPRB2010}. Samples were imaged at 5K after cleaving at 120K (above the 100K at which the extra surface superstructure diffraction spots started to disappear, as discussed in the following section, but below the 200K at which the surface long range order has fully disappeared\cite{HsiehArxiv0812.2289,MasseePRB2009b}).  Zhang found that a $\sqrt{2}\times\sqrt{2}$ superstructure dominated at all dopings (figure~\ref{fig:122-Ba-topos}c). Theoretical simulations with a half-Ba terminating layer showed a strong feature at -200 meV, and a gross energy asymmetry with larger empty-state DOS, in good agreement with the spectra measured on all six samples (see figure~\ref{fig:high-energy-DOS}a). Zhang therefore concluded that this $\sqrt{2}\times\sqrt{2}$ structure was a half-Ba surface.  Note that the six dopings span the bulk orthorhombic to tetragonal transition (see figure~\ref{fig:phase-diagrams}c), so their ubiquitous $\sqrt{2}\times\sqrt{2}$ structure cannot be due solely to the bulk orthorhombic transition.  In only two samples out of many studied, they saw patches of the $2\times 1$ stripes.

The argument for the partial-Ba-terminated surface was further supported by measurements of the work function, $\varphi$\cite{MasseeThesis2011}.  Work functions for relevant elements are shown in table~\ref{table:work-function}, taken from Ref. \onlinecite{CRC2011}. In dozens of Ba122 samples studied by Massee \textit{et al}, representing $\sqrt{2}\times\sqrt{2}$, $2\times 1$, and other surface organizations, the work function was never found to be greater than 1.8 eV, and typically found to be $\sim$1.5 eV.  On Ca122, Massee also found a slightly higher work function, $\sim$1.9 eV, suggesting that the surface termination was not As, which would be expected to show the same, much higher work function in both Ba122 and Ca122.\\

\noindent \textit{\textbf{As Surface}} Nascimento \textit{et al} presented STM images of BaFe$_2$As$_2$, cleaved at $T$=80K, revealing a 5.6\AA\ square lattice\cite{NascimentoPRL2009}. This is the correct spacing for a $\sqrt{2} \times \sqrt{2}$ organization of a half Ba layer.  In contrast, the complete As layer would form a 3.8\AA, nearly square lattice. However, in combination with low energy electron diffraction (LEED) data (next section), Nascimento \textit{et al} hypothesized that half of the As were invisible due to a structural reconstruction. They pointed out that the surface is 1-6\% orthogonal (in comparison to the 0.7\% orthogonality of the bulk), and that this surface orthorhombicity persists across the superconducting dome\cite{PlummerPrivate2011}.

\begin{figure}[tbh]
\begin{center}
  {\includegraphics[width=0.95\columnwidth,clip]{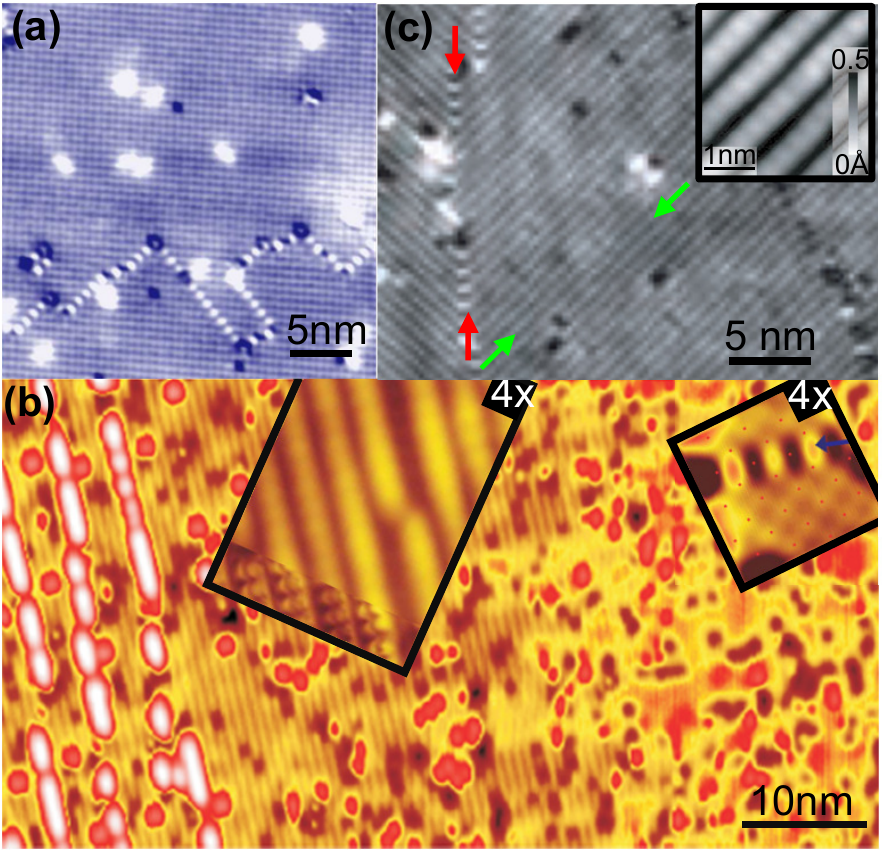}}
  \caption[As-terminated 122 surfaces]
    {\label{fig:122-As-topos}
    Topographic images of cleaved parent 122 surfaces, presumed to be As-terminated. Images are scaled to the same unit cell size. (a) BaFe$_2$As$_2$ cleaved at 80K, showing a $\sqrt{2}\times\sqrt{2}$ surface reconstruction with a domain boundary zigzagging across the bottom of the image\cite{LiArxiv1006.5907} (setup: $V_{\mathrm{sample}}=23\,\mathrm{mV}; I_{\mathrm{set}}=200\,\mathrm{pA}$). (b) SrFe$_2$As$_2$ showing $2\times 1$ reconstruction on the left, merging smoothly into $\sqrt{2}\times\sqrt{2}$ reconstruction on the right, without step edges\cite{NiestemskiArxiv0906.2761} ($R_J = 0.8\,\mathrm{G}\Omega$).  The left inset at $4\times$ magnification shows that at very low junction resistance (lower left strip, $R_J = 10\,\mathrm{M}\Omega$), the stripe appears to be two atoms wide. The right inset at $4\times$ magnification shows that half of the presumed $As$ atoms are invisible in the $\sqrt{2}\times\sqrt{2}$ reconstruction region ($R_J = 1.7\,\mathrm{G}\Omega$). (c) Non-superconducting Ca(Fe$_{0.97}$Co$_{0.03}$)$_2$As$_2$ showing a $2\times 1$ surface reconstruction, with a domain boundary in the lower left corner (setup: $V_{\mathrm{sample}}=-50\,\mathrm{mV}; I_{\mathrm{set}}=10\,\mathrm{pA}$), and $4\times$ magnified inset in the upper right corner (setup: $V_{\mathrm{sample}}=-5\,\mathrm{mV}; I_{\mathrm{set}}=100\,\mathrm{pA}$)\cite{ChuangScience2010}. }
\end{center}
\end{figure}

In a second study by the same group\cite{LiArxiv1006.5907}, more images were presented of the parent Ba122 $\sqrt{2}\times\sqrt{2}$ surface, sporting some white blobs which could be moved with the STM tip, which were identified as sparse remaining Ba (figure~\ref{fig:122-As-topos}a). Li \textit{et al} hypothesized here that half of the As atoms were invisible due to different spin environments.  Furthermore, they imaged domain walls, as seen zigzagging across the bottom half of figure~\ref{fig:122-As-topos}a.  Larger images showed that these domain walls always formed closed loops.  On either side of a domain wall, the orthorhombic $a$ and $b$ axes maintained their orientation, indicating no orthorhombic twin boundaries. But the visible and invisible atoms did switch, suggesting an antiphase boundary of the the spin state. These domain walls were similar to those seen by Massee, shown in figure~\ref{fig:122-Ba-topos}f. (The prevalence of such antiphase boundaries may explain the discrepancy between the large Fe spin moments computed from density functional theory, and the small spin moments measured experimentally\cite{MazinNatPhys2008}.)

Niestemski \textit{et al} studied SrFe$_2$As$_2$, cleaved both at room temperature and at $T=77$K\cite{NiestemskiArxiv0906.2761}.  Like Massee, they observed a $\sqrt{2}\times\sqrt{2}$ structure in some areas, which merged seamlessly into a $2\times 1$ structure in other areas, with no intervening step edges (figure~\ref{fig:122-As-topos}b).  They concluded that this was a complete As top layer, with half of the As atoms invisible under standard imaging conditions ($R_J \sim 0.8\,\mathrm{G}\Omega$).  They gave two arguments, as follows.  First, in the $2\times 1$ regions, very low junction resistance imaging allowed visualization of the expected number of atoms in a complete As or Sr layer (see bottom left edge of first inset in figure \ref{fig:122-As-topos}b).  Second, unlike Boyer's previous work on Sr122\cite{BoyerArxiv0806.4400}, after an extensive search over large areas of the surface, Niestemski never found a step down to an As layer below, which would be occasionally expected if the terminating surface were Sr.  In fact, Niestemski found some steps up to additional rod-like structures, which were suggested to be the remnants of Sr on top of the As surface. Therefore, Niestemski concluded that both the $\sqrt{2}\times\sqrt{2}$ and $2\times 1$ observed structures were different reconstructions of a complete terminating As layer. Both Nascimento and Niestemski reported identical $dI/dV$ at visible and `invisible' atomic sites.

Pan \textit{et al} used STM to study a series of Ba(Fe$_{1-x}$Co$_x$)$_2$As$_2$, and reported that the undoped samples showed the $\sqrt{2}\times\sqrt{2}$ structure, but on increasing $x$, the $2 \times 1$ structure became dominant\cite{PanBAPS2009}.\\



\noindent \textit{\textbf{Other Experimental Evidence}} Other surface studies of 122 materials have been carried out using photoemission spectroscopy (PES), ARPES, and LEED.

Hsieh \textit{et al} presented ARPES data on Sr122, cleaved at 10K, which showed a strong feature at the BZ $X$ point\cite{HsiehArxiv0812.2289} (see figure \ref{fig:unit-cell}d for BZ reference).  This feature was not present in the LDA-calculated band structure, but would result from the band folding due to a $2\times1$ surface structure. When the sample was heated to 200K and cooled to 10K for remeasurement, the band-folding artifact at $X$ disappeared, indicating that the $2\times1$ order was totally lost by temperature cycling.

Van Heumen \textit{et al} also used ARPES to directly show a surface state around -200meV near the $\Gamma$ point in cold-cleaved Ba122, which disappeared after thermal cycling\cite{VanHeumenPRL2011}. This -200 meV feature was also seen in the STM measurements by Zhang in 6 different samples of Ba(Fe$_{1-x}$Co$_x$)$_2$As$_2$ from $x$=0 to 0.16\cite{ZhangPRB2010} (figure \ref{fig:high-energy-DOS}a), by Yin\cite{YinThesis2009} (figure \ref{fig:high-energy-DOS}b), by Niestemski in parent Sr122\cite{NiestemskiArxiv0906.2761} (figure \ref{fig:high-energy-DOS}c), and appeared in Zhang's theoretical calculations for a half-Ba-terminated surface.

De Jong \textit{et al} reported photoemission on the undoped parent compound BaFe$_2$As$_2$ (with 7\% atomic weight Sn impurities), cleaved and measured at room temperature. Variations in photon energy affect the depth probed: they used $h\nu=140 eV$ photons to probe a few Angstroms, and $h\nu=3keV$ photons to probe 10s of nm from the surface.  With low energy (shallow) photons, the binding energy peaks for both the Ba $4d$ and As $3d$ core states showed additional shifted shoulders due to surface contributions. These shoulders were much more pronounced for the Ba $4d$ states than for the As $3d$ states, suggesting that Ba sat at the surface. Furthermore, these Ba shoulders were much broader than the As shoulders, suggesting that the surface Ba atoms were more disordered than the near-surface As atoms.

De Jong \textit{et al} also compared their Ba analysis to analogous work on the Ba $4d$ core states in the cuprate superconductor YBa$_2$Cu$_3$O$_{7-x}$.  The surface carrier concentration of cleaved YBa$_2$Cu$_3$O$_{7-x}$ is known to deviate from the bulk, resulting in shift to higher binding energy for both cations (Ba$^{2+}$) and anions (O$^{2-}$). In contrast, for Ba122 the shift of the surface states was to higher binding energy for cations (Ba$^{2+}$) and to lower binding energy for anions (As$^{1-}$), suggesting that surface doping was not the leading cause of the surface state.  The opposite binding energy shifts for Ba and As in Ba122 are more reminiscent of GaAs, which is known to have a Madelung potential shift at the surface. A Madelung energy shift is likely to affect the localized, ionic electronic levels, rather than the itinerant near-$\varepsilon_F$ states.  Therefore, de Jong concluded that the near-$\varepsilon_F$ states at the surface of Ba122 likely did not differ significantly from the bulk.

\begin{figure}[tbh]
\begin{center}
  {\includegraphics[width=1.0\columnwidth,clip]{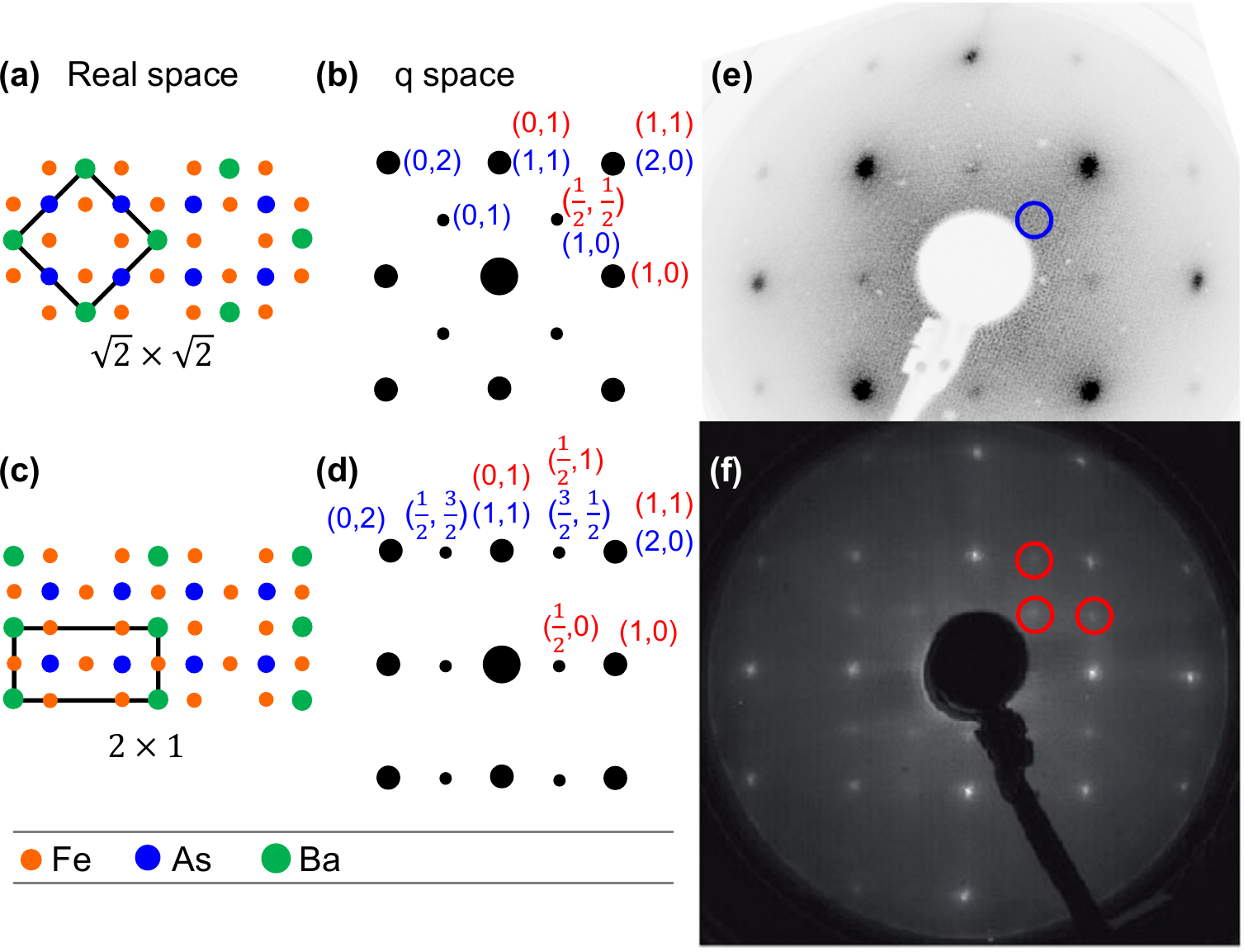}}
  \caption[LEED measurements of cleaved Ba122 surface.]
    {\label{fig:LEED}
    (a) Real space atomic configuration and (b) expected $q$ space LEED pattern, for a $\sqrt{2}\times \sqrt{2}$  arrangement of a 1/2-Ba layer on top of a complete As-Fe$_2$-As layer (the lower As atoms are not shown). (c) Real space atomic configuration and (d) expected $q$ space LEED pattern, for a $2\times 1$ arrangement of a 1/2-Ba layer on top of a complete As-Fe$_2$-As layer (the lower As atoms are not shown). In both (b) and (d), the LEED spots are labeled in the tetragonal notation (red) and the orthorhombic notation (blue). (e) LEED image of the surface of BaFe$_2$As$_2$, cleaved at room temperature and measured at $T$=20K\cite{NascimentoPRL2009}. Blue circle shows the missing (1,0) orthorhombic spot, which would be expected for the $\sqrt{2}\times\sqrt{2}$ surface observed by STM in the same paper. (f) LEED image of the surface of Ba(Fe$_{0.95}$Co$_{0.05}$)$_2$As$_2$ (in the bulk orthorhombic phase), cleaved and measured at $T<$20K\cite{VanHeumenPRL2011}. Red circles show diffraction spots corresponding to $\sqrt{2}\times\sqrt{2}$, $2\times 1$ and $1\times 2$ orders which coexist within the same macroscopic electron beam spot size. LEED image of Ba(Fe$_{0.915}$Co$_{0.085}$)$_2$As$_2$ in the bulk tetragonal phase (not shown here) similarly shows all $2\times1$ and $\sqrt{2}\times\sqrt{2}$ spots\cite{VanHeumenPRL2011}.}
\end{center}
\end{figure}

Massee \textit{et al} performed LEED on cold-cleaved Co-Ba122 surfaces at $T=17$K and observed extra spots corresponding to both the $\sqrt{2}\times\sqrt{2}$ and $2\times 1$ reconstructions (see schematic in figure \ref{fig:LEED}, a-d). When the sample was heated, the $\sqrt{2}\times\sqrt{2}$ and $2\times 1$ spots started to lose intensity at $T\sim100$K, and by $T\sim200$K they were gone (only tetragonal spots remained).  But when Massee recooled to 17K, none of the superstructure spots reappeared. This was consistent with both Nascimento\cite{NascimentoPRL2009}, who cleaved warm and never saw the $\sqrt{2}\times\sqrt{2}$ and $2\times 1$ orthorhombic spots to begin with, and with Hsieh\cite{HsiehArxiv0812.2289}, who also saw the low $T$ superstructure features in both STM and ARPES disappear upon warming.  Both Massee and Hsieh concluded that cleaving usually leaves some fraction of Ba layer.  If the material is cleaved while cold, then Ba may be stuck in any number of different metastable arrangements.  But if the material is cleaved while warm, or warmed up after cleaving, then the Ba revert to the lowest energy $\sqrt{2}\times\sqrt{2}$ configuration, but without long range order.  Massee speculated that invisibility of this $\sqrt{2}\times\sqrt{2}$ superstructure in room temperature LEED measurements may be explained by the proliferation of antiphase domain walls exemplified in figure~\ref{fig:122-Ba-topos}f.

Additional information can be gained by varying the energy of the incident electrons, and measuring the resultant variation in LEED spot intensity, in a technique called $IV$-LEED\cite{PendryBook1974,vanHoveBook1986}. By comparison to models of candidate structures, the $IV$-LEED data can be used to determine the terminating element. The Pendry factor is a measure of the reliability of a structural fit; it is considered good for $R_P\sim 0.2$, mediocre for $R_P\sim 0.3$ and bad for $R_P > 0.5$ \cite{PendryJPC1980}.

In contrast to Massee's LEED conclusions, an $IV$-LEED experiment performed by Nascimento \textit{et al} on warm-cleaved Ba122 led them to conclude that the surface termination is As\cite{NascimentoPRL2009}. In Nascimento's experiment, the LEED pattern reflected only the 3.8\AA\ tetragonal unit cell (figure~\ref{fig:LEED}e), with no hint of the 5.6\AA\ $\sqrt{2}\times\sqrt{2}$ structure observed by STM in the same material (figure~\ref{fig:122-As-topos}a).  From the orthorhombic (2,0) spot (tetragonal (1,1) spot), Nascimento calculated the Pendry factors for Ba, Fe, and As terminations, arriving at $R_P$= 0.57, 0.45, and 0.24,  respectively.




One caveat to Nasicmento's work is that the samples for LEED study were cleaved at room temperature, before cooling to 20K, so the structure from the low-$T$-cleaved surfaces studied by STM in the same paper was possibly very different from the room-$T$-cleaved surfaces studied by LEED.  This may explain the absence of the expected orthorhombic (1,0) spot in the LEED pattern.

Van Heumen \textit{et al} investigated the cold-cleaved surface of Ba(Fe$_{1-x}$Co$_x$)$_2$As$_2$ ($x$=0.05 and 0.085) via $IV$-LEED\cite{VanHeumenPRL2011}. The LEED images showed fractional spots corresponding to both $\sqrt{2}\times\sqrt{2}$ and $2\times 1$ surface structures (figure~\ref{fig:LEED}f), which can be explained by the coexistence of both types of domains.  Furthermore, the energy dependence of each of these spots was studied from 100 to 400 eV, and compared with theoretical simulations.  One set of simulations assumed a 1/2 Ba terminating layer, and allowed the top four layers to relax ($\frac{1}{2}$Ba-As-Fe$_2$-As), resulting in the low Pendry $R$ factors of $R_p=0.19$ for $\sqrt{2}\times\sqrt{2}$ and $R_p=0.29$ for $2\times 1$. A second set of simulations assumed a terminating As surface, also allowing the top three layers to relax, but this resulted in $R_p=0.42$ for $\sqrt{2}\times\sqrt{2}$ and $R_p=0.48$ for $2\times 1$.  Notably, in contrast to Nascimento's work, van Heumen's work compared calculated $IV$ curves only to the fractional spots, i.e. those spots corresponding directly and exclusively to the $\sqrt{2}\times\sqrt{2}$ or $2\times 1$ surface structures which were observed in most images by both surface termination camps.

Combining STM and these other measurements leads to a picture in which cleaving leaves $\sim$1/2 of the Ba/Sr/Ca atoms atop a reconstructed As layer.  Possible arrangements of the terminating Ba/Sr/Ca atoms are sketched in figure \ref{fig:Massee-surface-schematic}\cite{MasseePRB2009b}. \\

\noindent \textit{\textbf{Theory}} The experimentalists do not agree on the surface termination, so what do the theorists say?  Gao \textit{et al} presented electronic structure calculations on the (001) surface of three A122 compounds (A=Ba, Sr, or Ca)\cite{GaoPRB2010}.  They found that it is most energetically favorable for A to divide evenly between the two cleaved surfaces, and for the half-A terminating layers to organize into superstructures as shown in table \ref{table:Gao}.  They found that the electronic states imaged on surfaces with the $\sqrt{2}\times\sqrt{2}$ structure should be representative of the bulk states, while there may be some additional surface states near the Fermi energy in the $2 \times 1$ superstructure.  Gao commented specifically on the apparent dimerization of the stripes observed in Sr122 and given by Niestemski as a primary piece of evidence for a complete As terminating surface\cite{NiestemskiArxiv0906.2761}.  Gao instead ascribed this apparent dimerization to hybridization between a terminating half-Sr layer and a complete As layer beneath.

\begin{table}[h!]\footnotesize
\begin{center}
\begin{tabular}{ l | c | c }
 \hline
 & \multicolumn{2}{c}{Bulk structural phase} \\
 \hline
 & AF orthorhombic & tetragonal \\
 \hline \hline
 Ba122 & $\sqrt{2}\times\sqrt{2}$ & $\sqrt{2}\times\sqrt{2}$ \\
 Sr122 & $\sqrt{2}\times\sqrt{2}$ & $\sqrt{2}\times\sqrt{2}$ \\
 Ca122 & $2\times 1$ & $2\times 1$ \\
 \hline
\end{tabular}
\caption{\label{table:Gao} Energetically favorable arrangement of the 1/2-A terminating surface layer, as calculated by the plane-wave basis method\cite{GaoPRB2010}.}
\end{center}
\end{table}

Gao suggested that a metastable As termination may result from a fast cleave, so he investigated scenarios which might lead to half of the As atoms being invisible, as claimed in the $\sqrt{2}\times\sqrt{2}$ surface structure of parent Ba122 \cite{NascimentoPRL2009,LiArxiv1006.5907} and Sr122 \cite{NiestemskiArxiv0906.2761}.  Gao found that in the Ba122 orthorhombic phase, a small buckling of the As terminating surface does strongly affect the STM images, leading to half of the As being invisible.  However, he found that no such reconstruction occurs in the tetragonal phase of Ba122, or in either phase of Sr122.  Regardless of the presence or absence of As-buckling, Gao expected that the low energy electronic structure of any As-terminated surface would be strongly modified from the bulk. \\




\noindent \textit{\textbf{Wrap-up}} To round out this story of STM-measured surface structure, it is worth mentioning that STM images of very lightly Co-doped (non-superconducting) Ca122 also displayed a $2\times 1$ structure (figure \ref{fig:122-As-topos}c)\cite{ChuangScience2010}, matching Gao's prediction\cite{GaoPRB2010}.  In the Ca122 experimental work, the authors identified the terminating layer as As, based on solely on evidence by Nascimento and Niestemski.  However, given newer LEED data\cite{VanHeumenPRL2011}, the authors have revised their opinion and believe that their terminating surface is more likely Ca\cite{AllanPrivate2011}. It may also be worth noting that the unambiguously identified cleaved FeAs surface of LaFeAsO$_{1-x}$F$_x$ shows the $\sim$4\AA\ lattice of the full As layer with no reconstruction and no missing atoms\cite{ZhouPRL2011} (figure \ref{fig:LaFeAsO-QPI}a). Neither does the analogous Te/Se-terminated surface of FeTe$_{1-x}$Se$_x$ show any reconstruction or missing atoms\cite{MasseePRB2009b,KatoPRB2009,HanaguriScience2010,SongScience2011} (figure \ref{fig:Hanaguri-pnictide}c).

Table~\ref{table:A122structure} displays a summary of the many studies of 122 surfaces.  Even if we restrict ourselves to the interpretations of one of the two competing camps, we must conclude from the raw data (figures \ref{fig:122-Ba-topos} and \ref{fig:122-As-topos}) that there are many possible terminating surfaces, close in energy, which can result from very slight differences in cleaving and temperature history.  This conclusion, admittedly unsatisfying in its complexity, is supported by Gao's theory. Despite strong claims by both the half-A termination supporters and the As-termination supporters, we must ask: it worth debating about these surfaces at all? The only relevant question is: which, if any, of these surfaces are representative of the bulk electronic structure of the material?

\begin{table*}[tbh]\footnotesize
\begin{tabular}{ p{2.9cm} | l p{1.3cm} p{1.3cm} l p{2.5cm} p{2.5cm} l}
 \hline
    Material & Expt. & Cleave $T$ & Meas. $T$ & Term. & Structure & SC gap? & Ref. \\
 \hline \hline
    parent Ba122 & PES & 300K & 300K & Ba & disordered & - & de Jong\cite{DeJongPRB2009} \\

    parent Ba122 & $IV$-LEED & 300K & 20K & As & 3.8\AA\ tetragonal	 & -	& Nascimento\cite{NascimentoPRL2009} \\

    parent Ba122 & STM & 20K & 4.3K	& As & $\sqrt{2}\times\sqrt{2}$ & - & Nascimento\cite{NascimentoPRL2009} \\

    parent Ba122 & STM & 80K & 5K & As & $\sqrt{2}\times\sqrt{2}$ & - & Li\cite{LiArxiv1006.5907} \\

    parent Sr122 & STM & 300K & \raggedright 40K \newline \& 200K & Sr & disordered $2\times 1$	& - & Hsieh\cite{HsiehArxiv0812.2289} \\

    parent Sr122 & ARPES & 10K & 10K \newline \& 200K & Sr & $2\times 1$ & - & Hsieh\cite{HsiehArxiv0812.2289}\\

    parent Sr122 & STM & \raggedright warm \newline \& cold & 5K & As & $\sqrt{2}\times\sqrt{2}$, $2\times 1$	& - & Niestemski\cite{NiestemskiArxiv0906.2761} \\

    Ca(Fe$_{0.97}$Co$_{0.03}$)$_2$As$_2$ (``parent'') & STM & cold & 4.3K & ? & $2\times 1$ & - & Chuang\cite{ChuangScience2010} \\

    K$_x$Sr$_{1-x}$Fe$_2$As$_2$ ($T_c$=32K) & STM & 10K & 5.3K & Sr & $2\times 1$, disordered & \raggedright yes $2\times 1$, \newline no elsewhere & Boyer\cite{BoyerArxiv0806.4400} \\

    Ba(Fe$_{0.9}$Co$_{0.1}$)$_2$As$_2$ ($T_c$=25K) & STM & $\sim25$K & 6K & Ba	 & $2\times 1$ & yes $2\times 1$ & Yin\cite{YinPRL2009} \\

    Ba(Fe$_{1-x}$Co$_{x}$)$_2$As$_2$ ($x$=0,0.07) & STM & 300K & 4.2K & Ba & disordered $2\times 1$ & yes everywhere & Massee\cite{MasseePRB2009a} \\

    Ba(Fe$_{1-x}$Co$_{x}$)$_2$As$_2$ ($x$=0,0.07) & LEED & 300K & 4.2K & Ba	& disordered $2\times 1$ & - & Massee\cite{MasseePRB2009a} \\  

    Ba(Fe$_{0.93}$Co$_{0.07}$)$_2$As$_2$ & STM & \raggedright $<80$K \newline \& 300K & 4.2K & Ba & \raggedright $\sqrt{2}\times\sqrt{2}$, $2\times 1$, other & - & Massee\cite{MasseePRB2009b} \\

    Ba(Fe$_{0.93}$Co$_{0.07}$)$_2$As$_2$ & $IV$-LEED & \raggedright $<80$K \newline \& 300K & 4.2K & Ba & $2\times 1$ & - & Massee\cite{MasseePRB2009b} \\

    Ba(Fe$_{1-x}$Co$_{x}$)$_2$As$_2$ ($x$=0.04, 0.105) & STM & 80-100K & 4.5-24K & Ba & n/a	 & \raggedright yes except large corrugation & Massee\cite{MasseeEPL2010} \\

    Ba(Fe$_{1-x}$Co$_{x}$)$_2$As$_2$ ($x$=0.07) & STM & 300K & 4.5-24K & Ba & n/a	 & \raggedright yes except large corrugation & Massee\cite{MasseeEPL2010} \\

    Ba(Fe$_{1-x}$Co$_{x}$)$_2$As$_2$ & STM & ? & ? & ? & \raggedright $\sqrt{2}\times\sqrt{2} \newline \rightarrow 2\times 1$ as $x\uparrow$ & yes & Pan\cite{PanBAPS2009} \\

    Ba(Fe$_{1-x}$Co$_{x}$)$_2$As$_2$, ($x$=0-0.16) & STM & 120K & 5K & Ba & \raggedright mostly $\sqrt{2}\times\sqrt{2}$, rare $2\times 1$ & \raggedright yes $2\times 1$, \newline no elsewhere & Zhang\cite{ZhangPRB2010} \\

    Ba(Fe$_{1-x}$Co$_{x}$)$_2$As$_2$, ($x$=0.05,0.085) & $IV$-LEED & $<$20K & $<$20K & Ba & $\sqrt{2}\times\sqrt{2}$ and $2\times 1$ & - & van Heumen\cite{VanHeumenPRL2011} \\

    K$_x$Sr$_{1-x}$Fe$_2$As$_2$ ($T_c=38$K) & STM & cold & 2.5K & ? & no atomic res. & yes, 2 gaps & Shan\cite{ShanNatPhys2011, ShanPRB2011} \\

    Ba(Fe$_{1-x}$Co$_{x}$)$_2$As$_2$ ($x$=0.06,0.12; $T_c$=14,20K) & STM & \raggedright 300K \newline in Ar & 6K & ? & no atomic res. & yes, 2 gaps & Teague\cite{TeaguePRL2011} \\
 \hline
\end{tabular}
\caption{\label{table:A122structure} Summary of the surface structures reported for A122 compounds.}
\end{table*}

The penultimate column of table~\ref{table:A122structure} contains information about the purported superconducting gap (if applicable) observed by STM in the given experiment.  In the samples for which bulk superconductivity is expected, STM detected at least one gap of the expected magnitude on most surfaces. An appropriate gap was detected on every surface with an ordered $2\times 1$ structure, most surfaces with $\sqrt{2}\times\sqrt{2}$, and even most surfaces with disordered mixtures of these two structures. Massee explicitly observed ``identical superconducting gaps on both $\sqrt{2}\times\sqrt{2}$ as well as $2\times 1$ or other surface terminations''\cite{MasseeEPL2010}. However, extreme disorder does seem detrimental to the observation of a superconducting gap: Boyer\cite{BoyerArxiv0806.4400} observed no superconducting gap on very disordered surfaces which lacked atomic resolution, while Massee\cite{MasseeEPL2010} observed no superconducting gap on very disordered surfaces with corrugation $>$2\AA.

Therefore, according to the STM measurements of 6 different groups\cite{BoyerArxiv0806.4400, YinPRL2009, MasseeEPL2010, PanBAPS2009, TeaguePRL2011, ShanNatPhys2011}, Ba122 and Sr122 surfaces of nominally superconducting composition do display at least one superconducting gap of the expected magnitude, except in cases of extreme disorder.  The one puzzling exception is Zhang's failure to observe a superconducting gap on any of his superconducting samples with $\sqrt{2}\times\sqrt{2}$ surface structure.  It is also notable that none of the groups who have claimed As termination\cite{NascimentoPRL2009, NiestemskiArxiv0906.2761, ChuangScience2010}, have even looked at nominally superconducting samples.  Therefore, it is an open question as to whether an As-terminated surface, which is presumably polar and different in doping from the bulk, can support superconductivity.

We now close the section on STM of 122 materials, and look ahead to the sections on 11 and 111 materials, which will turn out to be much more conducive to surface studies, with reliable, well understood, and likely bulk-representative cleaved surfaces.  So we ask: is there anything more to be learned from STM on 122 materials?  Given the challenges associated with their surfaces, in comparison to the 11 and 111 materials which follow, it may be tempting to label the 122 materials a waste of time, and move on.  However, several new developments show that it is crucial that we understand the surfaces of the 122 materials, and when their properties can be trusted to be representative of the bulk.  Two of the most intriguing new Fe-based superconductors are of the 122 family: BaFe$_2$(As$_{1-x}$P$_x$)$_2$ and K$_{x}$Fe$_{2-y}$Se$_2$.

BaFe$_2$(As$_{1-x}$P$_x$)$_2$ becomes a superconductor without charge doping: isovalent chemical pressure from P disrupts the magnetic order and allows the material to become superconducting with $T_c$ as high as 31K\cite{KasaharaPRB2010}.  BaFe$_2$(As$_{1-x}$P$_x$)$_2$ is therefore an ideal material in which to isolate and study the causes and effects of the magnetic order, without the complications of changing carrier concentration, or disorder directly within the critical Fe plane. (This view has recently been questioned\cite{YeArxiv1105.5242}.) Furthermore, BaFe$_2$(As$_{1-x}$P$_x$)$_2$ is the highest-$T_c$ P-containing Fe-based superconductor, with $T_c$ comparable to hole-doped 122 materials, and higher than electron-doped 122 materials. This is especially surprising given the evidence for sign-changing gap nodes in the OP of this compound\cite{HashimotoPRB2010,NakaiPRB2010}. Such nodes are expected to significantly lower the $T_c$\cite{KurokiPRB2009}.

Additionally, BaFe$_2$(As$_{1-x}$P$_x$)$_2$ may actually be \textit{useful} to clarify the cleaved surface of the 122 compounds, because the P atoms are known to be smaller and therefore lower (closer to the Fe plane) than the As atoms\cite{RotterPRB2010}. If the cleaved surface is indeed the As$_{1-x}$P$_x$ layer, the expected concentration $x$ of P atoms should be clearly visible on the surface. In fact, Massee imaged the cleaved surface of BaFe$_2$(As$_{1-x}$P$_x$)$_2$ ($x$=0.32, $T_c$=31K) but he did not see any signature of $\sim 1/3$ of the surface atoms sitting lower than the others, as would be expected for a As$_{1-x}$P$_x$ top layer\cite{MasseeThesis2011}.

An even more intriguing newcomer is K$_{0.78}$Fe$_{1.7}$Se$_2$, with $T_c$=30K\cite{GuoPRB2010}, by far the highest in the Se subfamily of Fe-based superconductors.  More generally, this family (K,Tl,Rb,Cs)$_x$Fe$_{2-y}$Se$_2$ has been found to superconduct at ambient pressure with $T_c$ up to 40K\cite{FangEPL2011}.  The material is found to exist on the edge of an AF insulator phase which possibly originates from superlattice ordering of the Fe vacancies\cite{FangEPL2011}.  Remarkably, ARPES measurements show that K$_{0.8}$Fe$_{1.7}$Se$_2$ ($T_c$=30K) is strongly correlated (with band renormalization by a factor of 2.5 compared to LDA calculations) and lacks the hole-like FS at the $\Gamma$ point which is present in other high-$T_c$ Fe-based superconductors\cite{QianPRL2011}.  This compound therefore rules out nesting as the sole scenario to explain Fe-based high-$T_c$ superconductivity. This ARPES result is credible because the measured FS volume is found to satisfy Luttinger's theorem with the bulk carrier concentration; this is also very encouraging that STM studies of the surface this material will be fruitful windows into the bulk.

Very recently, K$_x$Fe$_{1-y}$Se$_2$ has been studied by STM by two groups.  First, Li \textit{et al} demonstrated MBE growth on a graphitized 6H-SiC(0001) substrate resulting in atomically flat (110)-oriented films of K$_x$Fe$_{1-y}$Se$_2$ \cite{LiNatPhys2011}. STM imaging of these films demonstrated phase separation into stoichiometric superconducting KFe$_2$Se$_2$, and insulating K$_x$Fe$_{1-y}$Se$_2$ in which the Fe vacancies tended to order into a $\sqrt{5}\times\sqrt{5}$ structure. In fact, a single isolated Fe vacancy in an otherwise stoichiometric region was shown to locally suppress superconductivity.

 Acomplementary STM study by Cai \textit{et al} of the (001) face of cleaved bulk K$_{0.73}$Fe$_{1.67}$Se$_2$ supported the idea of phase separation\cite{CaiArxiv1108.2798}. Cai \textit{et al} found that the cleaved surfaces of multiple samples were predominantly unstable and disordered due to a partial layer of mobile K. Occasionally, they found regions with an exposed, atomically ordered Se surface, and good superconducting spectra. However, despite extensive searching, they never found signatures of Fe vacancies in these regions, thus supporting Li's conclusion that the good superconducting regions are stoichiometric KFe$_2$Se$_2$.  Furthermore, in these `good' superconducting regions, Cai \textit{et al} found a weak $\sqrt{2}\times\sqrt{2}$ reconstruction, suggesting coexisting antiferromagnetism.

Both BaFe$_2$(As$_{1-x}$P$_x$)$_2$ and (K,Tl,Rb,Cs)$_x$Fe$_{2-y}$Se$_2$ are likely to provide insights into the mechanisms behind Fe-based high $T_c$ superconductivity.  As with any doped material, inhomogeneity must play a role, and therefore it is crucial to apply a local tool such as STM to resolve the mysteries of the DOS in these intriguing new compounds.  We hope that insights from the cleaved surfaces of Co- and K-doped Ba,Sr,Ca-122 can be extended to allow rapid characterization of surfaces, and deep insights about bulk DOS in these intriguing new materials.

\subsection{11 Materials}
\label{sec:Surface11}

The simplest family of Fe-based superconductors, the `11' materials, have actually been studied for decades, but their superconductivity was discovered only after the sudden attention on the iron-pnictides\cite{HsuPNAS2008}.  Because there is no intervening ionic layer between the Fe(Te,Se,S) planes, there is no polar surface catastrophe, so the 11's were the first of the Fe-based superconductors which were ideally suited for STM study.

Massee \textit{et al} first showed that the surfaces of Fe$_{1.07}$Te$_{0.55}$Se$_{0.45}$ appeared identical when cleaved at room temperature or low temperature, with every Se/Te atom visible, and no apparent surface reconstruction\cite{MasseePRB2009b}. Cold-cleaved LEED (17K) experiments showed only the 4\AA-periodic tetragonal spots with no extra reconstruction.  In this sample the 7\% interstitial Fe were seen as very bright spots exactly between four visible Se/Te atoms, suggesting that they sat beneath the surface but contributed significant extra local DOS.


Kato \textit{et al} studied Fe$_{1.05}$Te$_{0.85}$Se$_{0.15}$ with onset $T_c$=14K\cite{KatoPRB2009}.  They also saw the excess Fe as bright spots on the surface. Spectroscopy at $T$=4.2K showed $\Delta$=2.3 meV and $2\Delta/k_B T_c \sim$3.8, reasonable values which give no cause to suspect the surface is not representative of the bulk. The standard deviation $\sigma_{\Delta}$=0.23 meV gave $\sigma_{\Delta}/\overline{\Delta} \sim$ 10\%, in good agreement with $\sigma_{\Delta}/\overline{\Delta}$ in Ba(Fe$_{0.9}$Co$_{0.1}$)$_2$As$_2$ \cite{YinPRL2009}, and about half the relative gap variation reported in the cuprate superconductor Bi-2212\cite{McElroyPRL2005}.  Kato \textit{et al} did observe larger variations in the background DOS at higher energies, and speculated that this variation was due to the excess Fe.  They further speculated that this inhomogeneity at higher energies may be responsible for the apparent $\sim$10\% inhomogeneity in the superconducting $\Delta$, but they did not address this point directly using spectra at the same locations above and below $T_c$ for normalization.

Fridman \textit{et al} reported atomic resolution STM on FeTe$_{1-x}$Se$_x$, at nominal doping levels of $x$=0.3 and 0.5, with critical temperatures $T_c\sim$12K\cite{FridmanJPCS2010}.  The samples were warm cleaved in vacuum, then the spectra were measured at 300 mK.  Fridman confirmed the nanoscale inhomogeneity in the topography and spectroscopy.  Generally, the spectra displayed a broad, V-shaped background out to $\pm$10 mV, with a sharper gap structure $\Delta\sim$2-4 meV, which vanished above $T_c$.

He \textit{et al} reported STM on FeTe$_{1-x}$Se$_x$ with $x$=0 and 0.45, determined by energy dispersive x-ray spectroscopy ($T_c$=14K for the $x$=0.45 sample)\cite{HePRB2011}.  The samples were cleaved and imaged at $T$=80K. They claimed a small surface corrugation of 8 pm (but this would be dependent on the resolution and work function of the tip). They identified the ``bright'' atoms as Te and the ``dark'' ones as Se, and noted that the heights were bimodal, separated by $\Delta z = 44.7 \pm 12.0$pm, and that the fraction of bright and dark atoms matched the known bulk Te/Se composition ratio.  Both filled state and empty state topographic images showed the same $\Delta z$, supporting the conclusion that the apparent height difference was a real geometric effect, rather than a difference in DOS.  This STM-measured $\Delta z$ was almost a factor of two larger than the value measured by x-ray scattering\cite{TegelSSC2010}.  The larger value observed by STM could be a systematic experimental effect or a surface relaxation. He \textit{et al} observed nanoscale clustering of the Se and Te, but noted that the spectra were homogeneous (indistinguishable between Te and Se) out to $\pm$100 meV.  They speculated that the length scale of chemical inhomogeneity was likely shorter than the superconducting coherence length.  They further speculated that the apparent chemical inhomogeneity of the surface Se/Te may have been balanced by the opposite species on the opposite side of the Fe plane.


The previous experiments all studied the cleaved surfaces of bulk Fe$_{1+y}$Te$_{1-x}$Se$_x$. Better control over the crystal composition can be obtained in films grown by molecular beam epitaxy (MBE)\cite{SongPRB2011}. Song \textit{et al} grew and performed STM imaging on FeSe$_{1+z}$ films, demonstrating superconductivity for $z<2.5\%$. Sparse excess Se appeared as dimers and eventually, on increasing concentration, merged into an ordered $\sqrt{5}\times\sqrt{5}$ surface structure with a 0.5 eV insulating gap. By optimizing the substrate temperature during growth, Song was able to reduce the number of defects to fewer than 1 in 70,000 atoms, and reliably achieve superconductivity. A study of superconducting gap vs. film thickness showed that the $T_c$ scales inversely with the film thickness as $T_c(d) = T_{c0}(1 - d_c/d)$ where $T_{c0}$ is the bulk $T_c$ of 9.3K, and $d_c$ is the minimum thickness for superconductivity, estimated as 7\AA, approximately two Se-Fe$_2$-Se layers.

These experiments addressed the structure and chemistry of the surface of FeTe$_{1-x}$Se$_x$. Two of them observed a superconducting gap of $\sim$2-4 meV, but without sufficient energy resolution to make any concrete statements about pairing symmetry.  Recently, two different studies by Hanaguri \textit{et al}\cite{HanaguriScience2010} and Song \textit{et al}\cite{SongScience2011} addressed the superconducting pairing symmetry in the FeTe$_{1-x}$Se$_x$ family, to be discussed in section~\ref{sec:OP}.

\subsection{\label{sec:Surface111}111 Materials}

The `111' family of Fe-based superconductors is also ideally suited for STM study.  LiFeAs cleaves well with Li splitting evenly between cleaved surfaces.  Calculations show no surface states in LiFeAs\cite{LankauPRB2010}. The first (unpublished) work on the 111 family shows an unreconstructed, atomically resolved surface, on which 6 different types of impurity states, as well as magnetic vortices have been imaged\cite{HanaguriAspen2011}.

Very recent QPI imaging experiments on clean LiFeAs surfaces purport to demonstrated $p$-wave superconductivity\cite{HankeArxiv1106.4217}, or to resolve the in-plane anisotropy of several coexisting gaps\cite{DavisBCcolloquium2011}.

\subsection{\label{secSurfOther}Other distant cousins}

Other Fe-based superconductors include Sr$_2$VO$_3$FeAs (21311) with $T_c$=37.2 K\cite{ZhuPRB2009}, (Sr$_4$Sc$_2$O$_6$)(Fe$_2$P$_2$) (42622) with $T_c$=17K\cite{OginoSST2009}, and Sr$_3$Sc$_2$Fe$_2$As$_2$O$_5$ (32225) with a possible $T_c\sim$20K \cite{ChenSST2009}.  In addition, there are several series with increasing Fe$_2$As$_2$ interlayer separation: (Fe$_2$As$_2$)(Sr$_4$(Sc,Ti)$_3$O$_8$), (Fe$_2$As$_2$)(Ba$_4$Sc$_3$O$_{7.5}$), and (Fe$_2$As$_2$)(Ba$_3$Sc$_2$O$_5$) with $T_c$ up to 28K\cite{KawaguchiAPEX2010}; also (Fe$_2$As$_2$)(Ca$_{n+1}$(Sc,Ti)$_n$O$_y$ with $n$=3,4,5 and with $T_c$ up to 42K for $n$=5 \cite{OginoAPL2010}.  Many of these appear to have mirror planes which will be suitable for STM study, but so far they have not been successfully grown as large single crystals.  For example, the largest so far in the Sr$_2$VO$_3$FeAs family are ~300 $\mu$m~\cite{WenPrivate2011}, smaller than is convenient for STM study.

\section{\label{sec:OP}Superconducting Order Parameter}

A superconducting OP, which describes the symmetry of the pairing state, has both magnitude and phase.  We start this section with an overview of theory and early experiments on the OP in Fe-based superconductors.  We then explain the STM determination of the OP, starting with phase, which may be more relevant to the pairing mechanism, and proceeding to magnitude.

Within the first few weeks of Hosono's discovery, every imaginable superconducting OP was proposed.  Because all five Fe $d$ bands may cross the Fermi level, the number of possible OPs which can live on these five FSs is large. Some candidates were nicely summarized pictorially by Hicks \textit{et al}\cite{HicksArxiv0807.0467v1}.  Because of the presence of Fe and suspected magnetic ground state of the parents, and the early LDA calculations showing nesting between hole and electron FSs at the AF wavevector\cite{SinghPRL2008}, two early papers\cite{MazinPRL2008,KurokiPRL2008} convincingly argued for an $s\pm$ OP (sketched in figure~\ref{fig:Hanaguri-pnictide}a).  A spin-mediated pairing mechanism would require a sign flip between the phases of the paired carriers, forcing a plus sign on one nested FS and a minus sign on the other.  In this scenario, there would be no need for a sign change on a single FS, i.e.\ no need for gap nodes.

To cement a picture of spin-mediated $s\pm$ pairing, three pieces of evidence are necessary: (1) demonstration of the absence of nodes on each FS; (2) demonstration of a phase flip between FSs; and (3) demonstration of a spin resonance at the nesting wavevector connecting the two opposite-signed FSs.

Early ARPES studies did not show nodes in NdFeAsO$_{0.9}$F$_{0.1}$\cite{KondoPRL2008}, Ba$_{0.6}$K$_{0.4}$Fe$_2$As$_2$\cite{DingEPL2008} (see figure~\ref{fig:DingARPES}a), or Ba(Fe$_{1-x}$Co$_x$)$_2$As$_2$\cite{TerashimaPNAS2009} (see figure \ref{fig:DingARPES}, b and c). Many subsequent ARPES experiments have also failed to find evidence of nodes.  But ARPES experiments do not provide phase information.

Several phase-sensitive tunneling experiments have been performed. First, a search for half-flux-quantum vortices trapped in the native grain boundaries of polycrystalline NdFeAsO$_{0.94}$F$_{0.06}$ ($T_c$=48K) detected none, ruling out $d$-wave order in this compound\cite{HicksJPSJ2008}. However, a detection of half-quantum magnetic flux in a loop between NdFeAsO$_{0.88}$F$_{0.12}$ and conventional superconducting niobium\cite{ChenNatPhys2010} was explained by frustrated Josephson coupling due to a sign change in the OP\cite{BergPRL2011}.  In combination, these two experiments strongly suggested $s\pm$ symmetry in NdFeAsO$_{1-x}$F$_x$, but did not give proof for the pairing mechanism. A neutron scattering experiment on another 1111 FeAs superconductor, LaFeAsO$_{1-x}$F$_x$, detected a resonant spin fluctuation that peaked sharply in the superconducting state, at the nesting wave vector\cite{WakimotoJPSJ2010}.  (The resonant spin fluctuation had already been seen in Ba$_{0.6}$K$_{0.4}$Fe$_2$As$_2$.\cite{ChristiansonNature2008})

If we are willing to combine evidence from several compounds, a picture arises of a spin-mediated $s\pm$ OP in 1111 and 122 materials.  Nonetheless, other calculations have shown that various OPs with nodes on the FSs were not too different in energy, and may be obtained under some circumstances\cite{DaghoferPRL2008,KurokiPRB2009,ZhaiPRB2009}. Furthermore, there is a long list of experiments which showed evidence for nodal OPs in various Fe-based superconductors\cite{MuChinPhysLett2008,RenArxiv0804.1726, ShanEPL2008,LuetkensPRL2008,GrafePRL2008,MukudaJPSJ2008, KotegawaJPSJ2008,CheckelskyArxiv0811.4668,GordonPRL2009}.  Most of these experiments were sensitive only to the presence or absence of low-lying quasiparticles, thus impurities could give the same signature as a nodal OP.  In the early days, as sample quality was still rapidly improving, it was commonly believed that the elimination of impurities would eliminate the appearance of nodal superconductivity.

However, the 11 family refused to fit the mold.  In particular, the parent FeTe compound did not show the same spin ordering at the nesting wavevector, but rather spin ordering at an alternative non-nesting wavevector\cite{LiPRB2009}.  Therefore, the idea of spin-mediated $s\pm$ pairing in this material was called into question. As described in section~\ref{sec:QPI}, STM is an ideal tool to take on this question.  STM can simultaneously look for nodes (via the presence of low-lying quasiparticles in $dI/dV$ spectroscopy) and can perform a phase-sensitive test of the OP via QPI imaging. Several theoretical approaches predicted discriminating QPI patterns for expected OPs in the Fe-based superconductors\cite{WangEPL2009, ZhangPRB2009, PlamadealaPRB2010}.

\subsection{\label{sec:gap-symmetry}Gap Symmetry}

\begin{figure}[tbh]
\begin{center}
  {\includegraphics[width=1.0\columnwidth,clip]{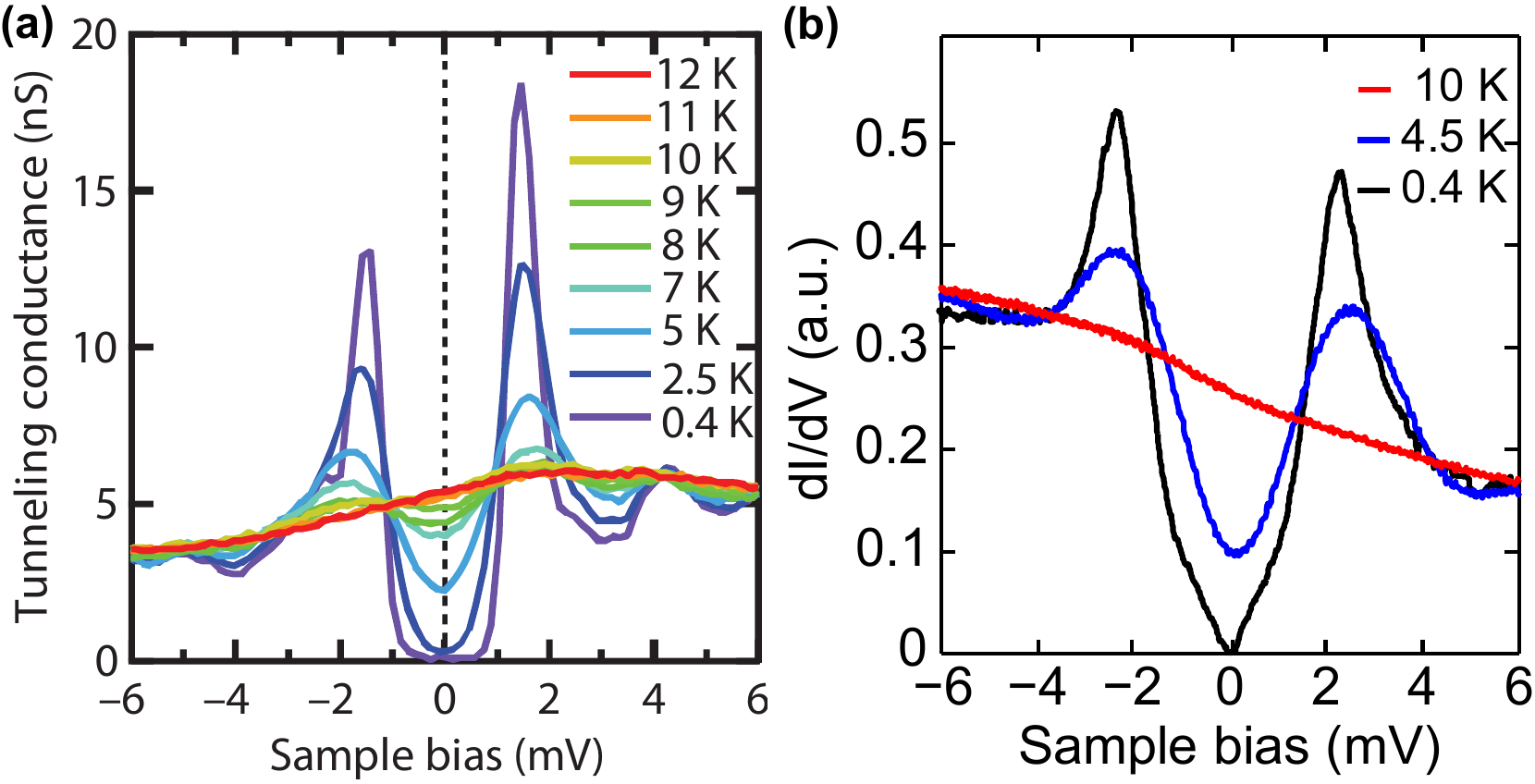}}
  \caption[DOS in FeTe$_{1-x}$Se$_x$.]
    {\label{fig:FeSe-spectra}
    (a) $dI/dV$ measured on FeTe$_{1-x}$Se$_x$ with $x\sim$0.4 and bulk $T_c$=14.5K (and apparent local $T_c\sim$11K). The lowest temperature $dI/dV$ is clearly fully gapped. (b) $dI/dV$ as a function of $T$ for FeSe with $T_c\sim$8K. Although topographic imaging shows fewer than 1 in 70,000 defects, the low $T$ $dI/dV$ is still V-shaped, indicating the presence of low energy quasiparticles. In both Hanaguri's and Song's spectroscopy, the bias modulation amplitude was set to 0.1 mV$_{\mathrm{rms}}$. }
\end{center}
\end{figure}

\noindent \textbf{\textit{$\bm{s\pm}$ order parameter in FeTe$_{0.6}$Se$_{0.4}$}}
Hanaguri \textit{et al} performed STM experiments on FeTe$_{1-x}$Se$_{x}$ with $x\sim0.4$ and bulk $T_c$ in the range 13K to 14.5K\cite{HanaguriScience2010}.  He first observed that the low temperature ($T$=400 mK) spectra were fully gapped, as shown in figure~\ref{fig:FeSe-spectra}a.  These fully-gapped spectra provide strong evidence against nodes in the OP.

\begin{figure}[tbh]
\begin{center}
  {\includegraphics[width=0.9\columnwidth,clip]{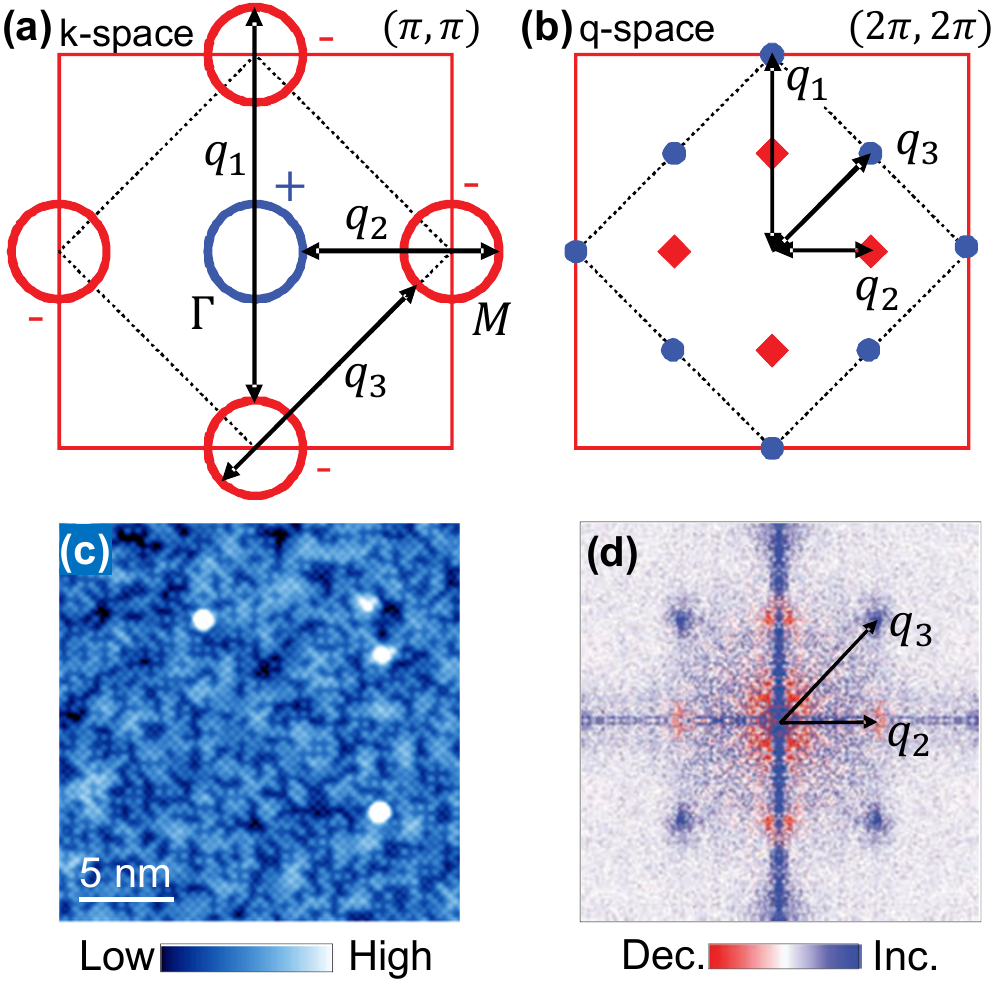}}
  \caption[QPI cuprate OP.]
    {\label{fig:Hanaguri-pnictide}
    (a) Schematic FSs of FeTe$_{0.6}$Se$_{0.4}$.  Red square denotes the unfolded BZ (one Fe per unit cell), while the dashed square at 45$^{\circ}$ denotes the tetragonal BZ. The hole FS at the $\Gamma$ point (blue) and the electron FS at the $M$ point (red) are thought to have different signs in the $s\pm$ pairing scenario. (b) QPI will be dominated by the scattering vectors $q_1$, $q_2$, and $q_3$ as shown. Blue circles represent sign-preserving scattering, whereas red diamonds represent sign-reversing scattering. (c) Topography, showing clean atomic resolution (setup: $V_{\mathrm{sample}}=-20\,\mathrm{mV}; I_{\mathrm{set}}=100\,\mathrm{pA}$).  Four bright spots correspond to extra Fe atoms. (d) Magnetic-field induced change in QPI intensities, $Z(\vec{q},E=1.0\mathrm{meV},B=10\mathrm{T})- Z(\vec{q},E=1.0\mathrm{meV},B=0\mathrm{T})$, supports the $s\pm$ scenario. Figure from Ref. \onlinecite{HanaguriScience2010}.}
\end{center}
\end{figure}

\begin{figure}[tbh]
\begin{center}
  {\includegraphics[width=0.95\columnwidth,clip]{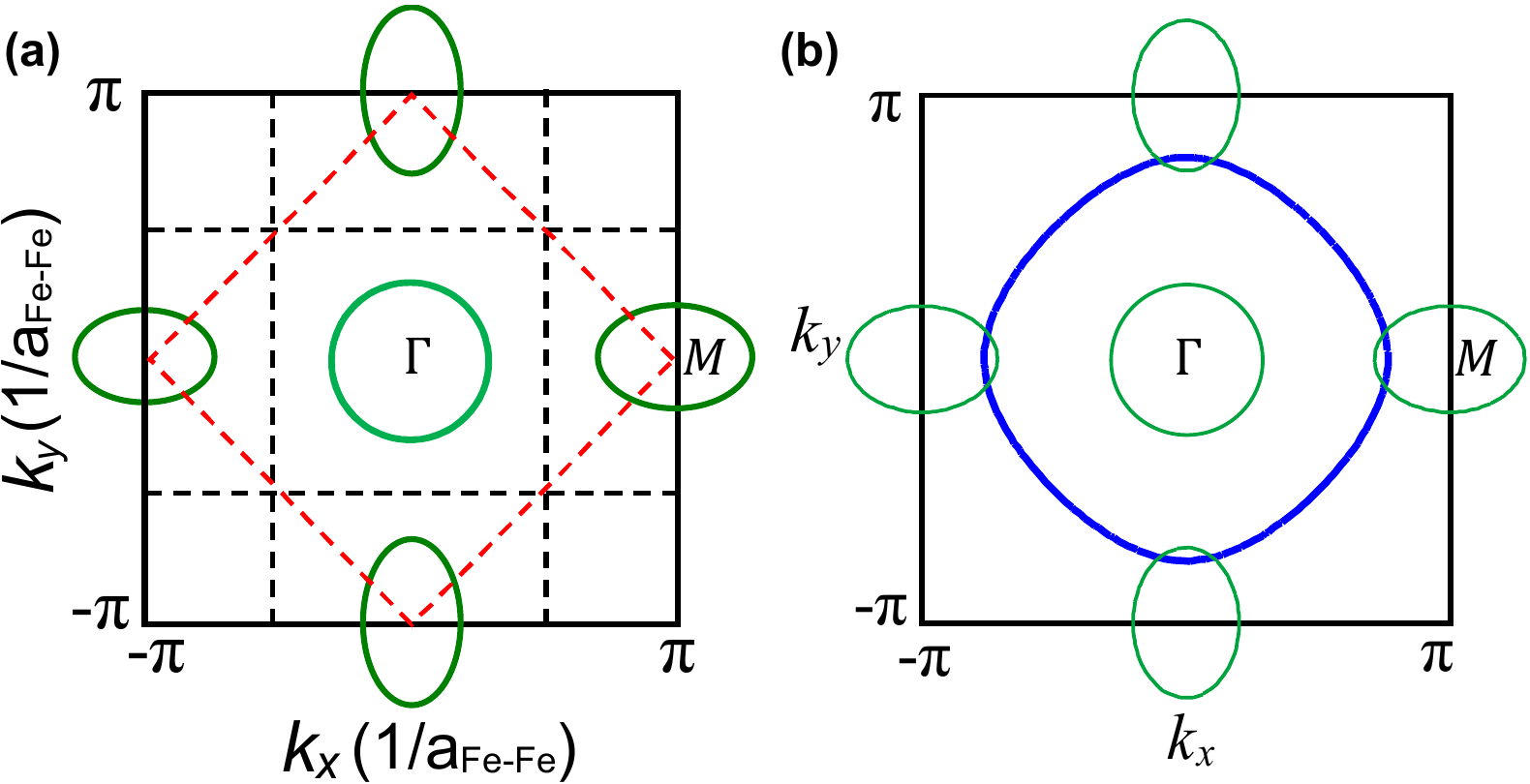}}
  \caption[FeSe BZ.]
    {\label{fig:FeSe-BZ}
    Schematic BZ of FeSe, in the unfolded scheme, with one Fe per unit cell. (a) Hole-like FS around the $\Gamma$ point, and electron-like FSs around the $M$ points are shown in green. Nodes of the $\cos k_x \cos k_y$ OP are shown in black. Nodes of the $\cos k_x + \cos k_y$ OP are shown in red. (b) Model showing the nodes of the combined $\Delta_1 \cos k_x \cos k_y + \Delta_2 (\cos k_x + \cos k_y)$ OPs with $\Delta_1/\Delta_2$=0.35. Increasing $\Delta_1/\Delta_2$ causes the blue line to move inwards, removing the intersections of the nodes with the $M$-centered electron FSs.}
\end{center}
\end{figure}

We address two possible arguments against the nodeless gap: (1) STM sensitivity to a quasiparticle state $\psi_k$ is exponentially suppressed with increasing in-plane momentum $\vec{k}$\cite{TersoffPRL1983}. One might therefore argue that STM lacks the sensitivity to detect quasiparticles arising from gap nodes far from the $\Gamma$ point, e.g. on $M$-centered FSs. Two counter-arguments follow. (a) The FT-STS images (e.g. figure~\ref{fig:Hanaguri-pnictide}d) show broad QPI peaks at $q_2$ and $q_3$, near the corners of the BZ, demonstrating the sensitivity of this STM and this particular tip to these large-$\vec{k}$ quasiparticles.  (b) STM spectroscopy at the same temperature, on a related material that is too clean to support impurity-induced quasiparticles, shows a V-shaped spectrum which could arise only from nodal quasiparticles\cite{SongScience2011}. These gap nodes are likely to live on the $M$-centered FSs (see figure~\ref{fig:FeSe-BZ}b), again suggesting that STM is generally sensitive to nodal quasiparticles even far from the $\Gamma$ point. (2) Fridman \textit{et al} found a V-shaped gap for FeTe$_{1-x}$Se$_x$ crystals of a similar composition, with nominal $x \sim 0.3$ (measured at at $T$=300 mK)\cite{FridmanJPCS2010}. However, Fridman's samples were cleaved at room temperature, allowing the possibility of significant surface contamination. In fact, their images did not achieve the same clean atomic resolution as shown by Hanaguri in figure~\ref{fig:Hanaguri-pnictide}c. It therefore seems likely that Fridman's V-shaped gap arose as a consequence of scattering from surface contamination.

Although Hanaguri's experiment provides strong evidence for a nodeless gap, it does not rule out the possibility of gap anisotropy.  Angle-resolved specific heat (ARSH) measurements on FeTe$_{0.55}$Se$_{0.45}$ suggest deep gap minima or nodes along the Fe-Fe direction ($\Gamma$-$M$ direction in the BZ of figure \ref{fig:unit-cell}d)\cite{ZengNatComm2010}. In comparison with Hanaguri's tunneling data, Zeng \textit{et al} speculated that the gap minima or nodes detected by ARSH must live on the inner electron pocket ($M$ pocket), i.e. at maximal $k$ values where the tunneling matrix element would be most severely suppressed, thus hiding the low energy quasiparticles from STM.  However, Hanaguri's QPI data suggests that quasiparticles are highly visible to STM even when located at the $M$ point, so reconciliation of ARSH with STM does not require that the nodes or gap minima live near the $M$ point.  Indeed, a closer look at Hanaguri's $T=0.4$K spectrum shows energy difference $\sim$0.75 meV between the lowest-energy quasiparticles, and the maximum of the coherence peak.  This difference greatly exceeds the expected thermal broadening of $\sim4k_B T = 0.14$ meV, suggesting that the observed broadening is due instead to gap anisotropy.  Therefore, Hanaguri's spectra support the likelihood of deep gap minima, but strongly argue against gap nodes.

In the $s\pm$ OP model, the electron and hole FSs have opposite signs, as sketched in figure~\ref{fig:Hanaguri-pnictide}a. The FSs are nested, with three possible nesting vectors sketched in figure~\ref{fig:Hanaguri-pnictide}b. Two of these nesting vectors, $q_1$ and $q_3$, are sign-preserving in the $s\pm$ model, while the third, $q_2$, is sign-changing. Upon application of a magnetic field, pinned vortices are expected to become a new source of sign-changing scattering\cite{MaltsevaPRB2009}. Furthermore, magnetic impurities, and non-magnetic resonant scatterers are both expected to enhance sign-changing scattering and suppress sign-preserving scattering upon application of a magnetic field\cite{SykoraPRB2011}.  In the $s\pm$ pairing model, all three mechanisms are therefore expected to enhance the coherence factors for $q_1$ and $q_3$ nesting, at the expense of $q_2$ nesting. This expectation is supported by the data shown in figure~\ref{fig:Hanaguri-pnictide}d.
The sign-changing $s\pm$ OP is therefore strongly supported by STM in FeTe$_{0.6}$Se$_{0.4}$.

A concern was raised by Mazin \textit{et al}\cite{MazinArxiv1007.0047}, who pointed out that all three $q$-vectors discussed by Hanaguri seem too sharp to be ascribed to QPI, and in fact each corresponds to long range structural or magnetic order in the crystal.  In particular (referring to the crystal lattice parameters defined in figures~\ref{fig:unit-cell}a and b), $q_1=2\pi/a_{\mathrm{Fe-Fe}}$, $q_2 = 2\pi/a_O$, and $q_3=2\pi/a_T$ would each correspond to long range crystalline order and give rise to sharp $q$-space peaks. In contrast, elastic scattering between each of the pairs of points on the FSs shown in figure~\ref{fig:Hanaguri-pnictide}a would result in broad $q$-space peaks, with diameters matching the diameters of the FSs, approximately 15-20\% of the BZ. Mazin suggested that the observed intensity vs. magnetic field trends for $q_2$ and $q_3$ could instead be explained by a field-induced suppression of both the superconductivity and the spin density wave (SDW) and its assumed concomitant surface reconstruction, which would enhance the structural Bragg peak at $q_3$ within the superconducting gap energy and suppress the SDW-induced peak at $q_2$.

Hanaguri countered\cite{HanaguriArxiv1007.0307} with linecuts through the $q_2$ and $q_3$ peaks which showed that each peak was made up of two components with distinct energy and field dependence. Each peak could be separated into a central sharp Bragg peak, and a broader QPI peak with the expected width $\sim20\%$ of the BZ. Hanaguri's $s\pm$ gap symmetry conclusions were drawn only from the broader QPI peaks. Hanaguri argued that the coexistence of sharper Bragg peaks does not negate these conclusions.\\

\noindent\textbf{\textit{Nodal order parameter in FeSe}}
There remain many credible claims of a nodal gap in Fe-based superconductors\cite{MuChinPhysLett2008,RenArxiv0804.1726, ShanEPL2008,LuetkensPRL2008,GrafePRL2008,MukudaJPSJ2008, KotegawaJPSJ2008,CheckelskyArxiv0811.4668,GordonPRL2009}.  However, from bulk studies, it can be hard to conclusively rule out impurity scattering as the source of apparent nodal quasiparticles.  Most recently, Song \textit{et al} performed a beautiful set of experiments on the single purest Fe-based material studied to date: MBE-grown FeSe, with fewer than one defect per 70,000 Se atoms (i.e. no visible impurities in a 100 nm square field of view)\cite{SongScience2011}.  Clearly, there should be no impurity-induced low-lying quasiparticles in a sample this clean.  Nonetheless, at the lowest temperatures (down to $T$=0.4K), Song found a V-shaped gap (figure~\ref{fig:FeSe-spectra}b), which provided clear evidence of nodal superconductivity.

How could the same 11 material have two different OPs at different dopings?  In fact, both $\cos k_x \cos k_y$ and $\cos k_x + \cos k_y$ pairing functions are consistent with the same point group symmetry, so they may naturally coexist\cite{SongScience2011}.  The gap function is given by
\begin{equation}
\Delta(\vec{k}) = \Delta_1 \cos k_x \cos k_y + \Delta_2 (\cos k_x + \cos k_y).
\end{equation}
Figure~\ref{fig:FeSe-BZ}a shows the nodal lines of these two terms in the unfolded BZ.  Depending on the relative magnitudes of $\Delta_1$ and $\Delta_2$, the nodal lines of the full function $\Delta(\vec{k})$ may or may not pass through the electron FSs (figure \ref{fig:FeSe-BZ}b). It has been shown by functional renormalization group calculations that when the interlayer coupling $t_z$ is weaker, the nodal lines are more likely to intersect with the electron FSs, giving a $d$-wave OP\cite{KurokiPRB2009,ZhaiPRB2009}.

In FeTe$_{1-x}$Se$_x$, the Se atoms lie closer to the Fe plane than the Te atoms\cite{TegelSSC2010,HePRB2011}, which may lead to reduced interlayer coupling $t_z$. It is therefore consistent that the FeSe studied by Song would have lower interlayer coupling than the FeTe$_{0.6}$Se$_{0.4}$ studied by Hanaguri, and would have greater tendency to $d$-wave order.  (It is also consistent with the finding that LaFePO, where P sits close to the Fe plane, has nodes\cite{FletcherPRL2009,HicksPRL2009}, whereas 1111 materials with As, which sits farther from the Fe plane, seem not to have nodes\cite{KondoPRL2008}).

Therefore, low temperature STM on clean samples has proven to be a sensitive test of both the presence of nodes, and the phase of the OP.  It is important to note again that STM is presumed to be more sensitive to the near-$\Gamma$ states\cite{LeePRL2009}. Low-$T$ spectroscopy on clean FeSe showed a V-shaped gap spectrum, although the gap nodes are thought to be on the $M$-centered FS. The fully-gapped spectra observed by STM at low $T$ are therefore likely truly indicative of no nodes on any FS\cite{HanaguriScience2010}. More generally, Song's work strongly suggests that STM is sensitive to the presence of nodes on any FS. \\

\noindent \textbf{\textit{$\bm{p}$-wave order parameter in LiFeAs}}
Recently, another QPI imaging study argued for $p$-wave superconducting gap symmetry in LiFeAs\cite{HankeArxiv1106.4217}. LiFeAs with $T_c$=18K was cleaved and imaged at low temperature, displaying an atomically resolved field of view with sparse impurities around which were centered clear interference patterns. The Fourier transforms of the interference patterns strongly resembled the band structure seen by ARPES\cite{BorisenkoPRL2010}, suggesting that the JDOS was dominated by a van Hove singularity at the $\Gamma$ point. The authors computed the expected $q$-space QPI patterns for $s\pm$, $d$, $p$, and $s++$ OP symmetries. From a qualitative comparison, they suggested that $p$-wave symmetry was the best match, but they did not rule out more exotic symmetries such as $s+id$.  However, their images showed some distortions, and were not consistent with other STM reports on LiFeAs\cite{HanaguriAspen2011,DavisBCcolloquium2011}.

\subsection{\label{sec:gap-magnitude}Gap magnitude}

The magnitude of the superconducting gap can be measured by bulk techniques such as heat capacity and electromagnetic absorption. The momentum-dependence of the gap can be directly measured by ARPES, which has so far been primarily limited to temperatures above 4.2K (with rare exceptions\cite{BorisenkoPRL2010}). ARPES measures only filled states, and cannot be used in a magnetic field. More precise energy resolution, as well as access to empty states and field-induced states, can be obtained through low $T$ tunneling measurements\cite{DynesPRL1978}. If a clean surface can be obtained, STS can determine the gap with atomic spatial resolution, and with energy resolution limited only by the measurement temperature, which has been pushed down to 10s of milliKelvin\cite{MarzRSI2010}.  STM energy resolution is $\sim4k_B T$, where the broadening stems from a smearing of DOS by $\sim k_B T$ on either side of the Fermi level, for both the sample and tip states. On some Fe-based superconductors, STS has been performed at temperatures down to 300 mK\cite{FridmanJPCS2010} or 400 mK\cite{HanaguriScience2010,SongScience2011}. Using QPI imaging\cite{McElroyNature2003} or clever fitting techniques\cite{PushpScience2009}, the $k$-dependence of the gap magnitude can also be quantified by STM, although applications of these techniques to the Fe-based superconductors have not yet been published\cite{DavisBCcolloquium2011}.

A typical superconducting $dI/dV$ spectrum measured by STM has one or more pairs of roughly particle-hole symmetric peaks in the DOS, which may be interpreted as the superconducting coherence peaks.  The simplest measure of the superconducting gap $\Delta$ is $\frac{1}{2}$ the energy between the two peaks. This technique has been used with success to analyze large $dI/dV$ datasets to generate `gapmaps' which show the local variation of the OP magnitude in cuprate superconductors\cite{LangNature2002} and in Fe-based superconductors\cite{YinPRL2009,MasseePRB2009a} (see figure~\ref{fig:Gapmaps}).

In some cases, the coherence peaks may be fit very accurately to an expected functional form, originally proposed by Dynes \textit{et al}\cite{DynesPRL1978}.  Dynes noted that the superconducting DOS $\rho_s(E)=|E|/\sqrt{E^2-\Delta^2}$ could be generalized to take into account a finite quasiparticle lifetime by writing
\begin{equation}
\label{eq:Dynes-rho}
\rho_s(E,\Gamma)=\mathrm{Re}\left(\frac{E-i\Gamma}{\sqrt{(E-i\Gamma)^2-\Delta^2}}\right),
\end{equation}
\noindent where $\Gamma$ is the inverse quasiparticle lifetime. Further generalizing to an angle-dependent gap $\Delta(\theta)$, plugging into equation~\ref{eq:Itotal}, and differentiating gives
\begin{eqnarray}\label{eq:Dynes-dIdV}
\lefteqn{ \!\!\!\!\!\!\!\!\!\!\!\!\!\!\!\!\!\!\!\!\!\!\!\! \frac{dI}{dV}\propto \int_{-\infty}^{\infty} dE \int_{0}^{2\pi} d\theta \left. \frac{df(\varepsilon)}{d\varepsilon} \right|_{\varepsilon=E-eV} } \nonumber \\
 & & \cdot \mathrm{Re} \left(\frac{E-i\Gamma}{\sqrt{(E-i\Gamma)^2-\Delta(\theta)^2}}\right) .
\end{eqnarray}

\begin{figure}[tb]
\begin{center}
  {\includegraphics[width=1.0\columnwidth,clip]{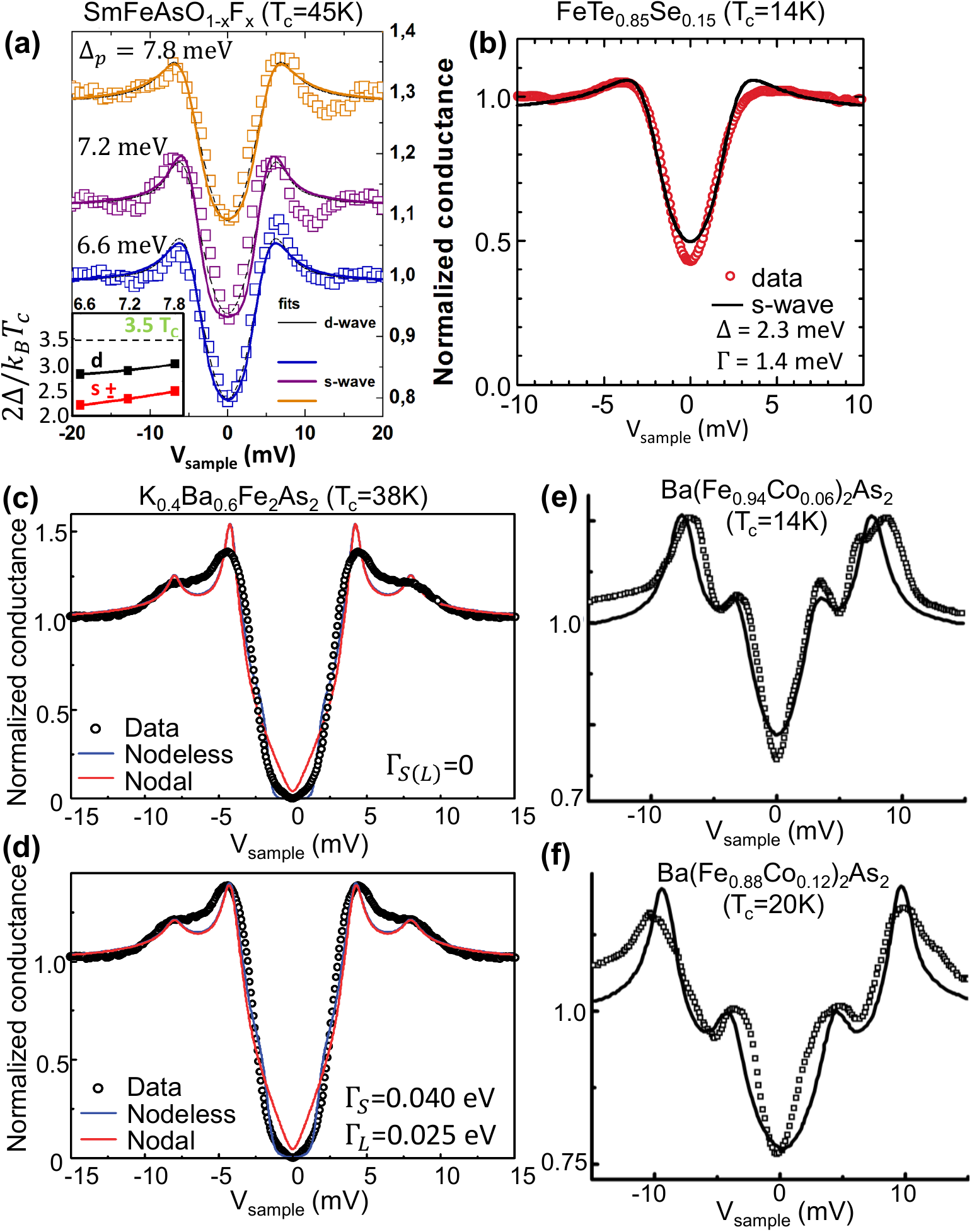}}
  \caption[Dynes fits of superconducting gaps.]
    {\label{fig:Dynes-gap-fit}
    Dynes fits to single- and double-gapped spectra. (a) $d$-wave and $s$-wave Dynes fits to 3 representative spectra from SmFeAsO$_{1-x}$F$_x$ with nominal $x$=0.2 and measured $T_c$=45K\cite{FasanoPRL2010}. Both $d$-wave and $s$-wave curves are indistinguishable by eye, yet result in very different values for the superconducting gap $\Delta$. Neither are visually satisfying fits, and both result in reduced gaps $2\Delta/k_B T_c$ more than 25\% too low for a BCS superconductor. (b) $s$-wave fit to FeTe$_{0.85}$Se$_{0.15}$\cite{KatoPRB2009} also fails the `chi-by-eye' test. (c) Fit using an extension of the Dynes form to two gaps, on Ba(Fe$_{1-x}$Co$_x$)$_2$As$_2$ with $x$=0.06 and $T_c$=14K\cite{TeaguePRL2011}. (d) Fit using an extension of the Dynes form to two gaps, on Ba(Fe$_{1-x}$Co$_x$)$_2$As$_2$ with $x$=0.12 and $T_c$=20K\cite{TeaguePRL2011}. (e) Fit using an extension of the Dynes form to two gaps, on Ba$_{1-x}$K$_x$Fe$_2$As$_2$ with $x$=0.4 and $T_c$=38K\cite{ShanNatPhys2011}.  Without introducing a finite Dynes broadening parameter $\Gamma$, neither nodal nor nodeless gaps produce good fits. (f) Fit to the same sample in (e), after introducing a finite Dynes broadening parameter $\Gamma$. Still, neither nodal nor nodeless gap fits pass the common sense `chi-by-eye' test. }
\end{center}
\end{figure}

However, in most cases, fits to equation~\ref{eq:Dynes-dIdV} fail the common sense `chi-by-eye' test (see figure \ref{fig:Dynes-gap-fit}).  Possible causes for this failure include unaccounted-for energy-dependence of the background DOS, energy or momentum variation of the tunneling matrix element, or energy-dependent limitations to the quasiparticle lifetime\cite{AlldredgeNatPhys2008}. The energies of the coherence peaks seem by eye to be reasonably well found by the various fits, but it must be noted that changes to the inverse quasiparticle lifetime $\Gamma$ (also sometimes called the `Dynes broadening parameter'\cite{DynesPRL1978}) can dramatically shift the fit value of $\Delta$. Additionally, thermal broadening causes the coherence peaks to appear to shift to larger energy, even as the gap $\Delta$ is actually closing upon increased temperature\cite{PanRSI1999}.  Therefore, in complex materials with multiple gaps and/or non-trivial momentum dependence, it is not clear whether careful fitting actually results in more accurate gap determination than the simple method of extracting half the distance between apparent coherence peak maxima (see figure~\ref{fig:Dynes-gap-fit}, a and b).

Fasano \textit{et al} studied the `as-grown' (air-exposed and uncleaved) surface of a $\sim$100 $\mu$m single crystal of SmFeAsO$_{1-x}$F$_x$ with nominal $x$=0.2 and measured $T_c$=45K\cite{FasanoPRL2010} (figure \ref{fig:pseudogap-DOS}a).  Spectra, acquired from base temperature 4.2K, showed a $V$-shaped gap that disappeared around the bulk $T_c$. The half-distance between `peaks' (or kinks) in the spectra, gave an average gap value $\overline{\Delta_p}$=7 meV, with standard deviation $\sim$6\% of the mean value. The reduced gap $2\Delta_p/k_B T_c \sim 3.6$ was in good agreement with point contact spectroscopy on the same compound. However, after averaging (binned by $\Delta_p$), and normalization (division by a smooth polynomial fit to the high energy part of the spectrum above the peaks), the spectra were fit to the Dynes equation~\ref{eq:Dynes-dIdV}, with both $s$ and $d$ wave functional forms for $\Delta(\theta)$, yielding $\Delta_{s}$ values from 4.2 to 4.8 meV ($2\Delta_{s}/k_B T_c \sim 2.2$), and $\Delta_{d}$ values from 5.1 to 5.8 meV ($2\Delta_d/k_B T_c \sim 3$), as shown in figure \ref{fig:Dynes-gap-fit}a.  Neither fit gave a reduced gap value close to the expected BCS values of $2\Delta_s/k_B T_c=3.5$\cite{BardeenPhysRev1957} or $2\Delta_d/k_B T_c=4.3$\cite{WonPRB1994}, nor did either fitting method give consistent values of the inverse quasiparticle lifetime $\Gamma$.  Because the gap values from $d$-wave fitting were only $\sim$22\% smaller than expected, in comparison to the $s$-wave fits which were $\sim$36\% smaller than expected, the authors concluded that their data supported $d$-wave pairing.

Fasano \textit{et al} also studied a `dip-hump' feature in the spectra, which occurred just outside the superconducting gap, and was interpreted as a signature of a collective mode. The energy $E_2$ of this feature had an even larger spread around its mean value of $\sim$ 15 to 20 meV.  Unlike the inner superconducting gap, the outer dip-hump feature was not symmetric about the Fermi level.  On the positive energy side (empty states), the energy of the presumed collective mode ($\Omega=E_{\mathrm{dip}}-\Delta_p$) was anti-correlated with the presumed pairing strength ($\Delta_p$).  Note that $\Omega$ ranged from 2 to 8 meV in Fasano's sample, giving $\Omega/k_B T_c \sim 0.5 - 2$.  This low energy of the purported `spin resonance' $\Omega$ in Fasano's work, may be cause for some concern, as the energy of the magnetic resonance has been shown to scale with $T_c$ in cuprates as $\Omega_r \sim 5k_B T_c$\cite{HufnerROPP2008}, in Ba122 as $\Omega_r = 4.3 k_B T_c$\cite{ChristiansonNature2008}, and in LaFeAsO$_{1-x}$F$_x$ with near-optimal $x=0.082$ and $T_c$=29 K as $\Omega_r = 4.4 k_B T_c$\cite{WakimotoJPSJ2010}. In summary, all of the energy scales reported by Fasano ($\Delta_{s\pm}$, $\Delta_d$, and $\Omega$) were alarmingly low compared to measured values from related materials. It is possible that surface contamination played a role in these results.

The expected reduced gaps $2\Delta/k_B T_c$ for weak-coupling $s$-wave and $d$-wave BCS superconductors are 3.5 and 4.3, respectively\cite{BardeenPhysRev1957,WonPRB1994}. In the cuprates, experiments which decoupled the pseudogap by coherent tunneling\cite{DeutscherRMP2005} or normalization\cite{BoyerNatPhys2007} have found reduced gap values between 6 and 10. Superconducting gaps measured by STM in Fe-based materials are shown in table~\ref{table:SC-gaps}.  There is a wide distribution of reduced gap values $2\Delta/k_B T_c$, which may arise from surface quality issues, unavoidable surface states even in clean surfaces, or method of determination of the gap, be it careful Dynes fitting or simple peak-to-peak measurement. In most cases, the reduced gap suggests that although Fe-based superconductors are in a strong coupling regime, they are not as strongly coupled as cuprates.

\begin{table*}[tbh]\footnotesize
\begin{center}
\begin{tabular}{ l | c c c c c c l}
 \hline
Material & $T_c$ (K) & $\overline{\Delta}$ (meV) & $\sigma_{\Delta}/\overline{\Delta}$ & $2\overline{\Delta}/k_B T_c$	 & Determined by & Reference \\
\hline
NdFeAsO$_{0.86}$F$_{0.14}$ & 48 & 9.3 & & 4.5 & Dynes fit (s) & Jin\cite{JinSST2010} \\
SmFeAsO$_{0.8}$F$_{0.2}$ & 45 & 7 & 0.06 & 3.6 & coh. pk. & Fasano\cite{FasanoPRL2010} \\
 & & & & 2.2 & Dynes fit ($s$) \\
 & & & & 3.0 & Dynes fit ($d$) \\
\hline	
Sr$_{1-x}$K$_x$Fe$_2$As$_2$ & 32 & 10 & & 7.2 & coh. pk. & Boyer\cite{BoyerArxiv0806.4400} \\
Ba$_{1-x}$K$_x$Fe$_2$As$_2$ & 37 & 15 & & 9.4 & coh. pk. & Wray\cite{WrayPRB2008} \\
Ba(Fe$_{0.9}$Co$_{0.1}$)$_2$As$_2$ & 25.3 & 6.25 & 0.12 & 5.7 & coh. pk. & Yin\cite{YinPRL2009} \\
Ba(Fe$_{0.96}$Co$_{0.04}$)$_2$As$_2$ & 14 & 4 & & 6.6 & coh. pk. & Massee\cite{MasseeEPL2010,MasseeThesis2011} \\
Ba(Fe$_{0.93}$Co$_{0.07}$)$_2$As$_2$ & 22 & 6.94 & 0.13 & 7.3 & coh. pk. & Massee\cite{MasseePRB2009a} \\
Ba(Fe$_{0.895}$Co$_{0.105}$)$_2$As$_2$ & 13 & 5.8 & & 10.3 & coh. pk.	 &
Massee\cite{MasseeEPL2010,MasseeThesis2011} \\
Ba(Fe$_{0.9}$Co$_{0.1}$)$_2$As$_2$ & 25 & 7 & & 6.5 & coh. pk. & Zhang\cite{ZhangPRB2010} \\
Ba$_{0.6}$K$_{0.4}$Fe$_2$As$_2$ & 38 & 3.6 & 0.12 & 2.2 & coh. pk.$^{(a)}$ & Shan\cite{ShanNatPhys2011} \\
& & 8.1 & 0.09 & 5.0 & \\
Ba$_{0.6}$K$_{0.4}$Fe$_2$As$_2$ & 37.2 & 3.32 & 0.28 & 2.1 & coh. pk.$^{(a)}$ & Shan\cite{ShanPRB2011} \\
& & 7.63 & 0.10 & 4.8 & \\
Ba(Fe$_{0.94}$Co$_{0.06}$)$_2$As$_2$ & 14 & 4 & 0.45 & 6.6 & Dynes fit ($s$) & Teague\cite{TeaguePRL2011} \\
& & 8 & 0.19 & 13.3 & \\		
Ba(Fe$_{0.88}$Co$_{0.12}$)$_2$As$_2$ & 20 & 5 & 0.34 & 5.8 & Dynes fit ($s$) & \\
& & 10 & 0.19 & 11.6 & \\
\hline		
FeTe$_{0.85}$Se$_{0.15}$ & 14 & 2.3 & & 3.8 & Dynes fit ($s$) & Kato\cite{KatoPRB2009} \\
FeTe$_{0.7}$Se$_{0.3}$ & $\sim12.8$ & $\sim2-4$ & & $\sim3.6-7.2$ & shoulder & Fridman\cite{FridmanJPCS2010} \\
FeTe$_{0.6}$Se$_{0.4}$ & 11 & 1.7 & & 3.6 & coh. pk. & Hanaguri\cite{HanaguriScience2010} \\
FeSe & 8 & 2.2 & & 6.4 & coh. pk. & Song\cite{SongScience2011} \\
\hline
LiFeAs & 16 & 2.5 & & 3.6 & coh. pk. & Hanaguri\cite{HanaguriAspen2011} \\
& & 5.7 & & 8.3 & & \\
LiFeAs & 18 & 5 & & 6.5 & coh. pk. & H\"{a}nke\cite{HankeArxiv1106.4217} \\
\hline
KFe$_2$Se$_2$ & $\sim25$ & 1 & & & coh. pk. & Li\cite{LiNatPhys2011} \\
& & 4 & & & & \\
K$_{0.73}$Fe$_{1.67}$Se$_2$ & $32$ & 7 & & 5.1 & coh. pk. & Cai\cite{CaiArxiv1108.2798} \\
\hline
\end{tabular}
\caption{\label{table:SC-gaps} Superconducting gaps, measured by STM.\\$^{(a)}$Average gap values were determined by observing the coherence peaks; individual spectra were checked with $s$-wave Dynes fitting.}
\end{center}
\end{table*}

\subsection{\label{sec:two-gaps}Two Gaps}

\begin{figure}[tb]
\begin{center}
  {\includegraphics[width=.95\columnwidth,clip]{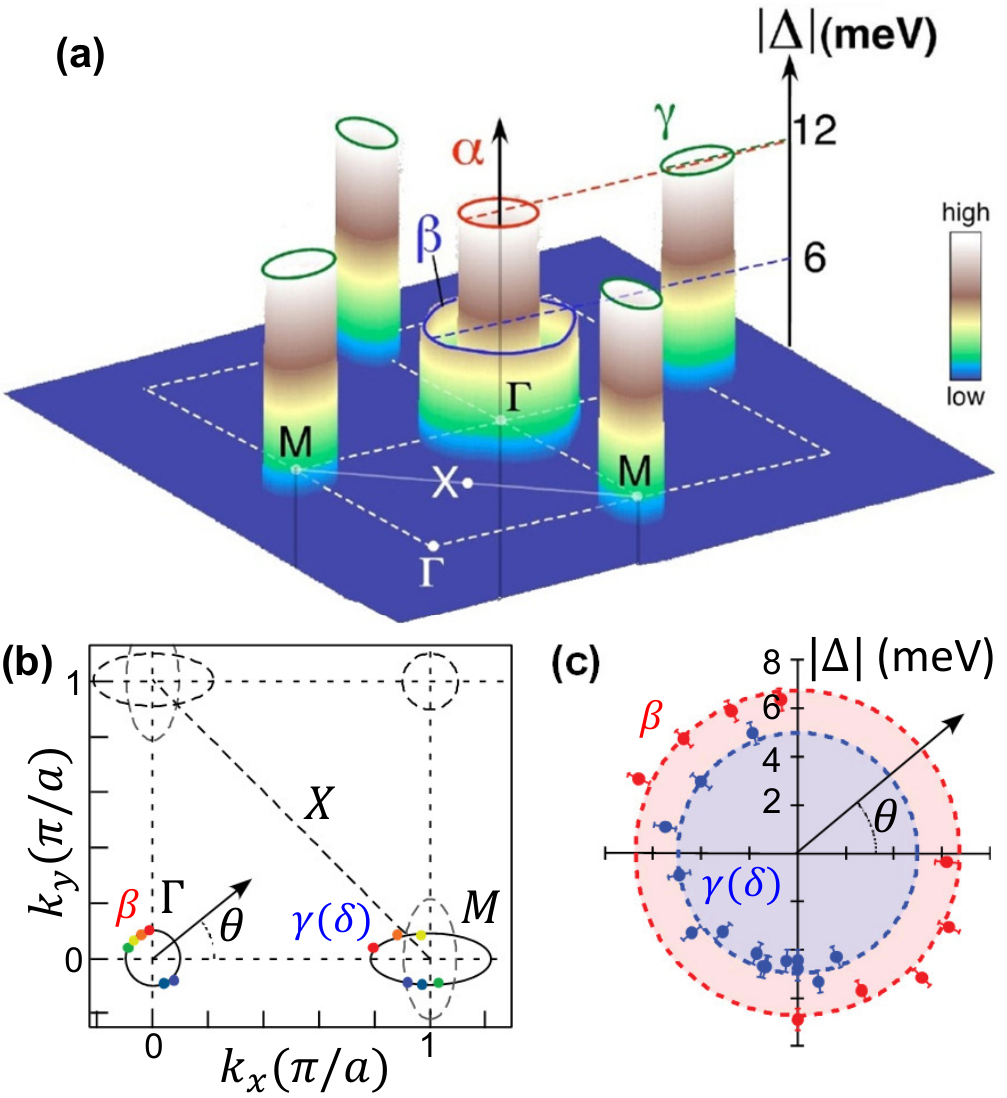}}
  \caption[ARPES on Ba$_{0.6}$K$_{0.4}$Fe$_2$As$_2$ and .]
    {\label{fig:DingARPES}
    Fermi sheets and superconducting gaps of Ba122. (a) Schematic of Fermi sheets and superconducting gaps of hole-doped Ba$_{0.6}$K$_{0.4}$Fe$_2$As$_2$ ($T_c$=37K), as measured by ARPES\cite{DingEPL2008}. (b) Schematic of Fermi sheets of electron-doped Ba(Fe$_{1.925}$Co$_{0.075}$)$_2$As$_2$ ($T_c$=25.5K), as measured by ARPES\cite{TerashimaPNAS2009}. (c) Isotropic superconducting gap values of the Fermi sheets shown in (b).}
\end{center}
\end{figure}

Early ARPES measurements detected multiple FSs for the 122 family, and showed that these FSs may support two gaps of rather different magnitudes\cite{DingEPL2008}. For example, in optimally hole-doped Ba$_{0.6}$K$_{0.4}$Fe$_2$As$_2$ ($T_c$=37K), the two gaps are $\Delta$=6meV and $\Delta$=12meV, as shown in figure~\ref{fig:DingARPES}a.  However, in optimally electron-doped Ba(Fe$_{1.925}$Co$_{0.075}$)$_2$As$_2$ ($T_c$=25.2K), these two gaps are much more closely spaced, at $\Delta$=6.6meV and $\Delta$=5meV\cite{TerashimaPNAS2009}, as shown in figure~\ref{fig:DingARPES}, b and c.

The tunneling rate along the $\hat{z}$ direction is expected to be strongly suppressed with increasing in-plane momentum $|\vec{k}_{||}|$\cite{TersoffPRL1983}. Therefore, the tunneling matrix element of the $M$-centered bands is expected to be strongly suppressed with respect to that of the $\Gamma$-centered bands\cite{LeePRL2009}.  In optimally electron-doped Ba(Fe$_{1.925}$Co$_{0.075}$)$_2$As$_2$, there is only one band centered at the $\Gamma$ point, with $\Delta$=6.6meV\cite{TerashimaPNAS2009}, as shown in figure~\ref{fig:DingARPES}, b and c.  Therefore, it is not surprising that in extensive STM studies of atomically resolved surfaces of electron doped Ba(Fe$_{1-x}$Co$_{x}$)$_2$As$_2$\cite{YinPRL2009,MasseePRB2009a}, the contribution from the outer bands has not been strong enough to show a clear second gap (with expected similar energy $\Delta_{\gamma,\delta}\sim$5meV) in addition to the observed gap of $\overline{\Delta}\sim$6.5meV.

However, two other very recent STM studies of non-atomically-resolved surfaces of superconducting Ba122 have shown two gaps. First, Shan \textit{et al} studied Ba$_{0.6}$K$_{0.4}$Fe$_2$As$_2$, the identical hole-doped compound pictured in figure~\ref{fig:DingARPES}a. In non-atomically-resolved regions, they found two gaps of $\overline{\Delta}\sim$3.6meV and $\overline{\Delta}\sim$8.1meV (fit values vary slightly with the choice of quasiparticle lifetime, as shown in figure~\ref{fig:Dynes-gap-fit}, c and d).

Second, Teague \textit{et al} resolved two gaps on the electron-doped Ba(Fe$_{1-x}$Co$_{x}$)$_2$As$_2$, as shown in figure~\ref{fig:Dynes-gap-fit}, e and f.  This study differed from all previous STM studies on the same material in that the sample was not cleaved in vacuum (rather in Ar at room $T$), and no atomic resolution was obtained, even after cooling to low $T$. This gives a clue that the imaged surface may be contaminated, so the tunneling electrons may have their momenta scrambled by passage through the contaminant layer.  This might allow enough mixing between bands to circumvent the rule of thumb that the tunneling matrix element for the $M$-centered bands will be suppressed\cite{LeePRL2009}.

Finally, a recent follow-up study on the same Ba$_{0.6}$K$_{0.4}$Fe$_2$As$_2$ sample\cite{ShanNatPhys2011} showed evidence of three gaps at $\sim$3.3 meV, $\sim$7.6 meV, and $\sim$10 meV, on the non-atomically resolved surface\cite{ShanPRB2011}. In conclusion, two or more gaps have been resolved only in STM measurements which lack atomic resolution, suggesting that surface disorder scrambles the quasiparticle momentum.


\subsection{\label{sec:gap-inhomogeneity}Gap Inhomogeneity}

\begin{figure}[tb]
\begin{center}
  {\includegraphics[width=.95\columnwidth,clip]{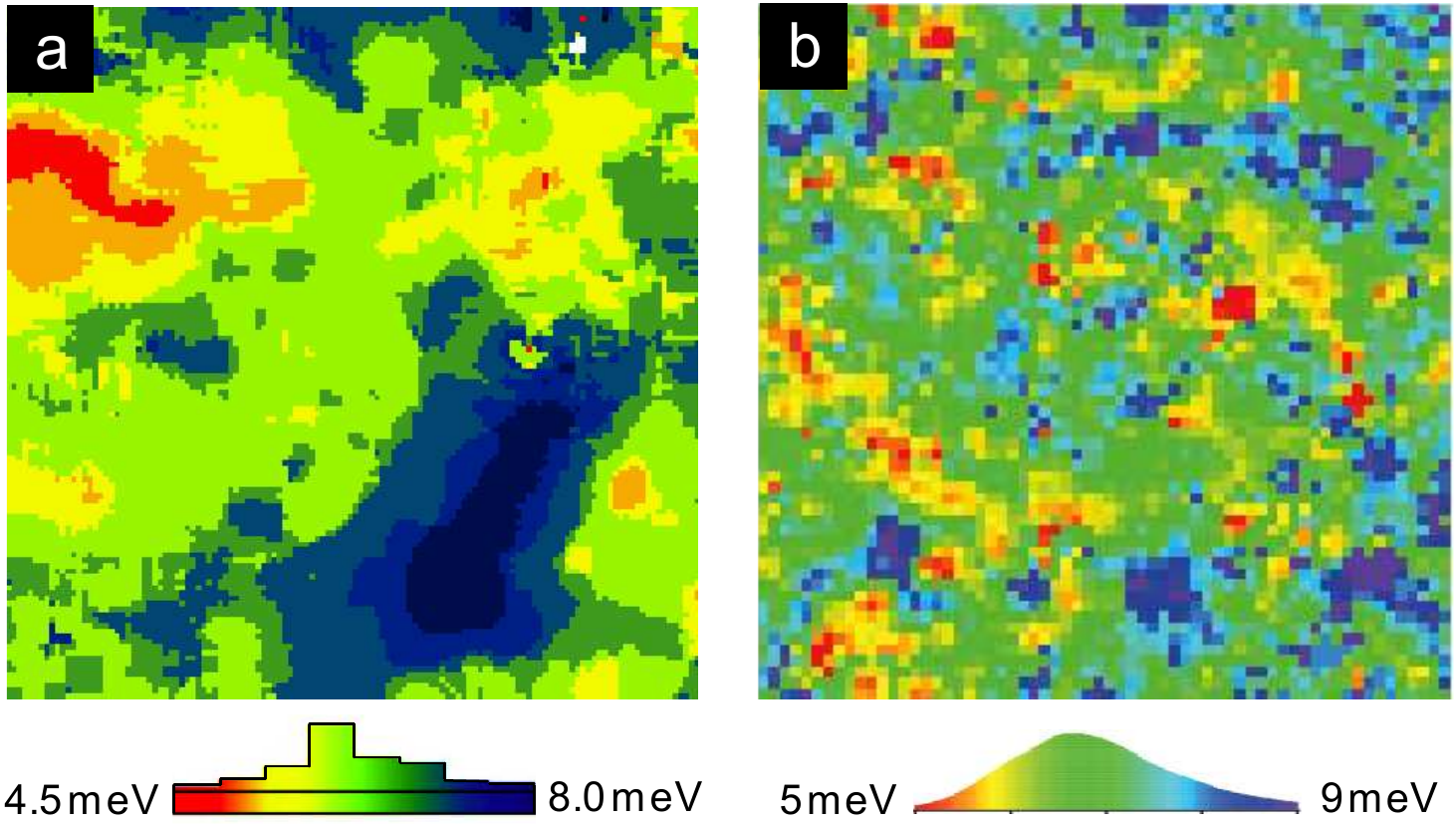}}
  \caption[Gapmaps from Ba(Fe$_{1-x}$Co$_x$)$_2$As$_2$]
    {\label{fig:Gapmaps}
    (a) Gap map recorded by Yin \textit{et al} over a 20$\times$20 nm$^2$ area of Ba(Fe$_{0.9}$Co$_{0.1}$)$_2$As$_2$ at 6.25 K\cite{YinPRL2009}; and (b) by Massee et al over a 18.9$\times$18.9 nm$^2$ area of Ba(Fe$_{0.93}$Co$_{0.07}$)$_2$As$_2$ at 4.2 K\cite{MasseePRB2009a}. In both cases, the reported gap values are computed as half the distance between coherence peaks.}
\end{center}
\end{figure}

A key advantage of STM over other gap measurement techniques is access to the gap variation on the nanoscale.  In most Fe-based superconductors, modest nanoscale variations have been found in the gap magnitude, with standard deviations amounting to $\sim$10\% of the average value $\overline{\Delta}$, as listed in table~\ref{table:SC-gaps} and exemplified in figure \ref{fig:Gapmaps}. In cuprates, superconducting gap variances of $\sigma_{\Delta}/\overline{\Delta}>$20\% have typically been reported\cite{McElroyPRL2005}.  More recently, it was shown that these gap variations in cuprates are heavily influenced by the pseudogap\cite{BoyerNatPhys2007}, but even after the pseudogap is removed by normalization or independent fitting, the superconducting gap variations in cuprates appear larger than those in the more weakly coupled Fe-based superconductors\cite{BoyerNatPhys2007,MaPRL2008}. A correlation between the superconducting gap variance $\sigma_{\Delta}/\overline{\Delta}$ and the reduced gap $2\Delta/k_B T_c$ has been suggested in Fe-based superconductors\cite{MasseeThesis2011}.

\section{\label{sec:Parent}Parent Compound}

Understanding the mechanism of superconductivity requires an understanding of the `normal state' of the parent compound out of which superconductivity arises. Here we discuss the spatial and energetic signatures of electronic ordered states of the parent compound. One spatial signature, detected by STM, is a rotational symmetry breaking from $C4 \rightarrow C2$. A much-debated energy signature in the parent compound is a depression in the DOS near the Fermi level, which has been framed in the language of a possible `pseudogap', in analogy to the cuprates.

\subsection{\label{sec:C4C2}\textit{C}4 $\rightarrow$ \textit{C}2 symmetry breaking}


Several theories predict $C4\rightarrow C2$ electronic symmetry breaking in the parent compounds of Fe-based superconductors, due to orbital ordering\cite{LvPRB2009,LeePRB2009,ChenPRB2010b,LvPRB2011} or a nematic state\cite{FangPRB2008,XuPRB2008,ZhaiPRB2009}.  A nematic state is a liquid which breaks long-range rotational symmetry without breaking long-range translational symmetry\cite{KivelsonNature1998}. Therefore, if the rotational symmetry is broken first by the orthorhombic crystal structure, the detection of $C2$ electronic order does not strictly constitute a nematic state.  However, if the crystal symmetry is broken by only $\sim$1\%, as is typical in the orthorhombic phase of these materials\cite{GoldmanPRB2008}, an electronic $C4\rightarrow C2$ symmetry breaking of $50-200\%$ or more has been argued to be too large to be solely due to a pre-existing structural effect, and has been ascribed instead to an independent electronic nematic state\cite{ChuangScience2010,SongScience2011}.

\begin{figure}[tbh]
\begin{center}
  {\includegraphics[width=0.8\columnwidth,clip]{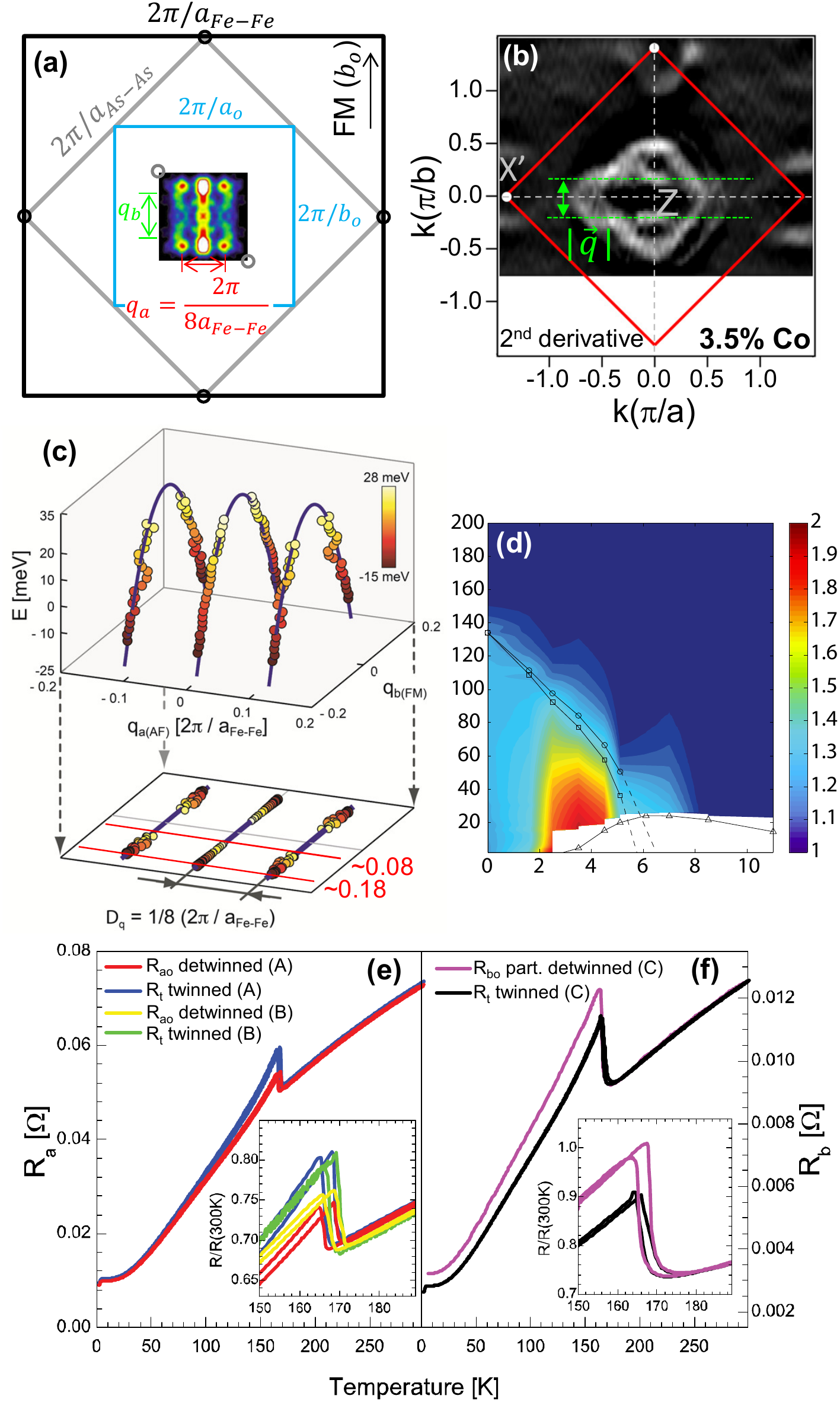}}
  \caption[$C2$ symmetry from several experiments.]
    {\label{fig:C2sym}
    STM, ARPES, and transport experiments have all provided evidence for $C2$ symmetry in 122 materials. (a) QPI images on 3\% Co-doped Ca122\cite{ChuangScience2010}. The dominant wavevector corresponds to length $\sim 8a_{\mathrm{Fe-Fe}}$ (shown in red). From inspection of twin boundaries, the authors determine this to be along the orthorhombic $a_O$ axis (the AF axis). (b) ARPES momentum space image (second derivative) from 3.5\% Co-doped Ca122, at 0meV binding energy\cite{DessauArxiv1009.0271}. There is a nesting vector between parallel FS sections at $\vec{q}=2\pi/8.3a_{\mathrm{Fe-Fe}}$ (shown in green). (c) QPI dispersion from Chuang\cite{ChuangScience2010}. (d) Transport anisotropy measured on detwinned Ba122\cite{ChuScience2010}. The high conductivity direction in Ba122 is the orthorhombic $a$ direction (the AF axis). (e-f) Transport anisotropy measured on detwinned Ca122\cite{TanatarPRB2010}. (e) The resistance decreases along the strain direction (with dominant $a$-oriented domains). (f) The resistance increases along the orthogonal direction (with dominant $b$-oriented domains). Taken together, these results confirm that in Ca122 as well, the high conductivity direction is the orthorhombic $a$ direction (the AF axis).}
\end{center}
\end{figure}


Some theories have predicted incipient magnetic ordering above the crystallographic orthorhombic transition at temperature $T_s$\cite{XuPRB2008,FangPRB2008} (see figure~\ref{fig:phase-diagrams}c for a representative phase diagram). According to these theories, above $T_s$, there are only fluctuations of the sublattice magnetic order, which will eventually, on cooling, lead first to the structural transition and then the full magnetic transition.  Below $T_{N}$, the AF magnetic order additionally breaks translational symmetry, so the phase is no longer strictly a nematic. Between $T_s$ and $T_{N}$, the $C4$ symmetry is broken but translational symmetry is not yet broken, fitting the definition of a true nematic phase\cite{FangPRB2008}.

Because as-grown crystals are composed of multiple orthogonally-oriented orthorhombic domains, intrinsic anisotropy is challenging to measure via bulk experiments. Local STM experiments were the first to show evidence for large $C4\rightarrow C2$ electronic state symmetry breaking\cite{ChuangScience2010}. Chuang \textit{et al} reported QPI measurements from cleaved Ca(Fe$_{0.97}$Co$_{0.03}$)$_2$As$_2$, at the underdoped edge of the superconducting dome (shown as a red $\star$ in the phase diagram of the similar Ba(Fe$_{1-x}$Co$_{x}$)$_2$As$_2$ system in figure~\ref{fig:phase-diagrams}c).  As expected at that doping, no signature of superconductivity was seen in STM spectroscopy at $T$=4.3 K.
In these non-superconducting parent compounds, deep in the orthorhombic phase, Chuang highlighted the radically different electronic behavior along the two orthorhombic axes (recall the orthorhombic unit cell in figure \ref{fig:unit-cell}b).

Along the $a$-axis (the longer, AF axis), the DOS showed self-similarity on a length scale $d\sim8a_{\mathrm{Fe-Fe}}$. The authors emphasized the existence of a `static' electronic nano-object of size $d$, which folded the band structure by a wavevector $q_a = 2\pi/d$\cite{AllanPreprint2011}, labeled in figure \ref{fig:C2sym}a.

Along the $b$-axis (the shorter, ferromagnetic axis), the authors noted significant QPI dispersion, labeled as wavevector $q_b$ in figure \ref{fig:C2sym}a.  The dispersion of $q_b$ along $b$, in contrast with the `static' wavevector $q_a\sim2\pi/8a_{\mathrm{Fe-Fe}}$ along $a$, led the authors to posit a nematic band, which `evolves along one axis only'.

Chuang showed excellent quantitative agreement between the $q_b$ dispersion along $b$, and the dispersion of strongly nested sections of the $\Gamma$-centered $\alpha_2$ hole band, observed by ARPES\cite{KondoPRB2010} in undoped Ca122 (a rigid band chemical potential shift was used to account for the different dopings in the STM and ARPES experiments). However, the ARPES measurement were carried out on twinned Ca122, giving the appearance of $C4$-symmetric nesting.

Given the $C2$-symmetric surface structures already demonstrated on various 122 materials\cite{BoyerArxiv0806.4400,YinPRL2009,MasseePRB2009b,NiestemskiArxiv0906.2761}, it is important to ask whether the STM-observed $C2$-symmetric QPI results from a surface or bulk phenomenon. Chuang did a meticulous job to separate the $C4\rightarrow C2$ symmetry breaking of the observed QPI from the more obvious $2\times 1$ reconstruction of the surface, which also breaks $C4$ symmetry, but at a 45$^{\circ}$ angle (see topography in figure~\ref{fig:122-As-topos}c).  To detect the very small 1\% length distortion and 1$^{\circ}$ angle distortion of the orthorhombic phase, the authors scanned the same field of view in two different directions, and found both surface reconstruction twin boundaries (see red arrows in figure~\ref{fig:122-As-topos}c), and orthorhombic twin boundaries (see green arrows in figure~\ref{fig:122-As-topos}c), at distinct locations.  The authors therefore empirically linked the QPI anisotropy to the bulk orthorhombic orientation, rather than the surface reconstruction orientation.  It is clear from this work that (a) STM can image the bulk electronic structure of 122 materials, and (b) bulk electronic structure is significantly more anisotropic than one might naively expect from the 1\% lattice orthorhombicity.

Although the experimental data gave clear evidence for bulk $C2$ symmetry, theoretical objections were raised to the claim of a nematic origin for this $C2$ symmetry\cite{MazinPRB2011}.  Mazin objected that by definition, nematic symmetry doesn't apply at Chuang's low temperature where the long range AF order breaks translation symmetry. Furthermore, Mazin insisted that a nematic band is not needed in order to explain Chuang's observations. Using a few empirical values to simply parameterize a plausible SDW-reconstructed FS, Knolle \textit{et al} calculated the QPI patterns due to scattering from pointlike non-magnetic impurities and magnetic impurities oriented along the $z$ axis\cite{KnollePRL2010}.  Knolle's calculations reproduced a $C2$-symmetric QPI pattern with wavevectors which were quantitatively similar to Chuang's measured $q_a$ and $q_b$, but unlike Chuang's data, the calculated QPI patterns dispersed along both orthogonal axes.  Mazin \textit{et al} used a different approach, starting from \textit{ab initio} FS (shown in figure \ref{fig:unit-cell}e)\cite{MazinPRB2011}. Mazin emphasized that no quantitative comparison should be possible without accurate knowledge of the surface doping and of the scatterers. He therefore made no assumptions about the scattering potential and computed only the JDOS rather than the full QPI pattern.  His calculated JDOS showed $C2$ symmetry, emphasizing a dispersive scattering vector along the $b$ axis, but lacking any features, dispersive or not, along the $a$ axis. Therefore, both Knolle and Mazin were able to reproduce some of the qualitative features of Chuang's measured QPI, in particular the $C2$ symmetry, without invoking a nematic band\cite{KnollePRL2010,MazinPRB2011}.

Following Chuang's local STM measurements, a breakthrough by Chu \textit{et al} allowed majority detwinning of the bulk orthorhombicity in 122 materials (for example, $\sim$5 MPa uniaxial pressure resulted in 86\% volume fraction of one orientation)\cite{ChuScience2010}. Chu reported that detwinned Ba(Fe$_{1-x}$Co$_x$)$_2$As$_2$ developed in-plane electronic transport anisotropy near the structural transition.  The resistivity along the shorter, ferromagnetic axis, $\rho_b$, became greater than $\rho_a$. The resistive anisotropy reached a maximum value of $\sim$2 for compositions near the underdoped edge of the superconducting dome (see figure~\ref{fig:C2sym}d). Even for temperatures well above the orthorhombic structural transition, uniaxial stress induced a resistivity anisotropy, suggesting a substantial nematic susceptibility.

Tanatar \textit{et al} verified the transport anisotropy in detwinned Ba122, and extended the measurement to detwinned Ca122, allowing direct comparison with Chuang's STM data\cite{TanatarPRB2010}. Tanatar showed that upon detwinning, resistivity decreased along the orthorhombic $a$ axis but increased along the orthorhombic $b$ axis in both compounds. In both materials Tanatar found that the resistive anisotropy
was largest at $T_s$ with $\rho_{b} /\rho_{a}\sim1.2$ in Ca122 and $\sim$1.5 in Ba122. For Ca122 this anisotropy was observed only below $T_s$, suggesting a first order transition, and diminished upon further cooling, reaching about 1.05 at $T\sim$50 K and remaining constant at lower temperatures. For Ba122 the anisotropy was observed both below and above $T_s$, suggesting a second order transition with nematic fluctuations above $T_{s}$.

Following Chu's demonstration of bulk detwinning, ARPES experiments on detwinned samples showed the anisotropic band structure of Ca122\cite{DessauArxiv1009.0271} and Ba122\cite{YiPNAS2011}. In Ca122, Wang \textit{et al} confirmed the presence of a strongly nested FS, with nesting vector approximately matching Chuang's QPI vector $q_b$ (figure \ref{fig:C2sym}b). Furthermore, the detwinned ARPES augmented Kondo's work\cite{KondoPRB2010} by showing that the nesting was $C2$-symmetric. In Ba122, the FS showed elongated regions of the same orientation and location, but without the strong nesting of Ca122, so a susceptibility to nematicity was emphasized only at temperatures above $T_s$\cite{YiPNAS2011}.

Some aspects of these experimental results from STM, transport, and ARPES seem counterintuitive, or even contradictory.
First, one expects that conductivity along the ferromagnetic direction ($b$) should be higher than along the AF direction ($a$). But transport experiments consistently show that $\rho_b > \rho_a$\cite{ChuScience2010,TanatarPRB2010}. Second, an $a$-oriented nematic band as posited by Chuang to explain the observed anisotropic QPI would be expected to lead to $\rho_a > \rho_b$ in contrast to the observed transport anisotropy. Third, given the dramatic anisotropy of the band structure, both from the experimentally measured strongly nested FSs\cite{DessauArxiv1009.0271}, and theoretically calculated fully broken symmetry between the Fe $xz$ and $yz$ orbitals\cite{TanatarPRB2010}, one might expect a dramatic electronic transport anisotropy. In fact, only weak to moderate anisotropy up to a factor of 2 is observed. Finally, the maximal electronic transport anisotropy in Ca122 was found right below the structural transition $T_s$ rather than at lower temperatures where the obvious $C4$ symmetry breaking OPs (either magnetic moment or structural orthorhombicity) would be expected to increase\cite{TanatarPRB2010}.


To qualitatively resolve some of these issues, Mazin noted that, while the calculated FSs completely break the $C4$ symmetry, which is fully reflected in Chuang's QPI images, the individual pockets are very three dimensional, so that the calculated conductivity is comparable for all three directions\cite{MazinPRB2011}, and the anisotropy may take either sign according to parameter details to be worked out\cite{TanatarPRB2010}.

Most recently, follow-up doping-dependent STM studies by Allan \textit{et al}\cite{AllanPreprint2011} have suggested a more complete and quantitative explanation for the counterintuitive and apparently contradictory results of STM, transport, and ARPES experiments. Allan \textit{et al} showed that in the presence of $C2$-symmetric electronic structure, Co dopants either establish or pin static electronic dimers, oriented along the $a$ axis, with a length scale of $d \sim 8a_{\mathrm{Fe-Fe}}$. He showed that these dimers may act as extended scattering objects which, in conjunction with the nested band structure measured by Wang \textit{et al}\cite{DessauArxiv1009.0271}, lead to QPI which is quantitatively consistent with that observed by Chuang, and to resistive anisotropy consistent with observations\cite{ChuScience2010,TanatarPRB2010}.

Several other parent compounds of the Fe-based superconductors have subsequently been studied by STM.  So far, all have shown evidence for a $C2$ ordered state.

Li \textit{et al}\cite{LiArxiv1006.5907} claimed evidence of $C4\rightarrow C2$ symmetry breaking in Ba122 from subtle differences between the shapes of $45^{\circ}$ and $-45^{\circ}$ domain walls, as shown in figure \ref{fig:122-As-topos}a.  However, unlike Chuang \textit{et al}, Li did not image a twin boundary in the bulk orthorhombic lattice, and thus could not demonstrate that the subtle domain wall orientation differences were coincident with the orthorhombic domain.  Therefore, it remains possible that this subtle observation may be due to surface state anisotropy or even to tip anisotropy.

\begin{figure}[tb]
\begin{center}
  {\includegraphics[width=.95\columnwidth,clip]{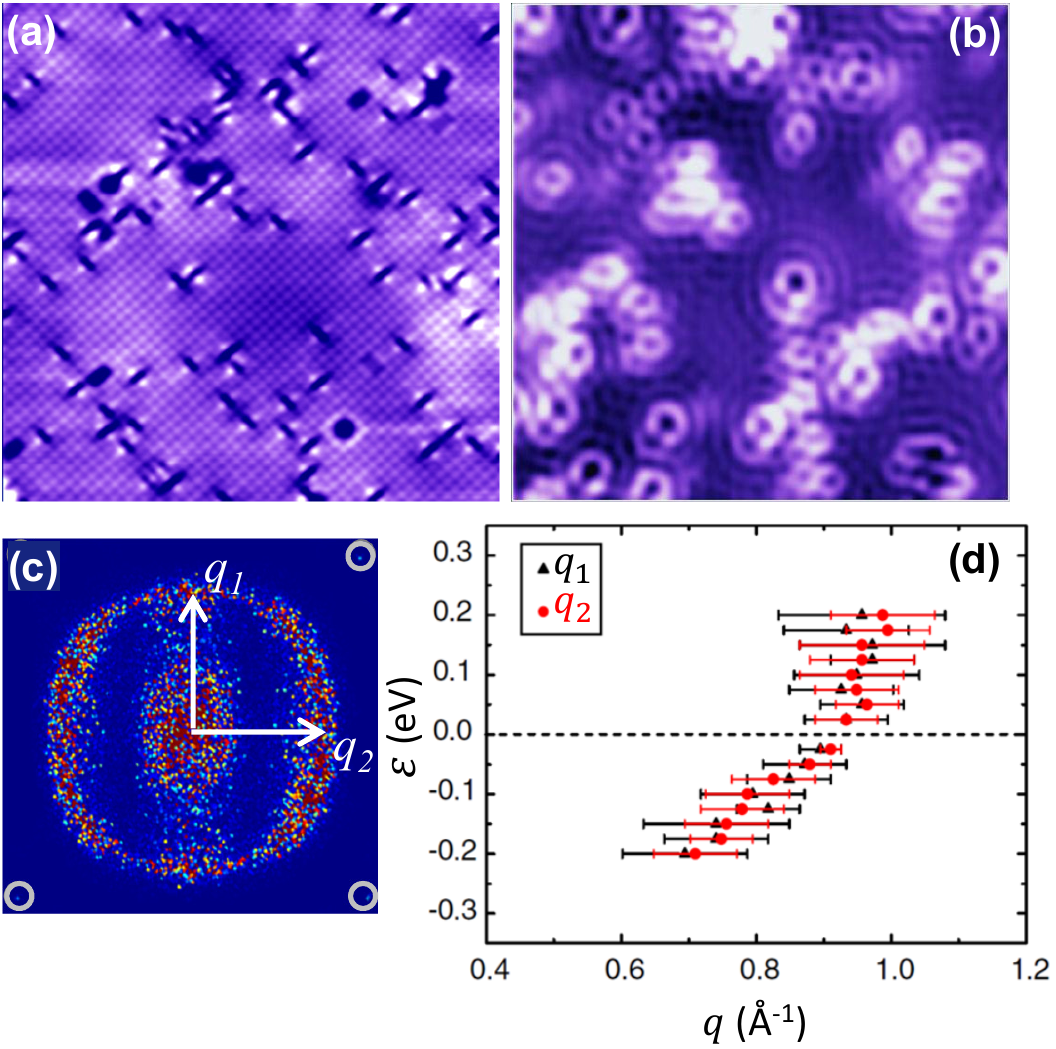}}
  \caption[$C2$-symmetric QPI from Ca122 and LaFeAsO.]
    {\label{fig:LaFeAsO-QPI}
    $C2$-symmetric QPI in LaFeAsO. (a-b) 20 nm square topographic images from the two inequivalent cleaved surfaces of LaFeAsO. (a) FeAs surface (setup: $V_{\mathrm{sample}}=100\,\mathrm{mV}; I_{\mathrm{set}}=50\,\mathrm{pA}$). (b) LaO surface (setup: $V_{\mathrm{sample}}=-100\,\mathrm{mV}; I_{\mathrm{set}}=40\,\mathrm{pA}$). (c) Fourier transform of a 40 nm square $dI/dV$ map of the LaO surface at -75 meV. The gray circles at the corners show the Bragg peaks of the La surface atoms, $2\pi/a_{\mathrm{La-La}}$ where $a_{\mathrm{La-La}}\sim$4\AA. (d) Dispersions of the QPI signatures at $q_1$ and $q_2$. Figures taken from Ref. \onlinecite{ZhouPRL2011}.
}
\end{center}
\end{figure}

STM images of the LaFeAsO parent compound also showed $C2$-symmetric QPI on one of the two inequivalent cleaved surfaces, shown in figure \ref{fig:LaFeAsO-QPI}, a and b\cite{ZhouPRL2011}. There are a number of significant differences in the acquisition and interpretation of the QPI data from Ca122 and LaFeAsO. The surface showing QPI in LaFeAsO was believed to be LaO (based on an observed increase in imaged surface impurity states on doping from parent LaFeAsO to LaFeAsO$_{0.9}$F$_{0.1}$), while the imaged surface in Ca122 was thought with less certainty to be FeAs\cite{AllanPrivate2011}.  The imaged QPI in LaFeAsO was believed to originate from a surface state, based on theoretical predictions\cite{EschrigPRB2010} and on the fact that the atomic topography was almost completely obscured.  The imaged QPI in Ca122 was shown to originate from the bulk, based on the fact that it changes orientation coincidently with a twin boundary in the bulk orthorhombic lattice but not the $2\times 1$ surface reconstruction.  It should be noted that in LaFeAsO, unlike in Ca122, there is no apparent reconstruction of surface atoms themselves into a $C2$ symmetric state.

Zhou observed QPI at long and short wavevectors, but chose to focus on the more-easily-quantifiable long wavevectors $q_1$ and $q_2$ (figure \ref{fig:LaFeAsO-QPI}c), because they are easier to distinguish from the long-wavelength inhomogeneity that results simply from the random impurity distribution.  Chuang removed the effects of this long-wavelength inhomogeneity by subtracting a Gaussian centered at the $q$-space origin.  Chuang focused on the shorter wavevectors, which are labeled $q_a$ and $q_b$ in figure~\ref{fig:C2sym}a.

Zhou found $q_1$ and $q_2$ to disperse over a large energy range from -200 to +200 meV, in perfect synchrony, as shown in figure \ref{fig:LaFeAsO-QPI}d.  In contrast, Chuang reported that $q_b$ dispersed strongly, while $q_a$ did not disperse at all, as shown in figure \ref{fig:C2sym}c.  Although Zhou's $q_1$ and $q_2$ wavevectors dispersed identically, the shape of their $q$-space weight distribution broke $C4$ symmetry. In Zhou's data, $q_1$ appeared as an isolated spot, whereas $q_2$ appeared elongated across almost the entire BZ.  Chuang's data also showed QPI peaks elongated along the $b$ axis. The main difference in the conclusions of the two papers stems from Chuang's claim that $q_a$ did not change with energy, in contrast to Zhou's changing $q_1$ and $q_2$ along both orthorhombic directions.  This claim of static $q_a$ led Chuang to posit a nematic band, in a melted stripes picture\cite{FradkinScience2010}. However, anisotropic QPI may arise simply from a scattering potential with $C2$ symmetric structure factors\cite{CapriottiPRB2003}, as shown empirically by Allan \textit{et al}\cite{AllanPreprint2011}.

Further investigations of $C4\rightarrow C2$ symmetry breaking were carried out on MBE-grown FeSe films with near-perfect Se surfaces (fewer than one Se vacancy per $\sim$70,000 sites)\cite{SongScience2011}. Song \textit{et al} found that magnetic vortices were elongated by a factor of $\sim$2 along one Fe-Fe direction with respect to the other. Furthermore, they imaged a twin boundary which separated two regions of orthogonally-elongated vortices, thus proving that the $C4$ symmetry breaking was not a tip effect.  Song also saw faint stripes in the background DOS, with wavelength $\sim16a_{\mathrm{Fe-Fe}}$ along the same direction as the vortex elongation.  These stripes are reminiscent of the $8a_{\mathrm{Fe-Fe}}$ stripes observed in Ca122\cite{ChuangScience2010}, but with approximately twice the wavelength.  Although Song did not investigate the energy dependence of these faint stripes, he speculated that they arose from intra-pocket nesting of the $\Gamma$-centered hole pocket, which may be distorted by orbital ordering\cite{KrugerPRB2009,LvPRB2009,LeePRB2009,ChenPRB2010b}.

Song also observed single atom impurities whose spatial resonances locally break $C4$ symmetry on two different length scales.  The short length scale 2-fold symmetric shapes (`atomic-scale dimers'), similar to those seen in LaFeAsO\cite{ZhouPRL2011} (figure \ref{fig:LaFeAsO-QPI}a), may be explained simply by the $\sim$50\% probability occupation of each of the two geometrically inequivalent Fe sites in the unit cell, without requiring a deeper mechanism of electronic symmetry breaking. Indeed, Song \textit{et al} show the atomic-scale dimers to be randomly oriented on both sides of their orthorhombic twin boundary. However, a longer length scale electronic dimer structure also surrounds the same impurities; this $\sim8\,\mathrm{nm}$ dimer is always oriented consistently with respect to the orthorhombic crystal direction. In comparison with the orientation of the $\sim8a_{\mathrm{Fe-Fe}}$ electronic dimers demonstrated by Allan \textit{et al}\cite{AllanPreprint2011}, it seems reasonable to identify the vortex elongation direction (longer $\xi$) as the crystalline $a$-axis (the AF axis).

\subsection{\label{sec:Pseudogap}Pseudogap}

Although the Fe-based superconductors are exciting materials in their own right, it must be admitted that a significant component of the interest is due to their role as a foil for the higher-$T_c$ cuprates.  One of the most puzzling unsolved problems in cuprates is the origin of the `pseudogap' - a suppression in the DOS near the Fermi level at temperatures far exceeding the superconducting $T_c$\cite{TimuskROPP1999}.  It is therefore a natural question: is there a pseudogap in the Fe-based superconductors?  The answer is still controversial, with pseudogap sightings claimed by NMR\cite{AhilanPRB2008}, ARPES\cite{XuNatComm2011}, femtosecond spectroscopy\cite{MerteljPRL2009}, and transport experiments\cite{HessEPL2009}.
In parent Sr122, the SDW transition occurs at $T_{SDW}\sim$190K, and opens a gap of $\sim$60 meV detected by optical spectroscopy\cite{HuPRL2008}.


The first question in evaluating these claims is, what do we mean by a `pseudogap'?  In the cuprates, the term arose to describe a mysterious gap of unknown origin, which appeared at temperatures far above the expected superconducting gap.  If we therefore take `pseudogap' to mean a `gap of unknown origin', then STM has measured many pseudogaps in the Fe-based superconductors.  If we restrict ourselves to discussing \textit{repeatable} gaps of unknown origin, then there does seem to be a frequently observed phenomena, manifesting as a sometimes-symmetric set of kinks around $\pm$ 20-50 meV, in a background V-shaped DOS.  However, there is not yet evidence from STM that this phenomenon turns on at any specific temperature higher than $T_c$. The few papers which did track the probable pseudogap energy range through the superconducting $T_c$ showed no evidence of a gap persisting above $T_c$\cite{JinSST2010,MasseeEPL2010}.

The situation in Fe-based superconductors is in some sense the inverse of the situation in cuprates.  In cuprates, there was a clear gap opening at temperatures far above $T_c$, but no obvious electronic ordered state, superconducting or otherwise, to take the blame for that gap.  This led to over a decade of speculation about electron correlation effects, numerous hidden OPs, and the possibility of phase incoherent precursor pairing far above the bulk superconducting transition\cite{TimuskROPP1999,HufnerROPP2008}.  However, in Fe-based superconductors, there are two clear phase transitions following a line in phase space which is suggestively similar to the $T^{*}$ pseudogap in line cuprates. First, there is a structural transition from tetragonal to orthorhombic at $T_s$.  Second (or in some cases simultaneously) there is a magnetic transition to a SDW state (typically an itinerant collinear antiferromagnet) at $T_N$.  This has launched a search for the `pseudogap' which should be associated with these transitions. This seems a bit of a misnomer, because the phase transitions are known, so if their gap is indeed detected, there will be nothing `pseudo' about it.

\begin{figure}[tb]
\begin{center}
  {\includegraphics[width=1\columnwidth,clip]{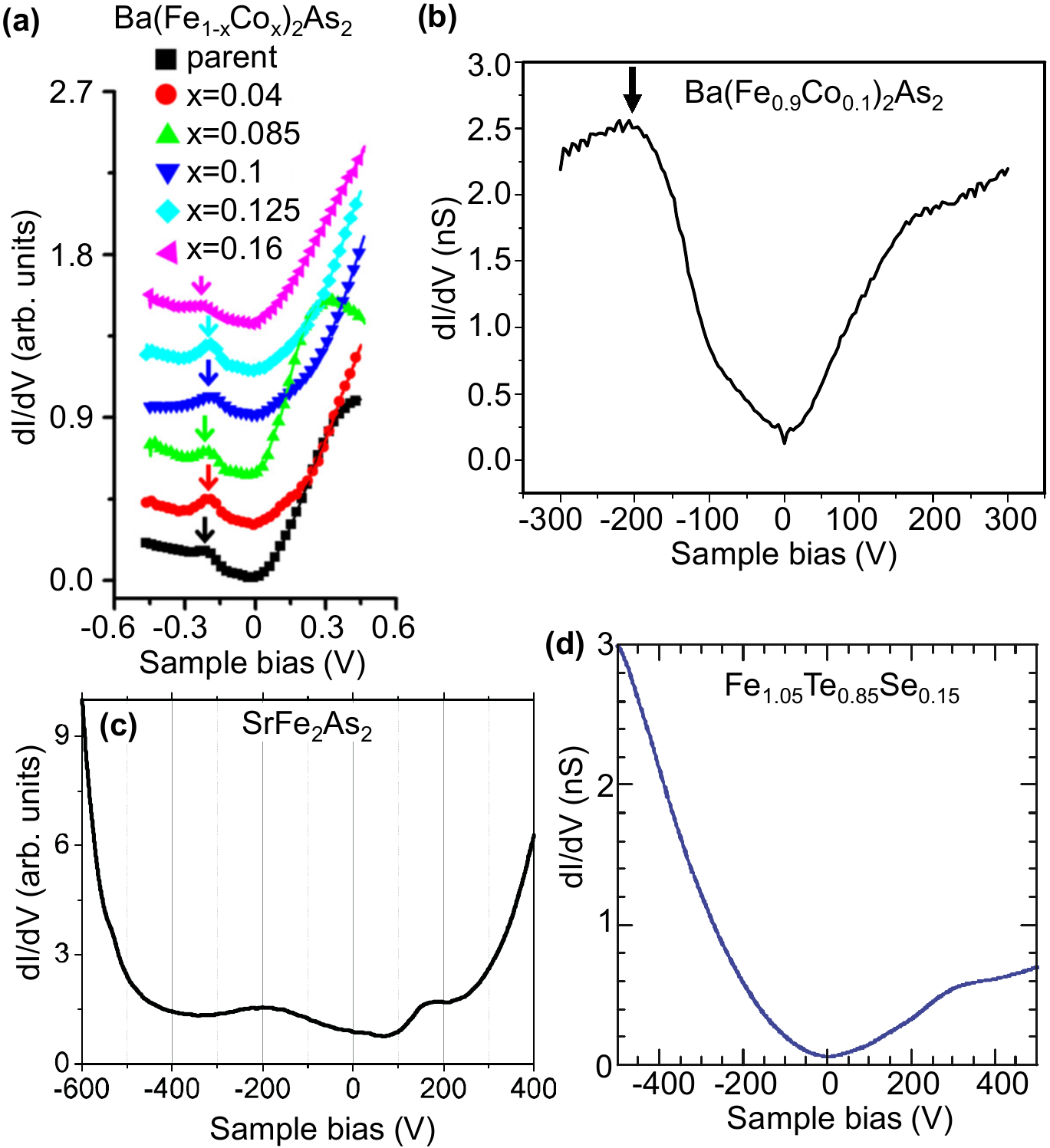}}
  \caption[High energy DOS.]
    {\label{fig:high-energy-DOS}
    High energy $dI/dV$ curves measured on several Fe-based superconductors.  All curves are consistent with a roughly compensated semimetallic DOS, displaying a broad miminum near the Fermi level. (a) $dI/dV$ from a series of 6 compounds Ba(Fe$_{1-x}$Co$_x$)$_2$As$_2$, where $x$ is the nominal concentration (from starting element ratio in the crystal growth process)\cite{ZhangPRB2010}. All curves show a pronounced feature at approximately -200 meV, which has been explained as a surface state, from calculations and ARPES measurements\cite{VanHeumenPRL2011}.  (b) A higher resolution spectrum from Ba(Fe$_{1-x}$Co$_x$)$_2$As$_2$ with nominal $x$=0.1 and $T_c$=25K shows the low-energy superconducting gap as well\cite{YinThesis2009}. (c) Spatially averaged DOS of the parent Sr122 compound\cite{NiestemskiArxiv0906.2761}. This average was taken from a region displaying the $\sqrt{2}\times\sqrt{2}$ structure, but is qualitatively similar to spectra from regions displaying the $2\times 1$ structure. (d) Spatially averaged DOS, measured at $T$=4.2K, from FeTe$_{0.85}$Se$_{0.15}$\cite{KatoPRB2009}. }
\end{center}
\end{figure}

In addition to the known SDW phase transition, there is an even more basic reason to expect a depressed DOS over a broad energy range close to the Fermi level.  The Fe-based superconductors are compensated semimetals, which means that the Fermi level is located just below the top edge of one band and just above the bottom edge of another, meaning that the DOS at the Fermi level itself will be low compared to the DOS farther into either of these bands.  So we should expect that all DOS measurements will be superimposed on a roughly V-shaped or U-shaped background, which will have a minimum roughly at the Fermi level for a compensated semimetal.
All high energy DOS curves have shown this background, as exemplified in figure~\ref{fig:high-energy-DOS}.

In contrast, a gap which opens up due to a new electronic ordered state (such as superconductivity or charge density wave (CDW) or SDW) will conserve states, pushing states out towards the gap edge, causing the appearance of gap edge peaks or coherence peaks.  So the `pseudogap' we seek should be more than just a depression in the DOS; it should open up at a particular $T$ and it should also show some pile-up of states at the gap edge.  Because it is superimposed on the semimetallic background DOS, the pile-up may manifest as a shoulder rather than a peak.

\begin{figure}[tbh]
\begin{center}
  {\includegraphics[width=0.9\columnwidth,clip]{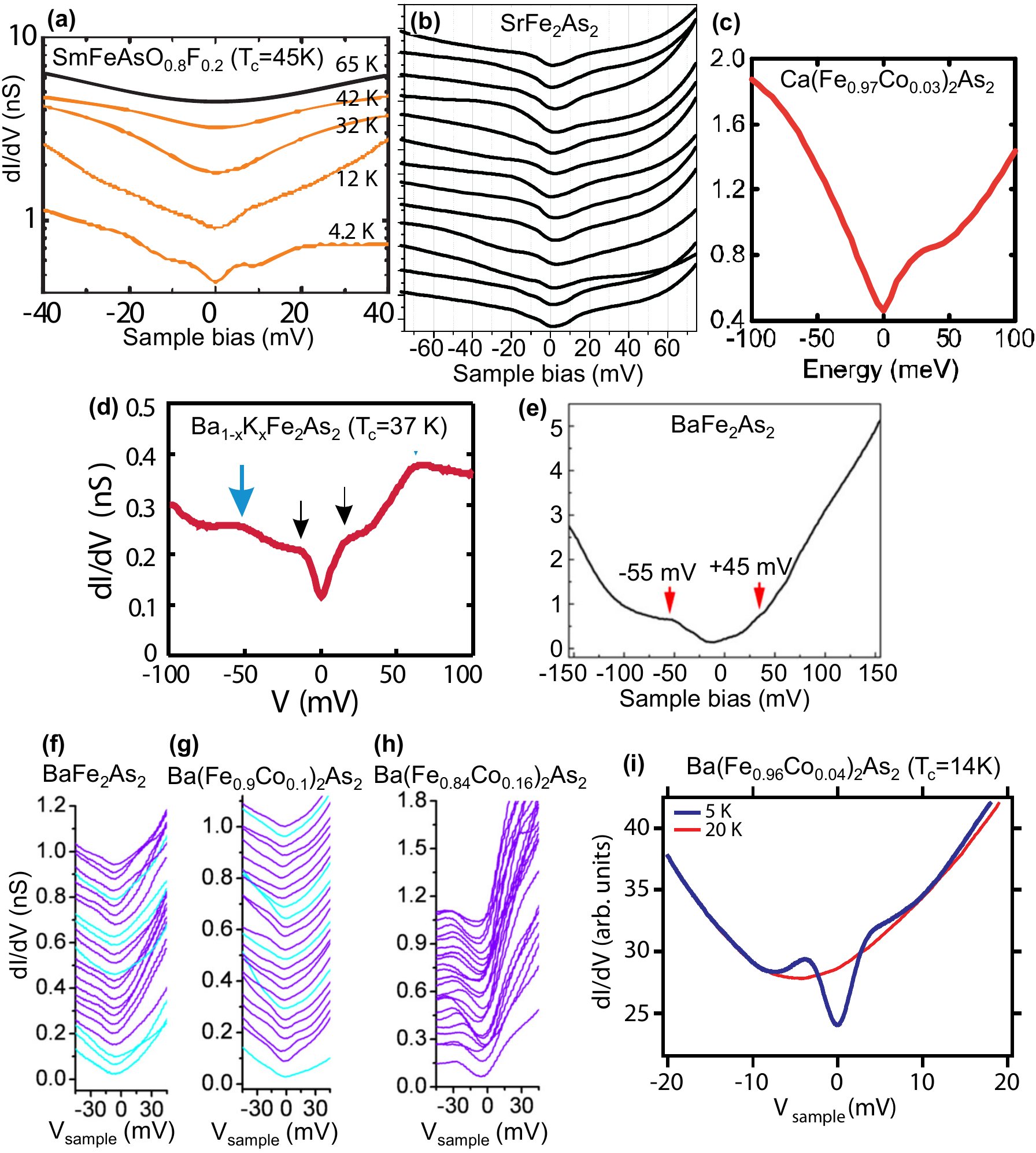}}
  \caption[Pseudogap claims.]
    {\label{fig:pseudogap-DOS}
    Spectra showing pseudogap-like features. (a) Spatially averaged DOS as a function of temperature from SmFeAsO$_{1-x}$F$_{x}$ with nominal $x=0.2$ and $T_c$=45K \cite{FasanoPRL2010}. Low energy feature is identified as the superconducting gap; higher energy feature is identified as a collective mode. (The phase diagram of SmFeAsO$_{1-x}$F$_x$ shows no overlap between the SDW and superconducting states, as shown in figure \ref{fig:phase-diagrams}a from Ref. \onlinecite{KamiharaNJP2010}). (b) A linecut along the parent Sr122 shows a ubiquitous feature at around $\pm$15 meV \cite{NiestemskiArxiv0906.2761}. (c) An average spectrum from Ca(Fe$_{1-x}$Co$_x$)$_2$As$_2$ with measured $x=0.03$\cite{ChuangScience2010}.  This sample is expected to be in the orthorhombic, non-superconducting SDW state. (d) Hole doped Ba$_{1-x}$K$_x$Fe$_2$As$_2$ with $T_c$=37K shows a superconducting gap marked by black arrows, and a second higher energy gap-like feature marked by blue arrows\cite{WrayPRB2008}. (e) BaFe$_2$As$_2$ shows asymmetric features of unknown origin \cite{NascimentoPRL2009}. (f-h) Ba(Fe$_{1-x}$Co$_x$)$_2$As$_2$ from Ref. \onlinecite{ZhangPRB2010}. (f) Parent BaFe$_2$As$_2$ in the orthorhombic state. (g) Ba(Fe$_{1-x}$Co$_x$)$_2$As$_2$ with nominal $x=0.1$ and measured $T_c$=25K. The measured $T_c$ indicates that this sample is not in the orthorhombic SDW phase (see Fig~\ref{fig:phase-diagrams}c and Ref. \onlinecite{NandiPRL2010}). (h) Ba(Fe$_{1-x}$Co$_x$)$_2$As$_2$ with nominal $x=0.16$ and measured $T_c$=9K, unambiguously not in the orthorhombic SDW phase. (i) Underdoped Ba(Fe$_{1-x}$Co$_x$)$_2$As$_2$ with measured $x=0.04$ and $T_c$=14K \cite{MasseeEPL2010}. This sample is expected to be in the orthorhombic SDW phase with $T_{SDW}\sim$70K.  Spectrum acquired at $T$=20K shows the expected disappearance of the superconducting gap, but no evidence of a `pseudogap' within the expected mean-field energy range for temperature $T_{SDW}$. }
\end{center}
\end{figure}

Several studies have shown evidence for a `pile-up' in the DOS, above the superconducting energy scale.  This higher-energy pileup has been interpreted as a second superconducting gap (figure~\ref{fig:Dynes-gap-fit}, c-f)\cite{ShanNatPhys2011,TeaguePRL2011}, the signature of a collective mode (figure~\ref{fig:pseudogap-DOS}a)\cite{FasanoPRL2010}, or a surface state (figure~\ref{fig:pseudogap-DOS}b)\cite{NiestemskiArxiv0906.2761}. In some cases, a higher energy pile-up of states has been observed, but not interpreted (for example, see figure~\ref{fig:FeSe-spectra}a)\cite{HanaguriScience2010}.

Some authors have explicitly labeled their higher energy features as `pseudogaps'. In Fe$_{1.05}$Te$_{0.85}$Se$_{0.15}$, Kato \textit{et al} saw a broad, asymmetric V-shaped DOS out to at least $\pm$500 meV, with a kink around $+300$ meV, which he identified as a `pseudogap feature' (figure~\ref{fig:high-energy-DOS}d)\cite{KatoPRB2009}. In lightly Co-doped Ca122, Chuang showed a V-shaped `pseudogap' spectrum out to $\pm$100 meV, with a metallic (non-zero) DOS at $E_F$ (figure~\ref{fig:pseudogap-DOS}c) \cite{ChuangScience2010}. In K-doped Ba122 with $T_c$=37K, Wray showed a `pseudogap' of $\sim$50 meV (figure~\ref{fig:pseudogap-DOS}d)\cite{WrayPRB2008}.
Other representative examples of DOS features at `pseudogap'-like energies can be seen in figure~\ref{fig:pseudogap-DOS}e-h.

Despite a suggestive title, ``Pseudogap-less high $T_c$ superconductivity in BaCo$_x$Fe$_{2-x}$As$_2$'', Massee did not definitively rule out a pseudogap in Co-doped Ba122\cite{MasseeEPL2010}.  What he did demonstrate was that, unlike in the cuprates\cite{BoyerNatPhys2007}, the apparent $\sim$10\% spatial variation in the energy of the superconducting gap cannot be blamed on the presence of a pseudogap. Using the same normalization technique as Boyer \textit{et al}\cite{BoyerNatPhys2007}, Massee divided the spectra below $T_c$ by the spectra at the same location just above $T_c$. Unlike Boyer, Massee found that the spatial variation in the apparent superconducting gap (defined as the distance between coherence peaks) was not removed by normalization.  However, all 3 samples studied, with $x$=0.08, 0.14, and 0.21, ranging from underdoped to optimally doped, did show a relative depression of DOS near $\varepsilon_F$ in the non-superconducting state spectra.  This depression was most pronounced in the most underdoped sample, as shown in figure \ref{fig:pseudogap-DOS}i.

The spin-density-wave transition in Ba(Fe$_{1-x}$Co$_x$)$_2$As$_2$ ranges from $T_{N}\sim$135K in the parent compound, down to $\sim$20K where the phase boundary enters the superconducting dome and actually folds back slightly at doping $x=0.06$ (see figure~\ref{fig:phase-diagrams}c)\cite{NandiPRL2010}. If the purported `pseudogap' in this system were a mean-field gap due to the SDW, then one would expect the pseudogap to range from $\sim$20 meV in the parent compound, down to $\sim$3meV where it cuts off at $x=0.06$.  In particular, the expected pseudogap would be $\sim$10 meV in Ba(Fe$_{0.96}$Co$_{0.04}$)$_2$As$_2$ studied by Massee, but he saw no evidence of a pseudogap edge out to $\pm$50 meV\cite{MasseeEPL2010}.

In the end, what matters about any DOS feature identified as a `pseudogap', is its relationship to superconductivity.  If it is the signature of a competing state, or a collaborating state, then it is worth understanding how it interacts with superconductivity in momentum space.  However, in these materials, particularly the 1111 and 122 materials which do not cleave nicely, there are likely numerous surface states which may also manifest as features in the DOS.  In our enthusiasm to draw connections with the cuprates, we must be cautious not to confuse these features with a pseudogap.

\section{\label{sec:Vortices}Vortex State}

Magnetic vortices are important both for their relevance to technological applications (vortices must be well-pinned in order to allow high critical current $J_c$), and for their scientific interest as nanoscale windows into the low $T$ normal state. The spatial shape and energetic signature of vortex core states may be used to determine the pairing symmetry. The size of vortex core states may be used to measure the superconducting coherence length $\xi$. STM has now imaged vortices in 122\cite{YinPRL2009,ShanNatPhys2011}, in 11\cite{HanaguriScience2010,SongScience2011}, and in 111 compounds\cite{HanaguriAspen2011}.

\subsection{\label{sec:pinning}Vortex Pinning}

One of the very exciting early findings in Fe-based superconductors was the unexpectedly high critical field $H_{c2}$\cite{HunteNature2008} and strong native vortex pinning\cite{YamamotoAPL2009}.  A summary of the technologically relevant properties of Fe-based superconductors, two years after their discovery, can be found in a review by Putti \cite{PuttiSST2010}.

Although critical current is a bulk property, measurable by bulk techniques, the quest to improve critical current relies on the pinning of vortices, typically by the judicious introduction of nanoscale defects which locally depress superconductivity.  The system may then find the lowest energy state in which vortex cores (which locally destroy superconductivity over the length scale of the coherence length $\xi$) are co-located with the defects, in order to avoid paying twice the energetic cost-of-destruction of superconductivity.
The mechanism of vortex pinning may be investigated in aggregate by bulk critical current measurements, but in the case of unknown distributions of pinning sites, a local probe may be used to investigate the pinning sites of individual vortices\cite{AuslaenderNatPhys2008}.

\begin{figure}[tbh]
\begin{center}
  {\includegraphics[width=.95\columnwidth,clip]{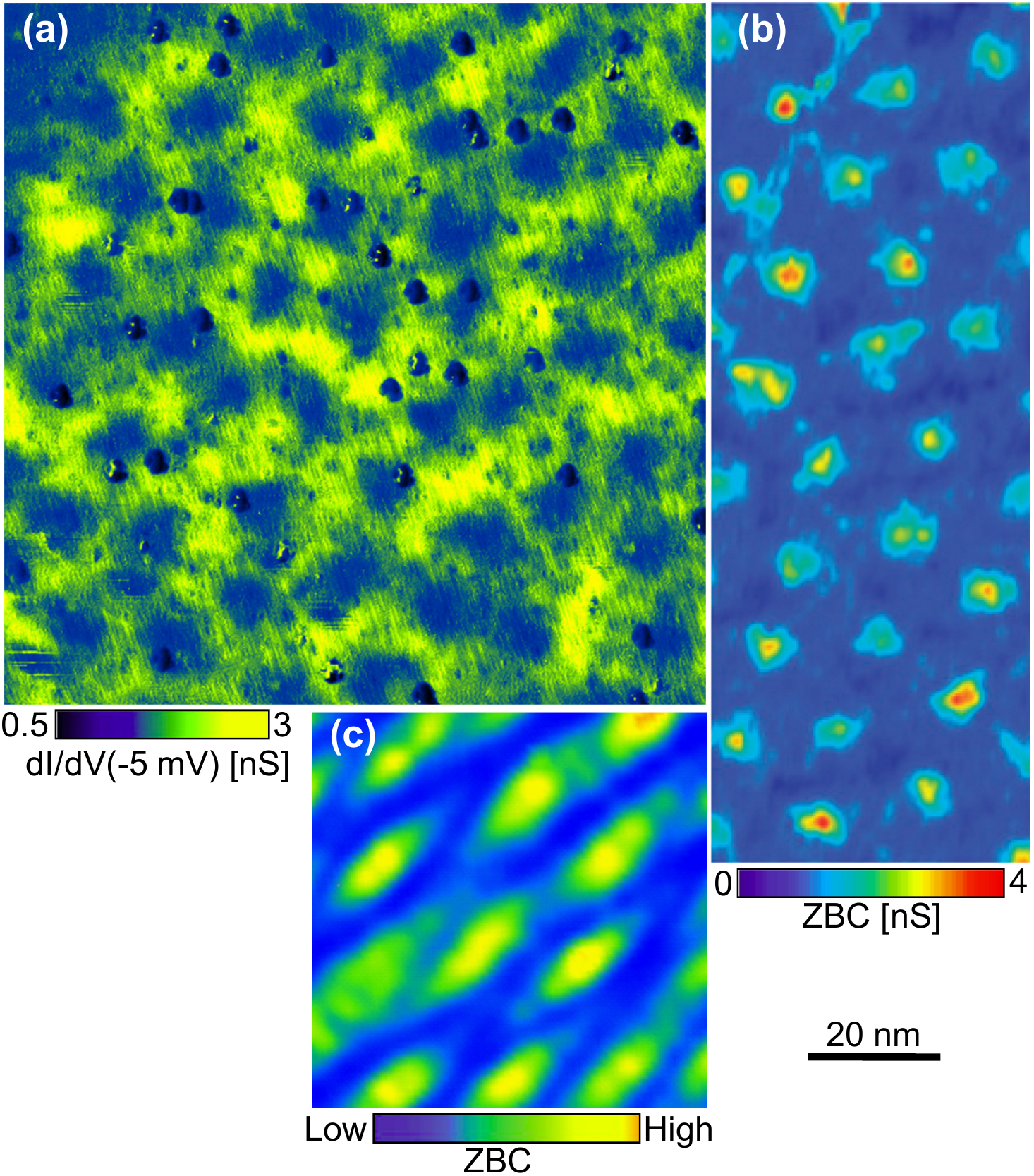}}
  \caption[Images of vortex arrays.]
    {\label{fig:vortex-images}
    (a) 106 nm square image of $dI/dV$ at the approximate coherence peak energy, -5 meV, showing vortices in Ba(Fe$_{1-x}$Co$_x$)$_2$As$_2$ with nominal $x=0.1$ and $T_c$=25K, as imaged at $H$=9T and $T$=6K (setup: $V_{\mathrm{sample}}=-5\,\mathrm{mV}; I_{\mathrm{set}}=10\,\mathrm{pA}$)\cite{YinPRL2009}.  Vortices are the broad blue regions, depressions in the DOS at the coherence peak energy.  Near-surface impurities are also visible as sharper, darker depressions in the DOS at the coherence peak energy. Vortices show no ordered lattice, and no correlation with the locations of surface impurities. (b) 130 nm $\times$ 50 nm image of zero bias conductance, showing vortices in Ba$_{0.6}$K$_{0.4}$Fe$_2$As$_2$ with $T_c$=38K, as imaged at $H$=9T and $T$=2K (setup: $V_{\mathrm{sample}}=100\,\mathrm{mV}; I_{\mathrm{set}}=200\,\mathrm{pA}$)\cite{ShanNatPhys2011}. In this case the vortices show a hexagonal lattice with slight disorder . (c) 60 nm square image of zero bias conductance, showing vortices in FeSe with $T_c\sim$8K, as imaged at $H$=8T and $T$=0.4K (setup: $V_{\mathrm{sample}}=10\,\mathrm{mV}; I_{\mathrm{set}}=100\,\mathrm{pA}$)\cite{SongScience2011}. Here vortices show a hexagonal lattice with slight disorder. Notably, they also show marked anisotropy, at least a factor of $2\times$ difference in size along the orthorhombic $a$ and $b$ axes, while the lattice distortion itself is only 0.5\%.}
\end{center}
\end{figure}

The first STM images of vortices in an Fe-based superconductor, optimally electron-doped Ba(Fe$_{0.9}$Co$_{0.1}$)$_2$As$_2$ (figure \ref{fig:vortex-images}a)\cite{YinPRL2009} showed a disordered array, indicating that vortex pinning forces trumped the inter-vortex interactions which would drive vortices to form an ordered lattice.  In the presence of strong pinning sites, the vortex arrangement with respect to those sites depends on the anisotropy of the material. In highly anisotropic superconductors, a one-dimensional vortex line may split like a stack of pancakes into point-like objects with the freedom to move independently in each superconducting layer\cite{ClemPRB1991}. In this scenario, pancakes may find pinning sites independently in each layer, resulting in a high correlation between observed vortex and pinning locations in any given layer. This scenario is realized in Bi2212\cite{PanPRL2000}. In a more isotropic superconductor with strong pinning sites, the vortices must remain as line objects, and can bend only slightly between layers to maximize their overlap with point impurities throughout the bulk. In this scenario, there may be very little observable correlation between vortex locations and impurities in any single layer. This latter scenario was observed by Yin in Ba(Fe$_{0.9}$Co$_{0.1}$)$_2$As$_2$ (figure~\ref{fig:vortex-images}a)\cite{YinPhysicaC2009}, giving evidence for strong pinning and low anisotropy in electron-doped Ba122.  In contrast, the vortices in hole-doped Ba122 were seen to form a hexagonal lattice, indicating weaker pinning in this material (figure~\ref{fig:vortex-images}b)\cite{ShanNatPhys2011}.  This supports the possibility that the Co atoms themselves, doped directly into the FeAs layer of Ba122 (unlike the K atoms which are doped into the Ba layer), may act as the strong pinning sites in electron-doped Ba122.

As would be expected, in clean FeSe (fewer than 1 defect in 70,000 Se sites), vortices were also seen to form a hexagonal lattice (figure~\ref{fig:vortex-images}c)\cite{SongScience2011}.

Unpublished conductance images by Hanaguri \textit{et al} on nominally stoichiometric LiFeAs show a disordered vortex arrangement, indicating strong pinning\cite{HanaguriAspen2011}.  However, the topographic images of Hanaguri's LiFeAs surface showed a native impurity concentration corresponding to $\sim2$\% of the Fe sites.

The superconducting coherence length $\xi$ is related to the upper critical field by $H_{c2}=\Phi_0/2\pi\xi^2$. In optimal electron-doped Ba122 ($T_c$=25K), Yin \textit{et al} measured a coherence length of $\xi=2.9$nm, corresponding to an upper critical field of $H_{c2}=\Phi_0/2\pi\xi^2=43$T\cite{YinPRL2009}.  In optimal hole-doped Ba122 ($T_c$=38K), Shan \textit{et al} measured a coherence length of $\xi=2.2$nm, corresponding to $H_{c2}=75$T\cite{ShanNatPhys2011}. Therefore, electron-doped Ba122 has stronger pinning at moderate fields, whereas hole-doped Ba122 seems to have a larger upper critical field.

\subsection{\label{sec:cores}Vortex Core States}

In a conventional $s$-wave superconductor, theory predicts and experiment confirms that the destruction of superconductivity in the vortex core will result in quasiparticle bound states at energy $\frac{1}{2}\Delta^2/\varepsilon_F$\cite{CaroliPhysLett1964}, which appear as a zero bias peak in the DOS\cite{HessPRL1989} that splits into two symmetric peaks and eventually merges into the coherence peaks on moving away from the vortex center\cite{HessPRL1990}.  In $d$-wave cuprate superconductors, particle-hole symmetric subgap states have been observed with energies approximately $\pm\Delta/4$\cite{PanPRL2000,Maggio-AprilePRL1995}.

Yin \textit{et al} found that in electron-doped Ba122, the vortex core destroyed the superconducting gap and coherence peaks, leaving behind a larger V-shaped background\cite{YinPRL2009}. But the cores lacked any of the sub-gap peaks predicted or observed in conventional $s$-wave or $d$-wave superconductors.  This experimental null result was verified by S.~H.~Pan, but has not been published (cited as a private communication in Ref.\onlinecite{ShanNatPhys2011}).

\begin{figure}[tb]
\begin{center}
  {\includegraphics[width=.95\columnwidth,clip]{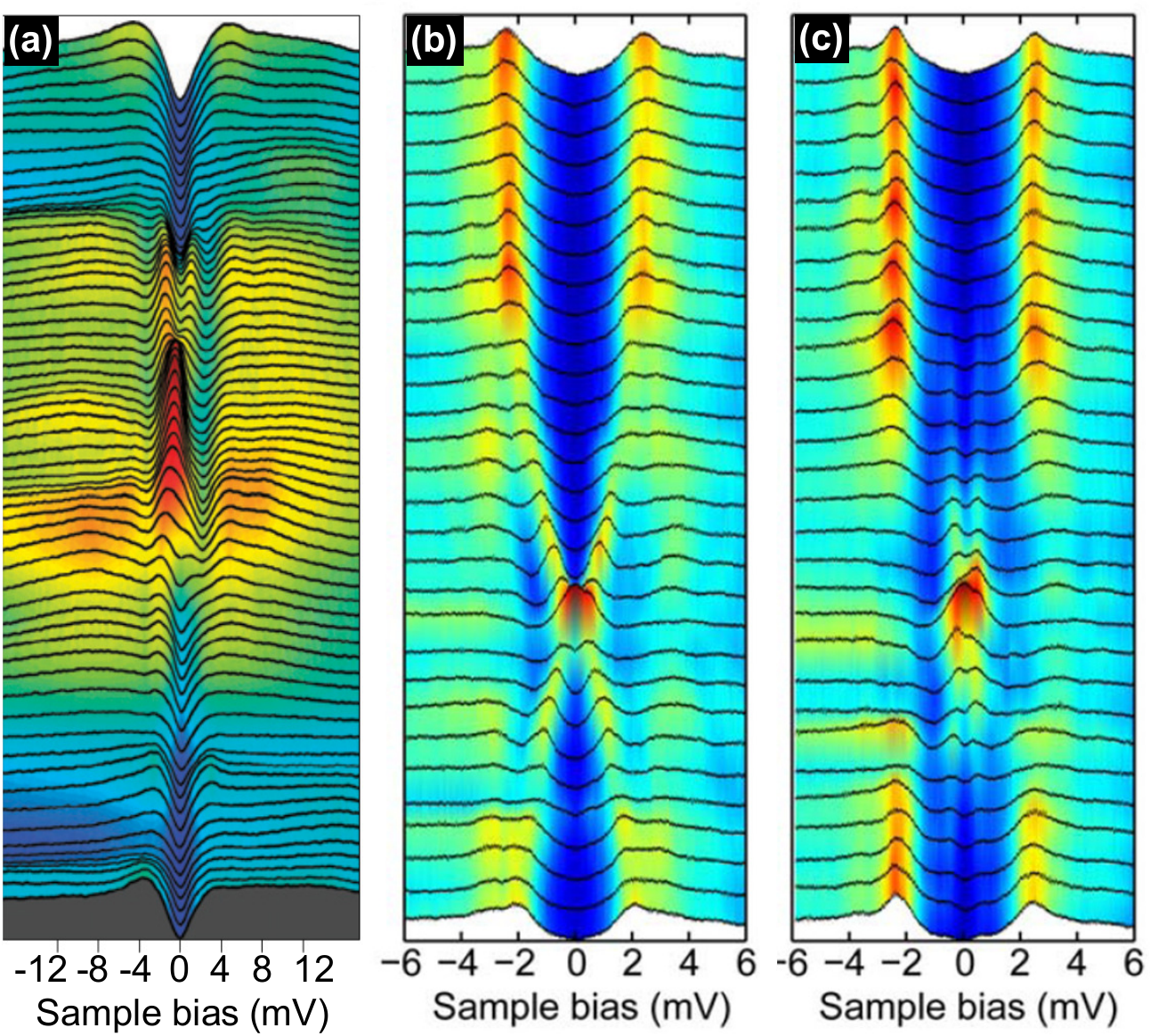}}
  \caption[Linecuts through vortices.]
    {\label{fig:vortex-linecuts}
    (a) $dI/dV$ spectra along a 104\AA\ trajectory through the center of a single vortex in Ba$_{0.6}$K$_{0.4}$Fe$_2$As$_2$, at $H$=9T, and $T$=2K\cite{ShanNatPhys2011}. (b) and (c) $dI/dV$ spectra, spaced 2nm apart on $a$-axis and $b$-axis trajectories through the center of a single vortex in FeSe at $H$=1T and $T$=0.4K\cite{SongScience2011}.}
\end{center}
\end{figure}

In contrast, STM studies of optimally hole-doped Ba122 showed a near-$\varepsilon_F$ peak in the DOS at the vortex center, which split and merged with the coherence peaks away from the vortex\cite{ShanNatPhys2011} (figure \ref{fig:vortex-linecuts}a). However, the vortex center peak was slightly offset from the Fermi level, as the authors suggest might be expected in the quantum limit in which $T/T_c \leq 1/(k_F \xi)$\cite{HayashiPRL1998}.  This phenomenology is very similar to that seen long ago in NbSe$_2$\cite{HessPRL1990}.

In unpublished work on LiFeAs\cite{HanaguriAspen2011}, Hanaguri also showed a vortex core peak in the DOS, offset from $\varepsilon_F$, which split and evolved towards the coherence peaks away from the vortex center.  Presumably, LiFeAs too was in the quantum limit at the 1.5K measurement temperature. Furthermore, the vortices in LiFeAs appeared star-shaped, with arms extending along either the Fe-Fe bond or diagonal direction, depending on the energy.

STM studies of FeSe showed a vortex core state exactly at the Fermi level. Along one orthorhombic axis, the peak split, moved to higher energy, and eventually merged into the coherence peaks on a linecut away from the vortex center, as shown in figure \ref{fig:vortex-linecuts}b. Behavior along the other orthorhombic axis was less straightforward, shown in figure \ref{fig:vortex-linecuts}c. In FeSe, the coherence length $\xi$ appeared much longer, so the measurements were probably not in the quantum limit where the vortex core state would be offset from zero energy.

A number of theoretical efforts have aimed to predict and explain vortex core states in Fe-based SCs, and particularly to discern whether the observed structure of the core states can be used to elucidate the pairing symmetry.

Vortex state calculations were performed on a 2-band model in the $d_{x^2-y^2}$ and $s_{x^2y^2}$ pairing states\cite{HuPRB2009}, and the Bogoliubov-deGennes equations were solved self-consistently by exact diagonalization. Resonance core states near the Fermi energy were found for both $d$ and $s$ pairing states. For the $d_{x^2-y^2}$ pairing, the states are localized while for the $s_{x^2y^2}$ pairing symmetry, they can evolve from sharp, energy-asymmetric localized states broad into extended ones with varying electron filling factor. To explain the non-observation of core states by Yin \textit{et al} in Ba(Fe$_{1-x}$Co$_x$)$_2$As$_2$, Hu \textit{et al} show that the resonant state is suppressed by an SDW, which may be a global coexisting phase (as in the underdoped materials) or may be locally induced by the vortex itself.

Similarly, Jiang \textit{et al} showed by self-consistently solving the Bogoliubov-de Gennes equations based on a two-orbital model, that the magnetic field can induce a SDW state\cite{JiangPRB2009} in and around the vortex cores.  Within the more favorable $s\pm$ pairing, Jiang \textit{et al} found that there are 2 kinds of vortices, those which induce SDW order, and those which don't, separated by a critical value of the Hund's coupling $J_c$. When present, the SDW state expands the vortex cores slightly, and suppresses the near-$\varepsilon_F$ core state.

Ara\'{u}jo \textit{et al} further considered the structure of a single vortex in two different 2-band models with $s\pm$ pairing\cite{AraujoNJP2009}. In a continuum model, they find a vortex core state which peaks at the Fermi level, while in a tight-binding model they find that the peak deviates from the Fermi level by an energy which depends on the band filling.  Interestingly, they find that an
impurity located outside the vortex core has little effect on the LDOS peak, but an
attractive impurity close to the vortex core can almost suppress the vortex core state and shift the peak to negative energy.

Wang \textit{et al} calculated the vortex-core electronic structure with both in-phase $s$-wave ($s_{++}$) and antiphase $s$-wave ($s_{+-}$) pairing, using four different tight-binding models\cite{WangPRB2010}.  They found a bound state LDOS peak at the core center, which splits away from
the center and eventually merges to the gap edges, in qualitative agreement with experimental data from Ba$_{0.6}$K$_{0.4}$Fe$_2$As$_2$\cite{ShanNatPhys2011}. They found that the sign of
the peak energy $\omega_p$ is positive/negative if the Fermi level is near
the bottom/top of the electron/hole bands, but that the peak energy is
insensitive to the relative phase of the pairing gaps.
The observed bound state in Ba$_{0.6}$K$_{0.4}$Fe$_2$As$_2$\cite{ShanNatPhys2011} is therefore compatible, although not exclusively, with $s_{+-}$ pairing.
According to their calculations, the bound state should also be observed in
Ba(Fe$_{1-x}$Co$_x$)$_2$As$_2$, but they attributed its non-observation\cite{YinPRL2009} to the Co dopants directly in the FeAs layers, which may act as in-plane scattering centers, smearing out the bound state.  (Later authors\cite{GaoPRL2011} noted that Wang's models were not self-consistent, and that the chemical potentials used for all 4 tight-binding models corresponded to undoped or electron-doped compounds.)

Gao \textit{et al} investigated the vortex states in Ba$_{1-x}$K$_{x}$Fe$_2$As$_2$ by solving Bogoliubov–de Gennes
equations based on a phenomenological 2-band model with extended $s$-wave
superconductivity and a competing SDW state\cite{GaoPRL2011}. Their result for the optimally doped compound without induced SDW is in qualitative agreement
with the STM data on Ba$_{0.6}$K$_{0.4}$Fe$_2$As$_2$\cite{ShanNatPhys2011}. Similar to Hu \textit{et al}\cite{HuPRB2009} and Jiang \textit{et al}\cite{JiangPRB2009}, they find that the SDW (present in underdoped samples) will reduce the intensity of the in-gap peak in the local DOS and transfer
the spectral weight to form additional peaks outside the gap. Furthermore, the SDW will slightly enlarge the vortex cores, and reduce the rotational symmetry of the states from $C4$ to $C2$. Like Jiang \textit{et al}, they speculate that an SDW state may be induced in the vortex cores of Ba(Fe$_{1-x}$Co$_x$)$_2$As$_2$, thereby suppressing the vortex core states.

In an effort to intepret Hanaguri's images of anisotropic vortex core states in LiFeAs\cite{HanaguriAspen2011}, the shape of the vortex core states was calculated by Wang \textit{et al}\cite{VekhterArxiv1111.0126}.  One expects the vortex core states to decay exponentially with length $\xi_0 = v_F/\pi\Delta$, where $v_F$ is the Fermi velocity and $\Delta$ is the gap amplitude. In the direction of a gap node, the decay will follow a power law. However, Wang found that the geometry of the vortex core state is strongly influenced by the normal state band structure. If the gap itself is not highly anisotropic, the anisotropy of $v_F$ dominates, preventing direct observation of superconducting gap features.

\section{\label{sec:Conclusion}Conclusions \& Future}

STM has made significant contributions to the understanding of Fe-based superconductors. Spectroscopy and QPI imaging have provided a phase-sensitive determination of the superconducting OP.  After several years of debate about the existence of nodes in the superconducting gap, STM has provided clarification by demonstrating the evolution of the OP from nodal to nodeless upon doping a single material\cite{SongScience2011,HanaguriScience2010}. STM has also shown a $C4\rightarrow C2$ symmetry breaking in the parent and superconducting states, highlighting the role of alternative and possibly competing electronic orders in these materials\cite{ChuangScience2010,ZhouPRL2011}.  Finally, STM has directly imaged the vortex state, as a route to better understanding and applying these exciting new superconductors\cite{YinPRL2009,ShanNatPhys2011,ShanPRB2011,SongScience2011,HanaguriAspen2011}. Following these accomplishments, there remain several important open questions, to which STM is an ideal tool to provide answers.

What is the nature of the $C2$ symmetry and its relation to superconductivity?  QPI imaging and analysis throughout the entire BZ should be coupled with careful determination of the local orthorhombic orientation\cite{ChuangScience2010}. Most Fe-based superconductors are non-stoichiometric, and therefore may have inhomogeneity in their electronic anisotropy and strength of the superconducting state. Song's elongated vortex cores give a local measure of coherence length anisotropy\cite{SongScience2011}. Real-space images of electronic `nematogens', and QPI-derived information about their scattering properties may give another measure of local anisotropy\cite{ChuangScience2010,AllanPreprint2011}. STM should be used to measure and correlate both the degree of local electronic anisotropy and the local superconducting energy gap.

How does superconductivity interact with spin order?  Spin-polarized STM (SP-STM) should be used to image local structures such as vortices or impurities which may pin or disrupt the spin order.
Measures of broken $C4$ symmetry should be combined with spin-polarized scanning tunneling spectroscopy, to understand the energy scales of this broken symmetry state.

How are vortices pinned in these materials?  Larger area studies will be needed, with better identification of local impurities.

We must also continue local studies of some of the most tantalizing new materials such as (K,Tl,Rb,Cs)$_x$Fe$_{2-y}$Se$_2$, which challenge three-year-old beliefs of Fe-based superconductivity, such as the necessity of nesting, and the deleterious effects of strong magnetic moments in the superconducting state.  Bulk investigations of this material have been confusing due to possible inhomogeneity and difficulty controlling the ordering of the non-stoichiometric Fe. Conflicting reports claim bulk coexistence of superconductivity with antiferromagnetism\cite{LiuEPL2011}, or phase separation\cite{FangEPL2011}.

In conclusion, Fe-based superconductors are an extraordinarily rich group of materials.  Compared to the cuprates, their metallic ground state, relative malleability and electronic isotropy (typically less than a factor of two), large upper critical field and strong native vortex pinning may all make them more useful for applications in bulk\cite{PuttiSST2010} or thin film form\cite{HiramatsuJPSJ2012}. Three years after the discovery of $T_c$=26K in LaFeAsO$_{1-x}$F$_x$, new materials are still being discovered at a rapid rate\cite{SahaArxiv1105.4798}, which may lead to higher $T_c$ yet in Fe-based materials. They are intriguingly similar to the higher-$T_c$ cuprates, and serve as a foil which has hastened the discovery of broader underlying principles across even farther flung families of unconventional superconductors\cite{Doiron-LeyraudArxiv0905.0964}.  With multiple pairing symmetries within the same subfamily, as well as multiple temperature scales for magnetic sublattice ordering, structural transition, AF ordering, and superconductivity within the same material, Fe-based superconductors are a complex playground in which to understand how various electronic and spin orders may support or inhibit superconductivity.

\textbf{\textit{Acknowledgements}} The author thanks Milan Allan, Jim Analytis, Dan Dessau, Eric Hudson, Freek Massee, Vidya Madhavan, Ivan Maggio-Aprile, Igor Mazin, Ward Plummer, Erik van Heumen, and Ming Yi for helpful discussions.  This work was supported by the Sloan Foundation, the National Science Foundation (DMR-0847433), and the Air Force Office of Scientific Research (FA9550-06-1-0531).


\begin{thebibliography}{229}%
\makeatletter
\providecommand \@ifxundefined [1]{%
 \@ifx{#1\undefined}
}%
\providecommand \@ifnum [1]{%
 \ifnum #1\expandafter \@firstoftwo
 \else \expandafter \@secondoftwo
 \fi
}%
\providecommand \@ifx [1]{%
 \ifx #1\expandafter \@firstoftwo
 \else \expandafter \@secondoftwo
 \fi
}%
\providecommand \natexlab [1]{#1}%
\providecommand \enquote  [1]{``#1''}%
\providecommand \bibnamefont  [1]{#1}%
\providecommand \bibfnamefont [1]{#1}%
\providecommand \citenamefont [1]{#1}%
\providecommand \href@noop [0]{\@secondoftwo}%
\providecommand \href [0]{\begingroup \@sanitize@url \@href}%
\providecommand \@href[1]{\@@startlink{#1}\@@href}%
\providecommand \@@href[1]{\endgroup#1\@@endlink}%
\providecommand \@sanitize@url [0]{\catcode `\\12\catcode `\$12\catcode
  `\&12\catcode `\#12\catcode `\^12\catcode `\_12\catcode `\%12\relax}%
\providecommand \@@startlink[1]{}%
\providecommand \@@endlink[0]{}%
\providecommand \url  [0]{\begingroup\@sanitize@url \@url }%
\providecommand \@url [1]{\endgroup\@href {#1}{\urlprefix }}%
\providecommand \urlprefix  [0]{URL }%
\providecommand \Eprint [0]{\href }%
\providecommand \doibase [0]{http://dx.doi.org/}%
\providecommand \selectlanguage [0]{\@gobble}%
\providecommand \bibinfo  [0]{\@secondoftwo}%
\providecommand \bibfield  [0]{\@secondoftwo}%
\providecommand \translation [1]{[#1]}%
\providecommand \BibitemOpen [0]{}%
\providecommand \bibitemStop [0]{}%
\providecommand \bibitemNoStop [0]{.\EOS\space}%
\providecommand \EOS [0]{\spacefactor3000\relax}%
\providecommand \BibitemShut  [1]{\csname bibitem#1\endcsname}%
\let\auto@bib@innerbib\@empty
\bibitem [{\citenamefont {Kamihara}\ \emph {et~al.}(2008)\citenamefont
  {Kamihara}, \citenamefont {Watanabe}, \citenamefont {Hirano},\ and\
  \citenamefont {Hosono}}]{KamiharaJACS2008}%
  \BibitemOpen
  \bibfield  {author} {\bibinfo {author} {\bibfnamefont {Y.}~\bibnamefont
  {Kamihara}}, \bibinfo {author} {\bibfnamefont {T.}~\bibnamefont {Watanabe}},
  \bibinfo {author} {\bibfnamefont {M.}~\bibnamefont {Hirano}}, \ and\ \bibinfo
  {author} {\bibfnamefont {H.}~\bibnamefont {Hosono}},\ }\href {\doibase
  10.1021/ja800073m} {\bibfield  {journal} {\bibinfo  {journal} {Journal of the
  American Chemical Society}\ }\textbf {\bibinfo {volume} {130}},\ \bibinfo
  {pages} {3296} (\bibinfo {year} {2008})}\BibitemShut {NoStop}%
\bibitem [{\citenamefont {Fischer}\ \emph {et~al.}(2007)\citenamefont
  {Fischer}, \citenamefont {Kugler}, \citenamefont {Maggio-Aprile},
  \citenamefont {Berthod},\ and\ \citenamefont {Renner}}]{FischerRMP2007}%
  \BibitemOpen
  \bibfield  {author} {\bibinfo {author} {\bibfnamefont {{\O}.}~\bibnamefont
  {Fischer}}, \bibinfo {author} {\bibfnamefont {M.}~\bibnamefont {Kugler}},
  \bibinfo {author} {\bibfnamefont {I.}~\bibnamefont {Maggio-Aprile}}, \bibinfo
  {author} {\bibfnamefont {C.}~\bibnamefont {Berthod}}, \ and\ \bibinfo
  {author} {\bibfnamefont {C.}~\bibnamefont {Renner}},\ }\href {\doibase
  10.1103/RevModPhys.79.353} {\bibfield  {journal} {\bibinfo  {journal}
  {Reviews of Modern Physics}\ }\textbf {\bibinfo {volume} {79}},\ \bibinfo
  {pages} {353} (\bibinfo {year} {2007})}\BibitemShut {NoStop}%
\bibitem [{\citenamefont {Yin}\ \emph {et~al.}(2009{\natexlab{a}})\citenamefont
  {Yin}, \citenamefont {Zech}, \citenamefont {Williams},\ and\ \citenamefont
  {Hoffman}}]{YinPhysicaC2009}%
  \BibitemOpen
  \bibfield  {author} {\bibinfo {author} {\bibfnamefont {Y.}~\bibnamefont
  {Yin}}, \bibinfo {author} {\bibfnamefont {M.}~\bibnamefont {Zech}}, \bibinfo
  {author} {\bibfnamefont {T.~L.}\ \bibnamefont {Williams}}, \ and\ \bibinfo
  {author} {\bibfnamefont {J.~E.}\ \bibnamefont {Hoffman}},\ }\href {\doibase
  10.1016/j.physc.2009.03.053} {\bibfield  {journal} {\bibinfo  {journal}
  {Physica C}\ }\textbf {\bibinfo {volume} {469}},\ \bibinfo {pages} {535}
  (\bibinfo {year} {2009}{\natexlab{a}})}\BibitemShut {NoStop}%
\bibitem [{\citenamefont {Johnston}(2010)}]{JohnstonAdvPhys2010}%
  \BibitemOpen
  \bibfield  {author} {\bibinfo {author} {\bibfnamefont {D.}~\bibnamefont
  {Johnston}},\ }\href {\doibase 10.1080/00018732.2010.513480} {\bibfield
  {journal} {\bibinfo  {journal} {Advances in Physics}\ }\textbf {\bibinfo
  {volume} {59}},\ \bibinfo {pages} {803} (\bibinfo {year} {2010})}\BibitemShut
  {NoStop}%
\bibitem [{\citenamefont {Stewart}(2011)}]{StewartRMP2011}%
  \BibitemOpen
  \bibfield  {author} {\bibinfo {author} {\bibfnamefont {G.}~\bibnamefont
  {Stewart}},\ }\href {\doibase 10.1103/RevModPhys.83.1589} {\bibfield
  {journal} {\bibinfo  {journal} {Reviews of Modern Physics}\ }\textbf
  {\bibinfo {volume} {83}},\ \bibinfo {pages} {1589} (\bibinfo {year}
  {2011})}\BibitemShut {NoStop}%
\bibitem [{\citenamefont {Binnig}\ \emph {et~al.}(1982)\citenamefont {Binnig},
  \citenamefont {Rohrer}, \citenamefont {Gerber},\ and\ \citenamefont
  {Weibel}}]{BinnigPRL1982}%
  \BibitemOpen
  \bibfield  {author} {\bibinfo {author} {\bibfnamefont {G.}~\bibnamefont
  {Binnig}}, \bibinfo {author} {\bibfnamefont {H.}~\bibnamefont {Rohrer}},
  \bibinfo {author} {\bibfnamefont {C.}~\bibnamefont {Gerber}}, \ and\ \bibinfo
  {author} {\bibfnamefont {E.}~\bibnamefont {Weibel}},\ }\href {\doibase
  10.1103/PhysRevLett.49.57} {\bibfield  {journal} {\bibinfo  {journal}
  {Physical Review Letters}\ }\textbf {\bibinfo {volume} {49}},\ \bibinfo
  {pages} {57} (\bibinfo {year} {1982})}\BibitemShut {NoStop}%
\bibitem [{\citenamefont {Chen}(2007)}]{ChenBook2007}%
  \BibitemOpen
  \bibfield  {author} {\bibinfo {author} {\bibfnamefont {C.~J.}\ \bibnamefont
  {Chen}},\ }\href@noop {} {\emph {\bibinfo {title} {{Introduction to Scanning
  Tunneling Microscopy}}}},\ \bibinfo {edition} {2nd}\ ed.\ (\bibinfo
  {publisher} {Oxford University Press},\ \bibinfo {year} {2007})\BibitemShut
  {NoStop}%
\bibitem [{\citenamefont {Stroscio}\ and\ \citenamefont
  {Kaiser}(1993)}]{StroscioBook1994}%
  \BibitemOpen
  \bibinfo {editor} {\bibfnamefont {J.}~\bibnamefont {Stroscio}}\ and\ \bibinfo
  {editor} {\bibfnamefont {W.}~\bibnamefont {Kaiser}},\ eds.,\ \href@noop {}
  {\emph {\bibinfo {title} {{Scanning Tunneling Microscopy}}}}\ (\bibinfo
  {publisher} {Academic Press},\ \bibinfo {year} {1993})\BibitemShut {NoStop}%
\bibitem [{\citenamefont {Wiesendanger}(1994)}]{WiesendangerBook1994}%
  \BibitemOpen
  \bibfield  {author} {\bibinfo {author} {\bibfnamefont {R.}~\bibnamefont
  {Wiesendanger}},\ }\href@noop {} {\emph {\bibinfo {title} {{Scanning probe
  microscopy and spectroscopy}}}}\ (\bibinfo  {publisher} {Cambridge University
  Press},\ \bibinfo {year} {1994})\BibitemShut {NoStop}%
\bibitem [{\citenamefont {Bardeen}(1961)}]{BardeenPRL1961}%
  \BibitemOpen
  \bibfield  {author} {\bibinfo {author} {\bibfnamefont {J.}~\bibnamefont
  {Bardeen}},\ }\href {\doibase 10.1103/PhysRevLett.6.57} {\bibfield  {journal}
  {\bibinfo  {journal} {Physical Review Letters}\ }\textbf {\bibinfo {volume}
  {6}},\ \bibinfo {pages} {57} (\bibinfo {year} {1961})}\BibitemShut {NoStop}%
\bibitem [{\citenamefont {Tersoff}\ and\ \citenamefont
  {Hamann}(1983)}]{TersoffPRL1983}%
  \BibitemOpen
  \bibfield  {author} {\bibinfo {author} {\bibfnamefont {J.}~\bibnamefont
  {Tersoff}}\ and\ \bibinfo {author} {\bibfnamefont {D.~R.}\ \bibnamefont
  {Hamann}},\ }\href {\doibase 10.1103/PhysRevLett.50.1998} {\bibfield
  {journal} {\bibinfo  {journal} {Physical Review Letters}\ }\textbf {\bibinfo
  {volume} {50}},\ \bibinfo {pages} {1998} (\bibinfo {year}
  {1983})}\BibitemShut {NoStop}%
\bibitem [{\citenamefont {Haynes}(2011)}]{CRC2011}%
  \BibitemOpen
  \bibinfo {editor} {\bibfnamefont {W.~M.}\ \bibnamefont {Haynes}},\ ed.,\
  \href {http://www.hbcpnetbase.com//articles/12\_21\_86.pdf} {\emph {\bibinfo
  {title} {{Handbook of Chemistry and Physics}}}},\ \bibinfo {edition} {92nd}\
  ed.\ (\bibinfo  {publisher} {CRC Press},\ \bibinfo {year} {2011})\ pp.\
  \bibinfo {pages} {12--114}\BibitemShut {NoStop}%
\bibitem [{\citenamefont {Hewson}(1993)}]{HewsonBook1993}%
  \BibitemOpen
  \bibfield  {author} {\bibinfo {author} {\bibfnamefont {A.~C.}\ \bibnamefont
  {Hewson}},\ }\href@noop {} {\emph {\bibinfo {title} {{The Kondo Problem to
  Heavy Fermions}}}},\ \bibinfo {edition} {2nd}\ ed.\ (\bibinfo  {publisher}
  {Cambridge University Press},\ \bibinfo {address} {Cambridge, UK},\ \bibinfo
  {year} {1993})\ pp.\ \bibinfo {pages} {4--8}\BibitemShut {NoStop}%
\bibitem [{\citenamefont {Crommie}\ \emph {et~al.}(1993)\citenamefont
  {Crommie}, \citenamefont {Lutz},\ and\ \citenamefont
  {Eigler}}]{CrommieNature1993}%
  \BibitemOpen
  \bibfield  {author} {\bibinfo {author} {\bibfnamefont {M.~F.}\ \bibnamefont
  {Crommie}}, \bibinfo {author} {\bibfnamefont {C.~P.}\ \bibnamefont {Lutz}}, \
  and\ \bibinfo {author} {\bibfnamefont {D.~M.}\ \bibnamefont {Eigler}},\
  }\href {\doibase 10.1038/363524a0} {\bibfield  {journal} {\bibinfo  {journal}
  {Nature}\ }\textbf {\bibinfo {volume} {363}},\ \bibinfo {pages} {524}
  (\bibinfo {year} {1993})}\BibitemShut {NoStop}%
\bibitem [{\citenamefont {Hasegawa}\ and\ \citenamefont
  {Avouris}(1993)}]{HasegawaPRL1993}%
  \BibitemOpen
  \bibfield  {author} {\bibinfo {author} {\bibfnamefont {Y.}~\bibnamefont
  {Hasegawa}}\ and\ \bibinfo {author} {\bibfnamefont {P.}~\bibnamefont
  {Avouris}},\ }\href {\doibase 10.1103/PhysRevLett.71.1071} {\bibfield
  {journal} {\bibinfo  {journal} {Physical Review Letters}\ }\textbf {\bibinfo
  {volume} {71}},\ \bibinfo {pages} {1071} (\bibinfo {year}
  {1993})}\BibitemShut {NoStop}%
\bibitem [{\citenamefont {Wittneven}\ \emph {et~al.}(1998)\citenamefont
  {Wittneven}, \citenamefont {Dombrowski}, \citenamefont {Morgenstern},\ and\
  \citenamefont {Wiesendanger}}]{WittnevenPRL1998}%
  \BibitemOpen
  \bibfield  {author} {\bibinfo {author} {\bibfnamefont {C.}~\bibnamefont
  {Wittneven}}, \bibinfo {author} {\bibfnamefont {R.}~\bibnamefont
  {Dombrowski}}, \bibinfo {author} {\bibfnamefont {M.}~\bibnamefont
  {Morgenstern}}, \ and\ \bibinfo {author} {\bibfnamefont {R.}~\bibnamefont
  {Wiesendanger}},\ }\href {\doibase 10.1103/PhysRevLett.81.5616} {\bibfield
  {journal} {\bibinfo  {journal} {Physical Review Letters}\ }\textbf {\bibinfo
  {volume} {81}},\ \bibinfo {pages} {5616} (\bibinfo {year}
  {1998})}\BibitemShut {NoStop}%
\bibitem [{\citenamefont {B\"{u}rgi}\ \emph {et~al.}(2000)\citenamefont
  {B\"{u}rgi}, \citenamefont {Brune}, \citenamefont {Jeandupeux},\ and\
  \citenamefont {Kern}}]{BurgiJESRP2000}%
  \BibitemOpen
  \bibfield  {author} {\bibinfo {author} {\bibfnamefont {L.}~\bibnamefont
  {B\"{u}rgi}}, \bibinfo {author} {\bibfnamefont {H.}~\bibnamefont {Brune}},
  \bibinfo {author} {\bibfnamefont {O.}~\bibnamefont {Jeandupeux}}, \ and\
  \bibinfo {author} {\bibfnamefont {K.}~\bibnamefont {Kern}},\ }\href {\doibase
  10.1016/S0368-2048(00)00105-5} {\bibfield  {journal} {\bibinfo  {journal}
  {Journal of Electron Spectroscopy and Related Phenomena}\ }\textbf {\bibinfo
  {volume} {109}},\ \bibinfo {pages} {33} (\bibinfo {year} {2000})}\BibitemShut
  {NoStop}%
\bibitem [{\citenamefont {Fujita}\ \emph {et~al.}(1999)\citenamefont {Fujita},
  \citenamefont {Amemiya}, \citenamefont {Yakabe}, \citenamefont {Nejoh},
  \citenamefont {Sato},\ and\ \citenamefont {Iwatsuki}}]{FujitaSurfSci1999}%
  \BibitemOpen
  \bibfield  {author} {\bibinfo {author} {\bibfnamefont {D.}~\bibnamefont
  {Fujita}}, \bibinfo {author} {\bibfnamefont {K.}~\bibnamefont {Amemiya}},
  \bibinfo {author} {\bibfnamefont {T.}~\bibnamefont {Yakabe}}, \bibinfo
  {author} {\bibfnamefont {H.}~\bibnamefont {Nejoh}}, \bibinfo {author}
  {\bibfnamefont {T.}~\bibnamefont {Sato}}, \ and\ \bibinfo {author}
  {\bibfnamefont {M.}~\bibnamefont {Iwatsuki}},\ }\href {\doibase
  10.1016/S0039-6028(98)00886-3} {\bibfield  {journal} {\bibinfo  {journal}
  {Surface Science}\ }\textbf {\bibinfo {volume} {423}},\ \bibinfo {pages}
  {160} (\bibinfo {year} {1999})}\BibitemShut {NoStop}%
\bibitem [{\citenamefont {Petersen}\ \emph {et~al.}(2000)\citenamefont
  {Petersen}, \citenamefont {Hofmann}, \citenamefont {Plummer},\ and\
  \citenamefont {Besenbacher}}]{PetersenJESRP2000}%
  \BibitemOpen
  \bibfield  {author} {\bibinfo {author} {\bibfnamefont {L.}~\bibnamefont
  {Petersen}}, \bibinfo {author} {\bibfnamefont {P.}~\bibnamefont {Hofmann}},
  \bibinfo {author} {\bibfnamefont {E.~W.}\ \bibnamefont {Plummer}}, \ and\
  \bibinfo {author} {\bibfnamefont {F.}~\bibnamefont {Besenbacher}},\ }\href
  {\doibase 10.1016/S0368-2048(00)00110-9} {\bibfield  {journal} {\bibinfo
  {journal} {Journal of Electron Spectroscopy and Related Phenomena}\ }\textbf
  {\bibinfo {volume} {109}},\ \bibinfo {pages} {97} (\bibinfo {year}
  {2000})}\BibitemShut {NoStop}%
\bibitem [{\citenamefont {Kanisawa}\ \emph {et~al.}(2001)\citenamefont
  {Kanisawa}, \citenamefont {Butcher}, \citenamefont {Yamaguchi},\ and\
  \citenamefont {Hirayama}}]{KanisawaPRL2001}%
  \BibitemOpen
  \bibfield  {author} {\bibinfo {author} {\bibfnamefont {K.}~\bibnamefont
  {Kanisawa}}, \bibinfo {author} {\bibfnamefont {M.}~\bibnamefont {Butcher}},
  \bibinfo {author} {\bibfnamefont {H.}~\bibnamefont {Yamaguchi}}, \ and\
  \bibinfo {author} {\bibfnamefont {Y.}~\bibnamefont {Hirayama}},\ }\href
  {\doibase 10.1103/PhysRevLett.86.3384} {\bibfield  {journal} {\bibinfo
  {journal} {Physical Review Letters}\ }\textbf {\bibinfo {volume} {86}},\
  \bibinfo {pages} {3384} (\bibinfo {year} {2001})}\BibitemShut {NoStop}%
\bibitem [{\citenamefont {Hoffman}\ \emph {et~al.}(2002)\citenamefont
  {Hoffman}, \citenamefont {McElroy}, \citenamefont {Lee}, \citenamefont
  {Lang}, \citenamefont {Eisaki}, \citenamefont {Uchida},\ and\ \citenamefont
  {Davis}}]{HoffmanScience2002b}%
  \BibitemOpen
  \bibfield  {author} {\bibinfo {author} {\bibfnamefont {J.~E.}\ \bibnamefont
  {Hoffman}}, \bibinfo {author} {\bibfnamefont {K.}~\bibnamefont {McElroy}},
  \bibinfo {author} {\bibfnamefont {D.-H.}\ \bibnamefont {Lee}}, \bibinfo
  {author} {\bibfnamefont {K.~M.}\ \bibnamefont {Lang}}, \bibinfo {author}
  {\bibfnamefont {H.}~\bibnamefont {Eisaki}}, \bibinfo {author} {\bibfnamefont
  {S.}~\bibnamefont {Uchida}}, \ and\ \bibinfo {author} {\bibfnamefont {J.~C.}\
  \bibnamefont {Davis}},\ }\href {\doibase 10.1126/science.1072640} {\bibfield
  {journal} {\bibinfo  {journal} {Science}\ }\textbf {\bibinfo {volume}
  {297}},\ \bibinfo {pages} {1148} (\bibinfo {year} {2002})}\BibitemShut
  {NoStop}%
\bibitem [{\citenamefont {Zhang}\ \emph
  {et~al.}(2009{\natexlab{a}})\citenamefont {Zhang}, \citenamefont {Brar},
  \citenamefont {Girit}, \citenamefont {Zettl},\ and\ \citenamefont
  {Crommie}}]{ZhangNatPhys2009}%
  \BibitemOpen
  \bibfield  {author} {\bibinfo {author} {\bibfnamefont {Y.}~\bibnamefont
  {Zhang}}, \bibinfo {author} {\bibfnamefont {V.~W.}\ \bibnamefont {Brar}},
  \bibinfo {author} {\bibfnamefont {C.}~\bibnamefont {Girit}}, \bibinfo
  {author} {\bibfnamefont {A.}~\bibnamefont {Zettl}}, \ and\ \bibinfo {author}
  {\bibfnamefont {M.~F.}\ \bibnamefont {Crommie}},\ }\href {\doibase
  10.1038/nphys1365} {\bibfield  {journal} {\bibinfo  {journal} {Nature
  Physics}\ }\textbf {\bibinfo {volume} {5}},\ \bibinfo {pages} {722} (\bibinfo
  {year} {2009}{\natexlab{a}})}\BibitemShut {NoStop}%
\bibitem [{\citenamefont {Roushan}\ \emph {et~al.}(2009)\citenamefont
  {Roushan}, \citenamefont {Seo}, \citenamefont {Parker}, \citenamefont {Hor},
  \citenamefont {Hsieh}, \citenamefont {Qian}, \citenamefont {Richardella},
  \citenamefont {Hasan}, \citenamefont {Cava},\ and\ \citenamefont
  {Yazdani}}]{RoushanNature2009}%
  \BibitemOpen
  \bibfield  {author} {\bibinfo {author} {\bibfnamefont {P.}~\bibnamefont
  {Roushan}}, \bibinfo {author} {\bibfnamefont {J.}~\bibnamefont {Seo}},
  \bibinfo {author} {\bibfnamefont {C.~V.}\ \bibnamefont {Parker}}, \bibinfo
  {author} {\bibfnamefont {Y.~S.}\ \bibnamefont {Hor}}, \bibinfo {author}
  {\bibfnamefont {D.}~\bibnamefont {Hsieh}}, \bibinfo {author} {\bibfnamefont
  {D.}~\bibnamefont {Qian}}, \bibinfo {author} {\bibfnamefont {A.}~\bibnamefont
  {Richardella}}, \bibinfo {author} {\bibfnamefont {M.~Z.}\ \bibnamefont
  {Hasan}}, \bibinfo {author} {\bibfnamefont {R.~J.}\ \bibnamefont {Cava}}, \
  and\ \bibinfo {author} {\bibfnamefont {A.}~\bibnamefont {Yazdani}},\ }\href
  {\doibase 10.1038/nature08308} {\bibfield  {journal} {\bibinfo  {journal}
  {Nature}\ }\textbf {\bibinfo {volume} {460}},\ \bibinfo {pages} {1106}
  (\bibinfo {year} {2009})}\BibitemShut {NoStop}%
\bibitem [{\citenamefont {Byers}\ \emph {et~al.}(1993)\citenamefont {Byers},
  \citenamefont {Flatt\'{e}},\ and\ \citenamefont {Scalapino}}]{ByersPRL1993}%
  \BibitemOpen
  \bibfield  {author} {\bibinfo {author} {\bibfnamefont {J.}~\bibnamefont
  {Byers}}, \bibinfo {author} {\bibfnamefont {M.}~\bibnamefont {Flatt\'{e}}}, \
  and\ \bibinfo {author} {\bibfnamefont {D.}~\bibnamefont {Scalapino}},\ }\href
  {\doibase 10.1103/PhysRevLett.71.3363} {\bibfield  {journal} {\bibinfo
  {journal} {Physical Review Letters}\ }\textbf {\bibinfo {volume} {71}},\
  \bibinfo {pages} {3363} (\bibinfo {year} {1993})}\BibitemShut {NoStop}%
\bibitem [{\citenamefont {Wang}\ \emph {et~al.}(2009)\citenamefont {Wang},
  \citenamefont {Zhai},\ and\ \citenamefont {Lee}}]{WangEPL2009}%
  \BibitemOpen
  \bibfield  {author} {\bibinfo {author} {\bibfnamefont {F.}~\bibnamefont
  {Wang}}, \bibinfo {author} {\bibfnamefont {H.}~\bibnamefont {Zhai}}, \ and\
  \bibinfo {author} {\bibfnamefont {D.-H.}\ \bibnamefont {Lee}},\ }\href
  {\doibase 10.1209/0295-5075/85/37005} {\bibfield  {journal} {\bibinfo
  {journal} {Europhysics Letters}\ }\textbf {\bibinfo {volume} {85}},\ \bibinfo
  {pages} {37005} (\bibinfo {year} {2009})}\BibitemShut {NoStop}%
\bibitem [{\citenamefont {Hanaguri}\ \emph {et~al.}(2009)\citenamefont
  {Hanaguri}, \citenamefont {Kohsaka}, \citenamefont {Ono}, \citenamefont
  {Maltseva}, \citenamefont {Coleman}, \citenamefont {Yamada}, \citenamefont
  {Azuma}, \citenamefont {Takano}, \citenamefont {Ohishi},\ and\ \citenamefont
  {Takagi}}]{HanaguriScience2009}%
  \BibitemOpen
  \bibfield  {author} {\bibinfo {author} {\bibfnamefont {T.}~\bibnamefont
  {Hanaguri}}, \bibinfo {author} {\bibfnamefont {Y.}~\bibnamefont {Kohsaka}},
  \bibinfo {author} {\bibfnamefont {M.}~\bibnamefont {Ono}}, \bibinfo {author}
  {\bibfnamefont {M.}~\bibnamefont {Maltseva}}, \bibinfo {author}
  {\bibfnamefont {P.}~\bibnamefont {Coleman}}, \bibinfo {author} {\bibfnamefont
  {I.}~\bibnamefont {Yamada}}, \bibinfo {author} {\bibfnamefont
  {M.}~\bibnamefont {Azuma}}, \bibinfo {author} {\bibfnamefont
  {M.}~\bibnamefont {Takano}}, \bibinfo {author} {\bibfnamefont
  {K.}~\bibnamefont {Ohishi}}, \ and\ \bibinfo {author} {\bibfnamefont
  {H.}~\bibnamefont {Takagi}},\ }\href {\doibase 10.1126/science.1166138}
  {\bibfield  {journal} {\bibinfo  {journal} {Science}\ }\textbf {\bibinfo
  {volume} {323}},\ \bibinfo {pages} {923} (\bibinfo {year}
  {2009})}\BibitemShut {NoStop}%
\bibitem [{\citenamefont {Hanaguri}\ \emph {et~al.}(2010)\citenamefont
  {Hanaguri}, \citenamefont {Niitaka}, \citenamefont {Kuroki},\ and\
  \citenamefont {Takagi}}]{HanaguriScience2010}%
  \BibitemOpen
  \bibfield  {author} {\bibinfo {author} {\bibfnamefont {T.}~\bibnamefont
  {Hanaguri}}, \bibinfo {author} {\bibfnamefont {S.}~\bibnamefont {Niitaka}},
  \bibinfo {author} {\bibfnamefont {K.}~\bibnamefont {Kuroki}}, \ and\ \bibinfo
  {author} {\bibfnamefont {H.}~\bibnamefont {Takagi}},\ }\href {\doibase
  10.1126/science.1187399} {\bibfield  {journal} {\bibinfo  {journal}
  {Science}\ }\textbf {\bibinfo {volume} {328}},\ \bibinfo {pages} {474}
  (\bibinfo {year} {2010})}\BibitemShut {NoStop}%
\bibitem [{\citenamefont {Hanaguri}\ \emph {et~al.}(2007)\citenamefont
  {Hanaguri}, \citenamefont {Kohsaka}, \citenamefont {Davis}, \citenamefont
  {Lupien}, \citenamefont {Yamada}, \citenamefont {Azuma}, \citenamefont
  {Takano}, \citenamefont {Ohishi}, \citenamefont {Ono},\ and\ \citenamefont
  {Takagi}}]{HanaguriNatPhys2007}%
  \BibitemOpen
  \bibfield  {author} {\bibinfo {author} {\bibfnamefont {T.}~\bibnamefont
  {Hanaguri}}, \bibinfo {author} {\bibfnamefont {Y.}~\bibnamefont {Kohsaka}},
  \bibinfo {author} {\bibfnamefont {J.~C.}\ \bibnamefont {Davis}}, \bibinfo
  {author} {\bibfnamefont {C.}~\bibnamefont {Lupien}}, \bibinfo {author}
  {\bibfnamefont {I.}~\bibnamefont {Yamada}}, \bibinfo {author} {\bibfnamefont
  {M.}~\bibnamefont {Azuma}}, \bibinfo {author} {\bibfnamefont
  {M.}~\bibnamefont {Takano}}, \bibinfo {author} {\bibfnamefont
  {K.}~\bibnamefont {Ohishi}}, \bibinfo {author} {\bibfnamefont
  {M.}~\bibnamefont {Ono}}, \ and\ \bibinfo {author} {\bibfnamefont
  {H.}~\bibnamefont {Takagi}},\ }\href {\doibase 10.1038/nphys753} {\bibfield
  {journal} {\bibinfo  {journal} {Nature Physics}\ }\textbf {\bibinfo {volume}
  {3}},\ \bibinfo {pages} {865} (\bibinfo {year} {2007})}\BibitemShut {NoStop}%
\bibitem [{\citenamefont {Damascelli}(2004)}]{DamascelliPhysScript2004}%
  \BibitemOpen
  \bibfield  {author} {\bibinfo {author} {\bibfnamefont {A.}~\bibnamefont
  {Damascelli}},\ }\href {\doibase 10.1238/Physica.Topical.109a00061}
  {\bibfield  {journal} {\bibinfo  {journal} {Physica Scripta}\ }\textbf
  {\bibinfo {volume} {T109}},\ \bibinfo {pages} {61} (\bibinfo {year}
  {2004})}\BibitemShut {NoStop}%
\bibitem [{\citenamefont {Capriotti}\ \emph {et~al.}(2003)\citenamefont
  {Capriotti}, \citenamefont {Scalapino},\ and\ \citenamefont
  {Sedgewick}}]{CapriottiPRB2003}%
  \BibitemOpen
  \bibfield  {author} {\bibinfo {author} {\bibfnamefont {L.}~\bibnamefont
  {Capriotti}}, \bibinfo {author} {\bibfnamefont {D.}~\bibnamefont
  {Scalapino}}, \ and\ \bibinfo {author} {\bibfnamefont {R.}~\bibnamefont
  {Sedgewick}},\ }\href {\doibase 10.1103/PhysRevB.68.014508} {\bibfield
  {journal} {\bibinfo  {journal} {Physical Review B}\ }\textbf {\bibinfo
  {volume} {68}},\ \bibinfo {pages} {014508} (\bibinfo {year}
  {2003})}\BibitemShut {NoStop}%
\bibitem [{\citenamefont {Hanke}\ \emph {et~al.}()\citenamefont {Hanke},
  \citenamefont {Sykora}, \citenamefont {Schlegel}, \citenamefont {Baumann},
  \citenamefont {Harnagea}, \citenamefont {Wurmehl}, \citenamefont {Daghofer},
  \citenamefont {B\"{u}chner}, \citenamefont {van~den Brink},\ and\
  \citenamefont {Hess}}]{HankeArxiv1106.4217}%
  \BibitemOpen
  \bibfield  {author} {\bibinfo {author} {\bibfnamefont {T.}~\bibnamefont
  {Hanke}}, \bibinfo {author} {\bibfnamefont {S.}~\bibnamefont {Sykora}},
  \bibinfo {author} {\bibfnamefont {R.}~\bibnamefont {Schlegel}}, \bibinfo
  {author} {\bibfnamefont {D.}~\bibnamefont {Baumann}}, \bibinfo {author}
  {\bibfnamefont {L.}~\bibnamefont {Harnagea}}, \bibinfo {author}
  {\bibfnamefont {S.}~\bibnamefont {Wurmehl}}, \bibinfo {author} {\bibfnamefont
  {M.}~\bibnamefont {Daghofer}}, \bibinfo {author} {\bibfnamefont
  {B.}~\bibnamefont {B\"{u}chner}}, \bibinfo {author} {\bibfnamefont
  {J.}~\bibnamefont {van~den Brink}}, \ and\ \bibinfo {author} {\bibfnamefont
  {C.}~\bibnamefont {Hess}},\ }\href@noop {} {\ }\Eprint
  {http://arxiv.org/abs/1106.4217} {arXiv:1106.4217} \BibitemShut {NoStop}%
\bibitem [{\citenamefont {Markiewicz}(2004)}]{MarkiewiczPRB2004}%
  \BibitemOpen
  \bibfield  {author} {\bibinfo {author} {\bibfnamefont {R.}~\bibnamefont
  {Markiewicz}},\ }\href {\doibase 10.1103/PhysRevB.69.214517} {\bibfield
  {journal} {\bibinfo  {journal} {Physical Review B}\ }\textbf {\bibinfo
  {volume} {69}},\ \bibinfo {pages} {214517} (\bibinfo {year}
  {2004})}\BibitemShut {NoStop}%
\bibitem [{\citenamefont {McElroy}\ \emph {et~al.}(2006)\citenamefont
  {McElroy}, \citenamefont {Gweon}, \citenamefont {Zhou}, \citenamefont {Graf},
  \citenamefont {Uchida}, \citenamefont {Eisaki}, \citenamefont {Takagi},
  \citenamefont {Sasagawa}, \citenamefont {Lee},\ and\ \citenamefont
  {Lanzara}}]{McElroyPRL2006}%
  \BibitemOpen
  \bibfield  {author} {\bibinfo {author} {\bibfnamefont {K.}~\bibnamefont
  {McElroy}}, \bibinfo {author} {\bibfnamefont {G.-H.}\ \bibnamefont {Gweon}},
  \bibinfo {author} {\bibfnamefont {S.}~\bibnamefont {Zhou}}, \bibinfo {author}
  {\bibfnamefont {J.}~\bibnamefont {Graf}}, \bibinfo {author} {\bibfnamefont
  {S.}~\bibnamefont {Uchida}}, \bibinfo {author} {\bibfnamefont
  {H.}~\bibnamefont {Eisaki}}, \bibinfo {author} {\bibfnamefont
  {H.}~\bibnamefont {Takagi}}, \bibinfo {author} {\bibfnamefont
  {T.}~\bibnamefont {Sasagawa}}, \bibinfo {author} {\bibfnamefont {D.-H.}\
  \bibnamefont {Lee}}, \ and\ \bibinfo {author} {\bibfnamefont
  {A.}~\bibnamefont {Lanzara}},\ }\href {\doibase
  10.1103/PhysRevLett.96.067005} {\bibfield  {journal} {\bibinfo  {journal}
  {Physical Review Letters}\ }\textbf {\bibinfo {volume} {96}},\ \bibinfo
  {pages} {067005} (\bibinfo {year} {2006})}\BibitemShut {NoStop}%
\bibitem [{\citenamefont {Chatterjee}\ \emph {et~al.}(2006)\citenamefont
  {Chatterjee}, \citenamefont {Shi}, \citenamefont {Kaminski}, \citenamefont
  {Kanigel}, \citenamefont {Fretwell}, \citenamefont {Terashima}, \citenamefont
  {Takahashi}, \citenamefont {Rosenkranz}, \citenamefont {Li}, \citenamefont
  {Raffy}, \citenamefont {Santander-Syro}, \citenamefont {Kadowaki},
  \citenamefont {Norman}, \citenamefont {Randeria},\ and\ \citenamefont
  {Campuzano}}]{ChatterjeePRL2006}%
  \BibitemOpen
  \bibfield  {author} {\bibinfo {author} {\bibfnamefont {U.}~\bibnamefont
  {Chatterjee}}, \bibinfo {author} {\bibfnamefont {M.}~\bibnamefont {Shi}},
  \bibinfo {author} {\bibfnamefont {A.}~\bibnamefont {Kaminski}}, \bibinfo
  {author} {\bibfnamefont {A.}~\bibnamefont {Kanigel}}, \bibinfo {author}
  {\bibfnamefont {H.}~\bibnamefont {Fretwell}}, \bibinfo {author}
  {\bibfnamefont {K.}~\bibnamefont {Terashima}}, \bibinfo {author}
  {\bibfnamefont {T.}~\bibnamefont {Takahashi}}, \bibinfo {author}
  {\bibfnamefont {S.}~\bibnamefont {Rosenkranz}}, \bibinfo {author}
  {\bibfnamefont {Z.}~\bibnamefont {Li}}, \bibinfo {author} {\bibfnamefont
  {H.}~\bibnamefont {Raffy}}, \bibinfo {author} {\bibfnamefont
  {A.}~\bibnamefont {Santander-Syro}}, \bibinfo {author} {\bibfnamefont
  {K.}~\bibnamefont {Kadowaki}}, \bibinfo {author} {\bibfnamefont
  {M.}~\bibnamefont {Norman}}, \bibinfo {author} {\bibfnamefont
  {M.}~\bibnamefont {Randeria}}, \ and\ \bibinfo {author} {\bibfnamefont
  {J.}~\bibnamefont {Campuzano}},\ }\href {\doibase
  10.1103/PhysRevLett.96.107006} {\bibfield  {journal} {\bibinfo  {journal}
  {Physical Review Letters}\ }\textbf {\bibinfo {volume} {96}},\ \bibinfo
  {pages} {107006} (\bibinfo {year} {2006})}\BibitemShut {NoStop}%
\bibitem [{\citenamefont {Hashimoto}\ \emph {et~al.}(2011)\citenamefont
  {Hashimoto}, \citenamefont {He}, \citenamefont {Testaud}, \citenamefont
  {Meevasana}, \citenamefont {Moore}, \citenamefont {Lu}, \citenamefont
  {Yoshida}, \citenamefont {Eisaki}, \citenamefont {Devereaux}, \citenamefont
  {Hussain},\ and\ \citenamefont {Shen}}]{HashimotoPRL2011}%
  \BibitemOpen
  \bibfield  {author} {\bibinfo {author} {\bibfnamefont {M.}~\bibnamefont
  {Hashimoto}}, \bibinfo {author} {\bibfnamefont {R.-H.}\ \bibnamefont {He}},
  \bibinfo {author} {\bibfnamefont {J.}~\bibnamefont {Testaud}}, \bibinfo
  {author} {\bibfnamefont {W.}~\bibnamefont {Meevasana}}, \bibinfo {author}
  {\bibfnamefont {R.}~\bibnamefont {Moore}}, \bibinfo {author} {\bibfnamefont
  {D.}~\bibnamefont {Lu}}, \bibinfo {author} {\bibfnamefont {Y.}~\bibnamefont
  {Yoshida}}, \bibinfo {author} {\bibfnamefont {H.}~\bibnamefont {Eisaki}},
  \bibinfo {author} {\bibfnamefont {T.}~\bibnamefont {Devereaux}}, \bibinfo
  {author} {\bibfnamefont {Z.}~\bibnamefont {Hussain}}, \ and\ \bibinfo
  {author} {\bibfnamefont {Z.-X.}\ \bibnamefont {Shen}},\ }\href {\doibase
  10.1103/PhysRevLett.106.167003} {\bibfield  {journal} {\bibinfo  {journal}
  {Physical Review Letters}\ }\textbf {\bibinfo {volume} {106}},\ \bibinfo
  {pages} {167003} (\bibinfo {year} {2011})}\BibitemShut {NoStop}%
\bibitem [{\citenamefont {Wang}\ and\ \citenamefont {Lee}(2003)}]{WangPRB2003}%
  \BibitemOpen
  \bibfield  {author} {\bibinfo {author} {\bibfnamefont {Q.-H.}\ \bibnamefont
  {Wang}}\ and\ \bibinfo {author} {\bibfnamefont {D.-H.}\ \bibnamefont {Lee}},\
  }\href {\doibase 10.1103/PhysRevB.67.020511} {\bibfield  {journal} {\bibinfo
  {journal} {Physical Review B}\ }\textbf {\bibinfo {volume} {67}},\ \bibinfo
  {pages} {020511} (\bibinfo {year} {2003})}\BibitemShut {NoStop}%
\bibitem [{\citenamefont {Zhang}\ and\ \citenamefont
  {Ting}(2003)}]{ZhangPRB2003}%
  \BibitemOpen
  \bibfield  {author} {\bibinfo {author} {\bibfnamefont {D.}~\bibnamefont
  {Zhang}}\ and\ \bibinfo {author} {\bibfnamefont {C.}~\bibnamefont {Ting}},\
  }\href {\doibase 10.1103/PhysRevB.67.100506} {\bibfield  {journal} {\bibinfo
  {journal} {Physical Review B}\ }\textbf {\bibinfo {volume} {67}},\ \bibinfo
  {pages} {100506} (\bibinfo {year} {2003})}\BibitemShut {NoStop}%
\bibitem [{\citenamefont {Pereg-Barnea}\ and\ \citenamefont
  {Franz}(2003)}]{Pereg-BarneaPRB2003}%
  \BibitemOpen
  \bibfield  {author} {\bibinfo {author} {\bibfnamefont {T.}~\bibnamefont
  {Pereg-Barnea}}\ and\ \bibinfo {author} {\bibfnamefont {M.}~\bibnamefont
  {Franz}},\ }\href {\doibase 10.1103/PhysRevB.68.180506} {\bibfield  {journal}
  {\bibinfo  {journal} {Physical Review B}\ }\textbf {\bibinfo {volume} {68}},\
  \bibinfo {pages} {180506} (\bibinfo {year} {2003})}\BibitemShut {NoStop}%
\bibitem [{\citenamefont {Zhu}\ \emph {et~al.}(2004)\citenamefont {Zhu},
  \citenamefont {Atkinson},\ and\ \citenamefont {Hirschfeld}}]{ZhuPRB2004}%
  \BibitemOpen
  \bibfield  {author} {\bibinfo {author} {\bibfnamefont {L.}~\bibnamefont
  {Zhu}}, \bibinfo {author} {\bibfnamefont {W.}~\bibnamefont {Atkinson}}, \
  and\ \bibinfo {author} {\bibfnamefont {P.}~\bibnamefont {Hirschfeld}},\
  }\href {\doibase 10.1103/PhysRevB.69.060503} {\bibfield  {journal} {\bibinfo
  {journal} {Physical Review B}\ }\textbf {\bibinfo {volume} {69}},\ \bibinfo
  {pages} {060503} (\bibinfo {year} {2004})}\BibitemShut {NoStop}%
\bibitem [{\citenamefont {Nunner}\ \emph {et~al.}(2006)\citenamefont {Nunner},
  \citenamefont {Chen}, \citenamefont {Andersen}, \citenamefont {Melikyan},\
  and\ \citenamefont {Hirschfeld}}]{NunnerPRB2006}%
  \BibitemOpen
  \bibfield  {author} {\bibinfo {author} {\bibfnamefont {T.}~\bibnamefont
  {Nunner}}, \bibinfo {author} {\bibfnamefont {W.}~\bibnamefont {Chen}},
  \bibinfo {author} {\bibfnamefont {B.}~\bibnamefont {Andersen}}, \bibinfo
  {author} {\bibfnamefont {A.}~\bibnamefont {Melikyan}}, \ and\ \bibinfo
  {author} {\bibfnamefont {P.}~\bibnamefont {Hirschfeld}},\ }\href {\doibase
  10.1103/PhysRevB.73.104511} {\bibfield  {journal} {\bibinfo  {journal}
  {Physical Review B}\ }\textbf {\bibinfo {volume} {73}},\ \bibinfo {pages}
  {104511} (\bibinfo {year} {2006})}\BibitemShut {NoStop}%
\bibitem [{\citenamefont {Maltseva}\ and\ \citenamefont
  {Coleman}(2009)}]{MaltsevaPRB2009}%
  \BibitemOpen
  \bibfield  {author} {\bibinfo {author} {\bibfnamefont {M.}~\bibnamefont
  {Maltseva}}\ and\ \bibinfo {author} {\bibfnamefont {P.}~\bibnamefont
  {Coleman}},\ }\href {\doibase 10.1103/PhysRevB.80.144514} {\bibfield
  {journal} {\bibinfo  {journal} {Physical Review B}\ }\textbf {\bibinfo
  {volume} {80}},\ \bibinfo {pages} {144514} (\bibinfo {year}
  {2009})}\BibitemShut {NoStop}%
\bibitem [{\citenamefont {Sykora}\ and\ \citenamefont
  {Coleman}(2011)}]{SykoraPRB2011}%
  \BibitemOpen
  \bibfield  {author} {\bibinfo {author} {\bibfnamefont {S.}~\bibnamefont
  {Sykora}}\ and\ \bibinfo {author} {\bibfnamefont {P.}~\bibnamefont
  {Coleman}},\ }\href {\doibase 10.1103/PhysRevB.84.054501} {\bibfield
  {journal} {\bibinfo  {journal} {Physical Review B}\ }\textbf {\bibinfo
  {volume} {84}},\ \bibinfo {pages} {054501} (\bibinfo {year}
  {2011})}\BibitemShut {NoStop}%
\bibitem [{\citenamefont {Tinkham}(1996)}]{TinkhamBook1996}%
  \BibitemOpen
  \bibfield  {author} {\bibinfo {author} {\bibfnamefont {M.}~\bibnamefont
  {Tinkham}},\ }\href@noop {} {\emph {\bibinfo {title} {{Introduction to
  Superconductivity}}}},\ \bibinfo {edition} {2nd}\ ed.\ (\bibinfo  {publisher}
  {Dover Publications},\ \bibinfo {year} {1996})\BibitemShut {NoStop}%
\bibitem [{\citenamefont {Wollman}\ \emph {et~al.}(1993)\citenamefont
  {Wollman}, \citenamefont {{Van Harlingen}}, \citenamefont {Lee},
  \citenamefont {Ginsberg},\ and\ \citenamefont {Leggett}}]{WollmanPRL1993}%
  \BibitemOpen
  \bibfield  {author} {\bibinfo {author} {\bibfnamefont {D.}~\bibnamefont
  {Wollman}}, \bibinfo {author} {\bibfnamefont {D.}~\bibnamefont {{Van
  Harlingen}}}, \bibinfo {author} {\bibfnamefont {W.}~\bibnamefont {Lee}},
  \bibinfo {author} {\bibfnamefont {D.}~\bibnamefont {Ginsberg}}, \ and\
  \bibinfo {author} {\bibfnamefont {A.}~\bibnamefont {Leggett}},\ }\href
  {\doibase 10.1103/PhysRevLett.71.2134} {\bibfield  {journal} {\bibinfo
  {journal} {Physical Review Letters}\ }\textbf {\bibinfo {volume} {71}},\
  \bibinfo {pages} {2134} (\bibinfo {year} {1993})}\BibitemShut {NoStop}%
\bibitem [{\citenamefont {{Van Harlingen}}(1995)}]{VanHarlingenRMP1995}%
  \BibitemOpen
  \bibfield  {author} {\bibinfo {author} {\bibfnamefont {D.}~\bibnamefont {{Van
  Harlingen}}},\ }\href {\doibase 10.1103/RevModPhys.67.515} {\bibfield
  {journal} {\bibinfo  {journal} {Reviews of Modern Physics}\ }\textbf
  {\bibinfo {volume} {67}},\ \bibinfo {pages} {515} (\bibinfo {year}
  {1995})}\BibitemShut {NoStop}%
\bibitem [{\citenamefont {McElroy}\ \emph {et~al.}(2003)\citenamefont
  {McElroy}, \citenamefont {Simmonds}, \citenamefont {Hoffman}, \citenamefont
  {Lee}, \citenamefont {Orenstein}, \citenamefont {Eisaki}, \citenamefont
  {Uchida},\ and\ \citenamefont {Davis}}]{McElroyNature2003}%
  \BibitemOpen
  \bibfield  {author} {\bibinfo {author} {\bibfnamefont {K.}~\bibnamefont
  {McElroy}}, \bibinfo {author} {\bibfnamefont {R.~W.}\ \bibnamefont
  {Simmonds}}, \bibinfo {author} {\bibfnamefont {J.~E.}\ \bibnamefont
  {Hoffman}}, \bibinfo {author} {\bibfnamefont {D.-H.}\ \bibnamefont {Lee}},
  \bibinfo {author} {\bibfnamefont {J.}~\bibnamefont {Orenstein}}, \bibinfo
  {author} {\bibfnamefont {H.}~\bibnamefont {Eisaki}}, \bibinfo {author}
  {\bibfnamefont {S.}~\bibnamefont {Uchida}}, \ and\ \bibinfo {author}
  {\bibfnamefont {J.~C.}\ \bibnamefont {Davis}},\ }\href {\doibase
  10.1038/nature01496} {\bibfield  {journal} {\bibinfo  {journal} {Nature}\
  }\textbf {\bibinfo {volume} {422}},\ \bibinfo {pages} {592} (\bibinfo {year}
  {2003})}\BibitemShut {NoStop}%
\bibitem [{\citenamefont {Huang}\ \emph
  {et~al.}(2008{\natexlab{a}})\citenamefont {Huang}, \citenamefont {Zhao},
  \citenamefont {Lynn}, \citenamefont {Chen}, \citenamefont {Luo},
  \citenamefont {Wang},\ and\ \citenamefont {Dai}}]{HuangPRB2008}%
  \BibitemOpen
  \bibfield  {author} {\bibinfo {author} {\bibfnamefont {Q.}~\bibnamefont
  {Huang}}, \bibinfo {author} {\bibfnamefont {J.}~\bibnamefont {Zhao}},
  \bibinfo {author} {\bibfnamefont {J.}~\bibnamefont {Lynn}}, \bibinfo {author}
  {\bibfnamefont {G.}~\bibnamefont {Chen}}, \bibinfo {author} {\bibfnamefont
  {J.}~\bibnamefont {Luo}}, \bibinfo {author} {\bibfnamefont {N.}~\bibnamefont
  {Wang}}, \ and\ \bibinfo {author} {\bibfnamefont {P.}~\bibnamefont {Dai}},\
  }\href {\doibase 10.1103/PhysRevB.78.054529} {\bibfield  {journal} {\bibinfo
  {journal} {Physical Review B}\ }\textbf {\bibinfo {volume} {78}},\ \bibinfo
  {pages} {054529} (\bibinfo {year} {2008}{\natexlab{a}})}\BibitemShut
  {NoStop}%
\bibitem [{\citenamefont {Huang}\ \emph
  {et~al.}(2008{\natexlab{b}})\citenamefont {Huang}, \citenamefont {Qiu},
  \citenamefont {Bao}, \citenamefont {Green}, \citenamefont {Lynn},
  \citenamefont {Gasparovic}, \citenamefont {Wu}, \citenamefont {Wu},\ and\
  \citenamefont {Chen}}]{HuangPRL2008}%
  \BibitemOpen
  \bibfield  {author} {\bibinfo {author} {\bibfnamefont {Q.}~\bibnamefont
  {Huang}}, \bibinfo {author} {\bibfnamefont {Y.}~\bibnamefont {Qiu}}, \bibinfo
  {author} {\bibfnamefont {W.}~\bibnamefont {Bao}}, \bibinfo {author}
  {\bibfnamefont {M.~A.}\ \bibnamefont {Green}}, \bibinfo {author}
  {\bibfnamefont {J.~W.}\ \bibnamefont {Lynn}}, \bibinfo {author}
  {\bibfnamefont {Y.~C.}\ \bibnamefont {Gasparovic}}, \bibinfo {author}
  {\bibfnamefont {T.}~\bibnamefont {Wu}}, \bibinfo {author} {\bibfnamefont
  {G.}~\bibnamefont {Wu}}, \ and\ \bibinfo {author} {\bibfnamefont {X.~H.}\
  \bibnamefont {Chen}},\ }\href {\doibase 10.1103/PhysRevLett.101.257003}
  {\bibfield  {journal} {\bibinfo  {journal} {Physical Review Letters}\
  }\textbf {\bibinfo {volume} {101}},\ \bibinfo {pages} {257003} (\bibinfo
  {year} {2008}{\natexlab{b}})}\BibitemShut {NoStop}%
\bibitem [{\citenamefont {Hanaguri}(2011)}]{HanaguriAspen2011}%
  \BibitemOpen
  \bibfield  {author} {\bibinfo {author} {\bibfnamefont {T.}~\bibnamefont
  {Hanaguri}},\ }in\ \href@noop {} {\emph {\bibinfo {booktitle} {Aspen Winter
  Conference on Condensed Matter Physics}}}\ (\bibinfo {year}
  {2011})\BibitemShut {NoStop}%
\bibitem [{\citenamefont {Song}\ \emph
  {et~al.}(2011{\natexlab{a}})\citenamefont {Song}, \citenamefont {Wang},
  \citenamefont {Jiang}, \citenamefont {Li}, \citenamefont {Wang},
  \citenamefont {He}, \citenamefont {Chen}, \citenamefont {Ma},\ and\
  \citenamefont {Xue}}]{SongPRB2011}%
  \BibitemOpen
  \bibfield  {author} {\bibinfo {author} {\bibfnamefont {C.-L.}\ \bibnamefont
  {Song}}, \bibinfo {author} {\bibfnamefont {Y.-L.}\ \bibnamefont {Wang}},
  \bibinfo {author} {\bibfnamefont {Y.-P.}\ \bibnamefont {Jiang}}, \bibinfo
  {author} {\bibfnamefont {Z.}~\bibnamefont {Li}}, \bibinfo {author}
  {\bibfnamefont {L.}~\bibnamefont {Wang}}, \bibinfo {author} {\bibfnamefont
  {K.}~\bibnamefont {He}}, \bibinfo {author} {\bibfnamefont {X.}~\bibnamefont
  {Chen}}, \bibinfo {author} {\bibfnamefont {X.-C.}\ \bibnamefont {Ma}}, \ and\
  \bibinfo {author} {\bibfnamefont {Q.-K.}\ \bibnamefont {Xue}},\ }\href
  {\doibase 10.1103/PhysRevB.84.020503} {\bibfield  {journal} {\bibinfo
  {journal} {Physical Review B}\ }\textbf {\bibinfo {volume} {84}},\ \bibinfo
  {pages} {020503} (\bibinfo {year} {2011}{\natexlab{a}})}\BibitemShut
  {NoStop}%
\bibitem [{\citenamefont {Kamihara}\ \emph {et~al.}(2010)\citenamefont
  {Kamihara}, \citenamefont {Nomura}, \citenamefont {Hirano}, \citenamefont
  {{Eun Kim}}, \citenamefont {Kato}, \citenamefont {Takata}, \citenamefont
  {Kobayashi}, \citenamefont {Kitao}, \citenamefont {Higashitaniguchi},
  \citenamefont {Yoda}, \citenamefont {Seto},\ and\ \citenamefont
  {Hosono}}]{KamiharaNJP2010}%
  \BibitemOpen
  \bibfield  {author} {\bibinfo {author} {\bibfnamefont {Y.}~\bibnamefont
  {Kamihara}}, \bibinfo {author} {\bibfnamefont {T.}~\bibnamefont {Nomura}},
  \bibinfo {author} {\bibfnamefont {M.}~\bibnamefont {Hirano}}, \bibinfo
  {author} {\bibfnamefont {J.}~\bibnamefont {{Eun Kim}}}, \bibinfo {author}
  {\bibfnamefont {K.}~\bibnamefont {Kato}}, \bibinfo {author} {\bibfnamefont
  {M.}~\bibnamefont {Takata}}, \bibinfo {author} {\bibfnamefont
  {Y.}~\bibnamefont {Kobayashi}}, \bibinfo {author} {\bibfnamefont
  {S.}~\bibnamefont {Kitao}}, \bibinfo {author} {\bibfnamefont
  {S.}~\bibnamefont {Higashitaniguchi}}, \bibinfo {author} {\bibfnamefont
  {Y.}~\bibnamefont {Yoda}}, \bibinfo {author} {\bibfnamefont {M.}~\bibnamefont
  {Seto}}, \ and\ \bibinfo {author} {\bibfnamefont {H.}~\bibnamefont
  {Hosono}},\ }\href {\doibase 10.1088/1367-2630/12/3/033005} {\bibfield
  {journal} {\bibinfo  {journal} {New Journal of Physics}\ }\textbf {\bibinfo
  {volume} {12}},\ \bibinfo {pages} {033005} (\bibinfo {year}
  {2010})}\BibitemShut {NoStop}%
\bibitem [{\citenamefont {Luetkens}\ \emph {et~al.}(2009)\citenamefont
  {Luetkens}, \citenamefont {Klauss}, \citenamefont {Kraken}, \citenamefont
  {Litterst}, \citenamefont {Dellmann}, \citenamefont {Klingeler},
  \citenamefont {Hess}, \citenamefont {Khasanov}, \citenamefont {Amato},
  \citenamefont {Baines}, \citenamefont {Kosmala}, \citenamefont {Schumann},
  \citenamefont {Braden}, \citenamefont {Hamann-Borrero}, \citenamefont {Leps},
  \citenamefont {Kondrat}, \citenamefont {Behr}, \citenamefont {Werner},\ and\
  \citenamefont {B\"{u}chner}}]{LuetkensNatMat2009}%
  \BibitemOpen
  \bibfield  {author} {\bibinfo {author} {\bibfnamefont {H.}~\bibnamefont
  {Luetkens}}, \bibinfo {author} {\bibfnamefont {H.-H.}\ \bibnamefont
  {Klauss}}, \bibinfo {author} {\bibfnamefont {M.}~\bibnamefont {Kraken}},
  \bibinfo {author} {\bibfnamefont {F.~J.}\ \bibnamefont {Litterst}}, \bibinfo
  {author} {\bibfnamefont {T.}~\bibnamefont {Dellmann}}, \bibinfo {author}
  {\bibfnamefont {R.}~\bibnamefont {Klingeler}}, \bibinfo {author}
  {\bibfnamefont {C.}~\bibnamefont {Hess}}, \bibinfo {author} {\bibfnamefont
  {R.}~\bibnamefont {Khasanov}}, \bibinfo {author} {\bibfnamefont
  {A.}~\bibnamefont {Amato}}, \bibinfo {author} {\bibfnamefont
  {C.}~\bibnamefont {Baines}}, \bibinfo {author} {\bibfnamefont
  {M.}~\bibnamefont {Kosmala}}, \bibinfo {author} {\bibfnamefont {O.~J.}\
  \bibnamefont {Schumann}}, \bibinfo {author} {\bibfnamefont {M.}~\bibnamefont
  {Braden}}, \bibinfo {author} {\bibfnamefont {J.}~\bibnamefont
  {Hamann-Borrero}}, \bibinfo {author} {\bibfnamefont {N.}~\bibnamefont
  {Leps}}, \bibinfo {author} {\bibfnamefont {A.}~\bibnamefont {Kondrat}},
  \bibinfo {author} {\bibfnamefont {G.}~\bibnamefont {Behr}}, \bibinfo {author}
  {\bibfnamefont {J.}~\bibnamefont {Werner}}, \ and\ \bibinfo {author}
  {\bibfnamefont {B.}~\bibnamefont {B\"{u}chner}},\ }\href {\doibase
  10.1038/nmat2397} {\bibfield  {journal} {\bibinfo  {journal} {Nature
  Materials}\ }\textbf {\bibinfo {volume} {8}},\ \bibinfo {pages} {305}
  (\bibinfo {year} {2009})}\BibitemShut {NoStop}%
\bibitem [{\citenamefont {Nandi}\ \emph {et~al.}(2010)\citenamefont {Nandi},
  \citenamefont {Kim}, \citenamefont {Kreyssig}, \citenamefont {Fernandes},
  \citenamefont {Pratt}, \citenamefont {Thaler}, \citenamefont {Ni},
  \citenamefont {Bud'ko}, \citenamefont {Canfield}, \citenamefont
  {Schmalian}, \citenamefont {McQueeney},\ and\ \citenamefont
  {Goldman}}]{NandiPRL2010}%
  \BibitemOpen
  \bibfield  {author} {\bibinfo {author} {\bibfnamefont {S.}~\bibnamefont
  {Nandi}}, \bibinfo {author} {\bibfnamefont {M.~G.}\ \bibnamefont {Kim}},
  \bibinfo {author} {\bibfnamefont {A.}~\bibnamefont {Kreyssig}}, \bibinfo
  {author} {\bibfnamefont {R.~M.}\ \bibnamefont {Fernandes}}, \bibinfo {author}
  {\bibfnamefont {D.~K.}\ \bibnamefont {Pratt}}, \bibinfo {author}
  {\bibfnamefont {A.}~\bibnamefont {Thaler}}, \bibinfo {author} {\bibfnamefont
  {N.}~\bibnamefont {Ni}}, \bibinfo {author} {\bibfnamefont {S.~L.}\
  \bibnamefont {Bud'ko}}, \bibinfo {author} {\bibfnamefont {P.~C.}\
  \bibnamefont {Canfield}}, \bibinfo {author} {\bibfnamefont {J.}~\bibnamefont
  {Schmalian}}, \bibinfo {author} {\bibfnamefont {R.~J.}\ \bibnamefont
  {McQueeney}}, \ and\ \bibinfo {author} {\bibfnamefont {A.~I.}\ \bibnamefont
  {Goldman}},\ }\href {\doibase 10.1103/PhysRevLett.104.057006} {\bibfield
  {journal} {\bibinfo  {journal} {Physical Review Letters}\ }\textbf {\bibinfo
  {volume} {104}},\ \bibinfo {pages} {057006} (\bibinfo {year}
  {2010})}\BibitemShut {NoStop}%
\bibitem [{\citenamefont {Katayama}\ \emph {et~al.}(2010)\citenamefont
  {Katayama}, \citenamefont {Ji}, \citenamefont {Louca}, \citenamefont {Lee},
  \citenamefont {Fujita}, \citenamefont {Sato}, \citenamefont {Wen},
  \citenamefont {Xu}, \citenamefont {Gu}, \citenamefont {Xu}, \citenamefont
  {Lin}, \citenamefont {Enoki}, \citenamefont {Chang}, \citenamefont {Yamada},\
  and\ \citenamefont {Tranquada}}]{KatayamaJPSJ2010}%
  \BibitemOpen
  \bibfield  {author} {\bibinfo {author} {\bibfnamefont {N.}~\bibnamefont
  {Katayama}}, \bibinfo {author} {\bibfnamefont {S.}~\bibnamefont {Ji}},
  \bibinfo {author} {\bibfnamefont {D.}~\bibnamefont {Louca}}, \bibinfo
  {author} {\bibfnamefont {S.}~\bibnamefont {Lee}}, \bibinfo {author}
  {\bibfnamefont {M.}~\bibnamefont {Fujita}}, \bibinfo {author} {\bibfnamefont
  {T.~J.}\ \bibnamefont {Sato}}, \bibinfo {author} {\bibfnamefont
  {J.}~\bibnamefont {Wen}}, \bibinfo {author} {\bibfnamefont {Z.}~\bibnamefont
  {Xu}}, \bibinfo {author} {\bibfnamefont {G.}~\bibnamefont {Gu}}, \bibinfo
  {author} {\bibfnamefont {G.}~\bibnamefont {Xu}}, \bibinfo {author}
  {\bibfnamefont {Z.}~\bibnamefont {Lin}}, \bibinfo {author} {\bibfnamefont
  {M.}~\bibnamefont {Enoki}}, \bibinfo {author} {\bibfnamefont
  {S.}~\bibnamefont {Chang}}, \bibinfo {author} {\bibfnamefont
  {K.}~\bibnamefont {Yamada}}, \ and\ \bibinfo {author} {\bibfnamefont {J.~M.}\
  \bibnamefont {Tranquada}},\ }\href {\doibase 10.1143/JPSJ.79.113702}
  {\bibfield  {journal} {\bibinfo  {journal} {Journal of the Physical Society
  of Japan}\ }\textbf {\bibinfo {volume} {79}},\ \bibinfo {pages} {113702}
  (\bibinfo {year} {2010})}\BibitemShut {NoStop}%
\bibitem [{\citenamefont {McQueen}\ \emph {et~al.}(2009)\citenamefont
  {McQueen}, \citenamefont {Williams}, \citenamefont {Stephens}, \citenamefont
  {Tao}, \citenamefont {Zhu}, \citenamefont {Ksenofontov}, \citenamefont
  {Casper}, \citenamefont {Felser},\ and\ \citenamefont
  {Cava}}]{McQueenPRL2009}%
  \BibitemOpen
  \bibfield  {author} {\bibinfo {author} {\bibfnamefont {T.}~\bibnamefont
  {McQueen}}, \bibinfo {author} {\bibfnamefont {A.}~\bibnamefont {Williams}},
  \bibinfo {author} {\bibfnamefont {P.}~\bibnamefont {Stephens}}, \bibinfo
  {author} {\bibfnamefont {J.}~\bibnamefont {Tao}}, \bibinfo {author}
  {\bibfnamefont {Y.}~\bibnamefont {Zhu}}, \bibinfo {author} {\bibfnamefont
  {V.}~\bibnamefont {Ksenofontov}}, \bibinfo {author} {\bibfnamefont
  {F.}~\bibnamefont {Casper}}, \bibinfo {author} {\bibfnamefont
  {C.}~\bibnamefont {Felser}}, \ and\ \bibinfo {author} {\bibfnamefont {R.~J.}\
  \bibnamefont {Cava}},\ }\href {\doibase 10.1103/PhysRevLett.103.057002}
  {\bibfield  {journal} {\bibinfo  {journal} {Physical Review Letters}\
  }\textbf {\bibinfo {volume} {103}},\ \bibinfo {pages} {057002} (\bibinfo
  {year} {2009})}\BibitemShut {NoStop}%
\bibitem [{\citenamefont {Chuang}\ \emph {et~al.}(2010)\citenamefont {Chuang},
  \citenamefont {Allan}, \citenamefont {Lee}, \citenamefont {Xie},
  \citenamefont {Ni}, \citenamefont {Bud'ko}, \citenamefont {Boebinger},
  \citenamefont {Canfield},\ and\ \citenamefont {Davis}}]{ChuangScience2010}%
  \BibitemOpen
  \bibfield  {author} {\bibinfo {author} {\bibfnamefont {T.-M.}\ \bibnamefont
  {Chuang}}, \bibinfo {author} {\bibfnamefont {M.~P.}\ \bibnamefont {Allan}},
  \bibinfo {author} {\bibfnamefont {J.}~\bibnamefont {Lee}}, \bibinfo {author}
  {\bibfnamefont {Y.}~\bibnamefont {Xie}}, \bibinfo {author} {\bibfnamefont
  {N.}~\bibnamefont {Ni}}, \bibinfo {author} {\bibfnamefont {S.~L.}\
  \bibnamefont {Bud'ko}}, \bibinfo {author} {\bibfnamefont {G.~S.}\
  \bibnamefont {Boebinger}}, \bibinfo {author} {\bibfnamefont {P.~C.}\
  \bibnamefont {Canfield}}, \ and\ \bibinfo {author} {\bibfnamefont {J.~C.}\
  \bibnamefont {Davis}},\ }\href {\doibase 10.1126/science.1181083} {\bibfield
  {journal} {\bibinfo  {journal} {Science}\ }\textbf {\bibinfo {volume}
  {327}},\ \bibinfo {pages} {181} (\bibinfo {year} {2010})}\BibitemShut
  {NoStop}%
\bibitem [{\citenamefont {Song}\ \emph
  {et~al.}(2011{\natexlab{b}})\citenamefont {Song}, \citenamefont {Wang},
  \citenamefont {Cheng}, \citenamefont {Jiang}, \citenamefont {Li},
  \citenamefont {Zhang}, \citenamefont {Li}, \citenamefont {He}, \citenamefont
  {Wang}, \citenamefont {Jia}, \citenamefont {Hung}, \citenamefont {Wu},
  \citenamefont {Ma}, \citenamefont {Chen},\ and\ \citenamefont
  {Xue}}]{SongScience2011}%
  \BibitemOpen
  \bibfield  {author} {\bibinfo {author} {\bibfnamefont {C.-L.}\ \bibnamefont
  {Song}}, \bibinfo {author} {\bibfnamefont {Y.-L.}\ \bibnamefont {Wang}},
  \bibinfo {author} {\bibfnamefont {P.}~\bibnamefont {Cheng}}, \bibinfo
  {author} {\bibfnamefont {Y.-P.}\ \bibnamefont {Jiang}}, \bibinfo {author}
  {\bibfnamefont {W.}~\bibnamefont {Li}}, \bibinfo {author} {\bibfnamefont
  {T.}~\bibnamefont {Zhang}}, \bibinfo {author} {\bibfnamefont
  {Z.}~\bibnamefont {Li}}, \bibinfo {author} {\bibfnamefont {K.}~\bibnamefont
  {He}}, \bibinfo {author} {\bibfnamefont {L.}~\bibnamefont {Wang}}, \bibinfo
  {author} {\bibfnamefont {J.-F.}\ \bibnamefont {Jia}}, \bibinfo {author}
  {\bibfnamefont {H.-H.}\ \bibnamefont {Hung}}, \bibinfo {author}
  {\bibfnamefont {C.}~\bibnamefont {Wu}}, \bibinfo {author} {\bibfnamefont
  {X.}~\bibnamefont {Ma}}, \bibinfo {author} {\bibfnamefont {X.}~\bibnamefont
  {Chen}}, \ and\ \bibinfo {author} {\bibfnamefont {Q.-K.}\ \bibnamefont
  {Xue}},\ }\href {\doibase 10.1126/science.1202226} {\bibfield  {journal}
  {\bibinfo  {journal} {Science}\ }\textbf {\bibinfo {volume} {332}},\ \bibinfo
  {pages} {1410} (\bibinfo {year} {2011}{\natexlab{b}})}\BibitemShut {NoStop}%
\bibitem [{\citenamefont {Singh}\ and\ \citenamefont
  {Du}(2008)}]{SinghPRL2008}%
  \BibitemOpen
  \bibfield  {author} {\bibinfo {author} {\bibfnamefont {D.~J.}\ \bibnamefont
  {Singh}}\ and\ \bibinfo {author} {\bibfnamefont {M.-H.}\ \bibnamefont {Du}},\
  }\href {\doibase 10.1103/PhysRevLett.100.237003} {\bibfield  {journal}
  {\bibinfo  {journal} {Physical Review Letters}\ }\textbf {\bibinfo {volume}
  {100}},\ \bibinfo {pages} {237003} (\bibinfo {year} {2008})}\BibitemShut
  {NoStop}%
\bibitem [{\citenamefont {Mazin}\ and\ \citenamefont
  {Schmalian}(2009)}]{MazinPhysicaC2009}%
  \BibitemOpen
  \bibfield  {author} {\bibinfo {author} {\bibfnamefont {I.~I.}\ \bibnamefont
  {Mazin}}\ and\ \bibinfo {author} {\bibfnamefont {J.}~\bibnamefont
  {Schmalian}},\ }\href {\doibase 10.1016/j.physc.2009.03.019} {\bibfield
  {journal} {\bibinfo  {journal} {Physica C}\ }\textbf {\bibinfo {volume}
  {469}},\ \bibinfo {pages} {614} (\bibinfo {year} {2009})}\BibitemShut
  {NoStop}%
\bibitem [{\citenamefont {Mazin}\ \emph {et~al.}(2011)\citenamefont {Mazin},
  \citenamefont {Kimber},\ and\ \citenamefont {Argyriou}}]{MazinPRB2011}%
  \BibitemOpen
  \bibfield  {author} {\bibinfo {author} {\bibfnamefont {I.~I.}\ \bibnamefont
  {Mazin}}, \bibinfo {author} {\bibfnamefont {S.}~\bibnamefont {Kimber}}, \
  and\ \bibinfo {author} {\bibfnamefont {D.}~\bibnamefont {Argyriou}},\ }\href
  {\doibase 10.1103/PhysRevB.83.052501} {\bibfield  {journal} {\bibinfo
  {journal} {Physical Review B}\ }\textbf {\bibinfo {volume} {83}},\ \bibinfo
  {pages} {052501} (\bibinfo {year} {2011})}\BibitemShut {NoStop}%
\bibitem [{\citenamefont {Kamihara}\ \emph {et~al.}(2006)\citenamefont
  {Kamihara}, \citenamefont {Hiramatsu}, \citenamefont {Hirano}, \citenamefont
  {Kawamura}, \citenamefont {Yanagi}, \citenamefont {Kamiya},\ and\
  \citenamefont {Hosono}}]{KamiharaJACS2006}%
  \BibitemOpen
  \bibfield  {author} {\bibinfo {author} {\bibfnamefont {Y.}~\bibnamefont
  {Kamihara}}, \bibinfo {author} {\bibfnamefont {H.}~\bibnamefont {Hiramatsu}},
  \bibinfo {author} {\bibfnamefont {M.}~\bibnamefont {Hirano}}, \bibinfo
  {author} {\bibfnamefont {R.}~\bibnamefont {Kawamura}}, \bibinfo {author}
  {\bibfnamefont {H.}~\bibnamefont {Yanagi}}, \bibinfo {author} {\bibfnamefont
  {T.}~\bibnamefont {Kamiya}}, \ and\ \bibinfo {author} {\bibfnamefont
  {H.}~\bibnamefont {Hosono}},\ }\href {\doibase 10.1021/ja063355c} {\bibfield
  {journal} {\bibinfo  {journal} {Journal of the American Chemical Society}\
  }\textbf {\bibinfo {volume} {128}},\ \bibinfo {pages} {10012} (\bibinfo
  {year} {2006})}\BibitemShut {NoStop}%
\bibitem [{\citenamefont {Chen}\ \emph
  {et~al.}(2008{\natexlab{a}})\citenamefont {Chen}, \citenamefont {Wu},
  \citenamefont {Wu}, \citenamefont {Liu}, \citenamefont {Chen},\ and\
  \citenamefont {Fang}}]{ChenNature2008}%
  \BibitemOpen
  \bibfield  {author} {\bibinfo {author} {\bibfnamefont {X.~H.}\ \bibnamefont
  {Chen}}, \bibinfo {author} {\bibfnamefont {T.}~\bibnamefont {Wu}}, \bibinfo
  {author} {\bibfnamefont {G.}~\bibnamefont {Wu}}, \bibinfo {author}
  {\bibfnamefont {R.~H.}\ \bibnamefont {Liu}}, \bibinfo {author} {\bibfnamefont
  {H.}~\bibnamefont {Chen}}, \ and\ \bibinfo {author} {\bibfnamefont {D.~F.}\
  \bibnamefont {Fang}},\ }\href {\doibase 10.1038/nature07045} {\bibfield
  {journal} {\bibinfo  {journal} {Nature}\ }\textbf {\bibinfo {volume} {453}},\
  \bibinfo {pages} {761} (\bibinfo {year} {2008}{\natexlab{a}})}\BibitemShut
  {NoStop}%
\bibitem [{\citenamefont {Chen}\ \emph
  {et~al.}(2008{\natexlab{b}})\citenamefont {Chen}, \citenamefont {Li},
  \citenamefont {Wu}, \citenamefont {Li}, \citenamefont {Hu}, \citenamefont
  {Dong}, \citenamefont {Zheng}, \citenamefont {Luo},\ and\ \citenamefont
  {Wang}}]{ChenPRL2008}%
  \BibitemOpen
  \bibfield  {author} {\bibinfo {author} {\bibfnamefont {G.~F.}\ \bibnamefont
  {Chen}}, \bibinfo {author} {\bibfnamefont {Z.}~\bibnamefont {Li}}, \bibinfo
  {author} {\bibfnamefont {D.}~\bibnamefont {Wu}}, \bibinfo {author}
  {\bibfnamefont {G.}~\bibnamefont {Li}}, \bibinfo {author} {\bibfnamefont
  {W.~Z.}\ \bibnamefont {Hu}}, \bibinfo {author} {\bibfnamefont
  {J.}~\bibnamefont {Dong}}, \bibinfo {author} {\bibfnamefont {P.}~\bibnamefont
  {Zheng}}, \bibinfo {author} {\bibfnamefont {J.~L.}\ \bibnamefont {Luo}}, \
  and\ \bibinfo {author} {\bibfnamefont {N.~L.}\ \bibnamefont {Wang}},\ }\href
  {\doibase 10.1103/PhysRevLett.100.247002} {\bibfield  {journal} {\bibinfo
  {journal} {Physical Review Letters}\ }\textbf {\bibinfo {volume} {100}},\
  \bibinfo {pages} {247002} (\bibinfo {year} {2008}{\natexlab{b}})}\BibitemShut
  {NoStop}%
\bibitem [{\citenamefont {Ren}\ \emph {et~al.}(2008{\natexlab{a}})\citenamefont
  {Ren}, \citenamefont {Yang}, \citenamefont {Lu}, \citenamefont {Yi},
  \citenamefont {Che}, \citenamefont {Dong}, \citenamefont {Sun},\ and\
  \citenamefont {Zhao}}]{RenMatResInn2008}%
  \BibitemOpen
  \bibfield  {author} {\bibinfo {author} {\bibfnamefont {Z.~A.}\ \bibnamefont
  {Ren}}, \bibinfo {author} {\bibfnamefont {J.}~\bibnamefont {Yang}}, \bibinfo
  {author} {\bibfnamefont {W.}~\bibnamefont {Lu}}, \bibinfo {author}
  {\bibfnamefont {W.}~\bibnamefont {Yi}}, \bibinfo {author} {\bibfnamefont
  {G.~C.}\ \bibnamefont {Che}}, \bibinfo {author} {\bibfnamefont {X.~L.}\
  \bibnamefont {Dong}}, \bibinfo {author} {\bibfnamefont {L.~L.}\ \bibnamefont
  {Sun}}, \ and\ \bibinfo {author} {\bibfnamefont {Z.~X.}\ \bibnamefont
  {Zhao}},\ }\href {\doibase 10.1179/143307508X333686} {\bibfield  {journal}
  {\bibinfo  {journal} {Materials Research Innovations}\ }\textbf {\bibinfo
  {volume} {12}},\ \bibinfo {pages} {105} (\bibinfo {year}
  {2008}{\natexlab{a}})}\BibitemShut {NoStop}%
\bibitem [{\citenamefont {Ren}\ \emph {et~al.}(2008{\natexlab{b}})\citenamefont
  {Ren}, \citenamefont {Yang}, \citenamefont {Lu}, \citenamefont {Yi},
  \citenamefont {Shen}, \citenamefont {Li}, \citenamefont {Che}, \citenamefont
  {Dong}, \citenamefont {Sun}, \citenamefont {Zhou},\ and\ \citenamefont
  {Zhao}}]{RenEPL2008}%
  \BibitemOpen
  \bibfield  {author} {\bibinfo {author} {\bibfnamefont {Z.-A.}\ \bibnamefont
  {Ren}}, \bibinfo {author} {\bibfnamefont {J.}~\bibnamefont {Yang}}, \bibinfo
  {author} {\bibfnamefont {W.}~\bibnamefont {Lu}}, \bibinfo {author}
  {\bibfnamefont {W.}~\bibnamefont {Yi}}, \bibinfo {author} {\bibfnamefont
  {X.-L.}\ \bibnamefont {Shen}}, \bibinfo {author} {\bibfnamefont {Z.-C.}\
  \bibnamefont {Li}}, \bibinfo {author} {\bibfnamefont {G.-C.}\ \bibnamefont
  {Che}}, \bibinfo {author} {\bibfnamefont {X.-L.}\ \bibnamefont {Dong}},
  \bibinfo {author} {\bibfnamefont {L.-L.}\ \bibnamefont {Sun}}, \bibinfo
  {author} {\bibfnamefont {F.}~\bibnamefont {Zhou}}, \ and\ \bibinfo {author}
  {\bibfnamefont {Z.-X.}\ \bibnamefont {Zhao}},\ }\href {\doibase
  10.1209/0295-5075/82/57002} {\bibfield  {journal} {\bibinfo  {journal}
  {Europhysics Letters}\ }\textbf {\bibinfo {volume} {82}},\ \bibinfo {pages}
  {57002} (\bibinfo {year} {2008}{\natexlab{b}})}\BibitemShut {NoStop}%
\bibitem [{\citenamefont {Bos}\ \emph {et~al.}(2008)\citenamefont {Bos},
  \citenamefont {Penny}, \citenamefont {Rodgers}, \citenamefont {Sokolov},
  \citenamefont {Huxley},\ and\ \citenamefont {Attfield}}]{BosChemComm2008}%
  \BibitemOpen
  \bibfield  {author} {\bibinfo {author} {\bibfnamefont {J.-W.~G.}\
  \bibnamefont {Bos}}, \bibinfo {author} {\bibfnamefont {G.~B.~S.}\
  \bibnamefont {Penny}}, \bibinfo {author} {\bibfnamefont {J.~A.}\ \bibnamefont
  {Rodgers}}, \bibinfo {author} {\bibfnamefont {D.~A.}\ \bibnamefont
  {Sokolov}}, \bibinfo {author} {\bibfnamefont {A.~D.}\ \bibnamefont {Huxley}},
  \ and\ \bibinfo {author} {\bibfnamefont {J.~P.}\ \bibnamefont {Attfield}},\
  }\href {\doibase 10.1039/b808474b} {\bibfield  {journal} {\bibinfo  {journal}
  {Chemical Communications}\ }\textbf {\bibinfo {volume} {31}},\ \bibinfo
  {pages} {3634} (\bibinfo {year} {2008})}\BibitemShut {NoStop}%
\bibitem [{\citenamefont {Wang}\ \emph {et~al.}(2008)\citenamefont {Wang},
  \citenamefont {Li}, \citenamefont {Chi}, \citenamefont {Zhu}, \citenamefont
  {Ren}, \citenamefont {Li}, \citenamefont {Wang}, \citenamefont {Lin},
  \citenamefont {Luo}, \citenamefont {Jiang}, \citenamefont {Xu}, \citenamefont
  {Cao},\ and\ \citenamefont {Xu}}]{WangEPL2008}%
  \BibitemOpen
  \bibfield  {author} {\bibinfo {author} {\bibfnamefont {C.}~\bibnamefont
  {Wang}}, \bibinfo {author} {\bibfnamefont {L.}~\bibnamefont {Li}}, \bibinfo
  {author} {\bibfnamefont {S.}~\bibnamefont {Chi}}, \bibinfo {author}
  {\bibfnamefont {Z.}~\bibnamefont {Zhu}}, \bibinfo {author} {\bibfnamefont
  {Z.}~\bibnamefont {Ren}}, \bibinfo {author} {\bibfnamefont {Y.}~\bibnamefont
  {Li}}, \bibinfo {author} {\bibfnamefont {Y.}~\bibnamefont {Wang}}, \bibinfo
  {author} {\bibfnamefont {X.}~\bibnamefont {Lin}}, \bibinfo {author}
  {\bibfnamefont {Y.}~\bibnamefont {Luo}}, \bibinfo {author} {\bibfnamefont
  {S.}~\bibnamefont {Jiang}}, \bibinfo {author} {\bibfnamefont
  {X.}~\bibnamefont {Xu}}, \bibinfo {author} {\bibfnamefont {G.}~\bibnamefont
  {Cao}}, \ and\ \bibinfo {author} {\bibfnamefont {Z.}~\bibnamefont {Xu}},\
  }\href {\doibase 10.1209/0295-5075/83/67006} {\bibfield  {journal} {\bibinfo
  {journal} {Europhysics Letters}\ }\textbf {\bibinfo {volume} {83}},\ \bibinfo
  {pages} {67006} (\bibinfo {year} {2008})}\BibitemShut {NoStop}%
\bibitem [{\citenamefont {Jia}\ \emph {et~al.}(2008)\citenamefont {Jia},
  \citenamefont {Cheng}, \citenamefont {Fang}, \citenamefont {Luo},
  \citenamefont {Yang}, \citenamefont {Ren}, \citenamefont {Shan},
  \citenamefont {Gu},\ and\ \citenamefont {Wen}}]{JiaAPL2008}%
  \BibitemOpen
  \bibfield  {author} {\bibinfo {author} {\bibfnamefont {Y.}~\bibnamefont
  {Jia}}, \bibinfo {author} {\bibfnamefont {P.}~\bibnamefont {Cheng}}, \bibinfo
  {author} {\bibfnamefont {L.}~\bibnamefont {Fang}}, \bibinfo {author}
  {\bibfnamefont {H.}~\bibnamefont {Luo}}, \bibinfo {author} {\bibfnamefont
  {H.}~\bibnamefont {Yang}}, \bibinfo {author} {\bibfnamefont {C.}~\bibnamefont
  {Ren}}, \bibinfo {author} {\bibfnamefont {L.}~\bibnamefont {Shan}}, \bibinfo
  {author} {\bibfnamefont {C.}~\bibnamefont {Gu}}, \ and\ \bibinfo {author}
  {\bibfnamefont {H.-H.}\ \bibnamefont {Wen}},\ }\href {\doibase
  10.1063/1.2963361} {\bibfield  {journal} {\bibinfo  {journal} {Applied
  Physics Letters}\ }\textbf {\bibinfo {volume} {93}},\ \bibinfo {pages}
  {032503} (\bibinfo {year} {2008})}\BibitemShut {NoStop}%
\bibitem [{\citenamefont {Kondo}\ \emph {et~al.}(2008)\citenamefont {Kondo},
  \citenamefont {Santander-Syro}, \citenamefont {Copie}, \citenamefont {Liu},
  \citenamefont {Tillman}, \citenamefont {Mun}, \citenamefont {Schmalian},
  \citenamefont {Bud'ko}, \citenamefont {Tanatar}, \citenamefont {Canfield},\
  and\ \citenamefont {Kaminski}}]{KondoPRL2008}%
  \BibitemOpen
  \bibfield  {author} {\bibinfo {author} {\bibfnamefont {T.}~\bibnamefont
  {Kondo}}, \bibinfo {author} {\bibfnamefont {A.~F.}\ \bibnamefont
  {Santander-Syro}}, \bibinfo {author} {\bibfnamefont {O.}~\bibnamefont
  {Copie}}, \bibinfo {author} {\bibfnamefont {C.}~\bibnamefont {Liu}}, \bibinfo
  {author} {\bibfnamefont {M.~E.}\ \bibnamefont {Tillman}}, \bibinfo {author}
  {\bibfnamefont {E.~D.}\ \bibnamefont {Mun}}, \bibinfo {author} {\bibfnamefont
  {J.}~\bibnamefont {Schmalian}}, \bibinfo {author} {\bibfnamefont {S.~L.}\
  \bibnamefont {Bud'ko}}, \bibinfo {author} {\bibfnamefont {M.~A.}\
  \bibnamefont {Tanatar}}, \bibinfo {author} {\bibfnamefont {P.~C.}\
  \bibnamefont {Canfield}}, \ and\ \bibinfo {author} {\bibfnamefont
  {A.}~\bibnamefont {Kaminski}},\ }\href {\doibase
  10.1103/PhysRevLett.101.147003} {\bibfield  {journal} {\bibinfo  {journal}
  {Physical Review Letters}\ }\textbf {\bibinfo {volume} {101}},\ \bibinfo
  {pages} {147003} (\bibinfo {year} {2008})}\BibitemShut {NoStop}%
\bibitem [{\citenamefont {Millo}\ \emph {et~al.}(2008)\citenamefont {Millo},
  \citenamefont {Asulin}, \citenamefont {Yuli}, \citenamefont {Felner},
  \citenamefont {Ren}, \citenamefont {Shen}, \citenamefont {Che},\ and\
  \citenamefont {Zhao}}]{MilloPRB2008}%
  \BibitemOpen
  \bibfield  {author} {\bibinfo {author} {\bibfnamefont {O.}~\bibnamefont
  {Millo}}, \bibinfo {author} {\bibfnamefont {I.}~\bibnamefont {Asulin}},
  \bibinfo {author} {\bibfnamefont {O.}~\bibnamefont {Yuli}}, \bibinfo {author}
  {\bibfnamefont {I.}~\bibnamefont {Felner}}, \bibinfo {author} {\bibfnamefont
  {Z.-A.}\ \bibnamefont {Ren}}, \bibinfo {author} {\bibfnamefont {X.-L.}\
  \bibnamefont {Shen}}, \bibinfo {author} {\bibfnamefont {G.-C.}\ \bibnamefont
  {Che}}, \ and\ \bibinfo {author} {\bibfnamefont {Z.-X.}\ \bibnamefont
  {Zhao}},\ }\href {\doibase 10.1103/PhysRevB.78.092505} {\bibfield  {journal}
  {\bibinfo  {journal} {Physical Review B}\ }\textbf {\bibinfo {volume} {78}},\
  \bibinfo {pages} {092505} (\bibinfo {year} {2008})}\BibitemShut {NoStop}%
\bibitem [{\citenamefont {Pan}\ \emph {et~al.}()\citenamefont {Pan},
  \citenamefont {He}, \citenamefont {Li}, \citenamefont {Wendelken},
  \citenamefont {Jin}, \citenamefont {Sefat}, \citenamefont {McGuire},
  \citenamefont {Sales}, \citenamefont {Mandrus},\ and\ \citenamefont
  {Plummer}}]{PanArxiv0808.0895}%
  \BibitemOpen
  \bibfield  {author} {\bibinfo {author} {\bibfnamefont {M.~H.}\ \bibnamefont
  {Pan}}, \bibinfo {author} {\bibfnamefont {X.~B.}\ \bibnamefont {He}},
  \bibinfo {author} {\bibfnamefont {G.~R.}\ \bibnamefont {Li}}, \bibinfo
  {author} {\bibfnamefont {J.~F.}\ \bibnamefont {Wendelken}}, \bibinfo {author}
  {\bibfnamefont {R.}~\bibnamefont {Jin}}, \bibinfo {author} {\bibfnamefont
  {A.~S.}\ \bibnamefont {Sefat}}, \bibinfo {author} {\bibfnamefont {M.~A.}\
  \bibnamefont {McGuire}}, \bibinfo {author} {\bibfnamefont {B.~C.}\
  \bibnamefont {Sales}}, \bibinfo {author} {\bibfnamefont {D.}~\bibnamefont
  {Mandrus}}, \ and\ \bibinfo {author} {\bibfnamefont {E.~W.}\ \bibnamefont
  {Plummer}},\ }\href@noop {} {\ }\Eprint {http://arxiv.org/abs/0808.0895}
  {arXiv:0808.0895} \BibitemShut {NoStop}%
\bibitem [{\citenamefont {Fasano}\ \emph {et~al.}(2010)\citenamefont {Fasano},
  \citenamefont {Maggio-Aprile}, \citenamefont {Zhigadlo}, \citenamefont
  {Katrych}, \citenamefont {Karpinski},\ and\ \citenamefont
  {Fischer}}]{FasanoPRL2010}%
  \BibitemOpen
  \bibfield  {author} {\bibinfo {author} {\bibfnamefont {Y.}~\bibnamefont
  {Fasano}}, \bibinfo {author} {\bibfnamefont {I.}~\bibnamefont
  {Maggio-Aprile}}, \bibinfo {author} {\bibfnamefont {N.}~\bibnamefont
  {Zhigadlo}}, \bibinfo {author} {\bibfnamefont {S.}~\bibnamefont {Katrych}},
  \bibinfo {author} {\bibfnamefont {J.}~\bibnamefont {Karpinski}}, \ and\
  \bibinfo {author} {\bibfnamefont {{\O}.}~\bibnamefont {Fischer}},\ }\href
  {\doibase 10.1103/PhysRevLett.105.167005} {\bibfield  {journal} {\bibinfo
  {journal} {Physical Review Letters}\ }\textbf {\bibinfo {volume} {105}},\
  \bibinfo {pages} {167005} (\bibinfo {year} {2010})}\BibitemShut {NoStop}%
\bibitem [{\citenamefont {Yan}\ \emph {et~al.}(2009)\citenamefont {Yan},
  \citenamefont {Nandi}, \citenamefont {Zarestky}, \citenamefont {Tian},
  \citenamefont {Kreyssig}, \citenamefont {Jensen}, \citenamefont {Kracher},
  \citenamefont {Dennis}, \citenamefont {McQueeney}, \citenamefont {Goldman},
  \citenamefont {McCallum},\ and\ \citenamefont {Lograsso}}]{YanAPL2009}%
  \BibitemOpen
  \bibfield  {author} {\bibinfo {author} {\bibfnamefont {J.-Q.}\ \bibnamefont
  {Yan}}, \bibinfo {author} {\bibfnamefont {S.}~\bibnamefont {Nandi}}, \bibinfo
  {author} {\bibfnamefont {J.~L.}\ \bibnamefont {Zarestky}}, \bibinfo {author}
  {\bibfnamefont {W.}~\bibnamefont {Tian}}, \bibinfo {author} {\bibfnamefont
  {A.}~\bibnamefont {Kreyssig}}, \bibinfo {author} {\bibfnamefont
  {B.}~\bibnamefont {Jensen}}, \bibinfo {author} {\bibfnamefont
  {A.}~\bibnamefont {Kracher}}, \bibinfo {author} {\bibfnamefont {K.~W.}\
  \bibnamefont {Dennis}}, \bibinfo {author} {\bibfnamefont {R.~J.}\
  \bibnamefont {McQueeney}}, \bibinfo {author} {\bibfnamefont {A.~I.}\
  \bibnamefont {Goldman}}, \bibinfo {author} {\bibfnamefont {R.~W.}\
  \bibnamefont {McCallum}}, \ and\ \bibinfo {author} {\bibfnamefont {T.~A.}\
  \bibnamefont {Lograsso}},\ }\href {\doibase 10.1063/1.3268435} {\bibfield
  {journal} {\bibinfo  {journal} {Applied Physics Letters}\ }\textbf {\bibinfo
  {volume} {95}},\ \bibinfo {pages} {222504} (\bibinfo {year}
  {2009})}\BibitemShut {NoStop}%
\bibitem [{\citenamefont {Chen}\ \emph
  {et~al.}(2010{\natexlab{a}})\citenamefont {Chen}, \citenamefont {Yuan},
  \citenamefont {Dong},\ and\ \citenamefont {Wang}}]{ChenPRB2010}%
  \BibitemOpen
  \bibfield  {author} {\bibinfo {author} {\bibfnamefont {Z.~G.}\ \bibnamefont
  {Chen}}, \bibinfo {author} {\bibfnamefont {R.~H.}\ \bibnamefont {Yuan}},
  \bibinfo {author} {\bibfnamefont {T.}~\bibnamefont {Dong}}, \ and\ \bibinfo
  {author} {\bibfnamefont {N.~L.}\ \bibnamefont {Wang}},\ }\href {\doibase
  10.1103/PhysRevB.81.100502} {\bibfield  {journal} {\bibinfo  {journal}
  {Physical Review B}\ }\textbf {\bibinfo {volume} {81}},\ \bibinfo {pages}
  {100502} (\bibinfo {year} {2010}{\natexlab{a}})}\BibitemShut {NoStop}%
\bibitem [{\citenamefont {Zhou}\ \emph {et~al.}(2011)\citenamefont {Zhou},
  \citenamefont {Ye}, \citenamefont {Cai}, \citenamefont {Wang}, \citenamefont
  {Chen},\ and\ \citenamefont {Wang}}]{ZhouPRL2011}%
  \BibitemOpen
  \bibfield  {author} {\bibinfo {author} {\bibfnamefont {X.}~\bibnamefont
  {Zhou}}, \bibinfo {author} {\bibfnamefont {C.}~\bibnamefont {Ye}}, \bibinfo
  {author} {\bibfnamefont {P.}~\bibnamefont {Cai}}, \bibinfo {author}
  {\bibfnamefont {X.}~\bibnamefont {Wang}}, \bibinfo {author} {\bibfnamefont
  {X.}~\bibnamefont {Chen}}, \ and\ \bibinfo {author} {\bibfnamefont
  {Y.}~\bibnamefont {Wang}},\ }\href {\doibase 10.1103/PhysRevLett.106.087001}
  {\bibfield  {journal} {\bibinfo  {journal} {Physical Review Letters}\
  }\textbf {\bibinfo {volume} {106}},\ \bibinfo {pages} {087001} (\bibinfo
  {year} {2011})}\BibitemShut {NoStop}%
\bibitem [{\citenamefont {Nakagawa}\ \emph {et~al.}(2006)\citenamefont
  {Nakagawa}, \citenamefont {Hwang},\ and\ \citenamefont
  {Muller}}]{NakagawaNatMat2006}%
  \BibitemOpen
  \bibfield  {author} {\bibinfo {author} {\bibfnamefont {N.}~\bibnamefont
  {Nakagawa}}, \bibinfo {author} {\bibfnamefont {H.~Y.}\ \bibnamefont {Hwang}},
  \ and\ \bibinfo {author} {\bibfnamefont {D.~A.}\ \bibnamefont {Muller}},\
  }\href {\doibase 10.1038/nmat1569} {\bibfield  {journal} {\bibinfo  {journal}
  {Nature Materials}\ }\textbf {\bibinfo {volume} {5}},\ \bibinfo {pages} {204}
  (\bibinfo {year} {2006})}\BibitemShut {NoStop}%
\bibitem [{\citenamefont {Hossain}\ \emph {et~al.}(2008)\citenamefont
  {Hossain}, \citenamefont {Mottershead}, \citenamefont {Fournier},
  \citenamefont {Bostwick}, \citenamefont {McChesney}, \citenamefont
  {Rotenberg}, \citenamefont {Liang}, \citenamefont {Hardy}, \citenamefont
  {Sawatzky}, \citenamefont {Elfimov}, \citenamefont {Bonn},\ and\
  \citenamefont {Damascelli}}]{HossainNatPhys2008}%
  \BibitemOpen
  \bibfield  {author} {\bibinfo {author} {\bibfnamefont {M.~A.}\ \bibnamefont
  {Hossain}}, \bibinfo {author} {\bibfnamefont {J.~D.~F.}\ \bibnamefont
  {Mottershead}}, \bibinfo {author} {\bibfnamefont {D.}~\bibnamefont
  {Fournier}}, \bibinfo {author} {\bibfnamefont {A.}~\bibnamefont {Bostwick}},
  \bibinfo {author} {\bibfnamefont {J.~L.}\ \bibnamefont {McChesney}}, \bibinfo
  {author} {\bibfnamefont {E.}~\bibnamefont {Rotenberg}}, \bibinfo {author}
  {\bibfnamefont {R.}~\bibnamefont {Liang}}, \bibinfo {author} {\bibfnamefont
  {W.~N.}\ \bibnamefont {Hardy}}, \bibinfo {author} {\bibfnamefont {G.~A.}\
  \bibnamefont {Sawatzky}}, \bibinfo {author} {\bibfnamefont {I.~S.}\
  \bibnamefont {Elfimov}}, \bibinfo {author} {\bibfnamefont {D.~A.}\
  \bibnamefont {Bonn}}, \ and\ \bibinfo {author} {\bibfnamefont
  {A.}~\bibnamefont {Damascelli}},\ }\href {\doibase 10.1038/nphys998}
  {\bibfield  {journal} {\bibinfo  {journal} {Nature Physics}\ }\textbf
  {\bibinfo {volume} {4}},\ \bibinfo {pages} {527} (\bibinfo {year}
  {2008})}\BibitemShut {NoStop}%
\bibitem [{\citenamefont {Hesper}\ \emph {et~al.}(2000)\citenamefont {Hesper},
  \citenamefont {Tjeng}, \citenamefont {Heeres},\ and\ \citenamefont
  {Sawatzky}}]{HesperPRB2000}%
  \BibitemOpen
  \bibfield  {author} {\bibinfo {author} {\bibfnamefont {R.}~\bibnamefont
  {Hesper}}, \bibinfo {author} {\bibfnamefont {L.}~\bibnamefont {Tjeng}},
  \bibinfo {author} {\bibfnamefont {A.}~\bibnamefont {Heeres}}, \ and\ \bibinfo
  {author} {\bibfnamefont {G.}~\bibnamefont {Sawatzky}},\ }\href {\doibase
  10.1103/PhysRevB.62.16046} {\bibfield  {journal} {\bibinfo  {journal}
  {Physical Review B}\ }\textbf {\bibinfo {volume} {62}},\ \bibinfo {pages}
  {16046} (\bibinfo {year} {2000})}\BibitemShut {NoStop}%
\bibitem [{\citenamefont {Eschrig}\ \emph {et~al.}(2010)\citenamefont
  {Eschrig}, \citenamefont {Lankau},\ and\ \citenamefont
  {Koepernik}}]{EschrigPRB2010}%
  \BibitemOpen
  \bibfield  {author} {\bibinfo {author} {\bibfnamefont {H.}~\bibnamefont
  {Eschrig}}, \bibinfo {author} {\bibfnamefont {A.}~\bibnamefont {Lankau}}, \
  and\ \bibinfo {author} {\bibfnamefont {K.}~\bibnamefont {Koepernik}},\ }\href
  {\doibase 10.1103/PhysRevB.81.155447} {\bibfield  {journal} {\bibinfo
  {journal} {Physical Review B}\ }\textbf {\bibinfo {volume} {81}},\ \bibinfo
  {pages} {155447} (\bibinfo {year} {2010})}\BibitemShut {NoStop}%
\bibitem [{\citenamefont {Lu}\ \emph {et~al.}(2008)\citenamefont {Lu},
  \citenamefont {Yi}, \citenamefont {Mo}, \citenamefont {Erickson},
  \citenamefont {Analytis}, \citenamefont {Chu}, \citenamefont {Singh},
  \citenamefont {Hussain}, \citenamefont {Geballe}, \citenamefont {Fisher},\
  and\ \citenamefont {Shen}}]{LuNature2008}%
  \BibitemOpen
  \bibfield  {author} {\bibinfo {author} {\bibfnamefont {D.~H.}\ \bibnamefont
  {Lu}}, \bibinfo {author} {\bibfnamefont {M.}~\bibnamefont {Yi}}, \bibinfo
  {author} {\bibfnamefont {S.-K.}\ \bibnamefont {Mo}}, \bibinfo {author}
  {\bibfnamefont {A.~S.}\ \bibnamefont {Erickson}}, \bibinfo {author}
  {\bibfnamefont {J.}~\bibnamefont {Analytis}}, \bibinfo {author}
  {\bibfnamefont {J.-H.}\ \bibnamefont {Chu}}, \bibinfo {author} {\bibfnamefont
  {D.~J.}\ \bibnamefont {Singh}}, \bibinfo {author} {\bibfnamefont
  {Z.}~\bibnamefont {Hussain}}, \bibinfo {author} {\bibfnamefont {T.~H.}\
  \bibnamefont {Geballe}}, \bibinfo {author} {\bibfnamefont {I.~R.}\
  \bibnamefont {Fisher}}, \ and\ \bibinfo {author} {\bibfnamefont {Z.-X.}\
  \bibnamefont {Shen}},\ }\href {\doibase 10.1038/nature07263} {\bibfield
  {journal} {\bibinfo  {journal} {Nature}\ }\textbf {\bibinfo {volume} {455}},\
  \bibinfo {pages} {81} (\bibinfo {year} {2008})}\BibitemShut {NoStop}%
\bibitem [{\citenamefont {Lu}\ \emph {et~al.}(2009)\citenamefont {Lu},
  \citenamefont {Yi}, \citenamefont {Mo}, \citenamefont {Analytis},
  \citenamefont {Chu}, \citenamefont {Erickson}, \citenamefont {Singh},
  \citenamefont {Hussain}, \citenamefont {Geballe},\ and\ \citenamefont
  {Fisher}}]{LuPhysicaC2009}%
  \BibitemOpen
  \bibfield  {author} {\bibinfo {author} {\bibfnamefont {D.~H.}\ \bibnamefont
  {Lu}}, \bibinfo {author} {\bibfnamefont {M.}~\bibnamefont {Yi}}, \bibinfo
  {author} {\bibfnamefont {S.-K.}\ \bibnamefont {Mo}}, \bibinfo {author}
  {\bibfnamefont {J.~G.}\ \bibnamefont {Analytis}}, \bibinfo {author}
  {\bibfnamefont {J.-H.}\ \bibnamefont {Chu}}, \bibinfo {author} {\bibfnamefont
  {A.~S.}\ \bibnamefont {Erickson}}, \bibinfo {author} {\bibfnamefont {D.~J.}\
  \bibnamefont {Singh}}, \bibinfo {author} {\bibfnamefont {Z.}~\bibnamefont
  {Hussain}}, \bibinfo {author} {\bibfnamefont {T.~H.}\ \bibnamefont
  {Geballe}}, \ and\ \bibinfo {author} {\bibfnamefont {I.~R.}\ \bibnamefont
  {Fisher}},\ }\href {\doibase 10.1016/j.physc.2009.03.044} {\bibfield
  {journal} {\bibinfo  {journal} {Physica C}\ }\textbf {\bibinfo {volume}
  {469}},\ \bibinfo {pages} {452} (\bibinfo {year} {2009})}\BibitemShut
  {NoStop}%
\bibitem [{\citenamefont {Liu}\ \emph {et~al.}(2010)\citenamefont {Liu},
  \citenamefont {Lee}, \citenamefont {Palczewski}, \citenamefont {Yan},
  \citenamefont {Kondo}, \citenamefont {Harmon}, \citenamefont {McCallum},
  \citenamefont {Lograsso},\ and\ \citenamefont {Kaminski}}]{LiuPRB2010}%
  \BibitemOpen
  \bibfield  {author} {\bibinfo {author} {\bibfnamefont {C.}~\bibnamefont
  {Liu}}, \bibinfo {author} {\bibfnamefont {Y.}~\bibnamefont {Lee}}, \bibinfo
  {author} {\bibfnamefont {A.}~\bibnamefont {Palczewski}}, \bibinfo {author}
  {\bibfnamefont {J.-Q.}\ \bibnamefont {Yan}}, \bibinfo {author} {\bibfnamefont
  {T.}~\bibnamefont {Kondo}}, \bibinfo {author} {\bibfnamefont
  {B.}~\bibnamefont {Harmon}}, \bibinfo {author} {\bibfnamefont
  {R.}~\bibnamefont {McCallum}}, \bibinfo {author} {\bibfnamefont
  {T.}~\bibnamefont {Lograsso}}, \ and\ \bibinfo {author} {\bibfnamefont
  {A.}~\bibnamefont {Kaminski}},\ }\href {\doibase 10.1103/PhysRevB.82.075135}
  {\bibfield  {journal} {\bibinfo  {journal} {Physical Review B}\ }\textbf
  {\bibinfo {volume} {82}},\ \bibinfo {pages} {075135} (\bibinfo {year}
  {2010})}\BibitemShut {NoStop}%
\bibitem [{\citenamefont {Rotter}\ \emph {et~al.}(2008)\citenamefont {Rotter},
  \citenamefont {Tegel},\ and\ \citenamefont {Johrendt}}]{RotterPRL2008}%
  \BibitemOpen
  \bibfield  {author} {\bibinfo {author} {\bibfnamefont {M.}~\bibnamefont
  {Rotter}}, \bibinfo {author} {\bibfnamefont {M.}~\bibnamefont {Tegel}}, \
  and\ \bibinfo {author} {\bibfnamefont {D.}~\bibnamefont {Johrendt}},\ }\href
  {\doibase 10.1103/PhysRevLett.101.107006} {\bibfield  {journal} {\bibinfo
  {journal} {Physical Review Letters}\ }\textbf {\bibinfo {volume} {101}},\
  \bibinfo {pages} {107006} (\bibinfo {year} {2008})}\BibitemShut {NoStop}%
\bibitem [{\citenamefont {Sefat}\ \emph {et~al.}(2008)\citenamefont {Sefat},
  \citenamefont {Jin}, \citenamefont {McGuire}, \citenamefont {Sales},
  \citenamefont {Singh},\ and\ \citenamefont {Mandrus}}]{SefatPRL2008}%
  \BibitemOpen
  \bibfield  {author} {\bibinfo {author} {\bibfnamefont {A.~S.}\ \bibnamefont
  {Sefat}}, \bibinfo {author} {\bibfnamefont {R.}~\bibnamefont {Jin}}, \bibinfo
  {author} {\bibfnamefont {M.~A.}\ \bibnamefont {McGuire}}, \bibinfo {author}
  {\bibfnamefont {B.~C.}\ \bibnamefont {Sales}}, \bibinfo {author}
  {\bibfnamefont {D.~J.}\ \bibnamefont {Singh}}, \ and\ \bibinfo {author}
  {\bibfnamefont {D.}~\bibnamefont {Mandrus}},\ }\href {\doibase
  10.1103/PhysRevLett.101.117004} {\bibfield  {journal} {\bibinfo  {journal}
  {Physical Review Letters}\ }\textbf {\bibinfo {volume} {101}},\ \bibinfo
  {pages} {117004} (\bibinfo {year} {2008})}\BibitemShut {NoStop}%
\bibitem [{\citenamefont {Li}\ \emph {et~al.}(2009{\natexlab{a}})\citenamefont
  {Li}, \citenamefont {Luo}, \citenamefont {Wang}, \citenamefont {Chen},
  \citenamefont {Ren}, \citenamefont {Tao}, \citenamefont {Li}, \citenamefont
  {Lin}, \citenamefont {He}, \citenamefont {Zhu}, \citenamefont {Cao},\ and\
  \citenamefont {Xu}}]{LiNJP2009}%
  \BibitemOpen
  \bibfield  {author} {\bibinfo {author} {\bibfnamefont {L.~J.}\ \bibnamefont
  {Li}}, \bibinfo {author} {\bibfnamefont {Y.~K.}\ \bibnamefont {Luo}},
  \bibinfo {author} {\bibfnamefont {Q.~B.}\ \bibnamefont {Wang}}, \bibinfo
  {author} {\bibfnamefont {H.}~\bibnamefont {Chen}}, \bibinfo {author}
  {\bibfnamefont {Z.}~\bibnamefont {Ren}}, \bibinfo {author} {\bibfnamefont
  {Q.}~\bibnamefont {Tao}}, \bibinfo {author} {\bibfnamefont {Y.~K.}\
  \bibnamefont {Li}}, \bibinfo {author} {\bibfnamefont {X.}~\bibnamefont
  {Lin}}, \bibinfo {author} {\bibfnamefont {M.}~\bibnamefont {He}}, \bibinfo
  {author} {\bibfnamefont {Z.~W.}\ \bibnamefont {Zhu}}, \bibinfo {author}
  {\bibfnamefont {G.~H.}\ \bibnamefont {Cao}}, \ and\ \bibinfo {author}
  {\bibfnamefont {Z.~A.}\ \bibnamefont {Xu}},\ }\href {\doibase
  10.1088/1367-2630/11/2/025008} {\bibfield  {journal} {\bibinfo  {journal}
  {New Journal of Physics}\ }\textbf {\bibinfo {volume} {11}},\ \bibinfo
  {pages} {025008} (\bibinfo {year} {2009}{\natexlab{a}})}\BibitemShut
  {NoStop}%
\bibitem [{\citenamefont {Kasahara}\ \emph {et~al.}(2010)\citenamefont
  {Kasahara}, \citenamefont {Shibauchi}, \citenamefont {Hashimoto},
  \citenamefont {Ikada}, \citenamefont {Tonegawa}, \citenamefont {Okazaki},
  \citenamefont {Shishido}, \citenamefont {Ikeda}, \citenamefont {Takeya},
  \citenamefont {Hirata}, \citenamefont {Terashima},\ and\ \citenamefont
  {Matsuda}}]{KasaharaPRB2010}%
  \BibitemOpen
  \bibfield  {author} {\bibinfo {author} {\bibfnamefont {S.}~\bibnamefont
  {Kasahara}}, \bibinfo {author} {\bibfnamefont {T.}~\bibnamefont {Shibauchi}},
  \bibinfo {author} {\bibfnamefont {K.}~\bibnamefont {Hashimoto}}, \bibinfo
  {author} {\bibfnamefont {K.}~\bibnamefont {Ikada}}, \bibinfo {author}
  {\bibfnamefont {S.}~\bibnamefont {Tonegawa}}, \bibinfo {author}
  {\bibfnamefont {R.}~\bibnamefont {Okazaki}}, \bibinfo {author} {\bibfnamefont
  {H.}~\bibnamefont {Shishido}}, \bibinfo {author} {\bibfnamefont
  {H.}~\bibnamefont {Ikeda}}, \bibinfo {author} {\bibfnamefont
  {H.}~\bibnamefont {Takeya}}, \bibinfo {author} {\bibfnamefont
  {K.}~\bibnamefont {Hirata}}, \bibinfo {author} {\bibfnamefont
  {T.}~\bibnamefont {Terashima}}, \ and\ \bibinfo {author} {\bibfnamefont
  {Y.}~\bibnamefont {Matsuda}},\ }\href {\doibase 10.1103/PhysRevB.81.184519}
  {\bibfield  {journal} {\bibinfo  {journal} {Physical Review B}\ }\textbf
  {\bibinfo {volume} {81}},\ \bibinfo {pages} {184519} (\bibinfo {year}
  {2010})}\BibitemShut {NoStop}%
\bibitem [{\citenamefont {Alireza}\ \emph {et~al.}(2009)\citenamefont
  {Alireza}, \citenamefont {Ko}, \citenamefont {Gillett}, \citenamefont
  {Petrone}, \citenamefont {Cole}, \citenamefont {Lonzarich},\ and\
  \citenamefont {Sebastian}}]{AlirezaJPCM2009}%
  \BibitemOpen
  \bibfield  {author} {\bibinfo {author} {\bibfnamefont {P.~L.}\ \bibnamefont
  {Alireza}}, \bibinfo {author} {\bibfnamefont {Y.~T.~C.}\ \bibnamefont {Ko}},
  \bibinfo {author} {\bibfnamefont {J.}~\bibnamefont {Gillett}}, \bibinfo
  {author} {\bibfnamefont {C.~M.}\ \bibnamefont {Petrone}}, \bibinfo {author}
  {\bibfnamefont {J.~M.}\ \bibnamefont {Cole}}, \bibinfo {author}
  {\bibfnamefont {G.~G.}\ \bibnamefont {Lonzarich}}, \ and\ \bibinfo {author}
  {\bibfnamefont {S.~E.}\ \bibnamefont {Sebastian}},\ }\href {\doibase
  10.1088/0953-8984/21/1/012208} {\bibfield  {journal} {\bibinfo  {journal}
  {Journal of Physics: Condensed Matter}\ }\textbf {\bibinfo {volume} {21}},\
  \bibinfo {pages} {012208} (\bibinfo {year} {2009})}\BibitemShut {NoStop}%
\bibitem [{\citenamefont {Tarascon}\ \emph {et~al.}(1987)\citenamefont
  {Tarascon}, \citenamefont {Greene}, \citenamefont {Barboux}, \citenamefont
  {McKinnon}, \citenamefont {Hull}, \citenamefont {Orlando}, \citenamefont
  {Delin}, \citenamefont {Foner},\ and\ \citenamefont
  {McNiff}}]{TarasconPRB1987}%
  \BibitemOpen
  \bibfield  {author} {\bibinfo {author} {\bibfnamefont {J.}~\bibnamefont
  {Tarascon}}, \bibinfo {author} {\bibfnamefont {L.}~\bibnamefont {Greene}},
  \bibinfo {author} {\bibfnamefont {P.}~\bibnamefont {Barboux}}, \bibinfo
  {author} {\bibfnamefont {W.}~\bibnamefont {McKinnon}}, \bibinfo {author}
  {\bibfnamefont {G.}~\bibnamefont {Hull}}, \bibinfo {author} {\bibfnamefont
  {T.}~\bibnamefont {Orlando}}, \bibinfo {author} {\bibfnamefont
  {K.}~\bibnamefont {Delin}}, \bibinfo {author} {\bibfnamefont
  {S.}~\bibnamefont {Foner}}, \ and\ \bibinfo {author} {\bibfnamefont
  {E.}~\bibnamefont {McNiff}},\ }\href {\doibase 10.1103/PhysRevB.36.8393}
  {\bibfield  {journal} {\bibinfo  {journal} {Physical Review B}\ }\textbf
  {\bibinfo {volume} {36}},\ \bibinfo {pages} {8393} (\bibinfo {year}
  {1987})}\BibitemShut {NoStop}%
\bibitem [{\citenamefont {Ni}\ \emph {et~al.}(2008)\citenamefont {Ni},
  \citenamefont {Bud'ko}, \citenamefont {Kreyssig}, \citenamefont {Nandi},
  \citenamefont {Rustan}, \citenamefont {Goldman}, \citenamefont {Gupta},
  \citenamefont {Corbett}, \citenamefont {Kracher},\ and\ \citenamefont
  {Canfield}}]{NiPRB2008}%
  \BibitemOpen
  \bibfield  {author} {\bibinfo {author} {\bibfnamefont {N.}~\bibnamefont
  {Ni}}, \bibinfo {author} {\bibfnamefont {S.~L.}\ \bibnamefont {Bud'ko}},
  \bibinfo {author} {\bibfnamefont {A.}~\bibnamefont {Kreyssig}}, \bibinfo
  {author} {\bibfnamefont {S.}~\bibnamefont {Nandi}}, \bibinfo {author}
  {\bibfnamefont {G.~E.}\ \bibnamefont {Rustan}}, \bibinfo {author}
  {\bibfnamefont {A.~I.}\ \bibnamefont {Goldman}}, \bibinfo {author}
  {\bibfnamefont {S.}~\bibnamefont {Gupta}}, \bibinfo {author} {\bibfnamefont
  {J.~D.}\ \bibnamefont {Corbett}}, \bibinfo {author} {\bibfnamefont
  {A.}~\bibnamefont {Kracher}}, \ and\ \bibinfo {author} {\bibfnamefont
  {P.~C.}\ \bibnamefont {Canfield}},\ }\href {\doibase
  10.1103/PhysRevB.78.014507} {\bibfield  {journal} {\bibinfo  {journal}
  {Physical Review B}\ }\textbf {\bibinfo {volume} {78}},\ \bibinfo {pages}
  {014507} (\bibinfo {year} {2008})}\BibitemShut {NoStop}%
\bibitem [{\citenamefont {Boyer}\ \emph {et~al.}()\citenamefont {Boyer},
  \citenamefont {Chatterjee}, \citenamefont {Wise}, \citenamefont {Chen},
  \citenamefont {Luo}, \citenamefont {Wang},\ and\ \citenamefont
  {Hudson}}]{BoyerArxiv0806.4400}%
  \BibitemOpen
  \bibfield  {author} {\bibinfo {author} {\bibfnamefont {M.~C.}\ \bibnamefont
  {Boyer}}, \bibinfo {author} {\bibfnamefont {K.}~\bibnamefont {Chatterjee}},
  \bibinfo {author} {\bibfnamefont {W.~D.}\ \bibnamefont {Wise}}, \bibinfo
  {author} {\bibfnamefont {G.~F.}\ \bibnamefont {Chen}}, \bibinfo {author}
  {\bibfnamefont {J.~L.}\ \bibnamefont {Luo}}, \bibinfo {author} {\bibfnamefont
  {N.~L.}\ \bibnamefont {Wang}}, \ and\ \bibinfo {author} {\bibfnamefont
  {E.~W.}\ \bibnamefont {Hudson}},\ }\href@noop {} {\ }\Eprint
  {http://arxiv.org/abs/0806.4400} {arXiv:0806.4400} \BibitemShut {NoStop}%
\bibitem [{\citenamefont {Yin}\ \emph {et~al.}(2009{\natexlab{b}})\citenamefont
  {Yin}, \citenamefont {Zech}, \citenamefont {Williams}, \citenamefont {Wang},
  \citenamefont {Wu}, \citenamefont {Chen},\ and\ \citenamefont
  {Hoffman}}]{YinPRL2009}%
  \BibitemOpen
  \bibfield  {author} {\bibinfo {author} {\bibfnamefont {Y.}~\bibnamefont
  {Yin}}, \bibinfo {author} {\bibfnamefont {M.}~\bibnamefont {Zech}}, \bibinfo
  {author} {\bibfnamefont {T.~L.}\ \bibnamefont {Williams}}, \bibinfo {author}
  {\bibfnamefont {X.}~\bibnamefont {Wang}}, \bibinfo {author} {\bibfnamefont
  {G.}~\bibnamefont {Wu}}, \bibinfo {author} {\bibfnamefont {X.~H.}\
  \bibnamefont {Chen}}, \ and\ \bibinfo {author} {\bibfnamefont {J.~E.}\
  \bibnamefont {Hoffman}},\ }\href {\doibase 10.1103/PhysRevLett.102.097002}
  {\bibfield  {journal} {\bibinfo  {journal} {Physical Review Letters}\
  }\textbf {\bibinfo {volume} {102}},\ \bibinfo {pages} {097002} (\bibinfo
  {year} {2009}{\natexlab{b}})}\BibitemShut {NoStop}%
\bibitem [{\citenamefont {Hsieh}\ \emph {et~al.}()\citenamefont {Hsieh},
  \citenamefont {Xia}, \citenamefont {Wray}, \citenamefont {Qian},
  \citenamefont {Gomes}, \citenamefont {Yazdani}, \citenamefont {Chen},
  \citenamefont {Luo}, \citenamefont {Wang},\ and\ \citenamefont
  {Hasan}}]{HsiehArxiv0812.2289}%
  \BibitemOpen
  \bibfield  {author} {\bibinfo {author} {\bibfnamefont {D.}~\bibnamefont
  {Hsieh}}, \bibinfo {author} {\bibfnamefont {Y.}~\bibnamefont {Xia}}, \bibinfo
  {author} {\bibfnamefont {L.}~\bibnamefont {Wray}}, \bibinfo {author}
  {\bibfnamefont {D.}~\bibnamefont {Qian}}, \bibinfo {author} {\bibfnamefont
  {K.~K.}\ \bibnamefont {Gomes}}, \bibinfo {author} {\bibfnamefont
  {A.}~\bibnamefont {Yazdani}}, \bibinfo {author} {\bibfnamefont {G.~F.}\
  \bibnamefont {Chen}}, \bibinfo {author} {\bibfnamefont {J.~L.}\ \bibnamefont
  {Luo}}, \bibinfo {author} {\bibfnamefont {N.~L.}\ \bibnamefont {Wang}}, \
  and\ \bibinfo {author} {\bibfnamefont {M.~Z.}\ \bibnamefont {Hasan}},\
  }\href@noop {} {\ }\Eprint {http://arxiv.org/abs/0812.2289} {arXiv:0812.2289}
  \BibitemShut {NoStop}%
\bibitem [{\citenamefont {Gomes}(2011)}]{GomesPrivate2011}%
  \BibitemOpen
  \bibfield  {author} {\bibinfo {author} {\bibfnamefont {K.~K.}\ \bibnamefont
  {Gomes}},\ }\href@noop {} {\bibfield  {journal} {\bibinfo  {journal} {private
  communication}\ } (\bibinfo {year} {2011})}\BibitemShut {NoStop}%
\bibitem [{\citenamefont {Zhang}\ \emph {et~al.}(2010)\citenamefont {Zhang},
  \citenamefont {Dai}, \citenamefont {Zhang}, \citenamefont {Qu}, \citenamefont
  {Ji}, \citenamefont {Wu}, \citenamefont {Wang}, \citenamefont {Chen},
  \citenamefont {Wang}, \citenamefont {Zeng}, \citenamefont {Yang},\ and\
  \citenamefont {Hou}}]{ZhangPRB2010}%
  \BibitemOpen
  \bibfield  {author} {\bibinfo {author} {\bibfnamefont {H.}~\bibnamefont
  {Zhang}}, \bibinfo {author} {\bibfnamefont {J.}~\bibnamefont {Dai}}, \bibinfo
  {author} {\bibfnamefont {Y.}~\bibnamefont {Zhang}}, \bibinfo {author}
  {\bibfnamefont {D.}~\bibnamefont {Qu}}, \bibinfo {author} {\bibfnamefont
  {H.}~\bibnamefont {Ji}}, \bibinfo {author} {\bibfnamefont {G.}~\bibnamefont
  {Wu}}, \bibinfo {author} {\bibfnamefont {X.~F.}\ \bibnamefont {Wang}},
  \bibinfo {author} {\bibfnamefont {X.~H.}\ \bibnamefont {Chen}}, \bibinfo
  {author} {\bibfnamefont {B.}~\bibnamefont {Wang}}, \bibinfo {author}
  {\bibfnamefont {C.}~\bibnamefont {Zeng}}, \bibinfo {author} {\bibfnamefont
  {J.}~\bibnamefont {Yang}}, \ and\ \bibinfo {author} {\bibfnamefont {J.~G.}\
  \bibnamefont {Hou}},\ }\href {\doibase 10.1103/PhysRevB.81.104520} {\bibfield
   {journal} {\bibinfo  {journal} {Physical Review B}\ }\textbf {\bibinfo
  {volume} {81}},\ \bibinfo {pages} {104520} (\bibinfo {year}
  {2010})}\BibitemShut {NoStop}%
\bibitem [{\citenamefont {Massee}\ \emph
  {et~al.}(2009{\natexlab{a}})\citenamefont {Massee}, \citenamefont {de~Jong},
  \citenamefont {Huang}, \citenamefont {Kaas}, \citenamefont {van Heumen},
  \citenamefont {Goedkoop},\ and\ \citenamefont {Golden}}]{MasseePRB2009b}%
  \BibitemOpen
  \bibfield  {author} {\bibinfo {author} {\bibfnamefont {F.}~\bibnamefont
  {Massee}}, \bibinfo {author} {\bibfnamefont {S.}~\bibnamefont {de~Jong}},
  \bibinfo {author} {\bibfnamefont {Y.}~\bibnamefont {Huang}}, \bibinfo
  {author} {\bibfnamefont {J.}~\bibnamefont {Kaas}}, \bibinfo {author}
  {\bibfnamefont {E.}~\bibnamefont {van Heumen}}, \bibinfo {author}
  {\bibfnamefont {J.}~\bibnamefont {Goedkoop}}, \ and\ \bibinfo {author}
  {\bibfnamefont {M.}~\bibnamefont {Golden}},\ }\href {\doibase
  10.1103/PhysRevB.80.140507} {\bibfield  {journal} {\bibinfo  {journal}
  {Physical Review B}\ }\textbf {\bibinfo {volume} {80}},\ \bibinfo {pages}
  {140507} (\bibinfo {year} {2009}{\natexlab{a}})}\BibitemShut {NoStop}%
\bibitem [{\citenamefont {Massee}(2011{\natexlab{a}})}]{MasseePrivate2011}%
  \BibitemOpen
  \bibfield  {author} {\bibinfo {author} {\bibfnamefont {F.}~\bibnamefont
  {Massee}},\ }\href@noop {} {\bibfield  {journal} {\bibinfo  {journal}
  {private communication}\ } (\bibinfo {year}
  {2011}{\natexlab{a}})}\BibitemShut {NoStop}%
\bibitem [{\citenamefont {Massee}(2011{\natexlab{b}})}]{MasseeThesis2011}%
  \BibitemOpen
  \bibfield  {author} {\bibinfo {author} {\bibfnamefont {F.}~\bibnamefont
  {Massee}},\ }\emph {\bibinfo {title} {{A tunneler's view on correlated
  oxides and iron based superconductors}}},\ \href
  {http://www.science.uva.nl/research/cmp/docs/Golden/thesis/PhD\_thesis\_Freek\_Massee\_small.pdf}
  {Ph.D. thesis},\ \bibinfo  {school} {University of Amsterdam} (\bibinfo
  {year} {2011}{\natexlab{b}})\BibitemShut {NoStop}%
\bibitem [{\citenamefont {Nascimento}\ \emph {et~al.}(2009)\citenamefont
  {Nascimento}, \citenamefont {Li}, \citenamefont {Jayasundara}, \citenamefont
  {Xuan}, \citenamefont {O'Neal}, \citenamefont {Pan}, \citenamefont {Chien},
  \citenamefont {Hu}, \citenamefont {He}, \citenamefont {Li}, \citenamefont
  {Sefat}, \citenamefont {McGuire}, \citenamefont {Sales}, \citenamefont
  {Mandrus}, \citenamefont {Pan}, \citenamefont {Zhang}, \citenamefont {Jin},\
  and\ \citenamefont {Plummer}}]{NascimentoPRL2009}%
  \BibitemOpen
  \bibfield  {author} {\bibinfo {author} {\bibfnamefont {V.}~\bibnamefont
  {Nascimento}}, \bibinfo {author} {\bibfnamefont {A.}~\bibnamefont {Li}},
  \bibinfo {author} {\bibfnamefont {D.}~\bibnamefont {Jayasundara}}, \bibinfo
  {author} {\bibfnamefont {Y.}~\bibnamefont {Xuan}}, \bibinfo {author}
  {\bibfnamefont {J.}~\bibnamefont {O'Neal}}, \bibinfo {author}
  {\bibfnamefont {S.}~\bibnamefont {Pan}}, \bibinfo {author} {\bibfnamefont
  {T.}~\bibnamefont {Chien}}, \bibinfo {author} {\bibfnamefont
  {B.}~\bibnamefont {Hu}}, \bibinfo {author} {\bibfnamefont {X.}~\bibnamefont
  {He}}, \bibinfo {author} {\bibfnamefont {G.}~\bibnamefont {Li}}, \bibinfo
  {author} {\bibfnamefont {A.}~\bibnamefont {Sefat}}, \bibinfo {author}
  {\bibfnamefont {M.}~\bibnamefont {McGuire}}, \bibinfo {author} {\bibfnamefont
  {B.}~\bibnamefont {Sales}}, \bibinfo {author} {\bibfnamefont
  {D.}~\bibnamefont {Mandrus}}, \bibinfo {author} {\bibfnamefont
  {M.}~\bibnamefont {Pan}}, \bibinfo {author} {\bibfnamefont {J.}~\bibnamefont
  {Zhang}}, \bibinfo {author} {\bibfnamefont {R.}~\bibnamefont {Jin}}, \ and\
  \bibinfo {author} {\bibfnamefont {E.~W.}\ \bibnamefont {Plummer}},\ }\href
  {\doibase 10.1103/PhysRevLett.103.076104} {\bibfield  {journal} {\bibinfo
  {journal} {Physical Review Letters}\ }\textbf {\bibinfo {volume} {103}},\
  \bibinfo {pages} {076104} (\bibinfo {year} {2009})}\BibitemShut {NoStop}%
\bibitem [{\citenamefont {Plummer}(2011)}]{PlummerPrivate2011}%
  \BibitemOpen
  \bibfield  {author} {\bibinfo {author} {\bibfnamefont {E.~W.}\ \bibnamefont
  {Plummer}},\ }\href@noop {} {\bibfield  {journal} {\bibinfo  {journal}
  {private communication}\ } (\bibinfo {year} {2011})}\BibitemShut {NoStop}%
\bibitem [{\citenamefont {Li}\ \emph {et~al.}()\citenamefont {Li},
  \citenamefont {He}, \citenamefont {Li}, \citenamefont {Pan}, \citenamefont
  {Zhang}, \citenamefont {Jin}, \citenamefont {Sefat}, \citenamefont {McGuire},
  \citenamefont {Mandrus}, \citenamefont {Sales},\ and\ \citenamefont
  {Plummer}}]{LiArxiv1006.5907}%
  \BibitemOpen
  \bibfield  {author} {\bibinfo {author} {\bibfnamefont {G.}~\bibnamefont
  {Li}}, \bibinfo {author} {\bibfnamefont {X.}~\bibnamefont {He}}, \bibinfo
  {author} {\bibfnamefont {A.}~\bibnamefont {Li}}, \bibinfo {author}
  {\bibfnamefont {S.~H.}\ \bibnamefont {Pan}}, \bibinfo {author} {\bibfnamefont
  {J.}~\bibnamefont {Zhang}}, \bibinfo {author} {\bibfnamefont
  {R.}~\bibnamefont {Jin}}, \bibinfo {author} {\bibfnamefont {A.~S.}\
  \bibnamefont {Sefat}}, \bibinfo {author} {\bibfnamefont {M.~A.}\ \bibnamefont
  {McGuire}}, \bibinfo {author} {\bibfnamefont {D.~G.}\ \bibnamefont
  {Mandrus}}, \bibinfo {author} {\bibfnamefont {B.~C.}\ \bibnamefont {Sales}},
  \ and\ \bibinfo {author} {\bibfnamefont {E.~W.}\ \bibnamefont {Plummer}},\
  }\href@noop {} {\ }\Eprint {http://arxiv.org/abs/1006.5907} {arXiv:1006.5907}
  \BibitemShut {NoStop}%
\bibitem [{\citenamefont {Mazin}\ and\ \citenamefont
  {Johannes}(2008)}]{MazinNatPhys2008}%
  \BibitemOpen
  \bibfield  {author} {\bibinfo {author} {\bibfnamefont {I.~I.}\ \bibnamefont
  {Mazin}}\ and\ \bibinfo {author} {\bibfnamefont {M.~D.}\ \bibnamefont
  {Johannes}},\ }\href {\doibase 10.1038/nphys1160} {\bibfield  {journal}
  {\bibinfo  {journal} {Nature Physics}\ }\textbf {\bibinfo {volume} {5}},\
  \bibinfo {pages} {141} (\bibinfo {year} {2008})}\BibitemShut {NoStop}%
\bibitem [{\citenamefont {Niestemski}\ \emph {et~al.}()\citenamefont
  {Niestemski}, \citenamefont {Nascimento}, \citenamefont {Hu}, \citenamefont
  {Plummer}, \citenamefont {Gillett}, \citenamefont {Sebastian}, \citenamefont
  {Wang},\ and\ \citenamefont {Madhavan}}]{NiestemskiArxiv0906.2761}%
  \BibitemOpen
  \bibfield  {author} {\bibinfo {author} {\bibfnamefont {F.~C.}\ \bibnamefont
  {Niestemski}}, \bibinfo {author} {\bibfnamefont {V.~B.}\ \bibnamefont
  {Nascimento}}, \bibinfo {author} {\bibfnamefont {B.}~\bibnamefont {Hu}},
  \bibinfo {author} {\bibfnamefont {E.~W.}\ \bibnamefont {Plummer}}, \bibinfo
  {author} {\bibfnamefont {J.}~\bibnamefont {Gillett}}, \bibinfo {author}
  {\bibfnamefont {S.~E.}\ \bibnamefont {Sebastian}}, \bibinfo {author}
  {\bibfnamefont {Z.}~\bibnamefont {Wang}}, \ and\ \bibinfo {author}
  {\bibfnamefont {V.}~\bibnamefont {Madhavan}},\ }\href@noop {} {\ }\Eprint
  {http://arxiv.org/abs/0906.2761} {arXiv:0906.2761} \BibitemShut {NoStop}%
\bibitem [{\citenamefont {Pan}\ \emph {et~al.}(2009)\citenamefont {Pan},
  \citenamefont {Li}, \citenamefont {Jayasundara}, \citenamefont {Xuan},
  \citenamefont {O'Neal}, \citenamefont {Jin}, \citenamefont {Plummer},
  \citenamefont {Sefat}, \citenamefont {McGuire}, \citenamefont {Sales},\ and\
  \citenamefont {Mandrus}}]{PanBAPS2009}%
  \BibitemOpen
  \bibfield  {author} {\bibinfo {author} {\bibfnamefont {S.~H.}\ \bibnamefont
  {Pan}}, \bibinfo {author} {\bibfnamefont {A.}~\bibnamefont {Li}}, \bibinfo
  {author} {\bibfnamefont {D.~R.}\ \bibnamefont {Jayasundara}}, \bibinfo
  {author} {\bibfnamefont {Y.}~\bibnamefont {Xuan}}, \bibinfo {author}
  {\bibfnamefont {J.~P.}\ \bibnamefont {O'Neal}}, \bibinfo {author}
  {\bibfnamefont {R.}~\bibnamefont {Jin}}, \bibinfo {author} {\bibfnamefont
  {E.~W.}\ \bibnamefont {Plummer}}, \bibinfo {author} {\bibfnamefont {A.~S.}\
  \bibnamefont {Sefat}}, \bibinfo {author} {\bibfnamefont {M.~A.}\ \bibnamefont
  {McGuire}}, \bibinfo {author} {\bibfnamefont {B.~C.}\ \bibnamefont {Sales}},
  \ and\ \bibinfo {author} {\bibfnamefont {D.}~\bibnamefont {Mandrus}},\ }\href
  {http://meetings.aps.org/Meeting/MAR09/Event/93579} {\bibfield  {journal}
  {\bibinfo  {journal} {Bulletin of the American Physical Society}\ }\textbf
  {\bibinfo {volume} {54}} (\bibinfo {year} {2009})}\BibitemShut {NoStop}%
\bibitem [{\citenamefont {van Heumen}\ \emph {et~al.}(2011)\citenamefont {van
  Heumen}, \citenamefont {Vuorinen}, \citenamefont {Koepernik}, \citenamefont
  {Massee}, \citenamefont {Huang}, \citenamefont {Shi}, \citenamefont {Klei},
  \citenamefont {Goedkoop}, \citenamefont {Lindroos}, \citenamefont {van~den
  Brink},\ and\ \citenamefont {Golden}}]{VanHeumenPRL2011}%
  \BibitemOpen
  \bibfield  {author} {\bibinfo {author} {\bibfnamefont {E.}~\bibnamefont {van
  Heumen}}, \bibinfo {author} {\bibfnamefont {J.}~\bibnamefont {Vuorinen}},
  \bibinfo {author} {\bibfnamefont {K.}~\bibnamefont {Koepernik}}, \bibinfo
  {author} {\bibfnamefont {F.}~\bibnamefont {Massee}}, \bibinfo {author}
  {\bibfnamefont {Y.}~\bibnamefont {Huang}}, \bibinfo {author} {\bibfnamefont
  {M.}~\bibnamefont {Shi}}, \bibinfo {author} {\bibfnamefont {J.}~\bibnamefont
  {Klei}}, \bibinfo {author} {\bibfnamefont {J.}~\bibnamefont {Goedkoop}},
  \bibinfo {author} {\bibfnamefont {M.}~\bibnamefont {Lindroos}}, \bibinfo
  {author} {\bibfnamefont {J.}~\bibnamefont {van~den Brink}}, \ and\ \bibinfo
  {author} {\bibfnamefont {M.}~\bibnamefont {Golden}},\ }\href {\doibase
  10.1103/PhysRevLett.106.027002} {\bibfield  {journal} {\bibinfo  {journal}
  {Physical Review Letters}\ }\textbf {\bibinfo {volume} {106}},\ \bibinfo
  {pages} {027002} (\bibinfo {year} {2011})}\BibitemShut {NoStop}%
\bibitem [{\citenamefont {Yin}(2009)}]{YinThesis2009}%
  \BibitemOpen
  \bibfield  {author} {\bibinfo {author} {\bibfnamefont {Y.}~\bibnamefont
  {Yin}},\ }\emph {\bibinfo {title} {{The Investigation of Scanning Tunneling
  Microscopy and Spectroscopy on High-Tc Superconductors: Cuprates and
  Pnictides}}},\ \href
  {http://hoffman.physics.harvard.edu/publications/YinThesis.pdf} {Ph.D.
  thesis},\ \bibinfo  {school} {Harvard University} (\bibinfo {year}
  {2009})\BibitemShut {NoStop}%
\bibitem [{\citenamefont {Pendry}(1974)}]{PendryBook1974}%
  \BibitemOpen
  \bibfield  {author} {\bibinfo {author} {\bibfnamefont {J.~B.}\ \bibnamefont
  {Pendry}},\ }\href@noop {} {\emph {\bibinfo {title} {{Low-Energy Electron
  Diffraction}}}}\ (\bibinfo  {publisher} {Academic Press Inc.},\ \bibinfo
  {address} {London},\ \bibinfo {year} {1974})\BibitemShut {NoStop}%
\bibitem [{\citenamefont {Hove}\ \emph {et~al.}(1986)\citenamefont {Hove},
  \citenamefont {Weinberg},\ and\ \citenamefont {Chan}}]{vanHoveBook1986}%
  \BibitemOpen
  \bibfield  {author} {\bibinfo {author} {\bibfnamefont {M.~V.}\ \bibnamefont
  {Hove}}, \bibinfo {author} {\bibfnamefont {W.}~\bibnamefont {Weinberg}}, \
  and\ \bibinfo {author} {\bibfnamefont {C.~M.}\ \bibnamefont {Chan}},\
  }\href@noop {} {\emph {\bibinfo {title} {{Low-Energy Electron
  Diffraction}}}}\ (\bibinfo  {publisher} {Springer-Verlag},\ \bibinfo
  {address} {Berlin Heidelberg New York},\ \bibinfo {year} {1986})\BibitemShut
  {NoStop}%
\bibitem [{\citenamefont {Pendry}(1980)}]{PendryJPC1980}%
  \BibitemOpen
  \bibfield  {author} {\bibinfo {author} {\bibfnamefont {J.~B.}\ \bibnamefont
  {Pendry}},\ }\href {\doibase 10.1088/0022-3719/13/5/024} {\bibfield
  {journal} {\bibinfo  {journal} {Journal of Physics C: Solid State Physics}\
  }\textbf {\bibinfo {volume} {13}},\ \bibinfo {pages} {937} (\bibinfo {year}
  {1980})}\BibitemShut {NoStop}%
\bibitem [{\citenamefont {Gao}\ \emph {et~al.}(2010)\citenamefont {Gao},
  \citenamefont {Ma}, \citenamefont {Lu},\ and\ \citenamefont
  {Xiang}}]{GaoPRB2010}%
  \BibitemOpen
  \bibfield  {author} {\bibinfo {author} {\bibfnamefont {M.}~\bibnamefont
  {Gao}}, \bibinfo {author} {\bibfnamefont {F.}~\bibnamefont {Ma}}, \bibinfo
  {author} {\bibfnamefont {Z.-Y.}\ \bibnamefont {Lu}}, \ and\ \bibinfo {author}
  {\bibfnamefont {T.}~\bibnamefont {Xiang}},\ }\href {\doibase
  10.1103/PhysRevB.81.193409} {\bibfield  {journal} {\bibinfo  {journal}
  {Physical Review B}\ }\textbf {\bibinfo {volume} {81}},\ \bibinfo {pages}
  {193409} (\bibinfo {year} {2010})}\BibitemShut {NoStop}%
\bibitem [{\citenamefont {Allan}(2011)}]{AllanPrivate2011}%
  \BibitemOpen
  \bibfield  {author} {\bibinfo {author} {\bibfnamefont {M.}~\bibnamefont
  {Allan}},\ }\href@noop {} {\bibfield  {journal} {\bibinfo  {journal} {private
  communication}\ } (\bibinfo {year} {2011})}\BibitemShut {NoStop}%
\bibitem [{\citenamefont {Kato}\ \emph {et~al.}(2009)\citenamefont {Kato},
  \citenamefont {Mizuguchi}, \citenamefont {Nakamura}, \citenamefont {Machida},
  \citenamefont {Sakata},\ and\ \citenamefont {Takano}}]{KatoPRB2009}%
  \BibitemOpen
  \bibfield  {author} {\bibinfo {author} {\bibfnamefont {T.}~\bibnamefont
  {Kato}}, \bibinfo {author} {\bibfnamefont {Y.}~\bibnamefont {Mizuguchi}},
  \bibinfo {author} {\bibfnamefont {H.}~\bibnamefont {Nakamura}}, \bibinfo
  {author} {\bibfnamefont {T.}~\bibnamefont {Machida}}, \bibinfo {author}
  {\bibfnamefont {H.}~\bibnamefont {Sakata}}, \ and\ \bibinfo {author}
  {\bibfnamefont {Y.}~\bibnamefont {Takano}},\ }\href {\doibase
  10.1103/PhysRevB.80.180507} {\bibfield  {journal} {\bibinfo  {journal}
  {Physical Review B}\ }\textbf {\bibinfo {volume} {80}},\ \bibinfo {pages}
  {180507} (\bibinfo {year} {2009})}\BibitemShut {NoStop}%
\bibitem [{\citenamefont {de~Jong}\ \emph {et~al.}(2009)\citenamefont
  {de~Jong}, \citenamefont {Huang}, \citenamefont {Huisman}, \citenamefont
  {Massee}, \citenamefont {Thirupathaiah}, \citenamefont {Gorgoi},
  \citenamefont {Schaefers}, \citenamefont {Follath}, \citenamefont
  {Goedkoop},\ and\ \citenamefont {Golden}}]{DeJongPRB2009}%
  \BibitemOpen
  \bibfield  {author} {\bibinfo {author} {\bibfnamefont {S.}~\bibnamefont
  {de~Jong}}, \bibinfo {author} {\bibfnamefont {Y.}~\bibnamefont {Huang}},
  \bibinfo {author} {\bibfnamefont {R.}~\bibnamefont {Huisman}}, \bibinfo
  {author} {\bibfnamefont {F.}~\bibnamefont {Massee}}, \bibinfo {author}
  {\bibfnamefont {S.}~\bibnamefont {Thirupathaiah}}, \bibinfo {author}
  {\bibfnamefont {M.}~\bibnamefont {Gorgoi}}, \bibinfo {author} {\bibfnamefont
  {F.}~\bibnamefont {Schaefers}}, \bibinfo {author} {\bibfnamefont
  {R.}~\bibnamefont {Follath}}, \bibinfo {author} {\bibfnamefont
  {J.}~\bibnamefont {Goedkoop}}, \ and\ \bibinfo {author} {\bibfnamefont
  {M.}~\bibnamefont {Golden}},\ }\href {\doibase 10.1103/PhysRevB.79.115125}
  {\bibfield  {journal} {\bibinfo  {journal} {Physical Review B}\ }\textbf
  {\bibinfo {volume} {79}},\ \bibinfo {pages} {115125} (\bibinfo {year}
  {2009})}\BibitemShut {NoStop}%
\bibitem [{\citenamefont {Massee}\ \emph
  {et~al.}(2009{\natexlab{b}})\citenamefont {Massee}, \citenamefont {Huang},
  \citenamefont {Huisman}, \citenamefont {de~Jong}, \citenamefont {Goedkoop},\
  and\ \citenamefont {Golden}}]{MasseePRB2009a}%
  \BibitemOpen
  \bibfield  {author} {\bibinfo {author} {\bibfnamefont {F.}~\bibnamefont
  {Massee}}, \bibinfo {author} {\bibfnamefont {Y.}~\bibnamefont {Huang}},
  \bibinfo {author} {\bibfnamefont {R.}~\bibnamefont {Huisman}}, \bibinfo
  {author} {\bibfnamefont {S.}~\bibnamefont {de~Jong}}, \bibinfo {author}
  {\bibfnamefont {J.}~\bibnamefont {Goedkoop}}, \ and\ \bibinfo {author}
  {\bibfnamefont {M.}~\bibnamefont {Golden}},\ }\href {\doibase
  10.1103/PhysRevB.79.220517} {\bibfield  {journal} {\bibinfo  {journal}
  {Physical Review B}\ }\textbf {\bibinfo {volume} {79}},\ \bibinfo {pages}
  {220517} (\bibinfo {year} {2009}{\natexlab{b}})}\BibitemShut {NoStop}%
\bibitem [{\citenamefont {Massee}\ \emph {et~al.}(2010)\citenamefont {Massee},
  \citenamefont {Huang}, \citenamefont {Kaas}, \citenamefont {van Heumen},
  \citenamefont {de~Jong}, \citenamefont {Huisman}, \citenamefont {Luigjes},
  \citenamefont {Goedkoop},\ and\ \citenamefont {Golden}}]{MasseeEPL2010}%
  \BibitemOpen
  \bibfield  {author} {\bibinfo {author} {\bibfnamefont {F.}~\bibnamefont
  {Massee}}, \bibinfo {author} {\bibfnamefont {Y.~K.}\ \bibnamefont {Huang}},
  \bibinfo {author} {\bibfnamefont {J.}~\bibnamefont {Kaas}}, \bibinfo {author}
  {\bibfnamefont {E.}~\bibnamefont {van Heumen}}, \bibinfo {author}
  {\bibfnamefont {S.}~\bibnamefont {de~Jong}}, \bibinfo {author} {\bibfnamefont
  {R.}~\bibnamefont {Huisman}}, \bibinfo {author} {\bibfnamefont
  {H.}~\bibnamefont {Luigjes}}, \bibinfo {author} {\bibfnamefont {J.~B.}\
  \bibnamefont {Goedkoop}}, \ and\ \bibinfo {author} {\bibfnamefont {M.~S.}\
  \bibnamefont {Golden}},\ }\href {\doibase 10.1209/0295-5075/92/57012}
  {\bibfield  {journal} {\bibinfo  {journal} {Europhysics Letters}\ }\textbf
  {\bibinfo {volume} {92}},\ \bibinfo {pages} {57012} (\bibinfo {year}
  {2010})}\BibitemShut {NoStop}%
\bibitem [{\citenamefont {Shan}\ \emph
  {et~al.}(2011{\natexlab{a}})\citenamefont {Shan}, \citenamefont {Wang},
  \citenamefont {Shen}, \citenamefont {Zeng}, \citenamefont {Huang},
  \citenamefont {Li}, \citenamefont {Wang}, \citenamefont {Yang}, \citenamefont
  {Ren}, \citenamefont {Wang}, \citenamefont {Pan},\ and\ \citenamefont
  {Wen}}]{ShanNatPhys2011}%
  \BibitemOpen
  \bibfield  {author} {\bibinfo {author} {\bibfnamefont {L.}~\bibnamefont
  {Shan}}, \bibinfo {author} {\bibfnamefont {Y.-L.}\ \bibnamefont {Wang}},
  \bibinfo {author} {\bibfnamefont {B.}~\bibnamefont {Shen}}, \bibinfo {author}
  {\bibfnamefont {B.}~\bibnamefont {Zeng}}, \bibinfo {author} {\bibfnamefont
  {Y.}~\bibnamefont {Huang}}, \bibinfo {author} {\bibfnamefont
  {A.}~\bibnamefont {Li}}, \bibinfo {author} {\bibfnamefont {D.}~\bibnamefont
  {Wang}}, \bibinfo {author} {\bibfnamefont {H.}~\bibnamefont {Yang}}, \bibinfo
  {author} {\bibfnamefont {C.}~\bibnamefont {Ren}}, \bibinfo {author}
  {\bibfnamefont {Q.-H.}\ \bibnamefont {Wang}}, \bibinfo {author}
  {\bibfnamefont {S.~H.}\ \bibnamefont {Pan}}, \ and\ \bibinfo {author}
  {\bibfnamefont {H.-H.}\ \bibnamefont {Wen}},\ }\href {\doibase
  10.1038/nphys1908} {\bibfield  {journal} {\bibinfo  {journal} {Nature
  Physics}\ }\textbf {\bibinfo {volume} {7}},\ \bibinfo {pages} {325} (\bibinfo
  {year} {2011}{\natexlab{a}})}\BibitemShut {NoStop}%
\bibitem [{\citenamefont {Shan}\ \emph
  {et~al.}(2011{\natexlab{b}})\citenamefont {Shan}, \citenamefont {Wang},
  \citenamefont {Gong}, \citenamefont {Shen}, \citenamefont {Huang},
  \citenamefont {Yang}, \citenamefont {Ren},\ and\ \citenamefont
  {Wen}}]{ShanPRB2011}%
  \BibitemOpen
  \bibfield  {author} {\bibinfo {author} {\bibfnamefont {L.}~\bibnamefont
  {Shan}}, \bibinfo {author} {\bibfnamefont {Y.-L.}\ \bibnamefont {Wang}},
  \bibinfo {author} {\bibfnamefont {J.}~\bibnamefont {Gong}}, \bibinfo {author}
  {\bibfnamefont {B.}~\bibnamefont {Shen}}, \bibinfo {author} {\bibfnamefont
  {Y.}~\bibnamefont {Huang}}, \bibinfo {author} {\bibfnamefont
  {H.}~\bibnamefont {Yang}}, \bibinfo {author} {\bibfnamefont {C.}~\bibnamefont
  {Ren}}, \ and\ \bibinfo {author} {\bibfnamefont {H.-H.}\ \bibnamefont
  {Wen}},\ }\href {\doibase 10.1103/PhysRevB.83.060510} {\bibfield  {journal}
  {\bibinfo  {journal} {Physical Review B}\ }\textbf {\bibinfo {volume} {83}},\
  \bibinfo {pages} {060510} (\bibinfo {year} {2011}{\natexlab{b}})}\BibitemShut
  {NoStop}%
\bibitem [{\citenamefont {Teague}\ \emph {et~al.}(2011)\citenamefont {Teague},
  \citenamefont {Drayna}, \citenamefont {Lockhart}, \citenamefont {Cheng},
  \citenamefont {Shen}, \citenamefont {Wen},\ and\ \citenamefont
  {Yeh}}]{TeaguePRL2011}%
  \BibitemOpen
  \bibfield  {author} {\bibinfo {author} {\bibfnamefont {M.}~\bibnamefont
  {Teague}}, \bibinfo {author} {\bibfnamefont {G.}~\bibnamefont {Drayna}},
  \bibinfo {author} {\bibfnamefont {G.}~\bibnamefont {Lockhart}}, \bibinfo
  {author} {\bibfnamefont {P.}~\bibnamefont {Cheng}}, \bibinfo {author}
  {\bibfnamefont {B.}~\bibnamefont {Shen}}, \bibinfo {author} {\bibfnamefont
  {H.-H.}\ \bibnamefont {Wen}}, \ and\ \bibinfo {author} {\bibfnamefont
  {N.-C.}\ \bibnamefont {Yeh}},\ }\href {\doibase
  10.1103/PhysRevLett.106.087004} {\bibfield  {journal} {\bibinfo  {journal}
  {Physical Review Letters}\ }\textbf {\bibinfo {volume} {106}},\ \bibinfo
  {pages} {087004} (\bibinfo {year} {2011})}\BibitemShut {NoStop}%
\bibitem [{\citenamefont {Ye}\ \emph {et~al.}()\citenamefont {Ye},
  \citenamefont {Zhang}, \citenamefont {Xu}, \citenamefont {Ge}, \citenamefont
  {Chen}, \citenamefont {Jiang}, \citenamefont {Xie}, \citenamefont {Hu},\ and\
  \citenamefont {Feng}}]{YeArxiv1105.5242}%
  \BibitemOpen
  \bibfield  {author} {\bibinfo {author} {\bibfnamefont {Z.~R.}\ \bibnamefont
  {Ye}}, \bibinfo {author} {\bibfnamefont {Y.}~\bibnamefont {Zhang}}, \bibinfo
  {author} {\bibfnamefont {M.}~\bibnamefont {Xu}}, \bibinfo {author}
  {\bibfnamefont {Q.~Q.}\ \bibnamefont {Ge}}, \bibinfo {author} {\bibfnamefont
  {F.}~\bibnamefont {Chen}}, \bibinfo {author} {\bibfnamefont {J.}~\bibnamefont
  {Jiang}}, \bibinfo {author} {\bibfnamefont {B.~P.}\ \bibnamefont {Xie}},
  \bibinfo {author} {\bibfnamefont {J.~P.}\ \bibnamefont {Hu}}, \ and\ \bibinfo
  {author} {\bibfnamefont {D.~L.}\ \bibnamefont {Feng}},\ }\href@noop {} {\
  }\Eprint {http://arxiv.org/abs/1105.5242} {arXiv:1105.5242} \BibitemShut
  {NoStop}%
\bibitem [{\citenamefont {Hashimoto}\ \emph {et~al.}(2010)\citenamefont
  {Hashimoto}, \citenamefont {Yamashita}, \citenamefont {Kasahara},
  \citenamefont {Senshu}, \citenamefont {Nakata}, \citenamefont {Tonegawa},
  \citenamefont {Ikada}, \citenamefont {Serafin}, \citenamefont {Carrington},
  \citenamefont {Terashima}, \citenamefont {Ikeda}, \citenamefont {Shibauchi},\
  and\ \citenamefont {Matsuda}}]{HashimotoPRB2010}%
  \BibitemOpen
  \bibfield  {author} {\bibinfo {author} {\bibfnamefont {K.}~\bibnamefont
  {Hashimoto}}, \bibinfo {author} {\bibfnamefont {M.}~\bibnamefont
  {Yamashita}}, \bibinfo {author} {\bibfnamefont {S.}~\bibnamefont {Kasahara}},
  \bibinfo {author} {\bibfnamefont {Y.}~\bibnamefont {Senshu}}, \bibinfo
  {author} {\bibfnamefont {N.}~\bibnamefont {Nakata}}, \bibinfo {author}
  {\bibfnamefont {S.}~\bibnamefont {Tonegawa}}, \bibinfo {author}
  {\bibfnamefont {K.}~\bibnamefont {Ikada}}, \bibinfo {author} {\bibfnamefont
  {A.}~\bibnamefont {Serafin}}, \bibinfo {author} {\bibfnamefont
  {A.}~\bibnamefont {Carrington}}, \bibinfo {author} {\bibfnamefont
  {T.}~\bibnamefont {Terashima}}, \bibinfo {author} {\bibfnamefont
  {H.}~\bibnamefont {Ikeda}}, \bibinfo {author} {\bibfnamefont
  {T.}~\bibnamefont {Shibauchi}}, \ and\ \bibinfo {author} {\bibfnamefont
  {Y.}~\bibnamefont {Matsuda}},\ }\href {\doibase 10.1103/PhysRevB.81.220501}
  {\bibfield  {journal} {\bibinfo  {journal} {Physical Review B}\ }\textbf
  {\bibinfo {volume} {81}},\ \bibinfo {pages} {220501} (\bibinfo {year}
  {2010})}\BibitemShut {NoStop}%
\bibitem [{\citenamefont {Nakai}\ \emph {et~al.}(2010)\citenamefont {Nakai},
  \citenamefont {Iye}, \citenamefont {Kitagawa}, \citenamefont {Ishida},
  \citenamefont {Kasahara}, \citenamefont {Shibauchi}, \citenamefont
  {Matsuda},\ and\ \citenamefont {Terashima}}]{NakaiPRB2010}%
  \BibitemOpen
  \bibfield  {author} {\bibinfo {author} {\bibfnamefont {Y.}~\bibnamefont
  {Nakai}}, \bibinfo {author} {\bibfnamefont {T.}~\bibnamefont {Iye}}, \bibinfo
  {author} {\bibfnamefont {S.}~\bibnamefont {Kitagawa}}, \bibinfo {author}
  {\bibfnamefont {K.}~\bibnamefont {Ishida}}, \bibinfo {author} {\bibfnamefont
  {S.}~\bibnamefont {Kasahara}}, \bibinfo {author} {\bibfnamefont
  {T.}~\bibnamefont {Shibauchi}}, \bibinfo {author} {\bibfnamefont
  {Y.}~\bibnamefont {Matsuda}}, \ and\ \bibinfo {author} {\bibfnamefont
  {T.}~\bibnamefont {Terashima}},\ }\href {\doibase 10.1103/PhysRevB.81.020503}
  {\bibfield  {journal} {\bibinfo  {journal} {Physical Review B}\ }\textbf
  {\bibinfo {volume} {81}},\ \bibinfo {pages} {020503} (\bibinfo {year}
  {2010})}\BibitemShut {NoStop}%
\bibitem [{\citenamefont {Kuroki}\ \emph {et~al.}(2009)\citenamefont {Kuroki},
  \citenamefont {Usui}, \citenamefont {Onari}, \citenamefont {Arita},\ and\
  \citenamefont {Aoki}}]{KurokiPRB2009}%
  \BibitemOpen
  \bibfield  {author} {\bibinfo {author} {\bibfnamefont {K.}~\bibnamefont
  {Kuroki}}, \bibinfo {author} {\bibfnamefont {H.}~\bibnamefont {Usui}},
  \bibinfo {author} {\bibfnamefont {S.}~\bibnamefont {Onari}}, \bibinfo
  {author} {\bibfnamefont {R.}~\bibnamefont {Arita}}, \ and\ \bibinfo {author}
  {\bibfnamefont {H.}~\bibnamefont {Aoki}},\ }\href {\doibase
  10.1103/PhysRevB.79.224511} {\bibfield  {journal} {\bibinfo  {journal}
  {Physical Review B}\ }\textbf {\bibinfo {volume} {79}},\ \bibinfo {pages}
  {224511} (\bibinfo {year} {2009})}\BibitemShut {NoStop}%
\bibitem [{\citenamefont {Rotter}\ \emph {et~al.}(2010)\citenamefont {Rotter},
  \citenamefont {Hieke},\ and\ \citenamefont {Johrendt}}]{RotterPRB2010}%
  \BibitemOpen
  \bibfield  {author} {\bibinfo {author} {\bibfnamefont {M.}~\bibnamefont
  {Rotter}}, \bibinfo {author} {\bibfnamefont {C.}~\bibnamefont {Hieke}}, \
  and\ \bibinfo {author} {\bibfnamefont {D.}~\bibnamefont {Johrendt}},\ }\href
  {\doibase 10.1103/PhysRevB.82.014513} {\bibfield  {journal} {\bibinfo
  {journal} {Physical Review B}\ }\textbf {\bibinfo {volume} {82}},\ \bibinfo
  {pages} {014513} (\bibinfo {year} {2010})}\BibitemShut {NoStop}%
\bibitem [{\citenamefont {Guo}\ \emph {et~al.}(2010)\citenamefont {Guo},
  \citenamefont {Jin}, \citenamefont {Wang}, \citenamefont {Wang},
  \citenamefont {Zhu}, \citenamefont {Zhou}, \citenamefont {He},\ and\
  \citenamefont {Chen}}]{GuoPRB2010}%
  \BibitemOpen
  \bibfield  {author} {\bibinfo {author} {\bibfnamefont {J.}~\bibnamefont
  {Guo}}, \bibinfo {author} {\bibfnamefont {S.}~\bibnamefont {Jin}}, \bibinfo
  {author} {\bibfnamefont {G.}~\bibnamefont {Wang}}, \bibinfo {author}
  {\bibfnamefont {S.}~\bibnamefont {Wang}}, \bibinfo {author} {\bibfnamefont
  {K.}~\bibnamefont {Zhu}}, \bibinfo {author} {\bibfnamefont {T.}~\bibnamefont
  {Zhou}}, \bibinfo {author} {\bibfnamefont {M.}~\bibnamefont {He}}, \ and\
  \bibinfo {author} {\bibfnamefont {X.}~\bibnamefont {Chen}},\ }\href {\doibase
  10.1103/PhysRevB.82.180520} {\bibfield  {journal} {\bibinfo  {journal}
  {Physical Review B}\ }\textbf {\bibinfo {volume} {82}},\ \bibinfo {pages}
  {180520} (\bibinfo {year} {2010})}\BibitemShut {NoStop}%
\bibitem [{\citenamefont {Fang}\ \emph {et~al.}(2011)\citenamefont {Fang},
  \citenamefont {Wang}, \citenamefont {Dong}, \citenamefont {Li}, \citenamefont
  {Feng}, \citenamefont {Chen},\ and\ \citenamefont {Yuan}}]{FangEPL2011}%
  \BibitemOpen
  \bibfield  {author} {\bibinfo {author} {\bibfnamefont {M.-H.}\ \bibnamefont
  {Fang}}, \bibinfo {author} {\bibfnamefont {H.-D.}\ \bibnamefont {Wang}},
  \bibinfo {author} {\bibfnamefont {C.-H.}\ \bibnamefont {Dong}}, \bibinfo
  {author} {\bibfnamefont {Z.-J.}\ \bibnamefont {Li}}, \bibinfo {author}
  {\bibfnamefont {C.-M.}\ \bibnamefont {Feng}}, \bibinfo {author}
  {\bibfnamefont {J.}~\bibnamefont {Chen}}, \ and\ \bibinfo {author}
  {\bibfnamefont {H.~Q.}\ \bibnamefont {Yuan}},\ }\href {\doibase
  10.1209/0295-5075/94/27009} {\bibfield  {journal} {\bibinfo  {journal}
  {Europhysics Letters}\ }\textbf {\bibinfo {volume} {94}},\ \bibinfo {pages}
  {27009} (\bibinfo {year} {2011})}\BibitemShut {NoStop}%
\bibitem [{\citenamefont {Qian}\ \emph {et~al.}(2011)\citenamefont {Qian},
  \citenamefont {Wang}, \citenamefont {Jin}, \citenamefont {Zhang},
  \citenamefont {Richard}, \citenamefont {Xu}, \citenamefont {Dai},
  \citenamefont {Fang}, \citenamefont {Guo}, \citenamefont {Chen},\ and\
  \citenamefont {Ding}}]{QianPRL2011}%
  \BibitemOpen
  \bibfield  {author} {\bibinfo {author} {\bibfnamefont {T.}~\bibnamefont
  {Qian}}, \bibinfo {author} {\bibfnamefont {X.-P.}\ \bibnamefont {Wang}},
  \bibinfo {author} {\bibfnamefont {W.-C.}\ \bibnamefont {Jin}}, \bibinfo
  {author} {\bibfnamefont {P.}~\bibnamefont {Zhang}}, \bibinfo {author}
  {\bibfnamefont {P.}~\bibnamefont {Richard}}, \bibinfo {author} {\bibfnamefont
  {G.}~\bibnamefont {Xu}}, \bibinfo {author} {\bibfnamefont {X.}~\bibnamefont
  {Dai}}, \bibinfo {author} {\bibfnamefont {Z.}~\bibnamefont {Fang}}, \bibinfo
  {author} {\bibfnamefont {J.-G.}\ \bibnamefont {Guo}}, \bibinfo {author}
  {\bibfnamefont {X.-L.}\ \bibnamefont {Chen}}, \ and\ \bibinfo {author}
  {\bibfnamefont {H.}~\bibnamefont {Ding}},\ }\href {\doibase
  10.1103/PhysRevLett.106.187001} {\bibfield  {journal} {\bibinfo  {journal}
  {Physical Review Letters}\ }\textbf {\bibinfo {volume} {106}},\ \bibinfo
  {pages} {187001} (\bibinfo {year} {2011})}\BibitemShut {NoStop}%
\bibitem [{\citenamefont {Li}\ \emph {et~al.}(2011)\citenamefont {Li},
  \citenamefont {Ding}, \citenamefont {Deng}, \citenamefont {Chang},
  \citenamefont {Song}, \citenamefont {He}, \citenamefont {Wang}, \citenamefont
  {Ma}, \citenamefont {Hu}, \citenamefont {Chen},\ and\ \citenamefont
  {Xue}}]{LiNatPhys2011}%
  \BibitemOpen
  \bibfield  {author} {\bibinfo {author} {\bibfnamefont {W.}~\bibnamefont
  {Li}}, \bibinfo {author} {\bibfnamefont {H.}~\bibnamefont {Ding}}, \bibinfo
  {author} {\bibfnamefont {P.}~\bibnamefont {Deng}}, \bibinfo {author}
  {\bibfnamefont {K.}~\bibnamefont {Chang}}, \bibinfo {author} {\bibfnamefont
  {C.-L.}\ \bibnamefont {Song}}, \bibinfo {author} {\bibfnamefont
  {K.}~\bibnamefont {He}}, \bibinfo {author} {\bibfnamefont {L.}~\bibnamefont
  {Wang}}, \bibinfo {author} {\bibfnamefont {X.}~\bibnamefont {Ma}}, \bibinfo
  {author} {\bibfnamefont {J.-P.}\ \bibnamefont {Hu}}, \bibinfo {author}
  {\bibfnamefont {X.}~\bibnamefont {Chen}}, \ and\ \bibinfo {author}
  {\bibfnamefont {Q.-K.}\ \bibnamefont {Xue}},\ }\href {\doibase
  10.1038/nphys2155} {\bibfield  {journal} {\bibinfo  {journal} {Nature
  Physics}\ } (\bibinfo {year} {2011}),\ 10.1038/nphys2155}\BibitemShut
  {NoStop}%
\bibitem [{\citenamefont {Cai}\ \emph {et~al.}()\citenamefont {Cai},
  \citenamefont {Ye}, \citenamefont {Ruan}, \citenamefont {Zhou}, \citenamefont
  {Wang}, \citenamefont {Zhang}, \citenamefont {Chen},\ and\ \citenamefont
  {Wang}}]{CaiArxiv1108.2798}%
  \BibitemOpen
  \bibfield  {author} {\bibinfo {author} {\bibfnamefont {P.}~\bibnamefont
  {Cai}}, \bibinfo {author} {\bibfnamefont {C.}~\bibnamefont {Ye}}, \bibinfo
  {author} {\bibfnamefont {W.}~\bibnamefont {Ruan}}, \bibinfo {author}
  {\bibfnamefont {X.}~\bibnamefont {Zhou}}, \bibinfo {author} {\bibfnamefont
  {A.}~\bibnamefont {Wang}}, \bibinfo {author} {\bibfnamefont {M.}~\bibnamefont
  {Zhang}}, \bibinfo {author} {\bibfnamefont {X.}~\bibnamefont {Chen}}, \ and\
  \bibinfo {author} {\bibfnamefont {Y.}~\bibnamefont {Wang}},\ }\href@noop {}
  {\ }\Eprint {http://arxiv.org/abs/1108.2798} {arXiv:1108.2798} \BibitemShut
  {NoStop}%
\bibitem [{\citenamefont {Hsu}\ \emph {et~al.}(2008)\citenamefont {Hsu},
  \citenamefont {Luo}, \citenamefont {Yeh}, \citenamefont {Chen}, \citenamefont
  {Huang}, \citenamefont {Wu}, \citenamefont {Lee}, \citenamefont {Huang},
  \citenamefont {Chu}, \citenamefont {Yan},\ and\ \citenamefont
  {Wu}}]{HsuPNAS2008}%
  \BibitemOpen
  \bibfield  {author} {\bibinfo {author} {\bibfnamefont {F.-C.}\ \bibnamefont
  {Hsu}}, \bibinfo {author} {\bibfnamefont {J.-Y.}\ \bibnamefont {Luo}},
  \bibinfo {author} {\bibfnamefont {K.-W.}\ \bibnamefont {Yeh}}, \bibinfo
  {author} {\bibfnamefont {T.-K.}\ \bibnamefont {Chen}}, \bibinfo {author}
  {\bibfnamefont {T.-W.}\ \bibnamefont {Huang}}, \bibinfo {author}
  {\bibfnamefont {P.~M.}\ \bibnamefont {Wu}}, \bibinfo {author} {\bibfnamefont
  {Y.-C.}\ \bibnamefont {Lee}}, \bibinfo {author} {\bibfnamefont {Y.-L.}\
  \bibnamefont {Huang}}, \bibinfo {author} {\bibfnamefont {Y.-Y.}\ \bibnamefont
  {Chu}}, \bibinfo {author} {\bibfnamefont {D.-C.}\ \bibnamefont {Yan}}, \ and\
  \bibinfo {author} {\bibfnamefont {M.-K.}\ \bibnamefont {Wu}},\ }\href
  {\doibase 10.1073/pnas.0807325105} {\bibfield  {journal} {\bibinfo  {journal}
  {Proceedings of the National Academy of Sciences of the United States of
  America}\ }\textbf {\bibinfo {volume} {105}},\ \bibinfo {pages} {14262}
  (\bibinfo {year} {2008})}\BibitemShut {NoStop}%
\bibitem [{\citenamefont {McElroy}\ \emph {et~al.}(2005)\citenamefont
  {McElroy}, \citenamefont {Lee}, \citenamefont {Hoffman}, \citenamefont
  {Lang}, \citenamefont {Lee}, \citenamefont {Hudson}, \citenamefont {Eisaki},
  \citenamefont {Uchida},\ and\ \citenamefont {Davis}}]{McElroyPRL2005}%
  \BibitemOpen
  \bibfield  {author} {\bibinfo {author} {\bibfnamefont {K.}~\bibnamefont
  {McElroy}}, \bibinfo {author} {\bibfnamefont {D.-H.}\ \bibnamefont {Lee}},
  \bibinfo {author} {\bibfnamefont {J.}~\bibnamefont {Hoffman}}, \bibinfo
  {author} {\bibfnamefont {K.}~\bibnamefont {Lang}}, \bibinfo {author}
  {\bibfnamefont {J.}~\bibnamefont {Lee}}, \bibinfo {author} {\bibfnamefont
  {E.}~\bibnamefont {Hudson}}, \bibinfo {author} {\bibfnamefont
  {H.}~\bibnamefont {Eisaki}}, \bibinfo {author} {\bibfnamefont
  {S.}~\bibnamefont {Uchida}}, \ and\ \bibinfo {author} {\bibfnamefont {J.~C.}\
  \bibnamefont {Davis}},\ }\href {\doibase 10.1103/PhysRevLett.94.197005}
  {\bibfield  {journal} {\bibinfo  {journal} {Physical Review Letters}\
  }\textbf {\bibinfo {volume} {94}},\ \bibinfo {pages} {197005} (\bibinfo
  {year} {2005})}\BibitemShut {NoStop}%
\bibitem [{\citenamefont {Fridman}\ \emph {et~al.}(2011)\citenamefont
  {Fridman}, \citenamefont {Yeh}, \citenamefont {Wu},\ and\ \citenamefont
  {Wei}}]{FridmanJPCS2010}%
  \BibitemOpen
  \bibfield  {author} {\bibinfo {author} {\bibfnamefont {I.}~\bibnamefont
  {Fridman}}, \bibinfo {author} {\bibfnamefont {K.-W.}\ \bibnamefont {Yeh}},
  \bibinfo {author} {\bibfnamefont {M.-K.}\ \bibnamefont {Wu}}, \ and\ \bibinfo
  {author} {\bibfnamefont {J.~Y.~T.}\ \bibnamefont {Wei}},\ }\href {\doibase
  10.1016/j.jpcs.2010.10.002} {\bibfield  {journal} {\bibinfo  {journal}
  {Journal of Physics and Chemistry of Solids}\ }\textbf {\bibinfo {volume}
  {72}},\ \bibinfo {pages} {483} (\bibinfo {year} {2011})}\BibitemShut
  {NoStop}%
\bibitem [{\citenamefont {He}\ \emph {et~al.}(2011)\citenamefont {He},
  \citenamefont {Li}, \citenamefont {Zhang}, \citenamefont {Karki},
  \citenamefont {Jin}, \citenamefont {Sales}, \citenamefont {Sefat},
  \citenamefont {McGuire}, \citenamefont {Mandrus},\ and\ \citenamefont
  {Plummer}}]{HePRB2011}%
  \BibitemOpen
  \bibfield  {author} {\bibinfo {author} {\bibfnamefont {X.}~\bibnamefont
  {He}}, \bibinfo {author} {\bibfnamefont {G.}~\bibnamefont {Li}}, \bibinfo
  {author} {\bibfnamefont {J.}~\bibnamefont {Zhang}}, \bibinfo {author}
  {\bibfnamefont {A.}~\bibnamefont {Karki}}, \bibinfo {author} {\bibfnamefont
  {R.}~\bibnamefont {Jin}}, \bibinfo {author} {\bibfnamefont {B.}~\bibnamefont
  {Sales}}, \bibinfo {author} {\bibfnamefont {A.}~\bibnamefont {Sefat}},
  \bibinfo {author} {\bibfnamefont {M.}~\bibnamefont {McGuire}}, \bibinfo
  {author} {\bibfnamefont {D.}~\bibnamefont {Mandrus}}, \ and\ \bibinfo
  {author} {\bibfnamefont {E.~W.}\ \bibnamefont {Plummer}},\ }\href {\doibase
  10.1103/PhysRevB.83.220502} {\bibfield  {journal} {\bibinfo  {journal}
  {Physical Review B}\ }\textbf {\bibinfo {volume} {83}},\ \bibinfo {pages}
  {220502} (\bibinfo {year} {2011})}\BibitemShut {NoStop}%
\bibitem [{\citenamefont {Tegel}\ \emph {et~al.}(2010)\citenamefont {Tegel},
  \citenamefont {L\"{o}hnert},\ and\ \citenamefont {Johrendt}}]{TegelSSC2010}%
  \BibitemOpen
  \bibfield  {author} {\bibinfo {author} {\bibfnamefont {M.}~\bibnamefont
  {Tegel}}, \bibinfo {author} {\bibfnamefont {C.}~\bibnamefont {L\"{o}hnert}},
  \ and\ \bibinfo {author} {\bibfnamefont {D.}~\bibnamefont {Johrendt}},\
  }\href {\doibase 10.1016/j.ssc.2010.01.002} {\bibfield  {journal} {\bibinfo
  {journal} {Solid State Communications}\ }\textbf {\bibinfo {volume} {150}},\
  \bibinfo {pages} {383} (\bibinfo {year} {2010})}\BibitemShut {NoStop}%
\bibitem [{\citenamefont {Lankau}\ \emph {et~al.}(2010)\citenamefont {Lankau},
  \citenamefont {Koepernik}, \citenamefont {Borisenko}, \citenamefont
  {Zabolotnyy}, \citenamefont {B\"{u}chner}, \citenamefont {van~den Brink},\
  and\ \citenamefont {Eschrig}}]{LankauPRB2010}%
  \BibitemOpen
  \bibfield  {author} {\bibinfo {author} {\bibfnamefont {A.}~\bibnamefont
  {Lankau}}, \bibinfo {author} {\bibfnamefont {K.}~\bibnamefont {Koepernik}},
  \bibinfo {author} {\bibfnamefont {S.}~\bibnamefont {Borisenko}}, \bibinfo
  {author} {\bibfnamefont {V.}~\bibnamefont {Zabolotnyy}}, \bibinfo {author}
  {\bibfnamefont {B.}~\bibnamefont {B\"{u}chner}}, \bibinfo {author}
  {\bibfnamefont {J.}~\bibnamefont {van~den Brink}}, \ and\ \bibinfo {author}
  {\bibfnamefont {H.}~\bibnamefont {Eschrig}},\ }\href {\doibase
  10.1103/PhysRevB.82.184518} {\bibfield  {journal} {\bibinfo  {journal}
  {Physical Review B}\ }\textbf {\bibinfo {volume} {82}},\ \bibinfo {pages}
  {184518} (\bibinfo {year} {2010})}\BibitemShut {NoStop}%
\bibitem [{\citenamefont {Davis}(2011)}]{DavisBCcolloquium2011}%
  \BibitemOpen
  \bibfield  {author} {\bibinfo {author} {\bibfnamefont {J.~C.}\ \bibnamefont
  {Davis}},\ }\href
  {http://www.bc.edu/content/bc/schools/cas/physics/news-and-events/seminars-colloquia/coll\_09-28-2011.html}
  {\bibfield  {journal} {\bibinfo  {journal} {colloquium}\ } (\bibinfo {year}
  {2011})}\BibitemShut {NoStop}%
\bibitem [{\citenamefont {Zhu}\ \emph {et~al.}(2009)\citenamefont {Zhu},
  \citenamefont {Han}, \citenamefont {Mu}, \citenamefont {Cheng}, \citenamefont
  {Shen}, \citenamefont {Zeng},\ and\ \citenamefont {Wen}}]{ZhuPRB2009}%
  \BibitemOpen
  \bibfield  {author} {\bibinfo {author} {\bibfnamefont {X.}~\bibnamefont
  {Zhu}}, \bibinfo {author} {\bibfnamefont {F.}~\bibnamefont {Han}}, \bibinfo
  {author} {\bibfnamefont {G.}~\bibnamefont {Mu}}, \bibinfo {author}
  {\bibfnamefont {P.}~\bibnamefont {Cheng}}, \bibinfo {author} {\bibfnamefont
  {B.}~\bibnamefont {Shen}}, \bibinfo {author} {\bibfnamefont {B.}~\bibnamefont
  {Zeng}}, \ and\ \bibinfo {author} {\bibfnamefont {H.-H.}\ \bibnamefont
  {Wen}},\ }\href {\doibase 10.1103/PhysRevB.79.220512} {\bibfield  {journal}
  {\bibinfo  {journal} {Physical Review B}\ }\textbf {\bibinfo {volume} {79}},\
  \bibinfo {pages} {220512} (\bibinfo {year} {2009})}\BibitemShut {NoStop}%
\bibitem [{\citenamefont {Ogino}\ \emph {et~al.}(2009)\citenamefont {Ogino},
  \citenamefont {Matsumura}, \citenamefont {Katsura}, \citenamefont {Ushiyama},
  \citenamefont {Horii}, \citenamefont {Kishio},\ and\ \citenamefont
  {Shimoyama}}]{OginoSST2009}%
  \BibitemOpen
  \bibfield  {author} {\bibinfo {author} {\bibfnamefont {H.}~\bibnamefont
  {Ogino}}, \bibinfo {author} {\bibfnamefont {Y.}~\bibnamefont {Matsumura}},
  \bibinfo {author} {\bibfnamefont {Y.}~\bibnamefont {Katsura}}, \bibinfo
  {author} {\bibfnamefont {K.}~\bibnamefont {Ushiyama}}, \bibinfo {author}
  {\bibfnamefont {S.}~\bibnamefont {Horii}}, \bibinfo {author} {\bibfnamefont
  {K.}~\bibnamefont {Kishio}}, \ and\ \bibinfo {author} {\bibfnamefont {J.-i.}\
  \bibnamefont {Shimoyama}},\ }\href {\doibase 10.1088/0953-2048/22/7/075008}
  {\bibfield  {journal} {\bibinfo  {journal} {Superconductor Science and
  Technology}\ }\textbf {\bibinfo {volume} {22}},\ \bibinfo {pages} {075008}
  (\bibinfo {year} {2009})}\BibitemShut {NoStop}%
\bibitem [{\citenamefont {Chen}\ \emph {et~al.}(2009)\citenamefont {Chen},
  \citenamefont {Xia}, \citenamefont {Yang}, \citenamefont {Li}, \citenamefont
  {Zheng}, \citenamefont {Luo},\ and\ \citenamefont {Wang}}]{ChenSST2009}%
  \BibitemOpen
  \bibfield  {author} {\bibinfo {author} {\bibfnamefont {G.~F.}\ \bibnamefont
  {Chen}}, \bibinfo {author} {\bibfnamefont {T.-L.}\ \bibnamefont {Xia}},
  \bibinfo {author} {\bibfnamefont {H.~X.}\ \bibnamefont {Yang}}, \bibinfo
  {author} {\bibfnamefont {J.~Q.}\ \bibnamefont {Li}}, \bibinfo {author}
  {\bibfnamefont {P.}~\bibnamefont {Zheng}}, \bibinfo {author} {\bibfnamefont
  {J.~L.}\ \bibnamefont {Luo}}, \ and\ \bibinfo {author} {\bibfnamefont
  {N.~L.}\ \bibnamefont {Wang}},\ }\href {\doibase
  10.1088/0953-2048/22/7/072001} {\bibfield  {journal} {\bibinfo  {journal}
  {Superconductor Science and Technology}\ }\textbf {\bibinfo {volume} {22}},\
  \bibinfo {pages} {072001} (\bibinfo {year} {2009})}\BibitemShut {NoStop}%
\bibitem [{\citenamefont {Kawaguchi}\ \emph {et~al.}(2010)\citenamefont
  {Kawaguchi}, \citenamefont {Ogino}, \citenamefont {Shimizu}, \citenamefont
  {Kishio},\ and\ \citenamefont {Shimoyama}}]{KawaguchiAPEX2010}%
  \BibitemOpen
  \bibfield  {author} {\bibinfo {author} {\bibfnamefont {N.}~\bibnamefont
  {Kawaguchi}}, \bibinfo {author} {\bibfnamefont {H.}~\bibnamefont {Ogino}},
  \bibinfo {author} {\bibfnamefont {Y.}~\bibnamefont {Shimizu}}, \bibinfo
  {author} {\bibfnamefont {K.}~\bibnamefont {Kishio}}, \ and\ \bibinfo {author}
  {\bibfnamefont {J.-i.}\ \bibnamefont {Shimoyama}},\ }\href {\doibase
  10.1143/APEX.3.063102} {\bibfield  {journal} {\bibinfo  {journal} {Applied
  Physics Express}\ }\textbf {\bibinfo {volume} {3}},\ \bibinfo {pages}
  {063102} (\bibinfo {year} {2010})}\BibitemShut {NoStop}%
\bibitem [{\citenamefont {Ogino}\ \emph {et~al.}(2010)\citenamefont {Ogino},
  \citenamefont {Sato}, \citenamefont {Kishio}, \citenamefont {Shimoyama},
  \citenamefont {Tohei},\ and\ \citenamefont {Ikuhara}}]{OginoAPL2010}%
  \BibitemOpen
  \bibfield  {author} {\bibinfo {author} {\bibfnamefont {H.}~\bibnamefont
  {Ogino}}, \bibinfo {author} {\bibfnamefont {S.}~\bibnamefont {Sato}},
  \bibinfo {author} {\bibfnamefont {K.}~\bibnamefont {Kishio}}, \bibinfo
  {author} {\bibfnamefont {J.-i.}\ \bibnamefont {Shimoyama}}, \bibinfo {author}
  {\bibfnamefont {T.}~\bibnamefont {Tohei}}, \ and\ \bibinfo {author}
  {\bibfnamefont {Y.}~\bibnamefont {Ikuhara}},\ }\href {\doibase
  10.1063/1.3478850} {\bibfield  {journal} {\bibinfo  {journal} {Applied
  Physics Letters}\ }\textbf {\bibinfo {volume} {97}},\ \bibinfo {pages}
  {072506} (\bibinfo {year} {2010})}\BibitemShut {NoStop}%
\bibitem [{\citenamefont {Wen}(2011)}]{WenPrivate2011}%
  \BibitemOpen
  \bibfield  {author} {\bibinfo {author} {\bibfnamefont {H.-H.}\ \bibnamefont
  {Wen}},\ }\href@noop {} {\bibfield  {journal} {\bibinfo  {journal} {private
  communication}\ } (\bibinfo {year} {2011})}\BibitemShut {NoStop}%
\bibitem [{\citenamefont {Hicks}\ \emph {et~al.}()\citenamefont {Hicks},
  \citenamefont {Lippman}, \citenamefont {Huber}, \citenamefont {Ren},
  \citenamefont {Zhao},\ and\ \citenamefont {Moler}}]{HicksArxiv0807.0467v1}%
  \BibitemOpen
  \bibfield  {author} {\bibinfo {author} {\bibfnamefont {C.~W.}\ \bibnamefont
  {Hicks}}, \bibinfo {author} {\bibfnamefont {T.~M.}\ \bibnamefont {Lippman}},
  \bibinfo {author} {\bibfnamefont {M.~E.}\ \bibnamefont {Huber}}, \bibinfo
  {author} {\bibfnamefont {Z.~A.}\ \bibnamefont {Ren}}, \bibinfo {author}
  {\bibfnamefont {Z.~X.}\ \bibnamefont {Zhao}}, \ and\ \bibinfo {author}
  {\bibfnamefont {K.~A.}\ \bibnamefont {Moler}},\ }\href@noop {} {\ }\Eprint
  {http://arxiv.org/abs/0807.0467v1} {arXiv:0807.0467v1} \BibitemShut {NoStop}%
\bibitem [{\citenamefont {Mazin}\ \emph {et~al.}(2008)\citenamefont {Mazin},
  \citenamefont {Singh}, \citenamefont {Johannes},\ and\ \citenamefont
  {Du}}]{MazinPRL2008}%
  \BibitemOpen
  \bibfield  {author} {\bibinfo {author} {\bibfnamefont {I.~I.}\ \bibnamefont
  {Mazin}}, \bibinfo {author} {\bibfnamefont {D.~J.}\ \bibnamefont {Singh}},
  \bibinfo {author} {\bibfnamefont {M.~D.}\ \bibnamefont {Johannes}}, \ and\
  \bibinfo {author} {\bibfnamefont {M.~H.}\ \bibnamefont {Du}},\ }\href
  {\doibase 10.1103/PhysRevLett.101.057003} {\bibfield  {journal} {\bibinfo
  {journal} {Physical Review Letters}\ }\textbf {\bibinfo {volume} {101}},\
  \bibinfo {pages} {057003} (\bibinfo {year} {2008})}\BibitemShut {NoStop}%
\bibitem [{\citenamefont {Kuroki}\ \emph {et~al.}(2008)\citenamefont {Kuroki},
  \citenamefont {Onari}, \citenamefont {Arita}, \citenamefont {Usui},
  \citenamefont {Tanaka}, \citenamefont {Kontani},\ and\ \citenamefont
  {Aoki}}]{KurokiPRL2008}%
  \BibitemOpen
  \bibfield  {author} {\bibinfo {author} {\bibfnamefont {K.}~\bibnamefont
  {Kuroki}}, \bibinfo {author} {\bibfnamefont {S.}~\bibnamefont {Onari}},
  \bibinfo {author} {\bibfnamefont {R.}~\bibnamefont {Arita}}, \bibinfo
  {author} {\bibfnamefont {H.}~\bibnamefont {Usui}}, \bibinfo {author}
  {\bibfnamefont {Y.}~\bibnamefont {Tanaka}}, \bibinfo {author} {\bibfnamefont
  {H.}~\bibnamefont {Kontani}}, \ and\ \bibinfo {author} {\bibfnamefont
  {H.}~\bibnamefont {Aoki}},\ }\href {\doibase 10.1103/PhysRevLett.101.087004}
  {\bibfield  {journal} {\bibinfo  {journal} {Physical Review Letters}\
  }\textbf {\bibinfo {volume} {101}},\ \bibinfo {pages} {087004} (\bibinfo
  {year} {2008})}\BibitemShut {NoStop}%
\bibitem [{\citenamefont {Ding}\ \emph {et~al.}(2008)\citenamefont {Ding},
  \citenamefont {Richard}, \citenamefont {Nakayama}, \citenamefont {Sugawara},
  \citenamefont {Arakane}, \citenamefont {Sekiba}, \citenamefont {Takayama},
  \citenamefont {Souma}, \citenamefont {Sato}, \citenamefont {Takahashi},
  \citenamefont {Wang}, \citenamefont {Dai}, \citenamefont {Fang},
  \citenamefont {Chen}, \citenamefont {Luo},\ and\ \citenamefont
  {Wang}}]{DingEPL2008}%
  \BibitemOpen
  \bibfield  {author} {\bibinfo {author} {\bibfnamefont {H.}~\bibnamefont
  {Ding}}, \bibinfo {author} {\bibfnamefont {P.}~\bibnamefont {Richard}},
  \bibinfo {author} {\bibfnamefont {K.}~\bibnamefont {Nakayama}}, \bibinfo
  {author} {\bibfnamefont {K.}~\bibnamefont {Sugawara}}, \bibinfo {author}
  {\bibfnamefont {T.}~\bibnamefont {Arakane}}, \bibinfo {author} {\bibfnamefont
  {Y.}~\bibnamefont {Sekiba}}, \bibinfo {author} {\bibfnamefont
  {A.}~\bibnamefont {Takayama}}, \bibinfo {author} {\bibfnamefont
  {S.}~\bibnamefont {Souma}}, \bibinfo {author} {\bibfnamefont
  {T.}~\bibnamefont {Sato}}, \bibinfo {author} {\bibfnamefont {T.}~\bibnamefont
  {Takahashi}}, \bibinfo {author} {\bibfnamefont {Z.}~\bibnamefont {Wang}},
  \bibinfo {author} {\bibfnamefont {X.}~\bibnamefont {Dai}}, \bibinfo {author}
  {\bibfnamefont {Z.}~\bibnamefont {Fang}}, \bibinfo {author} {\bibfnamefont
  {G.~F.}\ \bibnamefont {Chen}}, \bibinfo {author} {\bibfnamefont {J.~L.}\
  \bibnamefont {Luo}}, \ and\ \bibinfo {author} {\bibfnamefont {N.~L.}\
  \bibnamefont {Wang}},\ }\href {\doibase 10.1209/0295-5075/83/47001}
  {\bibfield  {journal} {\bibinfo  {journal} {Europhysics Letters}\ }\textbf
  {\bibinfo {volume} {83}},\ \bibinfo {pages} {47001} (\bibinfo {year}
  {2008})}\BibitemShut {NoStop}%
\bibitem [{\citenamefont {Terashima}\ \emph {et~al.}(2009)\citenamefont
  {Terashima}, \citenamefont {Sekiba}, \citenamefont {Bowen}, \citenamefont
  {Nakayama}, \citenamefont {Kawahara}, \citenamefont {Sato}, \citenamefont
  {Richard}, \citenamefont {Xu}, \citenamefont {Li}, \citenamefont {Cao},
  \citenamefont {Xu}, \citenamefont {Ding},\ and\ \citenamefont
  {Takahashi}}]{TerashimaPNAS2009}%
  \BibitemOpen
  \bibfield  {author} {\bibinfo {author} {\bibfnamefont {K.}~\bibnamefont
  {Terashima}}, \bibinfo {author} {\bibfnamefont {Y.}~\bibnamefont {Sekiba}},
  \bibinfo {author} {\bibfnamefont {J.~H.}\ \bibnamefont {Bowen}}, \bibinfo
  {author} {\bibfnamefont {K.}~\bibnamefont {Nakayama}}, \bibinfo {author}
  {\bibfnamefont {T.}~\bibnamefont {Kawahara}}, \bibinfo {author}
  {\bibfnamefont {T.}~\bibnamefont {Sato}}, \bibinfo {author} {\bibfnamefont
  {P.}~\bibnamefont {Richard}}, \bibinfo {author} {\bibfnamefont {Y.-M.}\
  \bibnamefont {Xu}}, \bibinfo {author} {\bibfnamefont {L.~J.}\ \bibnamefont
  {Li}}, \bibinfo {author} {\bibfnamefont {G.~H.}\ \bibnamefont {Cao}},
  \bibinfo {author} {\bibfnamefont {Z.-A.}\ \bibnamefont {Xu}}, \bibinfo
  {author} {\bibfnamefont {H.}~\bibnamefont {Ding}}, \ and\ \bibinfo {author}
  {\bibfnamefont {T.}~\bibnamefont {Takahashi}},\ }\href {\doibase
  10.1073/pnas.0900469106} {\bibfield  {journal} {\bibinfo  {journal}
  {Proceedings of the National Academy of Sciences of the United States of
  America}\ }\textbf {\bibinfo {volume} {106}},\ \bibinfo {pages} {7330}
  (\bibinfo {year} {2009})}\BibitemShut {NoStop}%
\bibitem [{\citenamefont {Hicks}\ \emph {et~al.}(2008)\citenamefont {Hicks},
  \citenamefont {Lippman}, \citenamefont {Huber}, \citenamefont {Ren},
  \citenamefont {Yang}, \citenamefont {Zhao},\ and\ \citenamefont
  {Moler}}]{HicksJPSJ2008}%
  \BibitemOpen
  \bibfield  {author} {\bibinfo {author} {\bibfnamefont {C.~W.}\ \bibnamefont
  {Hicks}}, \bibinfo {author} {\bibfnamefont {T.~M.}\ \bibnamefont {Lippman}},
  \bibinfo {author} {\bibfnamefont {M.~E.}\ \bibnamefont {Huber}}, \bibinfo
  {author} {\bibfnamefont {Z.-A.}\ \bibnamefont {Ren}}, \bibinfo {author}
  {\bibfnamefont {J.}~\bibnamefont {Yang}}, \bibinfo {author} {\bibfnamefont
  {Z.-X.}\ \bibnamefont {Zhao}}, \ and\ \bibinfo {author} {\bibfnamefont
  {K.~A.}\ \bibnamefont {Moler}},\ }\href {\doibase 10.1143/JPSJ.78.013708}
  {\bibfield  {journal} {\bibinfo  {journal} {Journal of the Physical Society
  of Japan}\ }\textbf {\bibinfo {volume} {78}},\ \bibinfo {pages} {013708}
  (\bibinfo {year} {2008})}\BibitemShut {NoStop}%
\bibitem [{\citenamefont {Chen}\ \emph
  {et~al.}(2010{\natexlab{b}})\citenamefont {Chen}, \citenamefont {Tsuei},
  \citenamefont {Ketchen}, \citenamefont {Ren},\ and\ \citenamefont
  {Zhao}}]{ChenNatPhys2010}%
  \BibitemOpen
  \bibfield  {author} {\bibinfo {author} {\bibfnamefont {C.-T.}\ \bibnamefont
  {Chen}}, \bibinfo {author} {\bibfnamefont {C.~C.}\ \bibnamefont {Tsuei}},
  \bibinfo {author} {\bibfnamefont {M.~B.}\ \bibnamefont {Ketchen}}, \bibinfo
  {author} {\bibfnamefont {Z.-A.}\ \bibnamefont {Ren}}, \ and\ \bibinfo
  {author} {\bibfnamefont {Z.~X.}\ \bibnamefont {Zhao}},\ }\href {\doibase
  10.1038/nphys1531} {\bibfield  {journal} {\bibinfo  {journal} {Nature
  Physics}\ }\textbf {\bibinfo {volume} {6}},\ \bibinfo {pages} {260} (\bibinfo
  {year} {2010}{\natexlab{b}})}\BibitemShut {NoStop}%
\bibitem [{\citenamefont {Berg}\ \emph {et~al.}(2011)\citenamefont {Berg},
  \citenamefont {Lindner},\ and\ \citenamefont {Pereg-Barnea}}]{BergPRL2011}%
  \BibitemOpen
  \bibfield  {author} {\bibinfo {author} {\bibfnamefont {E.}~\bibnamefont
  {Berg}}, \bibinfo {author} {\bibfnamefont {N.}~\bibnamefont {Lindner}}, \
  and\ \bibinfo {author} {\bibfnamefont {T.}~\bibnamefont {Pereg-Barnea}},\
  }\href {\doibase 10.1103/PhysRevLett.106.147003} {\bibfield  {journal}
  {\bibinfo  {journal} {Physical Review Letters}\ }\textbf {\bibinfo {volume}
  {106}},\ \bibinfo {pages} {147003} (\bibinfo {year} {2011})}\BibitemShut
  {NoStop}%
\bibitem [{\citenamefont {Wakimoto}\ \emph {et~al.}(2010)\citenamefont
  {Wakimoto}, \citenamefont {Kodama}, \citenamefont {Ishikado}, \citenamefont
  {Matsuda}, \citenamefont {Kajimoto}, \citenamefont {Arai}, \citenamefont
  {Kakurai}, \citenamefont {Esaka}, \citenamefont {Iyo}, \citenamefont {Kito},
  \citenamefont {Eisaki},\ and\ \citenamefont {Shamoto}}]{WakimotoJPSJ2010}%
  \BibitemOpen
  \bibfield  {author} {\bibinfo {author} {\bibfnamefont {S.}~\bibnamefont
  {Wakimoto}}, \bibinfo {author} {\bibfnamefont {K.}~\bibnamefont {Kodama}},
  \bibinfo {author} {\bibfnamefont {M.}~\bibnamefont {Ishikado}}, \bibinfo
  {author} {\bibfnamefont {M.}~\bibnamefont {Matsuda}}, \bibinfo {author}
  {\bibfnamefont {R.}~\bibnamefont {Kajimoto}}, \bibinfo {author}
  {\bibfnamefont {M.}~\bibnamefont {Arai}}, \bibinfo {author} {\bibfnamefont
  {K.}~\bibnamefont {Kakurai}}, \bibinfo {author} {\bibfnamefont
  {F.}~\bibnamefont {Esaka}}, \bibinfo {author} {\bibfnamefont
  {A.}~\bibnamefont {Iyo}}, \bibinfo {author} {\bibfnamefont {H.}~\bibnamefont
  {Kito}}, \bibinfo {author} {\bibfnamefont {H.}~\bibnamefont {Eisaki}}, \ and\
  \bibinfo {author} {\bibfnamefont {S.-i.}\ \bibnamefont {Shamoto}},\ }\href
  {\doibase 10.1143/JPSJ.79.074715} {\bibfield  {journal} {\bibinfo  {journal}
  {Journal of the Physical Society of Japan}\ }\textbf {\bibinfo {volume}
  {79}},\ \bibinfo {pages} {074715} (\bibinfo {year} {2010})}\BibitemShut
  {NoStop}%
\bibitem [{\citenamefont {Christianson}\ \emph {et~al.}(2008)\citenamefont
  {Christianson}, \citenamefont {Goremychkin}, \citenamefont {Osborn},
  \citenamefont {Rosenkranz}, \citenamefont {Lumsden}, \citenamefont
  {Malliakas}, \citenamefont {Todorov}, \citenamefont {Claus}, \citenamefont
  {Chung}, \citenamefont {Kanatzidis}, \citenamefont {Bewley},\ and\
  \citenamefont {Guidi}}]{ChristiansonNature2008}%
  \BibitemOpen
  \bibfield  {author} {\bibinfo {author} {\bibfnamefont {A.~D.}\ \bibnamefont
  {Christianson}}, \bibinfo {author} {\bibfnamefont {E.~A.}\ \bibnamefont
  {Goremychkin}}, \bibinfo {author} {\bibfnamefont {R.}~\bibnamefont {Osborn}},
  \bibinfo {author} {\bibfnamefont {S.}~\bibnamefont {Rosenkranz}}, \bibinfo
  {author} {\bibfnamefont {M.~D.}\ \bibnamefont {Lumsden}}, \bibinfo {author}
  {\bibfnamefont {C.~D.}\ \bibnamefont {Malliakas}}, \bibinfo {author}
  {\bibfnamefont {I.~S.}\ \bibnamefont {Todorov}}, \bibinfo {author}
  {\bibfnamefont {H.}~\bibnamefont {Claus}}, \bibinfo {author} {\bibfnamefont
  {D.~Y.}\ \bibnamefont {Chung}}, \bibinfo {author} {\bibfnamefont {M.~G.}\
  \bibnamefont {Kanatzidis}}, \bibinfo {author} {\bibfnamefont {R.~I.}\
  \bibnamefont {Bewley}}, \ and\ \bibinfo {author} {\bibfnamefont
  {T.}~\bibnamefont {Guidi}},\ }\href {\doibase 10.1038/nature07625} {\bibfield
   {journal} {\bibinfo  {journal} {Nature}\ }\textbf {\bibinfo {volume}
  {456}},\ \bibinfo {pages} {930} (\bibinfo {year} {2008})}\BibitemShut
  {NoStop}%
\bibitem [{\citenamefont {Daghofer}\ \emph {et~al.}(2008)\citenamefont
  {Daghofer}, \citenamefont {Moreo}, \citenamefont {Riera}, \citenamefont
  {Arrigoni}, \citenamefont {Scalapino},\ and\ \citenamefont
  {Dagotto}}]{DaghoferPRL2008}%
  \BibitemOpen
  \bibfield  {author} {\bibinfo {author} {\bibfnamefont {M.}~\bibnamefont
  {Daghofer}}, \bibinfo {author} {\bibfnamefont {A.}~\bibnamefont {Moreo}},
  \bibinfo {author} {\bibfnamefont {J.}~\bibnamefont {Riera}}, \bibinfo
  {author} {\bibfnamefont {E.}~\bibnamefont {Arrigoni}}, \bibinfo {author}
  {\bibfnamefont {D.}~\bibnamefont {Scalapino}}, \ and\ \bibinfo {author}
  {\bibfnamefont {E.}~\bibnamefont {Dagotto}},\ }\href {\doibase
  10.1103/PhysRevLett.101.237004} {\bibfield  {journal} {\bibinfo  {journal}
  {Physical Review Letters}\ }\textbf {\bibinfo {volume} {101}},\ \bibinfo
  {pages} {237004} (\bibinfo {year} {2008})}\BibitemShut {NoStop}%
\bibitem [{\citenamefont {Zhai}\ \emph {et~al.}(2009)\citenamefont {Zhai},
  \citenamefont {Wang},\ and\ \citenamefont {Lee}}]{ZhaiPRB2009}%
  \BibitemOpen
  \bibfield  {author} {\bibinfo {author} {\bibfnamefont {H.}~\bibnamefont
  {Zhai}}, \bibinfo {author} {\bibfnamefont {F.}~\bibnamefont {Wang}}, \ and\
  \bibinfo {author} {\bibfnamefont {D.-H.}\ \bibnamefont {Lee}},\ }\href
  {\doibase 10.1103/PhysRevB.80.064517} {\bibfield  {journal} {\bibinfo
  {journal} {Physical Review B}\ }\textbf {\bibinfo {volume} {80}},\ \bibinfo
  {pages} {064517} (\bibinfo {year} {2009})}\BibitemShut {NoStop}%
\bibitem [{\citenamefont {Mu}\ \emph {et~al.}(2008)\citenamefont {Mu},
  \citenamefont {Zhu}, \citenamefont {Fang}, \citenamefont {Shan},
  \citenamefont {Ren},\ and\ \citenamefont {Wen}}]{MuChinPhysLett2008}%
  \BibitemOpen
  \bibfield  {author} {\bibinfo {author} {\bibfnamefont {G.}~\bibnamefont
  {Mu}}, \bibinfo {author} {\bibfnamefont {X.-Y.}\ \bibnamefont {Zhu}},
  \bibinfo {author} {\bibfnamefont {L.}~\bibnamefont {Fang}}, \bibinfo {author}
  {\bibfnamefont {L.}~\bibnamefont {Shan}}, \bibinfo {author} {\bibfnamefont
  {C.}~\bibnamefont {Ren}}, \ and\ \bibinfo {author} {\bibfnamefont {H.-H.}\
  \bibnamefont {Wen}},\ }\href
  {http://cpl.iphy.ac.cn/qikan/epaper/zhaiyao.asp?bsid=9350} {\bibfield
  {journal} {\bibinfo  {journal} {Chinese Physics Letters}\ }\textbf {\bibinfo
  {volume} {25}},\ \bibinfo {pages} {2221} (\bibinfo {year}
  {2008})}\BibitemShut {NoStop}%
\bibitem [{\citenamefont {Ren}\ \emph {et~al.}()\citenamefont {Ren},
  \citenamefont {Wang}, \citenamefont {Yang}, \citenamefont {Zhu},
  \citenamefont {Fang}, \citenamefont {Mu}, \citenamefont {Shan},\ and\
  \citenamefont {Wen}}]{RenArxiv0804.1726}%
  \BibitemOpen
  \bibfield  {author} {\bibinfo {author} {\bibfnamefont {C.}~\bibnamefont
  {Ren}}, \bibinfo {author} {\bibfnamefont {Z.-S.}\ \bibnamefont {Wang}},
  \bibinfo {author} {\bibfnamefont {H.}~\bibnamefont {Yang}}, \bibinfo {author}
  {\bibfnamefont {X.}~\bibnamefont {Zhu}}, \bibinfo {author} {\bibfnamefont
  {L.}~\bibnamefont {Fang}}, \bibinfo {author} {\bibfnamefont {G.}~\bibnamefont
  {Mu}}, \bibinfo {author} {\bibfnamefont {L.}~\bibnamefont {Shan}}, \ and\
  \bibinfo {author} {\bibfnamefont {H.-H.}\ \bibnamefont {Wen}},\ }\href@noop
  {} {\ }\Eprint {http://arxiv.org/abs/0804.1726} {arXiv:0804.1726}
  \BibitemShut {NoStop}%
\bibitem [{\citenamefont {Shan}\ \emph {et~al.}(2008)\citenamefont {Shan},
  \citenamefont {Wang}, \citenamefont {Zhu}, \citenamefont {Mu}, \citenamefont
  {Fang}, \citenamefont {Ren},\ and\ \citenamefont {Wen}}]{ShanEPL2008}%
  \BibitemOpen
  \bibfield  {author} {\bibinfo {author} {\bibfnamefont {L.}~\bibnamefont
  {Shan}}, \bibinfo {author} {\bibfnamefont {Y.}~\bibnamefont {Wang}}, \bibinfo
  {author} {\bibfnamefont {X.}~\bibnamefont {Zhu}}, \bibinfo {author}
  {\bibfnamefont {G.}~\bibnamefont {Mu}}, \bibinfo {author} {\bibfnamefont
  {L.}~\bibnamefont {Fang}}, \bibinfo {author} {\bibfnamefont {C.}~\bibnamefont
  {Ren}}, \ and\ \bibinfo {author} {\bibfnamefont {H.-H.}\ \bibnamefont
  {Wen}},\ }\href {\doibase 10.1209/0295-5075/83/57004} {\bibfield  {journal}
  {\bibinfo  {journal} {Europhysics Letters}\ }\textbf {\bibinfo {volume}
  {83}},\ \bibinfo {pages} {57004} (\bibinfo {year} {2008})}\BibitemShut
  {NoStop}%
\bibitem [{\citenamefont {Luetkens}\ \emph {et~al.}(2008)\citenamefont
  {Luetkens}, \citenamefont {Klauss}, \citenamefont {Khasanov}, \citenamefont
  {Amato}, \citenamefont {Klingeler}, \citenamefont {Hellmann}, \citenamefont
  {Leps}, \citenamefont {Kondrat}, \citenamefont {Hess}, \citenamefont
  {K\"{o}hler}, \citenamefont {Behr}, \citenamefont {Werner},\ and\
  \citenamefont {B\"{u}chner}}]{LuetkensPRL2008}%
  \BibitemOpen
  \bibfield  {author} {\bibinfo {author} {\bibfnamefont {H.}~\bibnamefont
  {Luetkens}}, \bibinfo {author} {\bibfnamefont {H.-H.}\ \bibnamefont
  {Klauss}}, \bibinfo {author} {\bibfnamefont {R.}~\bibnamefont {Khasanov}},
  \bibinfo {author} {\bibfnamefont {A.}~\bibnamefont {Amato}}, \bibinfo
  {author} {\bibfnamefont {R.}~\bibnamefont {Klingeler}}, \bibinfo {author}
  {\bibfnamefont {I.}~\bibnamefont {Hellmann}}, \bibinfo {author}
  {\bibfnamefont {N.}~\bibnamefont {Leps}}, \bibinfo {author} {\bibfnamefont
  {A.}~\bibnamefont {Kondrat}}, \bibinfo {author} {\bibfnamefont
  {C.}~\bibnamefont {Hess}}, \bibinfo {author} {\bibfnamefont {A.}~\bibnamefont
  {K\"{o}hler}}, \bibinfo {author} {\bibfnamefont {G.}~\bibnamefont {Behr}},
  \bibinfo {author} {\bibfnamefont {J.}~\bibnamefont {Werner}}, \ and\ \bibinfo
  {author} {\bibfnamefont {B.}~\bibnamefont {B\"{u}chner}},\ }\href {\doibase
  10.1103/PhysRevLett.101.097009} {\bibfield  {journal} {\bibinfo  {journal}
  {Physical Review Letters}\ }\textbf {\bibinfo {volume} {101}},\ \bibinfo
  {pages} {097009} (\bibinfo {year} {2008})}\BibitemShut {NoStop}%
\bibitem [{\citenamefont {Grafe}\ \emph {et~al.}(2008)\citenamefont {Grafe},
  \citenamefont {Paar}, \citenamefont {Lang}, \citenamefont {Curro},
  \citenamefont {Behr}, \citenamefont {Werner}, \citenamefont {Hamann-Borrero},
  \citenamefont {Hess}, \citenamefont {Leps}, \citenamefont {Klingeler},\ and\
  \citenamefont {B\"{u}chner}}]{GrafePRL2008}%
  \BibitemOpen
  \bibfield  {author} {\bibinfo {author} {\bibfnamefont {H.-J.}\ \bibnamefont
  {Grafe}}, \bibinfo {author} {\bibfnamefont {D.}~\bibnamefont {Paar}},
  \bibinfo {author} {\bibfnamefont {G.}~\bibnamefont {Lang}}, \bibinfo {author}
  {\bibfnamefont {N.~J.}\ \bibnamefont {Curro}}, \bibinfo {author}
  {\bibfnamefont {G.}~\bibnamefont {Behr}}, \bibinfo {author} {\bibfnamefont
  {J.}~\bibnamefont {Werner}}, \bibinfo {author} {\bibfnamefont
  {J.}~\bibnamefont {Hamann-Borrero}}, \bibinfo {author} {\bibfnamefont
  {C.}~\bibnamefont {Hess}}, \bibinfo {author} {\bibfnamefont {N.}~\bibnamefont
  {Leps}}, \bibinfo {author} {\bibfnamefont {R.}~\bibnamefont {Klingeler}}, \
  and\ \bibinfo {author} {\bibfnamefont {B.}~\bibnamefont {B\"{u}chner}},\
  }\href {\doibase 10.1103/PhysRevLett.101.047003} {\bibfield  {journal}
  {\bibinfo  {journal} {Physical Review Letters}\ }\textbf {\bibinfo {volume}
  {101}},\ \bibinfo {pages} {047003} (\bibinfo {year} {2008})}\BibitemShut
  {NoStop}%
\bibitem [{\citenamefont {Mukuda}\ \emph {et~al.}(2008)\citenamefont {Mukuda},
  \citenamefont {Terasaki}, \citenamefont {Kinouchi}, \citenamefont {Yashima},
  \citenamefont {Kitaoka}, \citenamefont {Suzuki}, \citenamefont {Miyasaka},
  \citenamefont {Tajima}, \citenamefont {Miyazawa}, \citenamefont {Shirage},
  \citenamefont {Kito}, \citenamefont {Eisaki},\ and\ \citenamefont
  {Iyo}}]{MukudaJPSJ2008}%
  \BibitemOpen
  \bibfield  {author} {\bibinfo {author} {\bibfnamefont {H.}~\bibnamefont
  {Mukuda}}, \bibinfo {author} {\bibfnamefont {N.}~\bibnamefont {Terasaki}},
  \bibinfo {author} {\bibfnamefont {H.}~\bibnamefont {Kinouchi}}, \bibinfo
  {author} {\bibfnamefont {M.}~\bibnamefont {Yashima}}, \bibinfo {author}
  {\bibfnamefont {Y.}~\bibnamefont {Kitaoka}}, \bibinfo {author} {\bibfnamefont
  {S.}~\bibnamefont {Suzuki}}, \bibinfo {author} {\bibfnamefont
  {S.}~\bibnamefont {Miyasaka}}, \bibinfo {author} {\bibfnamefont
  {S.}~\bibnamefont {Tajima}}, \bibinfo {author} {\bibfnamefont
  {K.}~\bibnamefont {Miyazawa}}, \bibinfo {author} {\bibfnamefont
  {P.}~\bibnamefont {Shirage}}, \bibinfo {author} {\bibfnamefont
  {H.}~\bibnamefont {Kito}}, \bibinfo {author} {\bibfnamefont {H.}~\bibnamefont
  {Eisaki}}, \ and\ \bibinfo {author} {\bibfnamefont {A.}~\bibnamefont {Iyo}},\
  }\href {\doibase 10.1143/JPSJ.77.093704} {\bibfield  {journal} {\bibinfo
  {journal} {Journal of the Physical Society of Japan}\ }\textbf {\bibinfo
  {volume} {77}},\ \bibinfo {pages} {093704} (\bibinfo {year}
  {2008})}\BibitemShut {NoStop}%
\bibitem [{\citenamefont {Kotegawa}\ \emph {et~al.}(2008)\citenamefont
  {Kotegawa}, \citenamefont {Masaki}, \citenamefont {Awai}, \citenamefont
  {Tou}, \citenamefont {Mizuguchi},\ and\ \citenamefont
  {Takano}}]{KotegawaJPSJ2008}%
  \BibitemOpen
  \bibfield  {author} {\bibinfo {author} {\bibfnamefont {H.}~\bibnamefont
  {Kotegawa}}, \bibinfo {author} {\bibfnamefont {S.}~\bibnamefont {Masaki}},
  \bibinfo {author} {\bibfnamefont {Y.}~\bibnamefont {Awai}}, \bibinfo {author}
  {\bibfnamefont {H.}~\bibnamefont {Tou}}, \bibinfo {author} {\bibfnamefont
  {Y.}~\bibnamefont {Mizuguchi}}, \ and\ \bibinfo {author} {\bibfnamefont
  {Y.}~\bibnamefont {Takano}},\ }\href {\doibase 10.1143/JPSJ.77.113703}
  {\bibfield  {journal} {\bibinfo  {journal} {Journal of the Physical Society
  of Japan}\ }\textbf {\bibinfo {volume} {77}},\ \bibinfo {pages} {113703}
  (\bibinfo {year} {2008})}\BibitemShut {NoStop}%
\bibitem [{\citenamefont {Checkelsky}\ \emph {et~al.}()\citenamefont
  {Checkelsky}, \citenamefont {Li}, \citenamefont {Chen}, \citenamefont {Luo},
  \citenamefont {Wang},\ and\ \citenamefont {Ong}}]{CheckelskyArxiv0811.4668}%
  \BibitemOpen
  \bibfield  {author} {\bibinfo {author} {\bibfnamefont {J.~G.}\ \bibnamefont
  {Checkelsky}}, \bibinfo {author} {\bibfnamefont {L.}~\bibnamefont {Li}},
  \bibinfo {author} {\bibfnamefont {G.~F.}\ \bibnamefont {Chen}}, \bibinfo
  {author} {\bibfnamefont {J.~L.}\ \bibnamefont {Luo}}, \bibinfo {author}
  {\bibfnamefont {N.~L.}\ \bibnamefont {Wang}}, \ and\ \bibinfo {author}
  {\bibfnamefont {N.~P.}\ \bibnamefont {Ong}},\ }\href@noop {} {\ }\Eprint
  {http://arxiv.org/abs/0811.4668} {arXiv:0811.4668} \BibitemShut {NoStop}%
\bibitem [{\citenamefont {Gordon}\ \emph {et~al.}(2009)\citenamefont {Gordon},
  \citenamefont {Ni}, \citenamefont {Martin}, \citenamefont {Tanatar},
  \citenamefont {Vannette}, \citenamefont {Kim}, \citenamefont {Samolyuk},
  \citenamefont {Schmalian}, \citenamefont {Nandi}, \citenamefont {Kreyssig},
  \citenamefont {Goldman}, \citenamefont {Yan}, \citenamefont {Bud'ko},
  \citenamefont {Canfield},\ and\ \citenamefont {Prozorov}}]{GordonPRL2009}%
  \BibitemOpen
  \bibfield  {author} {\bibinfo {author} {\bibfnamefont {R.}~\bibnamefont
  {Gordon}}, \bibinfo {author} {\bibfnamefont {N.}~\bibnamefont {Ni}}, \bibinfo
  {author} {\bibfnamefont {C.}~\bibnamefont {Martin}}, \bibinfo {author}
  {\bibfnamefont {M.}~\bibnamefont {Tanatar}}, \bibinfo {author} {\bibfnamefont
  {M.}~\bibnamefont {Vannette}}, \bibinfo {author} {\bibfnamefont
  {H.}~\bibnamefont {Kim}}, \bibinfo {author} {\bibfnamefont {G.}~\bibnamefont
  {Samolyuk}}, \bibinfo {author} {\bibfnamefont {J.}~\bibnamefont {Schmalian}},
  \bibinfo {author} {\bibfnamefont {S.}~\bibnamefont {Nandi}}, \bibinfo
  {author} {\bibfnamefont {A.}~\bibnamefont {Kreyssig}}, \bibinfo {author}
  {\bibfnamefont {A.}~\bibnamefont {Goldman}}, \bibinfo {author} {\bibfnamefont
  {J.}~\bibnamefont {Yan}}, \bibinfo {author} {\bibfnamefont {S.~L.}\
  \bibnamefont {Bud'ko}}, \bibinfo {author} {\bibfnamefont {P.}~\bibnamefont
  {Canfield}}, \ and\ \bibinfo {author} {\bibfnamefont {R.}~\bibnamefont
  {Prozorov}},\ }\href {\doibase 10.1103/PhysRevLett.102.127004} {\bibfield
  {journal} {\bibinfo  {journal} {Physical Review Letters}\ }\textbf {\bibinfo
  {volume} {102}},\ \bibinfo {pages} {127004} (\bibinfo {year}
  {2009})}\BibitemShut {NoStop}%
\bibitem [{\citenamefont {Li}\ \emph {et~al.}(2009{\natexlab{b}})\citenamefont
  {Li}, \citenamefont {de~la Cruz}, \citenamefont {Huang}, \citenamefont
  {Chen}, \citenamefont {Lynn}, \citenamefont {Hu}, \citenamefont {Huang},
  \citenamefont {Hsu}, \citenamefont {Yeh}, \citenamefont {Wu},\ and\
  \citenamefont {Dai}}]{LiPRB2009}%
  \BibitemOpen
  \bibfield  {author} {\bibinfo {author} {\bibfnamefont {S.}~\bibnamefont
  {Li}}, \bibinfo {author} {\bibfnamefont {C.}~\bibnamefont {de~la Cruz}},
  \bibinfo {author} {\bibfnamefont {Q.}~\bibnamefont {Huang}}, \bibinfo
  {author} {\bibfnamefont {Y.}~\bibnamefont {Chen}}, \bibinfo {author}
  {\bibfnamefont {J.}~\bibnamefont {Lynn}}, \bibinfo {author} {\bibfnamefont
  {J.}~\bibnamefont {Hu}}, \bibinfo {author} {\bibfnamefont {Y.-L.}\
  \bibnamefont {Huang}}, \bibinfo {author} {\bibfnamefont {F.-C.}\ \bibnamefont
  {Hsu}}, \bibinfo {author} {\bibfnamefont {K.-W.}\ \bibnamefont {Yeh}},
  \bibinfo {author} {\bibfnamefont {M.-K.}\ \bibnamefont {Wu}}, \ and\ \bibinfo
  {author} {\bibfnamefont {P.}~\bibnamefont {Dai}},\ }\href {\doibase
  10.1103/PhysRevB.79.054503} {\bibfield  {journal} {\bibinfo  {journal}
  {Physical Review B}\ }\textbf {\bibinfo {volume} {79}},\ \bibinfo {pages}
  {054503} (\bibinfo {year} {2009}{\natexlab{b}})}\BibitemShut {NoStop}%
\bibitem [{\citenamefont {Zhang}\ \emph
  {et~al.}(2009{\natexlab{b}})\citenamefont {Zhang}, \citenamefont {Fang},
  \citenamefont {Zhou}, \citenamefont {Seo}, \citenamefont {Tsai},
  \citenamefont {Bernevig},\ and\ \citenamefont {Hu}}]{ZhangPRB2009}%
  \BibitemOpen
  \bibfield  {author} {\bibinfo {author} {\bibfnamefont {Y.-Y.}\ \bibnamefont
  {Zhang}}, \bibinfo {author} {\bibfnamefont {C.}~\bibnamefont {Fang}},
  \bibinfo {author} {\bibfnamefont {X.}~\bibnamefont {Zhou}}, \bibinfo {author}
  {\bibfnamefont {K.}~\bibnamefont {Seo}}, \bibinfo {author} {\bibfnamefont
  {W.-F.}\ \bibnamefont {Tsai}}, \bibinfo {author} {\bibfnamefont
  {B.}~\bibnamefont {Bernevig}}, \ and\ \bibinfo {author} {\bibfnamefont
  {J.}~\bibnamefont {Hu}},\ }\href {\doibase 10.1103/PhysRevB.80.094528}
  {\bibfield  {journal} {\bibinfo  {journal} {Physical Review B}\ }\textbf
  {\bibinfo {volume} {80}},\ \bibinfo {pages} {094528} (\bibinfo {year}
  {2009}{\natexlab{b}})}\BibitemShut {NoStop}%
\bibitem [{\citenamefont {Plamadeala}\ \emph {et~al.}(2010)\citenamefont
  {Plamadeala}, \citenamefont {Pereg-Barnea},\ and\ \citenamefont
  {Refael}}]{PlamadealaPRB2010}%
  \BibitemOpen
  \bibfield  {author} {\bibinfo {author} {\bibfnamefont {E.}~\bibnamefont
  {Plamadeala}}, \bibinfo {author} {\bibfnamefont {T.}~\bibnamefont
  {Pereg-Barnea}}, \ and\ \bibinfo {author} {\bibfnamefont {G.}~\bibnamefont
  {Refael}},\ }\href {\doibase 10.1103/PhysRevB.81.134513} {\bibfield
  {journal} {\bibinfo  {journal} {Physical Review B}\ }\textbf {\bibinfo
  {volume} {81}},\ \bibinfo {pages} {134513} (\bibinfo {year}
  {2010})}\BibitemShut {NoStop}%
\bibitem [{\citenamefont {Zeng}\ \emph {et~al.}(2010)\citenamefont {Zeng},
  \citenamefont {Mu}, \citenamefont {Luo}, \citenamefont {Xiang}, \citenamefont
  {Mazin}, \citenamefont {Yang}, \citenamefont {Shan}, \citenamefont {Ren},
  \citenamefont {Dai},\ and\ \citenamefont {Wen}}]{ZengNatComm2010}%
  \BibitemOpen
  \bibfield  {author} {\bibinfo {author} {\bibfnamefont {B.}~\bibnamefont
  {Zeng}}, \bibinfo {author} {\bibfnamefont {G.}~\bibnamefont {Mu}}, \bibinfo
  {author} {\bibfnamefont {H.~Q.}\ \bibnamefont {Luo}}, \bibinfo {author}
  {\bibfnamefont {T.}~\bibnamefont {Xiang}}, \bibinfo {author} {\bibfnamefont
  {I.~I.}\ \bibnamefont {Mazin}}, \bibinfo {author} {\bibfnamefont
  {H.}~\bibnamefont {Yang}}, \bibinfo {author} {\bibfnamefont {L.}~\bibnamefont
  {Shan}}, \bibinfo {author} {\bibfnamefont {C.}~\bibnamefont {Ren}}, \bibinfo
  {author} {\bibfnamefont {P.~C.}\ \bibnamefont {Dai}}, \ and\ \bibinfo
  {author} {\bibfnamefont {H.-H.}\ \bibnamefont {Wen}},\ }\href {\doibase
  10.1038/ncomms1115} {\bibfield  {journal} {\bibinfo  {journal} {Nature
  Communications}\ }\textbf {\bibinfo {volume} {1}},\ \bibinfo {pages} {112}
  (\bibinfo {year} {2010})}\BibitemShut {NoStop}%
\bibitem [{\citenamefont {Mazin}\ and\ \citenamefont
  {Singh}()}]{MazinArxiv1007.0047}%
  \BibitemOpen
  \bibfield  {author} {\bibinfo {author} {\bibfnamefont {I.~I.}\ \bibnamefont
  {Mazin}}\ and\ \bibinfo {author} {\bibfnamefont {D.~J.}\ \bibnamefont
  {Singh}},\ }\href@noop {} {\ }\Eprint {http://arxiv.org/abs/1007.0047}
  {arXiv:1007.0047} \BibitemShut {NoStop}%
\bibitem [{\citenamefont {Hanaguri}\ \emph {et~al.}()\citenamefont {Hanaguri},
  \citenamefont {Niitaka}, \citenamefont {Kuroki},\ and\ \citenamefont
  {Takagi}}]{HanaguriArxiv1007.0307}%
  \BibitemOpen
  \bibfield  {author} {\bibinfo {author} {\bibfnamefont {T.}~\bibnamefont
  {Hanaguri}}, \bibinfo {author} {\bibfnamefont {S.}~\bibnamefont {Niitaka}},
  \bibinfo {author} {\bibfnamefont {K.}~\bibnamefont {Kuroki}}, \ and\ \bibinfo
  {author} {\bibfnamefont {H.}~\bibnamefont {Takagi}},\ }\href@noop {} {\
  }\Eprint {http://arxiv.org/abs/1007.0307} {arXiv:1007.0307} \BibitemShut
  {NoStop}%
\bibitem [{\citenamefont {Fletcher}\ \emph {et~al.}(2009)\citenamefont
  {Fletcher}, \citenamefont {Serafin}, \citenamefont {Malone}, \citenamefont
  {Analytis}, \citenamefont {Chu}, \citenamefont {Erickson}, \citenamefont
  {Fisher},\ and\ \citenamefont {Carrington}}]{FletcherPRL2009}%
  \BibitemOpen
  \bibfield  {author} {\bibinfo {author} {\bibfnamefont {J.}~\bibnamefont
  {Fletcher}}, \bibinfo {author} {\bibfnamefont {A.}~\bibnamefont {Serafin}},
  \bibinfo {author} {\bibfnamefont {L.}~\bibnamefont {Malone}}, \bibinfo
  {author} {\bibfnamefont {J.}~\bibnamefont {Analytis}}, \bibinfo {author}
  {\bibfnamefont {J.-H.}\ \bibnamefont {Chu}}, \bibinfo {author} {\bibfnamefont
  {A.}~\bibnamefont {Erickson}}, \bibinfo {author} {\bibfnamefont {I.~R.}\
  \bibnamefont {Fisher}}, \ and\ \bibinfo {author} {\bibfnamefont
  {A.}~\bibnamefont {Carrington}},\ }\href {\doibase
  10.1103/PhysRevLett.102.147001} {\bibfield  {journal} {\bibinfo  {journal}
  {Physical Review Letters}\ }\textbf {\bibinfo {volume} {102}},\ \bibinfo
  {pages} {147001} (\bibinfo {year} {2009})}\BibitemShut {NoStop}%
\bibitem [{\citenamefont {Hicks}\ \emph {et~al.}(2009)\citenamefont {Hicks},
  \citenamefont {Lippman}, \citenamefont {Huber}, \citenamefont {Analytis},
  \citenamefont {Chu}, \citenamefont {Erickson}, \citenamefont {Fisher},\ and\
  \citenamefont {Moler}}]{HicksPRL2009}%
  \BibitemOpen
  \bibfield  {author} {\bibinfo {author} {\bibfnamefont {C.}~\bibnamefont
  {Hicks}}, \bibinfo {author} {\bibfnamefont {T.}~\bibnamefont {Lippman}},
  \bibinfo {author} {\bibfnamefont {M.}~\bibnamefont {Huber}}, \bibinfo
  {author} {\bibfnamefont {J.}~\bibnamefont {Analytis}}, \bibinfo {author}
  {\bibfnamefont {J.-H.}\ \bibnamefont {Chu}}, \bibinfo {author} {\bibfnamefont
  {A.}~\bibnamefont {Erickson}}, \bibinfo {author} {\bibfnamefont {I.~R.}\
  \bibnamefont {Fisher}}, \ and\ \bibinfo {author} {\bibfnamefont
  {K.}~\bibnamefont {Moler}},\ }\href {\doibase 10.1103/PhysRevLett.103.127003}
  {\bibfield  {journal} {\bibinfo  {journal} {Physical Review Letters}\
  }\textbf {\bibinfo {volume} {103}},\ \bibinfo {pages} {127003} (\bibinfo
  {year} {2009})}\BibitemShut {NoStop}%
\bibitem [{\citenamefont {Lee}\ and\ \citenamefont {Wu}(2009)}]{LeePRL2009}%
  \BibitemOpen
  \bibfield  {author} {\bibinfo {author} {\bibfnamefont {W.-C.}\ \bibnamefont
  {Lee}}\ and\ \bibinfo {author} {\bibfnamefont {C.}~\bibnamefont {Wu}},\
  }\href {\doibase 10.1103/PhysRevLett.103.176101} {\bibfield  {journal}
  {\bibinfo  {journal} {Physical Review Letters}\ }\textbf {\bibinfo {volume}
  {103}},\ \bibinfo {pages} {176101} (\bibinfo {year} {2009})}\BibitemShut
  {NoStop}%
\bibitem [{\citenamefont {Borisenko}\ \emph {et~al.}(2010)\citenamefont
  {Borisenko}, \citenamefont {Zabolotnyy}, \citenamefont {Evtushinsky},
  \citenamefont {Kim}, \citenamefont {Morozov}, \citenamefont {Yaresko},
  \citenamefont {Kordyuk}, \citenamefont {Behr}, \citenamefont {Vasiliev},
  \citenamefont {Follath},\ and\ \citenamefont
  {B\"{u}chner}}]{BorisenkoPRL2010}%
  \BibitemOpen
  \bibfield  {author} {\bibinfo {author} {\bibfnamefont {S.}~\bibnamefont
  {Borisenko}}, \bibinfo {author} {\bibfnamefont {V.}~\bibnamefont
  {Zabolotnyy}}, \bibinfo {author} {\bibfnamefont {D.}~\bibnamefont
  {Evtushinsky}}, \bibinfo {author} {\bibfnamefont {T.}~\bibnamefont {Kim}},
  \bibinfo {author} {\bibfnamefont {I.}~\bibnamefont {Morozov}}, \bibinfo
  {author} {\bibfnamefont {A.}~\bibnamefont {Yaresko}}, \bibinfo {author}
  {\bibfnamefont {A.}~\bibnamefont {Kordyuk}}, \bibinfo {author} {\bibfnamefont
  {G.}~\bibnamefont {Behr}}, \bibinfo {author} {\bibfnamefont {A.}~\bibnamefont
  {Vasiliev}}, \bibinfo {author} {\bibfnamefont {R.}~\bibnamefont {Follath}}, \
  and\ \bibinfo {author} {\bibfnamefont {B.}~\bibnamefont {B\"{u}chner}},\
  }\href {\doibase 10.1103/PhysRevLett.105.067002} {\bibfield  {journal}
  {\bibinfo  {journal} {Physical Review Letters}\ }\textbf {\bibinfo {volume}
  {105}},\ \bibinfo {pages} {067002} (\bibinfo {year} {2010})}\BibitemShut
  {NoStop}%
\bibitem [{\citenamefont {Dynes}\ \emph {et~al.}(1978)\citenamefont {Dynes},
  \citenamefont {Narayanamurti},\ and\ \citenamefont {Garno}}]{DynesPRL1978}%
  \BibitemOpen
  \bibfield  {author} {\bibinfo {author} {\bibfnamefont {R.}~\bibnamefont
  {Dynes}}, \bibinfo {author} {\bibfnamefont {V.}~\bibnamefont
  {Narayanamurti}}, \ and\ \bibinfo {author} {\bibfnamefont {J.}~\bibnamefont
  {Garno}},\ }\href {\doibase 10.1103/PhysRevLett.41.1509} {\bibfield
  {journal} {\bibinfo  {journal} {Physical Review Letters}\ }\textbf {\bibinfo
  {volume} {41}},\ \bibinfo {pages} {1509} (\bibinfo {year}
  {1978})}\BibitemShut {NoStop}%
\bibitem [{\citenamefont {Marz}\ \emph {et~al.}(2010)\citenamefont {Marz},
  \citenamefont {Goll},\ and\ \citenamefont {L\"{o}hneysen}}]{MarzRSI2010}%
  \BibitemOpen
  \bibfield  {author} {\bibinfo {author} {\bibfnamefont {M.}~\bibnamefont
  {Marz}}, \bibinfo {author} {\bibfnamefont {G.}~\bibnamefont {Goll}}, \ and\
  \bibinfo {author} {\bibfnamefont {H.~v.}\ \bibnamefont {L\"{o}hneysen}},\
  }\href {\doibase 10.1063/1.3328059} {\bibfield  {journal} {\bibinfo
  {journal} {Review of Scientific Instruments}\ }\textbf {\bibinfo {volume}
  {81}},\ \bibinfo {pages} {045102} (\bibinfo {year} {2010})}\BibitemShut
  {NoStop}%
\bibitem [{\citenamefont {Pushp}\ \emph {et~al.}(2009)\citenamefont {Pushp},
  \citenamefont {Parker}, \citenamefont {Pasupathy}, \citenamefont {Gomes},
  \citenamefont {Ono}, \citenamefont {Wen}, \citenamefont {Xu}, \citenamefont
  {Gu},\ and\ \citenamefont {Yazdani}}]{PushpScience2009}%
  \BibitemOpen
  \bibfield  {author} {\bibinfo {author} {\bibfnamefont {A.}~\bibnamefont
  {Pushp}}, \bibinfo {author} {\bibfnamefont {C.~V.}\ \bibnamefont {Parker}},
  \bibinfo {author} {\bibfnamefont {A.~N.}\ \bibnamefont {Pasupathy}}, \bibinfo
  {author} {\bibfnamefont {K.~K.}\ \bibnamefont {Gomes}}, \bibinfo {author}
  {\bibfnamefont {S.}~\bibnamefont {Ono}}, \bibinfo {author} {\bibfnamefont
  {J.~S.}\ \bibnamefont {Wen}}, \bibinfo {author} {\bibfnamefont
  {Z.}~\bibnamefont {Xu}}, \bibinfo {author} {\bibfnamefont {G.}~\bibnamefont
  {Gu}}, \ and\ \bibinfo {author} {\bibfnamefont {A.}~\bibnamefont {Yazdani}},\
  }\href {\doibase 10.1126/science.1174338} {\bibfield  {journal} {\bibinfo
  {journal} {Science}\ }\textbf {\bibinfo {volume} {324}},\ \bibinfo {pages}
  {1689} (\bibinfo {year} {2009})}\BibitemShut {NoStop}%
\bibitem [{\citenamefont {Lang}\ \emph {et~al.}(2002)\citenamefont {Lang},
  \citenamefont {Madhavan}, \citenamefont {Hoffman}, \citenamefont {Hudson},
  \citenamefont {Eisaki}, \citenamefont {Uchida},\ and\ \citenamefont
  {Davis}}]{LangNature2002}%
  \BibitemOpen
  \bibfield  {author} {\bibinfo {author} {\bibfnamefont {K.~M.}\ \bibnamefont
  {Lang}}, \bibinfo {author} {\bibfnamefont {V.}~\bibnamefont {Madhavan}},
  \bibinfo {author} {\bibfnamefont {J.~E.}\ \bibnamefont {Hoffman}}, \bibinfo
  {author} {\bibfnamefont {E.~W.}\ \bibnamefont {Hudson}}, \bibinfo {author}
  {\bibfnamefont {H.}~\bibnamefont {Eisaki}}, \bibinfo {author} {\bibfnamefont
  {S.}~\bibnamefont {Uchida}}, \ and\ \bibinfo {author} {\bibfnamefont {J.~C.}\
  \bibnamefont {Davis}},\ }\href {\doibase 10.1038/415412a} {\bibfield
  {journal} {\bibinfo  {journal} {Nature}\ }\textbf {\bibinfo {volume} {415}},\
  \bibinfo {pages} {412} (\bibinfo {year} {2002})}\BibitemShut {NoStop}%
\bibitem [{\citenamefont {Alldredge}\ \emph {et~al.}(2008)\citenamefont
  {Alldredge}, \citenamefont {Lee}, \citenamefont {McElroy}, \citenamefont
  {Wang}, \citenamefont {Fujita}, \citenamefont {Kohsaka}, \citenamefont
  {Taylor}, \citenamefont {Eisaki}, \citenamefont {Uchida}, \citenamefont
  {Hirschfeld},\ and\ \citenamefont {Davis}}]{AlldredgeNatPhys2008}%
  \BibitemOpen
  \bibfield  {author} {\bibinfo {author} {\bibfnamefont {J.~W.}\ \bibnamefont
  {Alldredge}}, \bibinfo {author} {\bibfnamefont {J.}~\bibnamefont {Lee}},
  \bibinfo {author} {\bibfnamefont {K.}~\bibnamefont {McElroy}}, \bibinfo
  {author} {\bibfnamefont {M.}~\bibnamefont {Wang}}, \bibinfo {author}
  {\bibfnamefont {K.}~\bibnamefont {Fujita}}, \bibinfo {author} {\bibfnamefont
  {Y.}~\bibnamefont {Kohsaka}}, \bibinfo {author} {\bibfnamefont
  {C.}~\bibnamefont {Taylor}}, \bibinfo {author} {\bibfnamefont
  {H.}~\bibnamefont {Eisaki}}, \bibinfo {author} {\bibfnamefont
  {S.}~\bibnamefont {Uchida}}, \bibinfo {author} {\bibfnamefont {P.~J.}\
  \bibnamefont {Hirschfeld}}, \ and\ \bibinfo {author} {\bibfnamefont {J.~C.}\
  \bibnamefont {Davis}},\ }\href {\doibase 10.1038/nphys917} {\bibfield
  {journal} {\bibinfo  {journal} {Nature Physics}\ }\textbf {\bibinfo {volume}
  {4}},\ \bibinfo {pages} {319} (\bibinfo {year} {2008})}\BibitemShut {NoStop}%
\bibitem [{\citenamefont {Pan}\ \emph {et~al.}(1999)\citenamefont {Pan},
  \citenamefont {Hudson},\ and\ \citenamefont {Davis}}]{PanRSI1999}%
  \BibitemOpen
  \bibfield  {author} {\bibinfo {author} {\bibfnamefont {S.~H.}\ \bibnamefont
  {Pan}}, \bibinfo {author} {\bibfnamefont {E.~W.}\ \bibnamefont {Hudson}}, \
  and\ \bibinfo {author} {\bibfnamefont {J.~C.}\ \bibnamefont {Davis}},\ }\href
  {\doibase 10.1063/1.1149605} {\bibfield  {journal} {\bibinfo  {journal}
  {Review of Scientific Instruments}\ }\textbf {\bibinfo {volume} {70}},\
  \bibinfo {pages} {1459} (\bibinfo {year} {1999})}\BibitemShut {NoStop}%
\bibitem [{\citenamefont {Bardeen}\ \emph {et~al.}(1957)\citenamefont
  {Bardeen}, \citenamefont {Cooper},\ and\ \citenamefont
  {Schrieffer}}]{BardeenPhysRev1957}%
  \BibitemOpen
  \bibfield  {author} {\bibinfo {author} {\bibfnamefont {J.}~\bibnamefont
  {Bardeen}}, \bibinfo {author} {\bibfnamefont {L.~N.}\ \bibnamefont {Cooper}},
  \ and\ \bibinfo {author} {\bibfnamefont {J.~R.}\ \bibnamefont {Schrieffer}},\
  }\href {\doibase 10.1103/PhysRev.108.1175} {\bibfield  {journal} {\bibinfo
  {journal} {Physical Review}\ }\textbf {\bibinfo {volume} {108}},\ \bibinfo
  {pages} {1175} (\bibinfo {year} {1957})}\BibitemShut {NoStop}%
\bibitem [{\citenamefont {Won}\ and\ \citenamefont {Maki}(1994)}]{WonPRB1994}%
  \BibitemOpen
  \bibfield  {author} {\bibinfo {author} {\bibfnamefont {H.}~\bibnamefont
  {Won}}\ and\ \bibinfo {author} {\bibfnamefont {K.}~\bibnamefont {Maki}},\
  }\href {\doibase 10.1103/PhysRevB.49.1397} {\bibfield  {journal} {\bibinfo
  {journal} {Physical Review B}\ }\textbf {\bibinfo {volume} {49}},\ \bibinfo
  {pages} {1397} (\bibinfo {year} {1994})}\BibitemShut {NoStop}%
\bibitem [{\citenamefont {H\"{u}fner}\ \emph {et~al.}(2008)\citenamefont
  {H\"{u}fner}, \citenamefont {Hossain}, \citenamefont {Damascelli},\ and\
  \citenamefont {Sawatzky}}]{HufnerROPP2008}%
  \BibitemOpen
  \bibfield  {author} {\bibinfo {author} {\bibfnamefont {S.}~\bibnamefont
  {H\"{u}fner}}, \bibinfo {author} {\bibfnamefont {M.~A.}\ \bibnamefont
  {Hossain}}, \bibinfo {author} {\bibfnamefont {A.}~\bibnamefont {Damascelli}},
  \ and\ \bibinfo {author} {\bibfnamefont {G.~A.}\ \bibnamefont {Sawatzky}},\
  }\href {\doibase 10.1088/0034-4885/71/6/062501} {\bibfield  {journal}
  {\bibinfo  {journal} {Reports on Progress in Physics}\ }\textbf {\bibinfo
  {volume} {71}},\ \bibinfo {pages} {062501} (\bibinfo {year}
  {2008})}\BibitemShut {NoStop}%
\bibitem [{\citenamefont {Deutscher}(2005)}]{DeutscherRMP2005}%
  \BibitemOpen
  \bibfield  {author} {\bibinfo {author} {\bibfnamefont {G.}~\bibnamefont
  {Deutscher}},\ }\href {\doibase 10.1103/RevModPhys.77.109} {\bibfield
  {journal} {\bibinfo  {journal} {Reviews of Modern Physics}\ }\textbf
  {\bibinfo {volume} {77}},\ \bibinfo {pages} {109} (\bibinfo {year}
  {2005})}\BibitemShut {NoStop}%
\bibitem [{\citenamefont {Boyer}\ \emph {et~al.}(2007)\citenamefont {Boyer},
  \citenamefont {Wise}, \citenamefont {Chatterjee}, \citenamefont {Yi},
  \citenamefont {Kondo}, \citenamefont {Takeuchi}, \citenamefont {Ikuta},\ and\
  \citenamefont {Hudson}}]{BoyerNatPhys2007}%
  \BibitemOpen
  \bibfield  {author} {\bibinfo {author} {\bibfnamefont {M.~C.}\ \bibnamefont
  {Boyer}}, \bibinfo {author} {\bibfnamefont {W.~D.}\ \bibnamefont {Wise}},
  \bibinfo {author} {\bibfnamefont {K.}~\bibnamefont {Chatterjee}}, \bibinfo
  {author} {\bibfnamefont {M.}~\bibnamefont {Yi}}, \bibinfo {author}
  {\bibfnamefont {T.}~\bibnamefont {Kondo}}, \bibinfo {author} {\bibfnamefont
  {T.}~\bibnamefont {Takeuchi}}, \bibinfo {author} {\bibfnamefont
  {H.}~\bibnamefont {Ikuta}}, \ and\ \bibinfo {author} {\bibfnamefont {E.~W.}\
  \bibnamefont {Hudson}},\ }\href {\doibase 10.1038/nphys725} {\bibfield
  {journal} {\bibinfo  {journal} {Nature Physics}\ }\textbf {\bibinfo {volume}
  {3}},\ \bibinfo {pages} {802} (\bibinfo {year} {2007})}\BibitemShut {NoStop}%
\bibitem [{\citenamefont {Jin}\ \emph {et~al.}(2010)\citenamefont {Jin},
  \citenamefont {Pan}, \citenamefont {He}, \citenamefont {Li}, \citenamefont
  {Li}, \citenamefont {Peng}, \citenamefont {Thompson}, \citenamefont {Sales},
  \citenamefont {Sefat}, \citenamefont {McGuire}, \citenamefont {Mandrus},
  \citenamefont {Wendelken}, \citenamefont {Keppens},\ and\ \citenamefont
  {Plummer}}]{JinSST2010}%
  \BibitemOpen
  \bibfield  {author} {\bibinfo {author} {\bibfnamefont {R.}~\bibnamefont
  {Jin}}, \bibinfo {author} {\bibfnamefont {M.~H.}\ \bibnamefont {Pan}},
  \bibinfo {author} {\bibfnamefont {X.~B.}\ \bibnamefont {He}}, \bibinfo
  {author} {\bibfnamefont {G.}~\bibnamefont {Li}}, \bibinfo {author}
  {\bibfnamefont {D.}~\bibnamefont {Li}}, \bibinfo {author} {\bibfnamefont
  {R.-w.}\ \bibnamefont {Peng}}, \bibinfo {author} {\bibfnamefont {J.~R.}\
  \bibnamefont {Thompson}}, \bibinfo {author} {\bibfnamefont {B.~C.}\
  \bibnamefont {Sales}}, \bibinfo {author} {\bibfnamefont {A.~S.}\ \bibnamefont
  {Sefat}}, \bibinfo {author} {\bibfnamefont {M.~A.}\ \bibnamefont {McGuire}},
  \bibinfo {author} {\bibfnamefont {D.}~\bibnamefont {Mandrus}}, \bibinfo
  {author} {\bibfnamefont {J.~F.}\ \bibnamefont {Wendelken}}, \bibinfo {author}
  {\bibfnamefont {V.}~\bibnamefont {Keppens}}, \ and\ \bibinfo {author}
  {\bibfnamefont {E.~W.}\ \bibnamefont {Plummer}},\ }\href {\doibase
  10.1088/0953-2048/23/5/054005} {\bibfield  {journal} {\bibinfo  {journal}
  {Superconductor Science and Technology}\ }\textbf {\bibinfo {volume} {23}},\
  \bibinfo {pages} {054005} (\bibinfo {year} {2010})}\BibitemShut {NoStop}%
\bibitem [{\citenamefont {Wray}\ \emph {et~al.}(2008)\citenamefont {Wray},
  \citenamefont {Qian}, \citenamefont {Hsieh}, \citenamefont {Xia},
  \citenamefont {Li}, \citenamefont {Checkelsky}, \citenamefont {Pasupathy},
  \citenamefont {Gomes}, \citenamefont {Parker}, \citenamefont {Fedorov},
  \citenamefont {Chen}, \citenamefont {Luo}, \citenamefont {Yazdani},
  \citenamefont {Ong}, \citenamefont {Wang},\ and\ \citenamefont
  {Hasan}}]{WrayPRB2008}%
  \BibitemOpen
  \bibfield  {author} {\bibinfo {author} {\bibfnamefont {L.}~\bibnamefont
  {Wray}}, \bibinfo {author} {\bibfnamefont {D.}~\bibnamefont {Qian}}, \bibinfo
  {author} {\bibfnamefont {D.}~\bibnamefont {Hsieh}}, \bibinfo {author}
  {\bibfnamefont {Y.}~\bibnamefont {Xia}}, \bibinfo {author} {\bibfnamefont
  {L.}~\bibnamefont {Li}}, \bibinfo {author} {\bibfnamefont {J.~G.}\
  \bibnamefont {Checkelsky}}, \bibinfo {author} {\bibfnamefont
  {A.}~\bibnamefont {Pasupathy}}, \bibinfo {author} {\bibfnamefont {K.~K.}\
  \bibnamefont {Gomes}}, \bibinfo {author} {\bibfnamefont {C.}~\bibnamefont
  {Parker}}, \bibinfo {author} {\bibfnamefont {A.}~\bibnamefont {Fedorov}},
  \bibinfo {author} {\bibfnamefont {G.~F.}\ \bibnamefont {Chen}}, \bibinfo
  {author} {\bibfnamefont {J.~L.}\ \bibnamefont {Luo}}, \bibinfo {author}
  {\bibfnamefont {A.}~\bibnamefont {Yazdani}}, \bibinfo {author} {\bibfnamefont
  {N.~P.}\ \bibnamefont {Ong}}, \bibinfo {author} {\bibfnamefont
  {N.}~\bibnamefont {Wang}}, \ and\ \bibinfo {author} {\bibfnamefont {M.~Z.}\
  \bibnamefont {Hasan}},\ }\href {\doibase 10.1103/PhysRevB.78.184508}
  {\bibfield  {journal} {\bibinfo  {journal} {Physical Review B}\ }\textbf
  {\bibinfo {volume} {78}},\ \bibinfo {pages} {184508} (\bibinfo {year}
  {2008})}\BibitemShut {NoStop}%
\bibitem [{\citenamefont {Ma}\ \emph {et~al.}(2008)\citenamefont {Ma},
  \citenamefont {Pan}, \citenamefont {Niestemski}, \citenamefont {Neupane},
  \citenamefont {Xu}, \citenamefont {Richard}, \citenamefont {Nakayama},
  \citenamefont {Sato}, \citenamefont {Takahashi}, \citenamefont {Luo},
  \citenamefont {Fang}, \citenamefont {Wen}, \citenamefont {Wang},
  \citenamefont {Ding},\ and\ \citenamefont {Madhavan}}]{MaPRL2008}%
  \BibitemOpen
  \bibfield  {author} {\bibinfo {author} {\bibfnamefont {J.-H.}\ \bibnamefont
  {Ma}}, \bibinfo {author} {\bibfnamefont {Z.-H.}\ \bibnamefont {Pan}},
  \bibinfo {author} {\bibfnamefont {F.~C.}\ \bibnamefont {Niestemski}},
  \bibinfo {author} {\bibfnamefont {M.}~\bibnamefont {Neupane}}, \bibinfo
  {author} {\bibfnamefont {Y.-M.}\ \bibnamefont {Xu}}, \bibinfo {author}
  {\bibfnamefont {P.}~\bibnamefont {Richard}}, \bibinfo {author} {\bibfnamefont
  {K.}~\bibnamefont {Nakayama}}, \bibinfo {author} {\bibfnamefont
  {T.}~\bibnamefont {Sato}}, \bibinfo {author} {\bibfnamefont {T.}~\bibnamefont
  {Takahashi}}, \bibinfo {author} {\bibfnamefont {H.-Q.}\ \bibnamefont {Luo}},
  \bibinfo {author} {\bibfnamefont {L.}~\bibnamefont {Fang}}, \bibinfo {author}
  {\bibfnamefont {H.-H.}\ \bibnamefont {Wen}}, \bibinfo {author} {\bibfnamefont
  {Z.}~\bibnamefont {Wang}}, \bibinfo {author} {\bibfnamefont {H.}~\bibnamefont
  {Ding}}, \ and\ \bibinfo {author} {\bibfnamefont {V.}~\bibnamefont
  {Madhavan}},\ }\href {\doibase 10.1103/PhysRevLett.101.207002} {\bibfield
  {journal} {\bibinfo  {journal} {Physical Review Letters}\ }\textbf {\bibinfo
  {volume} {101}},\ \bibinfo {pages} {207002} (\bibinfo {year}
  {2008})}\BibitemShut {NoStop}%
\bibitem [{\citenamefont {Lv}\ \emph {et~al.}(2009)\citenamefont {Lv},
  \citenamefont {Wu},\ and\ \citenamefont {Phillips}}]{LvPRB2009}%
  \BibitemOpen
  \bibfield  {author} {\bibinfo {author} {\bibfnamefont {W.}~\bibnamefont
  {Lv}}, \bibinfo {author} {\bibfnamefont {J.}~\bibnamefont {Wu}}, \ and\
  \bibinfo {author} {\bibfnamefont {P.}~\bibnamefont {Phillips}},\ }\href
  {\doibase 10.1103/PhysRevB.80.224506} {\bibfield  {journal} {\bibinfo
  {journal} {Physical Review B}\ }\textbf {\bibinfo {volume} {80}},\ \bibinfo
  {pages} {224506} (\bibinfo {year} {2009})}\BibitemShut {NoStop}%
\bibitem [{\citenamefont {Lee}\ \emph {et~al.}(2009)\citenamefont {Lee},
  \citenamefont {Yin},\ and\ \citenamefont {Ku}}]{LeePRB2009}%
  \BibitemOpen
  \bibfield  {author} {\bibinfo {author} {\bibfnamefont {C.-C.}\ \bibnamefont
  {Lee}}, \bibinfo {author} {\bibfnamefont {W.-G.}\ \bibnamefont {Yin}}, \ and\
  \bibinfo {author} {\bibfnamefont {W.}~\bibnamefont {Ku}},\ }\href {\doibase
  10.1103/PhysRevLett.103.267001} {\bibfield  {journal} {\bibinfo  {journal}
  {Physical Review Letters}\ }\textbf {\bibinfo {volume} {103}},\ \bibinfo
  {pages} {267001} (\bibinfo {year} {2009})}\BibitemShut {NoStop}%
\bibitem [{\citenamefont {Chen}\ \emph
  {et~al.}(2010{\natexlab{c}})\citenamefont {Chen}, \citenamefont {Maciejko},
  \citenamefont {Sorini}, \citenamefont {Moritz}, \citenamefont {Singh},\ and\
  \citenamefont {Devereaux}}]{ChenPRB2010b}%
  \BibitemOpen
  \bibfield  {author} {\bibinfo {author} {\bibfnamefont {C.-C.}\ \bibnamefont
  {Chen}}, \bibinfo {author} {\bibfnamefont {J.}~\bibnamefont {Maciejko}},
  \bibinfo {author} {\bibfnamefont {A.}~\bibnamefont {Sorini}}, \bibinfo
  {author} {\bibfnamefont {B.}~\bibnamefont {Moritz}}, \bibinfo {author}
  {\bibfnamefont {R.}~\bibnamefont {Singh}}, \ and\ \bibinfo {author}
  {\bibfnamefont {T.~P.}\ \bibnamefont {Devereaux}},\ }\href {\doibase
  10.1103/PhysRevB.82.100504} {\bibfield  {journal} {\bibinfo  {journal}
  {Physical Review B}\ }\textbf {\bibinfo {volume} {82}},\ \bibinfo {pages}
  {100504} (\bibinfo {year} {2010}{\natexlab{c}})}\BibitemShut {NoStop}%
\bibitem [{\citenamefont {Lv}\ and\ \citenamefont
  {Phillips}(2011)}]{LvPRB2011}%
  \BibitemOpen
  \bibfield  {author} {\bibinfo {author} {\bibfnamefont {W.}~\bibnamefont
  {Lv}}\ and\ \bibinfo {author} {\bibfnamefont {P.}~\bibnamefont {Phillips}},\
  }\href {\doibase 10.1103/PhysRevB.84.174512} {\bibfield  {journal} {\bibinfo
  {journal} {Physical Review B}\ }\textbf {\bibinfo {volume} {84}},\ \bibinfo
  {pages} {174512} (\bibinfo {year} {2011})}\BibitemShut {NoStop}%
\bibitem [{\citenamefont {Fang}\ \emph {et~al.}(2008)\citenamefont {Fang},
  \citenamefont {Yao}, \citenamefont {Tsai}, \citenamefont {Hu},\ and\
  \citenamefont {Kivelson}}]{FangPRB2008}%
  \BibitemOpen
  \bibfield  {author} {\bibinfo {author} {\bibfnamefont {C.}~\bibnamefont
  {Fang}}, \bibinfo {author} {\bibfnamefont {H.}~\bibnamefont {Yao}}, \bibinfo
  {author} {\bibfnamefont {W.-F.}\ \bibnamefont {Tsai}}, \bibinfo {author}
  {\bibfnamefont {J.}~\bibnamefont {Hu}}, \ and\ \bibinfo {author}
  {\bibfnamefont {S.~A.}\ \bibnamefont {Kivelson}},\ }\href {\doibase
  10.1103/PhysRevB.77.224509} {\bibfield  {journal} {\bibinfo  {journal}
  {Physical Review B}\ }\textbf {\bibinfo {volume} {77}},\ \bibinfo {pages}
  {224509} (\bibinfo {year} {2008})}\BibitemShut {NoStop}%
\bibitem [{\citenamefont {Xu}\ \emph {et~al.}(2008)\citenamefont {Xu},
  \citenamefont {M\"{u}ller},\ and\ \citenamefont {Sachdev}}]{XuPRB2008}%
  \BibitemOpen
  \bibfield  {author} {\bibinfo {author} {\bibfnamefont {C.}~\bibnamefont
  {Xu}}, \bibinfo {author} {\bibfnamefont {M.}~\bibnamefont {M\"{u}ller}}, \
  and\ \bibinfo {author} {\bibfnamefont {S.}~\bibnamefont {Sachdev}},\ }\href
  {\doibase 10.1103/PhysRevB.78.020501} {\bibfield  {journal} {\bibinfo
  {journal} {Physical Review B}\ }\textbf {\bibinfo {volume} {78}},\ \bibinfo
  {pages} {020501} (\bibinfo {year} {2008})}\BibitemShut {NoStop}%
\bibitem [{\citenamefont {Kivelson}\ \emph {et~al.}(1998)\citenamefont
  {Kivelson}, \citenamefont {Fradkin},\ and\ \citenamefont
  {Emery}}]{KivelsonNature1998}%
  \BibitemOpen
  \bibfield  {author} {\bibinfo {author} {\bibfnamefont {S.~A.}\ \bibnamefont
  {Kivelson}}, \bibinfo {author} {\bibfnamefont {E.}~\bibnamefont {Fradkin}}, \
  and\ \bibinfo {author} {\bibfnamefont {V.~J.}\ \bibnamefont {Emery}},\ }\href
  {\doibase 10.1038/31177} {\bibfield  {journal} {\bibinfo  {journal} {Nature}\
  }\textbf {\bibinfo {volume} {393}},\ \bibinfo {pages} {550} (\bibinfo {year}
  {1998})}\BibitemShut {NoStop}%
\bibitem [{\citenamefont {Goldman}\ \emph {et~al.}(2008)\citenamefont
  {Goldman}, \citenamefont {Argyriou}, \citenamefont {Ouladdiaf}, \citenamefont
  {Chatterji}, \citenamefont {Kreyssig}, \citenamefont {Nandi}, \citenamefont
  {Ni}, \citenamefont {Bud'ko}, \citenamefont {Canfield},\ and\ \citenamefont
  {McQueeney}}]{GoldmanPRB2008}%
  \BibitemOpen
  \bibfield  {author} {\bibinfo {author} {\bibfnamefont {A.}~\bibnamefont
  {Goldman}}, \bibinfo {author} {\bibfnamefont {D.}~\bibnamefont {Argyriou}},
  \bibinfo {author} {\bibfnamefont {B.}~\bibnamefont {Ouladdiaf}}, \bibinfo
  {author} {\bibfnamefont {T.}~\bibnamefont {Chatterji}}, \bibinfo {author}
  {\bibfnamefont {A.}~\bibnamefont {Kreyssig}}, \bibinfo {author}
  {\bibfnamefont {S.}~\bibnamefont {Nandi}}, \bibinfo {author} {\bibfnamefont
  {N.}~\bibnamefont {Ni}}, \bibinfo {author} {\bibfnamefont {S.~L.}\
  \bibnamefont {Bud'ko}}, \bibinfo {author} {\bibfnamefont {P.}~\bibnamefont
  {Canfield}}, \ and\ \bibinfo {author} {\bibfnamefont {R.}~\bibnamefont
  {McQueeney}},\ }\href {\doibase 10.1103/PhysRevB.78.100506} {\bibfield
  {journal} {\bibinfo  {journal} {Physical Review B}\ }\textbf {\bibinfo
  {volume} {78}},\ \bibinfo {pages} {100506} (\bibinfo {year}
  {2008})}\BibitemShut {NoStop}%
\bibitem [{\citenamefont {Wang}\ \emph {et~al.}\citenamefont
  {Wang}, \citenamefont {Sun}, \citenamefont {Rotenberg}, \citenamefont
  {Ronning}, \citenamefont {Bauer}, \citenamefont {Lin}, \citenamefont
  {Markiewicz}, \citenamefont {Lindroos}, \citenamefont {Barbiellini},
  \citenamefont {Bansil},\ and\ \citenamefont {Dessau}}]{DessauArxiv1009.0271}%
  \BibitemOpen
  \bibfield  {author} {\bibinfo {author} {\bibfnamefont {Q.}~\bibnamefont
  {Wang}}, \bibinfo {author} {\bibfnamefont {Z.}~\bibnamefont {Sun}}, \bibinfo
  {author} {\bibfnamefont {E.}~\bibnamefont {Rotenberg}}, \bibinfo {author}
  {\bibfnamefont {F.}~\bibnamefont {Ronning}}, \bibinfo {author} {\bibfnamefont
  {E.~D.}\ \bibnamefont {Bauer}}, \bibinfo {author} {\bibfnamefont
  {H.}~\bibnamefont {Lin}}, \bibinfo {author} {\bibfnamefont {R.~S.}\
  \bibnamefont {Markiewicz}}, \bibinfo {author} {\bibfnamefont
  {M.}~\bibnamefont {Lindroos}}, \bibinfo {author} {\bibfnamefont
  {B.}~\bibnamefont {Barbiellini}}, \bibinfo {author} {\bibfnamefont
  {A.}~\bibnamefont {Bansil}}, \ and\ \bibinfo {author} {\bibfnamefont {D.~S.}\
  \bibnamefont {Dessau}},\ }\href@noop {} {\ }\Eprint {http://arxiv.org/abs/1009.0271}
  {arXiv:1009.0271} \BibitemShut {NoStop}%
\bibitem [{\citenamefont {Chu}\ \emph {et~al.}(2010)\citenamefont {Chu},
  \citenamefont {Analytis}, \citenamefont {{De Greve}}, \citenamefont
  {McMahon}, \citenamefont {Islam}, \citenamefont {Yamamoto},\ and\
  \citenamefont {Fisher}}]{ChuScience2010}%
  \BibitemOpen
  \bibfield  {author} {\bibinfo {author} {\bibfnamefont {J.-H.}\ \bibnamefont
  {Chu}}, \bibinfo {author} {\bibfnamefont {J.~G.}\ \bibnamefont {Analytis}},
  \bibinfo {author} {\bibfnamefont {K.}~\bibnamefont {{De Greve}}}, \bibinfo
  {author} {\bibfnamefont {P.~L.}\ \bibnamefont {McMahon}}, \bibinfo {author}
  {\bibfnamefont {Z.}~\bibnamefont {Islam}}, \bibinfo {author} {\bibfnamefont
  {Y.}~\bibnamefont {Yamamoto}}, \ and\ \bibinfo {author} {\bibfnamefont
  {I.~R.}\ \bibnamefont {Fisher}},\ }\href {\doibase 10.1126/science.1190482}
  {\bibfield  {journal} {\bibinfo  {journal} {Science}\ }\textbf {\bibinfo
  {volume} {329}},\ \bibinfo {pages} {824} (\bibinfo {year}
  {2010})}\BibitemShut {NoStop}%
\bibitem [{\citenamefont {Tanatar}\ \emph {et~al.}(2010)\citenamefont
  {Tanatar}, \citenamefont {Blomberg}, \citenamefont {Kreyssig}, \citenamefont
  {Kim}, \citenamefont {Ni}, \citenamefont {Thaler}, \citenamefont {Bud'ko},
  \citenamefont {Canfield}, \citenamefont {Goldman}, \citenamefont {Mazin},\
  and\ \citenamefont {Prozorov}}]{TanatarPRB2010}%
  \BibitemOpen
  \bibfield  {author} {\bibinfo {author} {\bibfnamefont {M.~A.}\ \bibnamefont
  {Tanatar}}, \bibinfo {author} {\bibfnamefont {E.~C.}\ \bibnamefont
  {Blomberg}}, \bibinfo {author} {\bibfnamefont {A.}~\bibnamefont {Kreyssig}},
  \bibinfo {author} {\bibfnamefont {M.~G.}\ \bibnamefont {Kim}}, \bibinfo
  {author} {\bibfnamefont {N.}~\bibnamefont {Ni}}, \bibinfo {author}
  {\bibfnamefont {A.}~\bibnamefont {Thaler}}, \bibinfo {author} {\bibfnamefont
  {S.~L.}\ \bibnamefont {Bud'ko}}, \bibinfo {author} {\bibfnamefont {P.~C.}\
  \bibnamefont {Canfield}}, \bibinfo {author} {\bibfnamefont {A.~I.}\
  \bibnamefont {Goldman}}, \bibinfo {author} {\bibfnamefont {I.~I.}\
  \bibnamefont {Mazin}}, \ and\ \bibinfo {author} {\bibfnamefont
  {R.}~\bibnamefont {Prozorov}},\ }\href {\doibase 10.1103/PhysRevB.81.184508}
  {\bibfield  {journal} {\bibinfo  {journal} {Physical Review B}\ }\textbf
  {\bibinfo {volume} {81}},\ \bibinfo {pages} {184508} (\bibinfo {year}
  {2010})}\BibitemShut {NoStop}%
\bibitem [{\citenamefont {Allan}\ \emph {et~al.}(2011)\citenamefont {Allan},
  \citenamefont {Chuang}, \citenamefont {Xie}, \citenamefont {Lee},
  \citenamefont {Ni}, \citenamefont {Bud'ko}, \citenamefont {Boebinger},
  \citenamefont {Wang}, \citenamefont {Dessau}, \citenamefont {Canfield},\ and\
  \citenamefont {Davis}}]{AllanPreprint2011}%
  \BibitemOpen
  \bibfield  {author} {\bibinfo {author} {\bibfnamefont {M.~P.}\ \bibnamefont
  {Allan}}, \bibinfo {author} {\bibfnamefont {T.-M.}\ \bibnamefont {Chuang}},
  \bibinfo {author} {\bibfnamefont {Y.}~\bibnamefont {Xie}}, \bibinfo {author}
  {\bibfnamefont {J.}~\bibnamefont {Lee}}, \bibinfo {author} {\bibfnamefont
  {N.}~\bibnamefont {Ni}}, \bibinfo {author} {\bibfnamefont {S.~L.}\
  \bibnamefont {Bud'ko}}, \bibinfo {author} {\bibfnamefont {G.~S.}\
  \bibnamefont {Boebinger}}, \bibinfo {author} {\bibfnamefont {Q.}~\bibnamefont
  {Wang}}, \bibinfo {author} {\bibfnamefont {D.}~\bibnamefont {Dessau}},
  \bibinfo {author} {\bibfnamefont {P.~C.}\ \bibnamefont {Canfield}}, \ and\
  \bibinfo {author} {\bibfnamefont {J.~C.}\ \bibnamefont {Davis}},\ }\href@noop
  {} {\bibfield  {journal} {\bibinfo  {journal} {Preprint}\ } (\bibinfo {year}
  {2011})}\BibitemShut {NoStop}%
\bibitem [{\citenamefont {Kondo}\ \emph {et~al.}(2010)\citenamefont {Kondo},
  \citenamefont {Fernandes}, \citenamefont {Khasanov}, \citenamefont {Liu},
  \citenamefont {Palczewski}, \citenamefont {Ni}, \citenamefont {Shi},
  \citenamefont {Bostwick}, \citenamefont {Rotenberg}, \citenamefont
  {Schmalian}, \citenamefont {Bud'ko}, \citenamefont {Canfield},\ and\
  \citenamefont {Kaminski}}]{KondoPRB2010}%
  \BibitemOpen
  \bibfield  {author} {\bibinfo {author} {\bibfnamefont {T.}~\bibnamefont
  {Kondo}}, \bibinfo {author} {\bibfnamefont {R.~M.}\ \bibnamefont
  {Fernandes}}, \bibinfo {author} {\bibfnamefont {R.}~\bibnamefont {Khasanov}},
  \bibinfo {author} {\bibfnamefont {C.}~\bibnamefont {Liu}}, \bibinfo {author}
  {\bibfnamefont {A.~D.}\ \bibnamefont {Palczewski}}, \bibinfo {author}
  {\bibfnamefont {N.}~\bibnamefont {Ni}}, \bibinfo {author} {\bibfnamefont
  {M.}~\bibnamefont {Shi}}, \bibinfo {author} {\bibfnamefont {A.}~\bibnamefont
  {Bostwick}}, \bibinfo {author} {\bibfnamefont {E.}~\bibnamefont {Rotenberg}},
  \bibinfo {author} {\bibfnamefont {J.}~\bibnamefont {Schmalian}}, \bibinfo
  {author} {\bibfnamefont {S.~L.}\ \bibnamefont {Bud'ko}}, \bibinfo {author}
  {\bibfnamefont {P.~C.}\ \bibnamefont {Canfield}}, \ and\ \bibinfo {author}
  {\bibfnamefont {A.}~\bibnamefont {Kaminski}},\ }\href {\doibase
  10.1103/PhysRevB.81.060507} {\bibfield  {journal} {\bibinfo  {journal}
  {Physical Review B}\ }\textbf {\bibinfo {volume} {81}},\ \bibinfo {pages}
  {060507} (\bibinfo {year} {2010})}\BibitemShut {NoStop}%
\bibitem [{\citenamefont {Knolle}\ \emph {et~al.}(2010)\citenamefont {Knolle},
  \citenamefont {Eremin}, \citenamefont {Akbari},\ and\ \citenamefont
  {Moessner}}]{KnollePRL2010}%
  \BibitemOpen
  \bibfield  {author} {\bibinfo {author} {\bibfnamefont {J.}~\bibnamefont
  {Knolle}}, \bibinfo {author} {\bibfnamefont {I.}~\bibnamefont {Eremin}},
  \bibinfo {author} {\bibfnamefont {A.}~\bibnamefont {Akbari}}, \ and\ \bibinfo
  {author} {\bibfnamefont {R.}~\bibnamefont {Moessner}},\ }\href {\doibase
  10.1103/PhysRevLett.104.257001} {\bibfield  {journal} {\bibinfo  {journal}
  {Physical Review Letters}\ }\textbf {\bibinfo {volume} {104}},\ \bibinfo
  {pages} {257001} (\bibinfo {year} {2010})}\BibitemShut {NoStop}%
\bibitem [{\citenamefont {Yi}\ \emph {et~al.}(2011)\citenamefont {Yi},
  \citenamefont {Lu}, \citenamefont {Chu}, \citenamefont {Analytis},
  \citenamefont {Sorini}, \citenamefont {Kemper}, \citenamefont {Moritz},
  \citenamefont {Mo}, \citenamefont {Moore}, \citenamefont {Hashimoto},
  \citenamefont {Lee}, \citenamefont {Hussain}, \citenamefont {Devereaux},
  \citenamefont {Fisher},\ and\ \citenamefont {Shen}}]{YiPNAS2011}%
  \BibitemOpen
  \bibfield  {author} {\bibinfo {author} {\bibfnamefont {M.}~\bibnamefont
  {Yi}}, \bibinfo {author} {\bibfnamefont {D.}~\bibnamefont {Lu}}, \bibinfo
  {author} {\bibfnamefont {J.-H.}\ \bibnamefont {Chu}}, \bibinfo {author}
  {\bibfnamefont {J.~G.}\ \bibnamefont {Analytis}}, \bibinfo {author}
  {\bibfnamefont {A.~P.}\ \bibnamefont {Sorini}}, \bibinfo {author}
  {\bibfnamefont {A.~F.}\ \bibnamefont {Kemper}}, \bibinfo {author}
  {\bibfnamefont {B.}~\bibnamefont {Moritz}}, \bibinfo {author} {\bibfnamefont
  {S.-K.}\ \bibnamefont {Mo}}, \bibinfo {author} {\bibfnamefont {R.~G.}\
  \bibnamefont {Moore}}, \bibinfo {author} {\bibfnamefont {M.}~\bibnamefont
  {Hashimoto}}, \bibinfo {author} {\bibfnamefont {W.-S.}\ \bibnamefont {Lee}},
  \bibinfo {author} {\bibfnamefont {Z.}~\bibnamefont {Hussain}}, \bibinfo
  {author} {\bibfnamefont {T.~P.}\ \bibnamefont {Devereaux}}, \bibinfo {author}
  {\bibfnamefont {I.~R.}\ \bibnamefont {Fisher}}, \ and\ \bibinfo {author}
  {\bibfnamefont {Z.-X.}\ \bibnamefont {Shen}},\ }\href {\doibase
  10.1073/pnas.1015572108} {\bibfield  {journal} {\bibinfo  {journal}
  {Proceedings of the National Academy of Sciences of the United States of
  America}\ }\textbf {\bibinfo {volume} {108}},\ \bibinfo {pages} {6878}
  (\bibinfo {year} {2011})}\BibitemShut {NoStop}%
\bibitem [{\citenamefont {Fradkin}\ and\ \citenamefont
  {Kivelson}(2010)}]{FradkinScience2010}%
  \BibitemOpen
  \bibfield  {author} {\bibinfo {author} {\bibfnamefont {E.}~\bibnamefont
  {Fradkin}}\ and\ \bibinfo {author} {\bibfnamefont {S.~A.}\ \bibnamefont
  {Kivelson}},\ }\href {\doibase 10.1126/science.1183464} {\bibfield  {journal}
  {\bibinfo  {journal} {Science}\ }\textbf {\bibinfo {volume} {327}},\ \bibinfo
  {pages} {155} (\bibinfo {year} {2010})}\BibitemShut {NoStop}%
\bibitem [{\citenamefont {Kr\"{u}ger}\ \emph {et~al.}(2009)\citenamefont
  {Kr\"{u}ger}, \citenamefont {Kumar}, \citenamefont {Zaanen},\ and\
  \citenamefont {van~den Brink}}]{KrugerPRB2009}%
  \BibitemOpen
  \bibfield  {author} {\bibinfo {author} {\bibfnamefont {F.}~\bibnamefont
  {Kr\"{u}ger}}, \bibinfo {author} {\bibfnamefont {S.}~\bibnamefont {Kumar}},
  \bibinfo {author} {\bibfnamefont {J.}~\bibnamefont {Zaanen}}, \ and\ \bibinfo
  {author} {\bibfnamefont {J.}~\bibnamefont {van~den Brink}},\ }\href {\doibase
  10.1103/PhysRevB.79.054504} {\bibfield  {journal} {\bibinfo  {journal}
  {Physical Review B}\ }\textbf {\bibinfo {volume} {79}},\ \bibinfo {pages}
  {054504} (\bibinfo {year} {2009})}\BibitemShut {NoStop}%
\bibitem [{\citenamefont {Timusk}\ and\ \citenamefont
  {Statt}(1999)}]{TimuskROPP1999}%
  \BibitemOpen
  \bibfield  {author} {\bibinfo {author} {\bibfnamefont {T.}~\bibnamefont
  {Timusk}}\ and\ \bibinfo {author} {\bibfnamefont {B.}~\bibnamefont {Statt}},\
  }\href {\doibase 10.1088/0034-4885/62/1/002} {\bibfield  {journal} {\bibinfo
  {journal} {Reports on Progress in Physics}\ }\textbf {\bibinfo {volume}
  {62}},\ \bibinfo {pages} {61} (\bibinfo {year} {1999})}\BibitemShut {NoStop}%
\bibitem [{\citenamefont {Ahilan}\ \emph {et~al.}(2008)\citenamefont {Ahilan},
  \citenamefont {Ning}, \citenamefont {Imai}, \citenamefont {Sefat},
  \citenamefont {Jin}, \citenamefont {McGuire}, \citenamefont {Sales},\ and\
  \citenamefont {Mandrus}}]{AhilanPRB2008}%
  \BibitemOpen
  \bibfield  {author} {\bibinfo {author} {\bibfnamefont {K.}~\bibnamefont
  {Ahilan}}, \bibinfo {author} {\bibfnamefont {F.}~\bibnamefont {Ning}},
  \bibinfo {author} {\bibfnamefont {T.}~\bibnamefont {Imai}}, \bibinfo {author}
  {\bibfnamefont {A.}~\bibnamefont {Sefat}}, \bibinfo {author} {\bibfnamefont
  {R.}~\bibnamefont {Jin}}, \bibinfo {author} {\bibfnamefont {M.}~\bibnamefont
  {McGuire}}, \bibinfo {author} {\bibfnamefont {B.}~\bibnamefont {Sales}}, \
  and\ \bibinfo {author} {\bibfnamefont {D.}~\bibnamefont {Mandrus}},\ }\href
  {\doibase 10.1103/PhysRevB.78.100501} {\bibfield  {journal} {\bibinfo
  {journal} {Physical Review B}\ }\textbf {\bibinfo {volume} {78}},\ \bibinfo
  {pages} {100501} (\bibinfo {year} {2008})}\BibitemShut {NoStop}%
\bibitem [{\citenamefont {Xu}\ \emph {et~al.}(2011)\citenamefont {Xu},
  \citenamefont {Richard}, \citenamefont {Nakayama}, \citenamefont {Kawahara},
  \citenamefont {Sekiba}, \citenamefont {Qian}, \citenamefont {Neupane},
  \citenamefont {Souma}, \citenamefont {Sato}, \citenamefont {Takahashi},
  \citenamefont {Luo}, \citenamefont {Wen}, \citenamefont {Chen}, \citenamefont
  {Wang}, \citenamefont {Wang}, \citenamefont {Fang}, \citenamefont {Dai},\
  and\ \citenamefont {Ding}}]{XuNatComm2011}%
  \BibitemOpen
  \bibfield  {author} {\bibinfo {author} {\bibfnamefont {Y.-M.}\ \bibnamefont
  {Xu}}, \bibinfo {author} {\bibfnamefont {P.}~\bibnamefont {Richard}},
  \bibinfo {author} {\bibfnamefont {K.}~\bibnamefont {Nakayama}}, \bibinfo
  {author} {\bibfnamefont {T.}~\bibnamefont {Kawahara}}, \bibinfo {author}
  {\bibfnamefont {Y.}~\bibnamefont {Sekiba}}, \bibinfo {author} {\bibfnamefont
  {T.}~\bibnamefont {Qian}}, \bibinfo {author} {\bibfnamefont {M.}~\bibnamefont
  {Neupane}}, \bibinfo {author} {\bibfnamefont {S.}~\bibnamefont {Souma}},
  \bibinfo {author} {\bibfnamefont {T.}~\bibnamefont {Sato}}, \bibinfo {author}
  {\bibfnamefont {T.}~\bibnamefont {Takahashi}}, \bibinfo {author}
  {\bibfnamefont {H.-Q.}\ \bibnamefont {Luo}}, \bibinfo {author} {\bibfnamefont
  {H.-H.}\ \bibnamefont {Wen}}, \bibinfo {author} {\bibfnamefont {G.-F.}\
  \bibnamefont {Chen}}, \bibinfo {author} {\bibfnamefont {N.-L.}\ \bibnamefont
  {Wang}}, \bibinfo {author} {\bibfnamefont {Z.}~\bibnamefont {Wang}}, \bibinfo
  {author} {\bibfnamefont {Z.}~\bibnamefont {Fang}}, \bibinfo {author}
  {\bibfnamefont {X.}~\bibnamefont {Dai}}, \ and\ \bibinfo {author}
  {\bibfnamefont {H.}~\bibnamefont {Ding}},\ }\href {\doibase
  10.1038/ncomms1394} {\bibfield  {journal} {\bibinfo  {journal} {Nature
  Communications}\ }\textbf {\bibinfo {volume} {2}},\ \bibinfo {pages} {394}
  (\bibinfo {year} {2011})}\BibitemShut {NoStop}%
\bibitem [{\citenamefont {Mertelj}\ \emph {et~al.}(2009)\citenamefont
  {Mertelj}, \citenamefont {Kabanov}, \citenamefont {Gadermaier}, \citenamefont
  {Zhigadlo}, \citenamefont {Katrych}, \citenamefont {Karpinski},\ and\
  \citenamefont {Mihailovic}}]{MerteljPRL2009}%
  \BibitemOpen
  \bibfield  {author} {\bibinfo {author} {\bibfnamefont {T.}~\bibnamefont
  {Mertelj}}, \bibinfo {author} {\bibfnamefont {V.}~\bibnamefont {Kabanov}},
  \bibinfo {author} {\bibfnamefont {C.}~\bibnamefont {Gadermaier}}, \bibinfo
  {author} {\bibfnamefont {N.}~\bibnamefont {Zhigadlo}}, \bibinfo {author}
  {\bibfnamefont {S.}~\bibnamefont {Katrych}}, \bibinfo {author} {\bibfnamefont
  {J.}~\bibnamefont {Karpinski}}, \ and\ \bibinfo {author} {\bibfnamefont
  {D.}~\bibnamefont {Mihailovic}},\ }\href {\doibase
  10.1103/PhysRevLett.102.117002} {\bibfield  {journal} {\bibinfo  {journal}
  {Physical Review Letters}\ }\textbf {\bibinfo {volume} {102}},\ \bibinfo
  {pages} {117002} (\bibinfo {year} {2009})}\BibitemShut {NoStop}%
\bibitem [{\citenamefont {Hess}\ \emph {et~al.}(2009)\citenamefont {Hess},
  \citenamefont {Kondrat}, \citenamefont {Narduzzo}, \citenamefont
  {Hamann-Borrero}, \citenamefont {Klingeler}, \citenamefont {Werner},
  \citenamefont {Behr},\ and\ \citenamefont {B\"{u}chner}}]{HessEPL2009}%
  \BibitemOpen
  \bibfield  {author} {\bibinfo {author} {\bibfnamefont {C.}~\bibnamefont
  {Hess}}, \bibinfo {author} {\bibfnamefont {A.}~\bibnamefont {Kondrat}},
  \bibinfo {author} {\bibfnamefont {A.}~\bibnamefont {Narduzzo}}, \bibinfo
  {author} {\bibfnamefont {J.~E.}\ \bibnamefont {Hamann-Borrero}}, \bibinfo
  {author} {\bibfnamefont {R.}~\bibnamefont {Klingeler}}, \bibinfo {author}
  {\bibfnamefont {J.}~\bibnamefont {Werner}}, \bibinfo {author} {\bibfnamefont
  {G.}~\bibnamefont {Behr}}, \ and\ \bibinfo {author} {\bibfnamefont
  {B.}~\bibnamefont {B\"{u}chner}},\ }\href {\doibase
  10.1209/0295-5075/87/17005} {\bibfield  {journal} {\bibinfo  {journal}
  {Europhysics Letters}\ }\textbf {\bibinfo {volume} {87}},\ \bibinfo {pages}
  {17005} (\bibinfo {year} {2009})}\BibitemShut {NoStop}%
\bibitem [{\citenamefont {Hu}\ \emph {et~al.}(2008)\citenamefont {Hu},
  \citenamefont {Dong}, \citenamefont {Li}, \citenamefont {Li}, \citenamefont
  {Zheng}, \citenamefont {Chen}, \citenamefont {Luo},\ and\ \citenamefont
  {Wang}}]{HuPRL2008}%
  \BibitemOpen
  \bibfield  {author} {\bibinfo {author} {\bibfnamefont {W.}~\bibnamefont
  {Hu}}, \bibinfo {author} {\bibfnamefont {J.}~\bibnamefont {Dong}}, \bibinfo
  {author} {\bibfnamefont {G.}~\bibnamefont {Li}}, \bibinfo {author}
  {\bibfnamefont {Z.}~\bibnamefont {Li}}, \bibinfo {author} {\bibfnamefont
  {P.}~\bibnamefont {Zheng}}, \bibinfo {author} {\bibfnamefont
  {G.}~\bibnamefont {Chen}}, \bibinfo {author} {\bibfnamefont {J.}~\bibnamefont
  {Luo}}, \ and\ \bibinfo {author} {\bibfnamefont {N.}~\bibnamefont {Wang}},\
  }\href {\doibase 10.1103/PhysRevLett.101.257005} {\bibfield  {journal}
  {\bibinfo  {journal} {Physical Review Letters}\ }\textbf {\bibinfo {volume}
  {101}},\ \bibinfo {pages} {257005} (\bibinfo {year} {2008})}\BibitemShut
  {NoStop}%
\bibitem [{\citenamefont {Hunte}\ \emph {et~al.}(2008)\citenamefont {Hunte},
  \citenamefont {Jaroszynski}, \citenamefont {Gurevich}, \citenamefont
  {Larbalestier}, \citenamefont {Jin}, \citenamefont {Sefat}, \citenamefont
  {McGuire}, \citenamefont {Sales}, \citenamefont {Christen},\ and\
  \citenamefont {Mandrus}}]{HunteNature2008}%
  \BibitemOpen
  \bibfield  {author} {\bibinfo {author} {\bibfnamefont {F.}~\bibnamefont
  {Hunte}}, \bibinfo {author} {\bibfnamefont {J.}~\bibnamefont {Jaroszynski}},
  \bibinfo {author} {\bibfnamefont {A.}~\bibnamefont {Gurevich}}, \bibinfo
  {author} {\bibfnamefont {D.~C.}\ \bibnamefont {Larbalestier}}, \bibinfo
  {author} {\bibfnamefont {R.}~\bibnamefont {Jin}}, \bibinfo {author}
  {\bibfnamefont {A.~S.}\ \bibnamefont {Sefat}}, \bibinfo {author}
  {\bibfnamefont {M.~A.}\ \bibnamefont {McGuire}}, \bibinfo {author}
  {\bibfnamefont {B.~C.}\ \bibnamefont {Sales}}, \bibinfo {author}
  {\bibfnamefont {D.~K.}\ \bibnamefont {Christen}}, \ and\ \bibinfo {author}
  {\bibfnamefont {D.}~\bibnamefont {Mandrus}},\ }\href {\doibase
  10.1038/nature07058} {\bibfield  {journal} {\bibinfo  {journal} {Nature}\
  }\textbf {\bibinfo {volume} {453}},\ \bibinfo {pages} {903} (\bibinfo {year}
  {2008})}\BibitemShut {NoStop}%
\bibitem [{\citenamefont {Yamamoto}\ \emph {et~al.}(2009)\citenamefont
  {Yamamoto}, \citenamefont {Jaroszynski}, \citenamefont {Tarantini},
  \citenamefont {Balicas}, \citenamefont {Jiang}, \citenamefont {Gurevich},
  \citenamefont {Larbalestier}, \citenamefont {Jin}, \citenamefont {Sefat},
  \citenamefont {McGuire}, \citenamefont {Sales}, \citenamefont {Christen},\
  and\ \citenamefont {Mandrus}}]{YamamotoAPL2009}%
  \BibitemOpen
  \bibfield  {author} {\bibinfo {author} {\bibfnamefont {A.}~\bibnamefont
  {Yamamoto}}, \bibinfo {author} {\bibfnamefont {J.}~\bibnamefont
  {Jaroszynski}}, \bibinfo {author} {\bibfnamefont {C.}~\bibnamefont
  {Tarantini}}, \bibinfo {author} {\bibfnamefont {L.}~\bibnamefont {Balicas}},
  \bibinfo {author} {\bibfnamefont {J.}~\bibnamefont {Jiang}}, \bibinfo
  {author} {\bibfnamefont {A.}~\bibnamefont {Gurevich}}, \bibinfo {author}
  {\bibfnamefont {D.~C.}\ \bibnamefont {Larbalestier}}, \bibinfo {author}
  {\bibfnamefont {R.}~\bibnamefont {Jin}}, \bibinfo {author} {\bibfnamefont
  {A.~S.}\ \bibnamefont {Sefat}}, \bibinfo {author} {\bibfnamefont {M.~A.}\
  \bibnamefont {McGuire}}, \bibinfo {author} {\bibfnamefont {B.~C.}\
  \bibnamefont {Sales}}, \bibinfo {author} {\bibfnamefont {D.~K.}\ \bibnamefont
  {Christen}}, \ and\ \bibinfo {author} {\bibfnamefont {D.}~\bibnamefont
  {Mandrus}},\ }\href {\doibase 10.1063/1.3081455} {\bibfield  {journal}
  {\bibinfo  {journal} {Applied Physics Letters}\ }\textbf {\bibinfo {volume}
  {94}},\ \bibinfo {pages} {062511} (\bibinfo {year} {2009})}\BibitemShut
  {NoStop}%
\bibitem [{\citenamefont {Putti}\ \emph {et~al.}(2010)\citenamefont {Putti},
  \citenamefont {Pallecchi}, \citenamefont {Bellingeri}, \citenamefont
  {Cimberle}, \citenamefont {Tropeano}, \citenamefont {Ferdeghini},
  \citenamefont {Palenzona}, \citenamefont {Tarantini}, \citenamefont
  {Yamamoto}, \citenamefont {Jiang}, \citenamefont {Jaroszynski}, \citenamefont
  {Kametani}, \citenamefont {Abraimov}, \citenamefont {Polyanskii},
  \citenamefont {Weiss}, \citenamefont {Hellstrom}, \citenamefont {Gurevich},
  \citenamefont {Larbalestier}, \citenamefont {Jin}, \citenamefont {Sales},
  \citenamefont {Sefat}, \citenamefont {McGuire}, \citenamefont {Mandrus},
  \citenamefont {Cheng}, \citenamefont {Jia}, \citenamefont {Wen},
  \citenamefont {Lee},\ and\ \citenamefont {Eom}}]{PuttiSST2010}%
  \BibitemOpen
  \bibfield  {author} {\bibinfo {author} {\bibfnamefont {M.}~\bibnamefont
  {Putti}}, \bibinfo {author} {\bibfnamefont {I.}~\bibnamefont {Pallecchi}},
  \bibinfo {author} {\bibfnamefont {E.}~\bibnamefont {Bellingeri}}, \bibinfo
  {author} {\bibfnamefont {M.~R.}\ \bibnamefont {Cimberle}}, \bibinfo {author}
  {\bibfnamefont {M.}~\bibnamefont {Tropeano}}, \bibinfo {author}
  {\bibfnamefont {C.}~\bibnamefont {Ferdeghini}}, \bibinfo {author}
  {\bibfnamefont {A.}~\bibnamefont {Palenzona}}, \bibinfo {author}
  {\bibfnamefont {C.}~\bibnamefont {Tarantini}}, \bibinfo {author}
  {\bibfnamefont {A.}~\bibnamefont {Yamamoto}}, \bibinfo {author}
  {\bibfnamefont {J.}~\bibnamefont {Jiang}}, \bibinfo {author} {\bibfnamefont
  {J.}~\bibnamefont {Jaroszynski}}, \bibinfo {author} {\bibfnamefont
  {F.}~\bibnamefont {Kametani}}, \bibinfo {author} {\bibfnamefont
  {D.}~\bibnamefont {Abraimov}}, \bibinfo {author} {\bibfnamefont
  {A.}~\bibnamefont {Polyanskii}}, \bibinfo {author} {\bibfnamefont {J.~D.}\
  \bibnamefont {Weiss}}, \bibinfo {author} {\bibfnamefont {E.~E.}\ \bibnamefont
  {Hellstrom}}, \bibinfo {author} {\bibfnamefont {A.}~\bibnamefont {Gurevich}},
  \bibinfo {author} {\bibfnamefont {D.~C.}\ \bibnamefont {Larbalestier}},
  \bibinfo {author} {\bibfnamefont {R.}~\bibnamefont {Jin}}, \bibinfo {author}
  {\bibfnamefont {B.~C.}\ \bibnamefont {Sales}}, \bibinfo {author}
  {\bibfnamefont {A.~S.}\ \bibnamefont {Sefat}}, \bibinfo {author}
  {\bibfnamefont {M.~A.}\ \bibnamefont {McGuire}}, \bibinfo {author}
  {\bibfnamefont {D.}~\bibnamefont {Mandrus}}, \bibinfo {author} {\bibfnamefont
  {P.}~\bibnamefont {Cheng}}, \bibinfo {author} {\bibfnamefont
  {Y.}~\bibnamefont {Jia}}, \bibinfo {author} {\bibfnamefont {H.-H.}\
  \bibnamefont {Wen}}, \bibinfo {author} {\bibfnamefont {S.}~\bibnamefont
  {Lee}}, \ and\ \bibinfo {author} {\bibfnamefont {C.~B.}\ \bibnamefont
  {Eom}},\ }\href {\doibase 10.1088/0953-2048/23/3/034003} {\bibfield
  {journal} {\bibinfo  {journal} {Superconductor Science and Technology}\
  }\textbf {\bibinfo {volume} {23}},\ \bibinfo {pages} {034003} (\bibinfo
  {year} {2010})}\BibitemShut {NoStop}%
\bibitem [{\citenamefont {Auslaender}\ \emph {et~al.}(2008)\citenamefont
  {Auslaender}, \citenamefont {Luan}, \citenamefont {Straver}, \citenamefont
  {Hoffman}, \citenamefont {Koshnick}, \citenamefont {Zeldov}, \citenamefont
  {Bonn}, \citenamefont {Liang}, \citenamefont {Hardy},\ and\ \citenamefont
  {Moler}}]{AuslaenderNatPhys2008}%
  \BibitemOpen
  \bibfield  {author} {\bibinfo {author} {\bibfnamefont {O.~M.}\ \bibnamefont
  {Auslaender}}, \bibinfo {author} {\bibfnamefont {L.}~\bibnamefont {Luan}},
  \bibinfo {author} {\bibfnamefont {E.~W.~J.}\ \bibnamefont {Straver}},
  \bibinfo {author} {\bibfnamefont {J.~E.}\ \bibnamefont {Hoffman}}, \bibinfo
  {author} {\bibfnamefont {N.~C.}\ \bibnamefont {Koshnick}}, \bibinfo {author}
  {\bibfnamefont {E.}~\bibnamefont {Zeldov}}, \bibinfo {author} {\bibfnamefont
  {D.~A.}\ \bibnamefont {Bonn}}, \bibinfo {author} {\bibfnamefont
  {R.}~\bibnamefont {Liang}}, \bibinfo {author} {\bibfnamefont {W.~N.}\
  \bibnamefont {Hardy}}, \ and\ \bibinfo {author} {\bibfnamefont {K.~A.}\
  \bibnamefont {Moler}},\ }\href {\doibase 10.1038/nphys1127} {\bibfield
  {journal} {\bibinfo  {journal} {Nature Physics}\ }\textbf {\bibinfo {volume}
  {5}},\ \bibinfo {pages} {35} (\bibinfo {year} {2008})}\BibitemShut {NoStop}%
\bibitem [{\citenamefont {Clem}(1991)}]{ClemPRB1991}%
  \BibitemOpen
  \bibfield  {author} {\bibinfo {author} {\bibfnamefont {J.}~\bibnamefont
  {Clem}},\ }\href {\doibase 10.1103/PhysRevB.43.7837} {\bibfield  {journal}
  {\bibinfo  {journal} {Physical Review B}\ }\textbf {\bibinfo {volume} {43}},\
  \bibinfo {pages} {7837} (\bibinfo {year} {1991})}\BibitemShut {NoStop}%
\bibitem [{\citenamefont {Pan}\ \emph {et~al.}(2000)\citenamefont {Pan},
  \citenamefont {Hudson}, \citenamefont {Gupta}, \citenamefont {Ng},
  \citenamefont {Eisaki}, \citenamefont {Uchida},\ and\ \citenamefont
  {Davis}}]{PanPRL2000}%
  \BibitemOpen
  \bibfield  {author} {\bibinfo {author} {\bibfnamefont {S.-H.}\ \bibnamefont
  {Pan}}, \bibinfo {author} {\bibfnamefont {E.~W.}\ \bibnamefont {Hudson}},
  \bibinfo {author} {\bibfnamefont {A.~K.}\ \bibnamefont {Gupta}}, \bibinfo
  {author} {\bibfnamefont {K.-W.}\ \bibnamefont {Ng}}, \bibinfo {author}
  {\bibfnamefont {H.}~\bibnamefont {Eisaki}}, \bibinfo {author} {\bibfnamefont
  {S.}~\bibnamefont {Uchida}}, \ and\ \bibinfo {author} {\bibfnamefont {J.~C.}\
  \bibnamefont {Davis}},\ }\href {\doibase 10.1103/PhysRevLett.85.1536}
  {\bibfield  {journal} {\bibinfo  {journal} {Physical Review Letters}\
  }\textbf {\bibinfo {volume} {85}},\ \bibinfo {pages} {1536} (\bibinfo {year}
  {2000})}\BibitemShut {NoStop}%
\bibitem [{\citenamefont {Caroli}\ \emph {et~al.}(1964)\citenamefont {Caroli},
  \citenamefont {{De Gennes}},\ and\ \citenamefont
  {Matricon}}]{CaroliPhysLett1964}%
  \BibitemOpen
  \bibfield  {author} {\bibinfo {author} {\bibfnamefont {C.}~\bibnamefont
  {Caroli}}, \bibinfo {author} {\bibfnamefont {P.~G.}\ \bibnamefont {{De
  Gennes}}}, \ and\ \bibinfo {author} {\bibfnamefont {J.}~\bibnamefont
  {Matricon}},\ }\href {\doibase 10.1016/0031-9163(64)90375-0} {\bibfield
  {journal} {\bibinfo  {journal} {Physics Letters}\ }\textbf {\bibinfo {volume}
  {9}},\ \bibinfo {pages} {307} (\bibinfo {year} {1964})}\BibitemShut {NoStop}%
\bibitem [{\citenamefont {Hess}\ \emph {et~al.}(1989)\citenamefont {Hess},
  \citenamefont {Robinson}, \citenamefont {Dynes}, \citenamefont {Valles},\
  and\ \citenamefont {Waszczak}}]{HessPRL1989}%
  \BibitemOpen
  \bibfield  {author} {\bibinfo {author} {\bibfnamefont {H.}~\bibnamefont
  {Hess}}, \bibinfo {author} {\bibfnamefont {R.}~\bibnamefont {Robinson}},
  \bibinfo {author} {\bibfnamefont {R.}~\bibnamefont {Dynes}}, \bibinfo
  {author} {\bibfnamefont {J.}~\bibnamefont {Valles}}, \ and\ \bibinfo {author}
  {\bibfnamefont {J.}~\bibnamefont {Waszczak}},\ }\href {\doibase
  10.1103/PhysRevLett.62.214} {\bibfield  {journal} {\bibinfo  {journal}
  {Physical Review Letters}\ }\textbf {\bibinfo {volume} {62}},\ \bibinfo
  {pages} {214} (\bibinfo {year} {1989})}\BibitemShut {NoStop}%
\bibitem [{\citenamefont {Hess}\ \emph {et~al.}(1990)\citenamefont {Hess},
  \citenamefont {Robinson},\ and\ \citenamefont {Waszczak}}]{HessPRL1990}%
  \BibitemOpen
  \bibfield  {author} {\bibinfo {author} {\bibfnamefont {H.}~\bibnamefont
  {Hess}}, \bibinfo {author} {\bibfnamefont {R.}~\bibnamefont {Robinson}}, \
  and\ \bibinfo {author} {\bibfnamefont {J.}~\bibnamefont {Waszczak}},\ }\href
  {\doibase 10.1103/PhysRevLett.64.2711} {\bibfield  {journal} {\bibinfo
  {journal} {Physical Review Letters}\ }\textbf {\bibinfo {volume} {64}},\
  \bibinfo {pages} {2711} (\bibinfo {year} {1990})}\BibitemShut {NoStop}%
\bibitem [{\citenamefont {Maggio-Aprile}\ \emph {et~al.}(1995)\citenamefont
  {Maggio-Aprile}, \citenamefont {Renner}, \citenamefont {Erb}, \citenamefont
  {Walker},\ and\ \citenamefont {Fischer}}]{Maggio-AprilePRL1995}%
  \BibitemOpen
  \bibfield  {author} {\bibinfo {author} {\bibfnamefont {I.}~\bibnamefont
  {Maggio-Aprile}}, \bibinfo {author} {\bibfnamefont {C.}~\bibnamefont
  {Renner}}, \bibinfo {author} {\bibfnamefont {A.}~\bibnamefont {Erb}},
  \bibinfo {author} {\bibfnamefont {E.}~\bibnamefont {Walker}}, \ and\ \bibinfo
  {author} {\bibfnamefont {{\O}.}~\bibnamefont {Fischer}},\ }\href {\doibase
  10.1103/PhysRevLett.75.2754} {\bibfield  {journal} {\bibinfo  {journal}
  {Physical Review Letters}\ }\textbf {\bibinfo {volume} {75}},\ \bibinfo
  {pages} {2754} (\bibinfo {year} {1995})}\BibitemShut {NoStop}%
\bibitem [{\citenamefont {Hayashi}\ \emph {et~al.}(1998)\citenamefont
  {Hayashi}, \citenamefont {Isoshima}, \citenamefont {Ichioka},\ and\
  \citenamefont {Machida}}]{HayashiPRL1998}%
  \BibitemOpen
  \bibfield  {author} {\bibinfo {author} {\bibfnamefont {N.}~\bibnamefont
  {Hayashi}}, \bibinfo {author} {\bibfnamefont {T.}~\bibnamefont {Isoshima}},
  \bibinfo {author} {\bibfnamefont {M.}~\bibnamefont {Ichioka}}, \ and\
  \bibinfo {author} {\bibfnamefont {K.}~\bibnamefont {Machida}},\ }\href
  {\doibase 10.1103/PhysRevLett.80.2921} {\bibfield  {journal} {\bibinfo
  {journal} {Physical Review Letters}\ }\textbf {\bibinfo {volume} {80}},\
  \bibinfo {pages} {2921} (\bibinfo {year} {1998})}\BibitemShut {NoStop}%
\bibitem [{\citenamefont {Hu}\ \emph {et~al.}(2009)\citenamefont {Hu},
  \citenamefont {Ting},\ and\ \citenamefont {Zhu}}]{HuPRB2009}%
  \BibitemOpen
  \bibfield  {author} {\bibinfo {author} {\bibfnamefont {X.}~\bibnamefont
  {Hu}}, \bibinfo {author} {\bibfnamefont {C.}~\bibnamefont {Ting}}, \ and\
  \bibinfo {author} {\bibfnamefont {J.-X.}\ \bibnamefont {Zhu}},\ }\href
  {\doibase 10.1103/PhysRevB.80.014523} {\bibfield  {journal} {\bibinfo
  {journal} {Physical Review B}\ }\textbf {\bibinfo {volume} {80}},\ \bibinfo
  {pages} {9} (\bibinfo {year} {2009})}\BibitemShut {NoStop}%
\bibitem [{\citenamefont {Jiang}\ \emph {et~al.}(2009)\citenamefont {Jiang},
  \citenamefont {Li},\ and\ \citenamefont {Wang}}]{JiangPRB2009}%
  \BibitemOpen
  \bibfield  {author} {\bibinfo {author} {\bibfnamefont {H.-M.}\ \bibnamefont
  {Jiang}}, \bibinfo {author} {\bibfnamefont {J.-X.}\ \bibnamefont {Li}}, \
  and\ \bibinfo {author} {\bibfnamefont {Z.}~\bibnamefont {Wang}},\ }\href
  {\doibase 10.1103/PhysRevB.80.134505} {\bibfield  {journal} {\bibinfo
  {journal} {Physical Review B}\ }\textbf {\bibinfo {volume} {80}},\ \bibinfo
  {pages} {134505} (\bibinfo {year} {2009})}\BibitemShut {NoStop}%
\bibitem [{\citenamefont {Ara\'{u}jo}\ \emph {et~al.}(2009)\citenamefont
  {Ara\'{u}jo}, \citenamefont {Cardoso},\ and\ \citenamefont
  {Sacramento}}]{AraujoNJP2009}%
  \BibitemOpen
  \bibfield  {author} {\bibinfo {author} {\bibfnamefont {M.~A.~N.}\
  \bibnamefont {Ara\'{u}jo}}, \bibinfo {author} {\bibfnamefont
  {M.}~\bibnamefont {Cardoso}}, \ and\ \bibinfo {author} {\bibfnamefont
  {P.~D.}\ \bibnamefont {Sacramento}},\ }\href {\doibase
  10.1088/1367-2630/11/11/113008} {\bibfield  {journal} {\bibinfo  {journal}
  {New Journal of Physics}\ }\textbf {\bibinfo {volume} {11}},\ \bibinfo
  {pages} {113008} (\bibinfo {year} {2009})}\BibitemShut {NoStop}%
\bibitem [{\citenamefont {Wang}\ \emph {et~al.}(2010)\citenamefont {Wang},
  \citenamefont {Xu}, \citenamefont {Xiang},\ and\ \citenamefont
  {Wang}}]{WangPRB2010}%
  \BibitemOpen
  \bibfield  {author} {\bibinfo {author} {\bibfnamefont {D.}~\bibnamefont
  {Wang}}, \bibinfo {author} {\bibfnamefont {J.}~\bibnamefont {Xu}}, \bibinfo
  {author} {\bibfnamefont {Y.-Y.}\ \bibnamefont {Xiang}}, \ and\ \bibinfo
  {author} {\bibfnamefont {Q.-H.}\ \bibnamefont {Wang}},\ }\href {\doibase
  10.1103/PhysRevB.82.184519} {\bibfield  {journal} {\bibinfo  {journal}
  {Physical Review B}\ }\textbf {\bibinfo {volume} {82}},\ \bibinfo {pages}
  {184519} (\bibinfo {year} {2010})}\BibitemShut {NoStop}%
\bibitem [{\citenamefont {Gao}\ \emph {et~al.}(2011)\citenamefont {Gao},
  \citenamefont {Huang}, \citenamefont {Chen}, \citenamefont {Ting},\ and\
  \citenamefont {Su}}]{GaoPRL2011}%
  \BibitemOpen
  \bibfield  {author} {\bibinfo {author} {\bibfnamefont {Y.}~\bibnamefont
  {Gao}}, \bibinfo {author} {\bibfnamefont {H.-X.}\ \bibnamefont {Huang}},
  \bibinfo {author} {\bibfnamefont {C.}~\bibnamefont {Chen}}, \bibinfo {author}
  {\bibfnamefont {C.}~\bibnamefont {Ting}}, \ and\ \bibinfo {author}
  {\bibfnamefont {W.-P.}\ \bibnamefont {Su}},\ }\href {\doibase
  10.1103/PhysRevLett.106.027004} {\bibfield  {journal} {\bibinfo  {journal}
  {Physical Review Letters}\ }\textbf {\bibinfo {volume} {106}},\ \bibinfo
  {pages} {027004} (\bibinfo {year} {2011})}\BibitemShut {NoStop}%
\bibitem [{\citenamefont {Wang}\ \emph {et~al.}()\citenamefont
  {Wang}, \citenamefont {Hirschfeld},\ and\ \citenamefont
  {Vekhter}}]{VekhterArxiv1111.0126}%
  \BibitemOpen
  \bibfield  {author} {\bibinfo {author} {\bibfnamefont {Y.}~\bibnamefont
  {Wang}}, \bibinfo {author} {\bibfnamefont {P.~J.}\ \bibnamefont
  {Hirschfeld}}, \ and\ \bibinfo {author} {\bibfnamefont {I.}~\bibnamefont
  {Vekhter}},\ }\href@noop {} 
  {\  }\Eprint {http://arxiv.org/abs/1111.0126} {arXiv:1111.0126} \BibitemShut
  {NoStop}%
\bibitem [{\citenamefont {Liu}\ \emph {et~al.}(2011)\citenamefont {Liu},
  \citenamefont {Luo}, \citenamefont {Zhang}, \citenamefont {Wang},
  \citenamefont {Ying}, \citenamefont {Wang}, \citenamefont {Yan},
  \citenamefont {Xiang}, \citenamefont {Cheng}, \citenamefont {Ye},
  \citenamefont {Li},\ and\ \citenamefont {Chen}}]{LiuEPL2011}%
  \BibitemOpen
  \bibfield  {author} {\bibinfo {author} {\bibfnamefont {R.~H.}\ \bibnamefont
  {Liu}}, \bibinfo {author} {\bibfnamefont {X.~G.}\ \bibnamefont {Luo}},
  \bibinfo {author} {\bibfnamefont {M.}~\bibnamefont {Zhang}}, \bibinfo
  {author} {\bibfnamefont {A.~F.}\ \bibnamefont {Wang}}, \bibinfo {author}
  {\bibfnamefont {J.~J.}\ \bibnamefont {Ying}}, \bibinfo {author}
  {\bibfnamefont {X.~F.}\ \bibnamefont {Wang}}, \bibinfo {author}
  {\bibfnamefont {Y.~J.}\ \bibnamefont {Yan}}, \bibinfo {author} {\bibfnamefont
  {Z.~J.}\ \bibnamefont {Xiang}}, \bibinfo {author} {\bibfnamefont
  {P.}~\bibnamefont {Cheng}}, \bibinfo {author} {\bibfnamefont {G.~J.}\
  \bibnamefont {Ye}}, \bibinfo {author} {\bibfnamefont {Z.~Y.}\ \bibnamefont
  {Li}}, \ and\ \bibinfo {author} {\bibfnamefont {X.~H.}\ \bibnamefont
  {Chen}},\ }\href {\doibase 10.1209/0295-5075/94/27008} {\bibfield  {journal}
  {\bibinfo  {journal} {Europhysics Letters}\ }\textbf {\bibinfo {volume}
  {94}},\ \bibinfo {pages} {27008} (\bibinfo {year} {2011})}\BibitemShut
  {NoStop}%
\bibitem [{\citenamefont {Hiramatsu}\ \emph {et~al.}(2011)\citenamefont
  {Hiramatsu}, \citenamefont {Katase}, \citenamefont {Kamiya},\ and\
  \citenamefont {Hosono}}]{HiramatsuJPSJ2012}%
  \BibitemOpen
  \bibfield  {author} {\bibinfo {author} {\bibfnamefont {H.}~\bibnamefont
  {Hiramatsu}}, \bibinfo {author} {\bibfnamefont {T.}~\bibnamefont {Katase}},
  \bibinfo {author} {\bibfnamefont {T.}~\bibnamefont {Kamiya}}, \ and\ \bibinfo
  {author} {\bibfnamefont {H.}~\bibnamefont {Hosono}},\ }\href {\doibase
  10.1143/JPSJ.81.011011} {\bibfield  {journal} {\bibinfo  {journal} {Journal
  of the Physical Society of Japan}\ }\textbf {\bibinfo {volume} {81}},\
  \bibinfo {pages} {011011} (\bibinfo {year} {2011})},\ \Eprint
  {http://arxiv.org/abs/1111.0358} {arXiv:1111.0358} \BibitemShut {NoStop}%
\bibitem [{\citenamefont {Saha}\ \emph {et~al.}()\citenamefont {Saha},
  \citenamefont {Butch}, \citenamefont {Drye}, \citenamefont {Magill},
  \citenamefont {Ziemak}, \citenamefont {Kirshenbaum}, \citenamefont {Zavalij},
  \citenamefont {Lynn},\ and\ \citenamefont {Paglione}}]{SahaArxiv1105.4798}%
  \BibitemOpen
  \bibfield  {author} {\bibinfo {author} {\bibfnamefont {S.~R.}\ \bibnamefont
  {Saha}}, \bibinfo {author} {\bibfnamefont {N.~P.}\ \bibnamefont {Butch}},
  \bibinfo {author} {\bibfnamefont {T.}~\bibnamefont {Drye}}, \bibinfo {author}
  {\bibfnamefont {J.}~\bibnamefont {Magill}}, \bibinfo {author} {\bibfnamefont
  {S.}~\bibnamefont {Ziemak}}, \bibinfo {author} {\bibfnamefont
  {K.}~\bibnamefont {Kirshenbaum}}, \bibinfo {author} {\bibfnamefont {P.~Y.}\
  \bibnamefont {Zavalij}}, \bibinfo {author} {\bibfnamefont {J.~W.}\
  \bibnamefont {Lynn}}, \ and\ \bibinfo {author} {\bibfnamefont
  {J.}~\bibnamefont {Paglione}},\ }\href@noop {} {\ }\Eprint
  {http://arxiv.org/abs/1105.4798} {arXiv:1105.4798} \BibitemShut {NoStop}%
\bibitem [{\citenamefont {Doiron-Leyraud}\ \emph {et~al.}()\citenamefont
  {Doiron-Leyraud}, \citenamefont {Auban-Senzier}, \citenamefont {de~Cotret},
  \citenamefont {Sedeki}, \citenamefont {Bourbonnais}, \citenamefont {Jerome},
  \citenamefont {Bechgaard},\ and\ \citenamefont
  {Taillefer}}]{Doiron-LeyraudArxiv0905.0964}%
  \BibitemOpen
  \bibfield  {author} {\bibinfo {author} {\bibfnamefont {N.}~\bibnamefont
  {Doiron-Leyraud}}, \bibinfo {author} {\bibfnamefont {P.}~\bibnamefont
  {Auban-Senzier}}, \bibinfo {author} {\bibfnamefont {S.~R.}\ \bibnamefont
  {de~Cotret}}, \bibinfo {author} {\bibfnamefont {A.}~\bibnamefont {Sedeki}},
  \bibinfo {author} {\bibfnamefont {C.}~\bibnamefont {Bourbonnais}}, \bibinfo
  {author} {\bibfnamefont {D.}~\bibnamefont {Jerome}}, \bibinfo {author}
  {\bibfnamefont {K.}~\bibnamefont {Bechgaard}}, \ and\ \bibinfo {author}
  {\bibfnamefont {L.}~\bibnamefont {Taillefer}},\ }\href@noop {} {\ }\Eprint
  {http://arxiv.org/abs/0905.0964} {arXiv:0905.0964} \BibitemShut {NoStop}%
\end{thebibliography}
\end{document}